\documentclass[11pt]{report}
\usepackage[utf8]{inputenc}
\usepackage[T1]{fontenc}
\usepackage{lmodern}

\usepackage[polish, english]{babel}

\usepackage{graphicx}
\usepackage{amsmath}
\usepackage{slashed}
\usepackage{geometry}
\usepackage{fancyhdr}
\usepackage{datetime}
\usepackage{cite}
\usepackage[hidelinks]{hyperref}
\usepackage[bottom]{footmisc}

\usepackage{enumitem}
\usepackage{fdsymbol}
\usepackage{physics}
\usepackage{subcaption}
\usepackage{xspace} 

\newdate{date}{\THEYEAR}{\THEMONTH}{}

\newcommand{\whizard}{\textsc{Whizard}\xspace}

\newcommand{\xbar}{\ensuremath{\bar{x}}\xspace}
\newcommand{\ptmax}{\ensuremath{p_{\perp,\text{max}}}\xspace}
\newcommand{\xmin}{\ensuremath{x_\text{min}}\xspace}

\newcommand{\epem}{e$^{_+}$e$^{_-}$\xspace}

\newcommand{\MidLarge}{\fontsize{15.5}{19}\selectfont}

\newlength{\figwidth}
\newlength{\figskip}
\usepackage[math]{cellspace}
\begin{document}
\pagenumbering{gobble}
\begin{titlepage}
    \centering
    \includegraphics[width=0.4\textwidth]{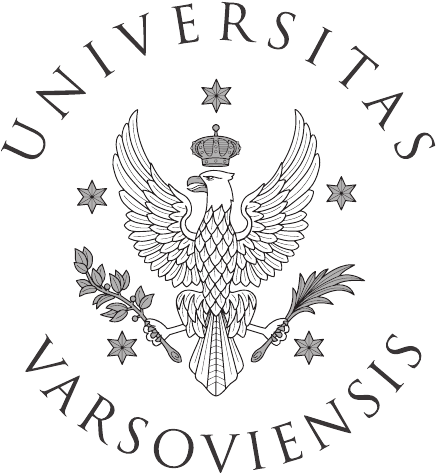}\par\vspace{1cm}
    
    {\scshape\Large University of Warsaw \par}
    \vspace{1.5cm}
    
    {\MidLarge\bfseries Modelling electroweak interactions at lepton colliders 
    
    --- from the Z pole to the highest energies\par}
    \vspace{1.5cm}
    
    {\Large Krzysztof Paweł Mękała\par}
    \vspace{1cm}
    
    {\Large Doctoral thesis written under the supervision of \\Professor Aleksander Filip Żarnecki\\
    and Doctor J\"urgen Reuter\par}
    \vfill
    
    {\large \monthname~\the\year\par}
    
\end{titlepage}

\chapter*{Abstract}

Despite the Standard Model’s ability to describe collider data from all experiments conducted around the world, several questions about the history of the Universe, including the hierarchy problem, the baryon asymmetry, and the dark matter density, remain open. To address these unresolved issues, the particle physics community agrees on the need to construct a next-generation collider that could probe the Standard Model with higher accuracy and potentially reveal deviations from the predictions of this fundamental theory. Future lepton colliders, operating at energies from the Z pole up to the multi-TeV scale, could probe the electroweak sector in various ways. In this thesis, the influence of radiative corrections on the modelling of electroweak processes is investigated.

The thesis first introduces the necessary concepts, including the construction of the Standard Model, future collider proposals and their physics capabilities, and Monte Carlo generators. It then examines three topics. First, a method to measure the electroweak couplings of light quarks is presented as an example showing how processes with photon radiation should be modelled for experimental analyses. Projected results for different facilities operating at the Z pole improve current precision by an order of magnitude. Second, the framework of collinear factorisation is introduced, and its application in the form of the Equivalent Vector Boson Approximation for multi-TeV collisions is re-evaluated. It is shown under what conditions the approximated framework provides a good description of the full process. Third, an extension of the factorised approach that resums large logarithmic contributions is introduced within the framework of Electroweak Parton Distribution Functions, together with its implementation in the Monte Carlo generator Whizard. Several examples are presented, suggesting that this approach may be a promising way to speed up computations and improve the precision of measurements at future colliders.
\chapter*{Streszczenie}

Chociaż Model Standardowy doskonale opisuje wyniki pomiarów w zderzaczach cząstek na całym świecie, wiele pytań dotyczących historii Wszechświata --- na przykład problem hierarchii, asymetria barionowa czy gęstość ciemnej materii --- pozostaje otwartych. Aby się z nimi zmierzyć, społeczność fizyków cząstek zgadza się co do zasadności budowy zderzacza nowej generacji, który mógłby testować Model Standardowy z większą precyzją i potencjalnie odkryć odstępstwa od tej teorii. Przyszłe zderzacze leptonów, działające przy energiach od rezonansu Z aż po skalę teraelektronowoltów, mogłyby badać oddziaływania elektrosłabe na różne sposoby. Niniejsza rozprawa doktorska omawia wpływ poprawek radiacyjnych na modelowanie procesów elektrosłabych.

Na początku rozprawa przedstawia niezbędne zagadnienia, w tym strukturę Model Standardowego, przyszłe zderzacze i ich możliwości fizyczne oraz generatory Monte Carlo, a następnie omawia trzy tematy. Po pierwsze, praca prezentuje metodę pomiaru elektrosłabych sprzężeń kwarków jako przykład pokazujący, jak należy modelować promie\-niowanie fotonów w analizach doświadczalnych. Oczekuje się, że przyszłe maszyny operujące przy rezonansie Z mogłyby poprawić obecną dokładność tego pomiaru o rząd wielkości. Po drugie, praca wprowadza formalizm faktoryzacji kolinearnej oraz analizuje jego zastosowanie w kontekście równoważnego przybliżenia bozonów wektorowych (ang. Equivalent Vector Boson Approximation) dla zderzeń w skali teraelektronowoltów. Rozprawa omawia także warunki, pod jakimi podejście to zadowalająco opisuje procesy fuzji bozonów. Po trzecie, praca przedstawia rozszerzenie tego formalizmu, które resumuje duże wkłady logarytmiczne, w ramach elektrosłabych funkcji rozkładu partonów (ang. Electroweak Parton Distribution Functions), a także jego techniczną implementację w generatorze Monte Carlo Whizard. Kilka omówionych dalej przykładów wskazuje, że podejście to ma szansę przyspieszyć obliczenia i poprawić precyzję pomiarów przy przyszłych zderzaczach.
\chapter*{Acknowledgments}
This thesis could not have been written without the help of many people who contributed to my academic development. First of all, I would like to thank Prof. Aleksander Filip Żarnecki, who has supervised my research for the last seven years. He introduced me to the world of future colliders, which has remained my main research topic. I cannot express in words how much I have learned from him, not only about physics, but also about academic work and life. I am deeply grateful for all his support and help on many levels, often provided even before I asked for it. I could not imagine a better supervisor to learn from and collaborate with.

Equal thanks go to my second supervisor, Dr. J\"urgen Reuter. I am grateful to him for introducing me to the world of MC generators and for his support during my time in Hamburg. I could always ask any question and receive an answer quickly, often within just a few minutes. I am happy that I could learn so much about particle theory from him. It is impressive how many ideas for new research projects he always has, and I am grateful for all the collaborations we started thanks to his extensive network of connections.

I would also like to thank all the collaborators who directly contributed to the research presented in this thesis: Tao Han, Daniel Jeans, Max L\"oschner, Yang Ma, Panos Stylianou, Junping Tian, and Keping Xie. I appreciate your work and expertise, which made this research meaningful. I extend this gratitude to all other collaborators I have worked with throughout my academic path on various topics, including the lepton collider collaborations of which I have been a member. Special thanks go to further members of the \whizard collaboration: Pia, Wolfgang, Nils, Thorsten, and Tobias. I am grateful for our work together and I look forward to future projects. I would also like to thank the DESY Summer and Winter Students whom I had the opportunity to partially supervise in recent years: Benjamin, Jan, Kateryna, Leonard, and Mariia.

I would like to express my sincere gratitude to my colleagues in Warsaw. First of all, I want to thank Janek K. for all our scientific discussions and business trips across four continents. I extend these thanks to my office mates, Gosia and Mateusz F., for their welcoming attitude whenever I appeared at the office. Special thanks go to all the friends I made during my studies at the University of Warsaw, especially Bartek, Janek Ch., Mateusz Z., Stanisław K, and Stanisław Ż. I am also grateful to all my colleagues in Hamburg who made my time at DESY more pleasant.

Finally, I am deeply grateful to my parents and my siblings, Anna and Szymon, for all the support I have received throughout my life. I would also like to thank all my friends for being such good companions, always ready to listen to me and help comfort my worries.
\newpage
\pagenumbering{arabic} 
\setcounter{page}{1}
\cleardoublepage
\tableofcontents
\cleardoublepage

\chapter{Introduction}

Since antiquity, humans have sought to understand what the Universe is made of. Over two thousand years ago, the first ideas of atoms appeared: small, indivisible particles differing in shape, size, and arrangement, giving rise to the diversity of materials. Since then, humanity has made considerable progress towards the modern understanding of matter and its fundamental interactions. Today, it is known that an atom consists of a nucleus, made of protons and neutrons, which are bound states of quarks held together by gluons, the carriers of the strong force. The nucleus is surrounded by a cloud of electrons. This structure is stabilised by the electromagnetic force, mediated by photons, while potential nuclear beta decays are governed by the weak force, carried by $W$ and $Z$ bosons.

Six types of quarks have been discovered, along with three charged leptons --- the lightest being the electron --- and three corresponding neutral leptons, called neutrinos. Each of these particles has an antiparticle with the same mass but opposite charges. The picture is completed by the recently discovered Higgs boson, responsible for the mechanism of mass generation. The \textit{Standard Model} is a modern theory that incorporates this knowledge.

Even though the mathematical framework of the Standard Model perfectly describes collider experiments around the world, several questions remain unanswered. Among these are the incorporation of the gravitational force into the theory, the observed distribution of dark matter in the Universe, and the origin of neutrino masses, all of which pose significant challenges to the Standard Model's viability. To reveal potential solutions to these outstanding problems, new experiments --- often involving the construction of next-generation collider facilities --- are being discussed. The proposals span a wide range of energies, from the $Z$-pole run up to the multi-TeV regime. Although these machines can pursue the goal of uncovering the hidden mysteries of the Universe in different ways, a common strategy is to measure the Standard Model as precisely as possible.

This doctoral thesis discusses how various future colliders can precisely probe the electroweak sector of the Standard Model using radiative processes in different incarnations, and how such processes can be simulated. To this end, Chapter~\ref{chap:SM} introduces the principles of the Standard Model and its known limitations. Chapter~\ref{chap:colliders} describes the landscape of future colliders, while Chapter~\ref{chap:futurephysics} highlights their physics research programmes. Chapter~\ref{chap:MonteCarlo} introduces the concept of a Monte Carlo generator and elucidates its importance to modern particle physics. The following chapters present the main scientific achievements of the thesis towards improving the precision in electroweak measurements: a method to measure $Z$ boson couplings to quarks at the $Z$-pole run of future $e^+e^-$ facilities (Chapter~\ref{chap:Zcouplings}), a revalidation of the collinear factorisation framework and the Equivalent Vector Boson Approximation at multi-TeV colliders (Chapter~\ref{chap:EVA}), and a novel implementation of the Electroweak Parton Distribution Functions into the Monte Carlo generator \textsc{Whizard} (Chapter~\ref{chap:EWPDF}). Further work planned in these directions is described in Chapter~\ref{chap:prospects}. The major findings of the thesis are summarised in Chapter~\ref{chap:summary}.

The main results of this thesis are based on the following publications: ``Determination of the first-generation quark couplings at the Z-pole''~\cite{Mekala:2025rlk}, ``EVAluation of the Equivalent Vector boson Approximation at highest energy colliders''~\cite{Dahlen:2025udl}, and a manuscript on Electroweak Parton Distribution Functions, currently being prepared. The Author of this thesis has made significant contributions to all of these publications. For the first analysis, the Author carried out all major stages of the work, including the literature review, the design and execution of the event-generation procedure, the application of the statistical framework, and the preparation of the final plots. In the second publication, the Author contributed to the revalidation of the Monte Carlo implementation of the formalism in \textsc{Whizard}, generated all Monte Carlo samples, investigated the kinematics of the processes under study, and assessed the level of agreement between results obtained using the approximation and those based on full matrix-element calculations for selected cases. The third publication was entirely driven by the Author, who implemented the formalism in \textsc{Whizard}, performed its validation, generated all event samples, prepared all figures, and discussed the application of the framework across a range of scenarios. Moreover, the new physics examples in Chapter~\ref{chap:futurephysics} rely substantially on the results obtained by the Author~\cite{Mekala:2022cmm,Mekala:2023diu,Mekala:2023kzo,Kalinowski:2021tyr,Korshynska:2024suh,Kling:2025zsb}. A wide range of other publications was consulted in the preparation of this thesis, and each has been cited accordingly. All academic content --- including the theoretical background, analytical methods, results, and conclusions --- was conceived and written entirely by the Author. AI-based tools such as \textsc{ChatGPT}, \textsc{DeepSeek}, \textsc{Grammarly}, and \textsc{Writefull} were used for language refinement.
\chapter{The Standard Model of particle physics}
\label{chap:SM}

All known particles and their interactions are incorporated in the mathematical framework of the Standard Model (SM), a modern theory that summarises the current understanding of particle physics. In this chapter, the structure of the SM is presented, and its limitations and open questions are discussed, providing foundations for the research in this thesis.

\section{Formulation of the Standard Model}
\label{sec:formSM}
The key object to understand the theory of particle physics is the Lagrangian of the SM,
which compactly encodes all known non-gravitational interactions between the fundamental
particles~\cite{Altarelli:2013tya}:
\[
\begin{aligned}
\mathcal{L}_{\text{SM}} =\; & 
-\frac{1}{4} G_{\mu\nu}^a G^{a\mu\nu}
-\frac{1}{4} W_{\mu\nu}^i W^{i\mu\nu}
-\frac{1}{4} B_{\mu\nu} B^{\mu\nu} \\[6pt]
& + \sum_{\text{fermions}} \bar{\psi}^f\, i \gamma^\mu D_\mu \psi^f
+ (D_\mu \phi)^\dagger (D^\mu \phi)
- V(\phi) \\[6pt]
& - \sum_{u,d,e} \left( \bar{\psi}^f_L \, Y_F \, \Phi_f \, \psi^f_R + \text{h.c.} \right)\,.
\end{aligned}
\]
Although somewhat intricate at first glance, this expression represents one of the greatest
triumphs of modern theoretical physics. It successfully describes an enormous range of
experimental results obtained over the past decades.

The SM is built upon the non-simple gauge symmetry group
\[
G_{\text{SM}} = SU(3)_C \times SU(2)_L \times U(1)_Y\,,
\]
where each factor corresponds to a fundamental interaction associated with a distinct type
of charge: colour for $SU(3)_C$, weak isospin for $SU(2)_L$, and hypercharge for $U(1)_Y$.
The first three terms in the Lagrangian describe the field strengths of the corresponding
gauge bosons, $G_\mu^a$, $W_\mu^i$, and $B_\mu$, respectively:
\begin{itemize}
    \item non-Abelian $SU(3)_C$: $G_{\mu\nu}^a \;=\; \partial_\mu G_\nu^a - \partial_\nu G_\mu^a + g_s\, f^{abc} G_\mu^b G_\nu^c, \qquad (a,b,c=1,\dots,8)$
    \item non-Abelian $SU(2)_L$: $W_{\mu\nu}^i \;=\; \partial_\mu W_\nu^i - \partial_\nu W_\mu^i
+ g\,\varepsilon^{ijk} W_\mu^j W_\nu^k, \qquad (i,j,k=1,2,3)$
    \item Abelian $U(1)_Y$: $B_{\mu\nu} \;=\; \partial_\mu B_\nu - \partial_\nu B_\mu\,.$
\end{itemize}
$g_s$ is the strong coupling constant, $g$ is the weak-isospin coupling constant, and $f^{abc}$ and $\epsilon^{ijk}$ are the structure constants of the $SU(3)_C$ and $SU(2)_L$ groups, respectively. The fields $G^a$ correspond to the gluons, while $W^i$ and $B$ combine to form the photon, the $Z$ boson and the $W^\pm$ bosons after electroweak symmetry breaking.

The next term, with the sum running over all fermions, i.e. quarks and leptons, is the fermion kinetic term. Not only does this term describe the fermion field propagation through spacetime, but it also encodes fermion-boson interactions via the covariant derivative:
\begin{equation*}
    \bar{\psi} \, i \, \gamma^\mu D_\mu \psi
=
\bar{\psi} \, i \, \gamma^\mu \left( \partial_\mu
- i g_s T^a G^a_\mu
- i g \frac{\tau^i}{2} W^i_\mu
- i g' \frac{Y}{2} B_\mu \right) \psi\,.
\end{equation*}
$T^a$ are the $SU(3)_C$ generators, $\tau^i$ are the $SU(2)_L$ generators, $g'$ is the hypercharge coupling constant, and $Y$ is the hypercharge of the fermion.

The next two terms, involving the Higgs doublet field $\phi$, are responsible for the dynamics of the Higgs field and electroweak symmetry breaking. The SM potential is given by 
\begin{equation*}
    V(\phi) = \mu^2 \, \phi^\dagger \phi + \lambda \, (\phi^\dagger \phi)^2\,,
\end{equation*}
where $\mu^2 < 0$ triggers spontaneous symmetry breaking and $\lambda > 0$ ensures that the potential is bounded from below, which is necessary for a stable vacuum. This combination of parameters leads to the famous ``Mexican hat'' shape of the potential, which makes the minimum occur at a non-zero field value.

The last term corresponds to the Yukawa interaction, connected to the mass generation mechanism for fermions. For three generations of fermions, $Y_f$ are the matrices of the Yukawa couplings that must be determined experimentally.

\section{Calculations in the Standard Model}
\label{sec:calculationsSM}
Even though the structure of the SM Lagrangian is relatively simple, deriving experimentally measurable predictions is non-trivial. First, one must consider electroweak symmetry breaking to get physical particles --- in contrast to, for instance, the unbroken bosonic gauge fields $W_\mu^i$ and $B_\mu$~\cite{Denner:1991kt,Peskin:1995ev,Bohm:2001yx,Denner:2019vbn}. In addition, for perturbative quantisation, an appropriate gauge fixing must be introduced to eliminate unphysical degrees of freedom and to define well-behaved propagators for the gauge fields. The resulting terms in the Lagrangian can be classified into two categories: quadratic terms, which describe field propagators, and cubic and quartic terms, which encode particle interactions.

A convenient method for calculating amplitudes for specific processes is the well-known Feynman-diagram approach. The expressions associated with the basic building blocks --- propagators and vertices --- can be derived from the Lagrangian using the \textit{Feynman rules}, which, in simplified terms, involve stripping off the fields, multiplying the corresponding SM coefficients by an imaginary unit, and including symmetry factors which account for identical fields. Although this procedure may appear to be a ``cookbook recipe'', the formalism is firmly grounded in the underlying path-integral formulation of quantum field theory.

\begin{figure}[t]
  \centering
  \begin{subfigure}[b]{0.23\textwidth}
    \centering
    \includegraphics[width=\textwidth]{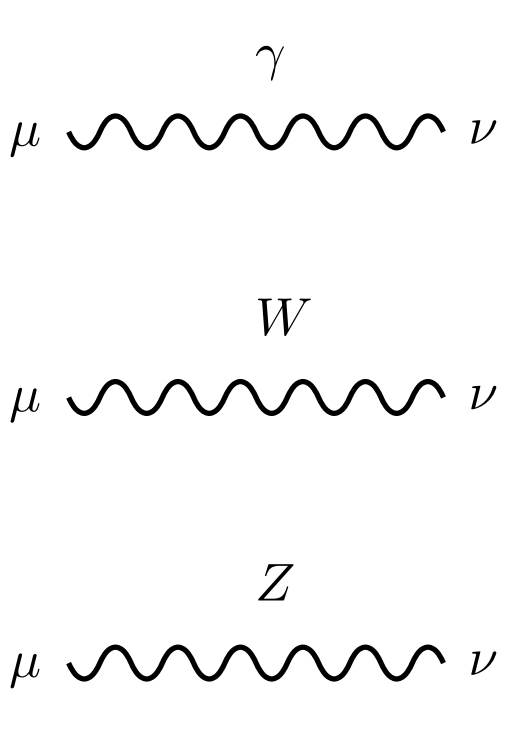}
    \caption{Propagators}
  \end{subfigure}\hfill
  \begin{subfigure}[b]{0.23\textwidth}
    \centering
    \includegraphics[width=\textwidth]{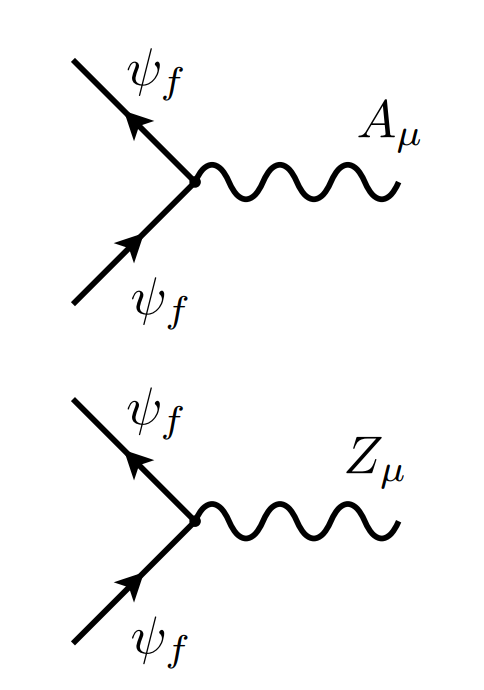}
    \caption{Neutral current}
  \end{subfigure}\hfill
  \begin{subfigure}[b]{0.41\textwidth}
    \centering
    \includegraphics[width=\textwidth]{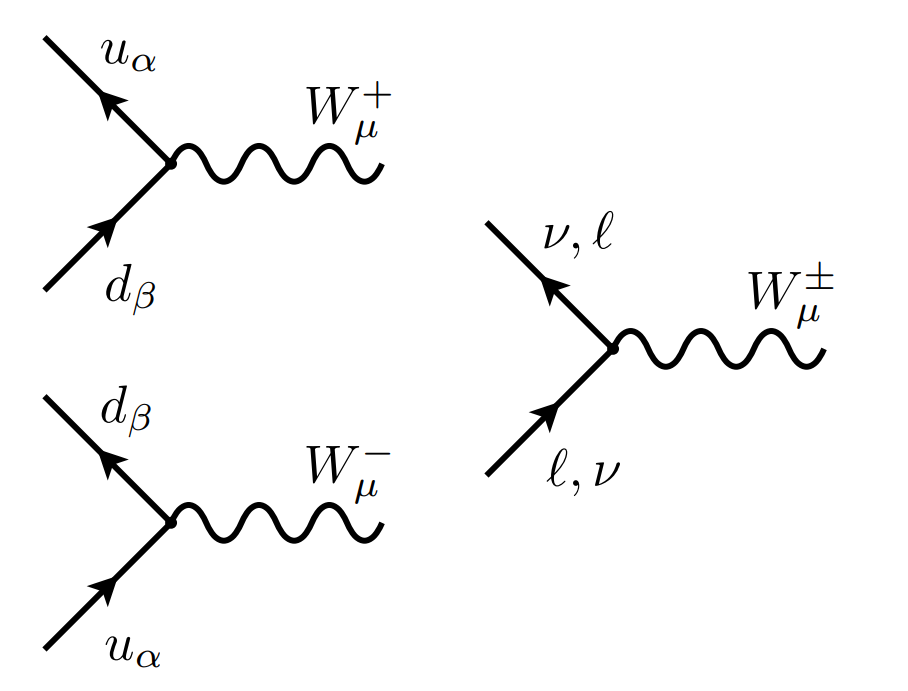}
    \caption{Charged current}
  \end{subfigure}

  \caption{Building blocks for Feynman diagrams of the electroweak theory: propagators, neutral-current vertices and charged-current vertices~\cite{Romao:2012pq}. Selected items are shown, relevant for further discussion in the thesis.}
  \label{fig:feynman_all}
\end{figure}

Figure~\ref{fig:feynman_all} shows selected building blocks of Feynman diagrams within electroweak theory~\cite{Romao:2012pq}, which will be used further in the thesis. The photon propagator is given by
\begin{equation*}
    -i \left[ \frac{g_{\mu \nu}}{k^2 + i \epsilon}
    - (1-\xi_A) \frac{k_\mu k_\nu}{(k^2)^2}\right]\,, 
\end{equation*}
the $W$ boson propagator by
\begin{equation*}
    \frac{-i}{k^2 -m^2_W + i \epsilon}
    \left[ g_{\mu \nu}- (1-\xi_W) \frac{k_\mu k_\nu}{k^2 -\xi_W m^2_W} \right] \,,
\end{equation*}
and the $Z$ boson propagator by
\begin{equation*}
    \frac{-i}{k^2 -m^2_Z + i \epsilon} 
  \left[ g_{\mu \nu}- (1-\xi_Z) \frac{k_\mu k_\nu}{k^2-\xi_Z m^2_Z} \right]\,.
\end{equation*}
Here, $k$ denotes the boson momentum, $g_{\mu\nu}$ is the metric tensor, $m_W$ stands for the $W$ boson mass, $m_Z$ for the $Z$ boson mass, and $\xi_X$ are the gauge fixing parameters in the $R_\xi$ gauge~\cite{tHooft:1971qjg,tHooft:1972tcz}.

The so-called \textit{neutral current} interactions can be mediated either by the photon or the $Z$ boson. In the former case, the vertex is associated with
\begin{equation*}
    ie Q_f \gamma^{\mu}\,,
\end{equation*}
while in the latter
\begin{equation*}
    i\frac{g}{\cos\theta_W} \gamma^{\mu}\left(g_V^f-g_A^f \gamma^5\right)\,.
\end{equation*}
$e$ is the elementary charge, $Q_f$ is the charge of a given fermion in units of the elementary charge, and $g_V^f$ ($g_A^f$) is the fermion vector (axial) coupling given by the weak isospin $T^f_3$, the charge, and the weak mixing angle $\theta_W$:
\begin{equation*}
g_V^f = T^f_3 - 2Q_f \sin^2\theta_W, \quad
g_A^f = T^f_3 \, .
\end{equation*}
Conventionally, the couplings can also be expressed in terms of the left- and right-handed couplings, $g_L^f$ and $g_R^f$, associated with the left- and right-handed projectors, $P_L = \frac{1-\gamma_5}{2}$ and $P_R = \frac{1+\gamma_5}{2}$, respectively.

The \textit{charged current} interaction is mediated by the $W$ bosons. An important feature of this interaction in the quark sector is the presence of the Cabibbo-Kobayashi-Maskawa (CKM) matrix~\cite{Cabibbo:1963yz, Kobayashi:1973fv}, which describes quark flavour mixing, encoding the fact that weak interaction eigenstates are not aligned with mass eigenstates. For the $W^+$ boson, the quark contribution is associated with
\begin{equation*}
    i \frac{g}{\sqrt{2}} V_{ij} \gamma^\mu P_L\,,
\end{equation*}
while for the $W^-$ boson, with
\begin{equation*}
    i \frac{g}{\sqrt{2}} V^*_{ij} \gamma^\mu P_L\,.
\end{equation*}
$V_{ij}$ are the elements of the CKM matrix. Similar mixing is observed in the lepton sector and is described by the Pontecorvo-Maki-Nakagawa-Sakata (PMNS) matrix~\cite{Pontecorvo:1957qd, Maki:1962mu}. This matrix is responsible for neutrino oscillations, which will be discussed later, and the mixing is generally larger than that observed in the quark sector. However, since neutrinos are hard to detect at colliders, the effects of the PMNS matrix are irrelevant for the analyses presented in this thesis and will be neglected. Following this assumption, the vertex involving leptons contributes as
\begin{equation*}
    i \frac{g}{\sqrt{2}} \gamma^\mu P_L\,.
\end{equation*}
This explicit form shows that only the left-handed fermions and right-handed antifermions take part in the charged-current interactions.

External lines also need to be taken into account, since they play an important role in considerations related to factorisation, as elaborated later in this thesis. An incoming fermion with mass $m$, momentum $p$, and helicity\footnote{In this thesis, $s=\pm 1$ is assumed to remain consistent with the notation for gauge bosons, for which $\lambda$ denotes the helicity. The helicity corresponds to a spin quantisation axis in the direction of the particle momentum and is a conserved quantum number only for massless particles. Typically, $s$ is associated with the fermion spin, i.e. $s= \pm \frac{1}{2}$.} $s$ is represented by the Dirac spinor $u(p,s)$, satisfying $(\slashed{p} - m)u(p,s) = 0$, while an incoming antifermion is represented by $\bar{v}(p,s)$, with $(\slashed{p} + m)v(p,s) = 0$. Outgoing fermions and antifermions are associated with $\bar{u}(p,s)$ and $v(p,s)$, respectively. Assuming $p^\mu = (E, p_x, p_y, p_z)$, for massless fermions in the Weyl representation, one can write
\begin{equation*}
 u(p,s) =  \left(\begin{matrix}
 \sqrt{E-s\vert\vec{p}\vert} \,\chi(p,s)
\\ 
\sqrt{E+s\vert\vec{p}\vert} \,\chi(p,s)
\end{matrix}\right)
\end{equation*}
and
\begin{equation*}
v(p,s) = \left(\begin{matrix}
-s\sqrt{E+s\vert\vec{p}\vert} \,\chi(p,-s)
\\ 
s\sqrt{E-s\vert\vec{p}\vert} \,\chi(p,-s)
\end{matrix}\right)\,,
\end{equation*}
where the two-component helicity eigenstates $\chi$ are given by
\begin{equation*}
 \chi(p,+1)=\frac{1}{\sqrt{2\vert\vec{p}\vert(\vert\vec{p}\vert+p_z)}}
\begin{pmatrix} \vert\vec{p}\vert+p_z \\ p_x + ip_y \end{pmatrix}
\end{equation*}
and
\begin{equation*}
\chi(p,-1)=\frac{1}{\sqrt{2\vert\vec{p}\vert(\vert\vec{p}\vert+p_z)}}
\begin{pmatrix} -p_x + ip_y \\ \vert\vec{p}\vert+p_z \end{pmatrix}\,.
\end{equation*}

Gauge bosons are included via their polarisation vectors: $\varepsilon_\mu(k,\lambda)$ for an incoming line and $\varepsilon_\mu^*(k,\lambda)$ for an outgoing one, where $\lambda$ denotes the boson helicity. Taking $k^\mu = (E_V, k_x, k_y, k_z)$, the polarisation vectors for a massive gauge boson with a mass $m_V$ can be written as\footnote{Different conventions exist for polarisation vectors, including relative complex phases. While these do not affect cross-section calculations, which depend only on the squared amplitudes, care must be taken when considering amplitude-level quantities. In this thesis, the convention used in \whizard is adopted.}
\begin{equation}
\label{eq:polpl}
\varepsilon^\mu(k,+1) 
= 
\frac{1}{\sqrt{2(k_x^2 + k_y^2)}}
\left(
0,\;
\frac{k_x k_z}{|\vec{k}|} - i k_y,\;
\frac{k_y k_z}{|\vec{k}|} + i k_x,\;
- \frac{k_x^2 + k_y^2}{|\vec{k}|}
\right),
\end{equation}
\begin{equation}
\label{eq:polmn}
\varepsilon^\mu(k, -1) 
= 
\frac{1}{\sqrt{2(k_x^2 + k_y^2)}}
\left(
0,\;
\frac{k_x k_z}{|\vec{k}|} + i k_y,\;
\frac{k_y k_z}{|\vec{k}|} - i k_x,\;
-\frac{k_x^2 + k_y^2}{|\vec{k}|}
\right),
\end{equation}
\begin{equation}
\label{eq:pollong}
\varepsilon^\mu(k, 0)
=
\frac{E_V}{m_V |\vec{k}|}
\left(
\frac{|\vec{k}|^2}{E_V},\;
k_x,\;
k_y,\;
k_z
\right).
\end{equation}

\section{Open questions in the Standard Model}
\label{sec:openquestions}

Although the SM provides an excellent framework for describing particle interactions at the fundamental level, it does not answer all the questions about the nature of matter and the evolution of the Universe. Some of these open issues are discussed in this section.

\subsection{Gravity}

The SM accounts for three of the four known fundamental interactions: the strong, weak, and electromagnetic forces. Gravity, however, is not included. Its modern theoretical description, general relativity, is a classical field theory and differs fundamentally from the quantum field theory framework used in the SM~\cite{Misner:1973prb}. This mismatch makes the unification of all four forces challenging.

From the practical perspective, incorporating gravity into the description of physics at colliders is unnecessary because its strength at accessible energy scales is negligible. Similarly, in most astrophysical contexts, particle-level forces play only a secondary role; gravitational interactions alone successfully explain the majority of observed phenomena.

Nevertheless, unifying all interactions is not merely an aesthetic goal. A consistent theory combining gravity with the other fundamental forces would have profound implications for the understanding of the early Universe, where microscopic quantum processes and macroscopic gravitational dynamics were tightly interconnected. Without such a unified framework, the description of the Universe’s earliest moments remains incomplete.

Another issue related to gravity is the so-called \textit{hierarchy problem}. In a natural theory, one expects dimensionless parameters to be of comparable order. It is therefore puzzling that gravity is many orders of magnitude weaker than the weak interaction. This disparity is closely related to the question of why the Higgs boson mass is so much smaller than the Planck scale.

\subsection{Unification of forces}

The unification of gravity with the quantum-level forces is only one of the challenges in understanding the interplay of fundamental interactions. The SM provides a partial unification by combining the electromagnetic and weak forces into the electroweak interaction~\cite{Glashow:1961tr,Weinberg:1967tq,Salam:1968rm}. Below the electroweak scale, these forces appear distinct due to spontaneous symmetry breaking; above this scale, they merge into a single interaction.  

The coupling constants of the strong, weak, and electromagnetic interactions evolve with energy, which is known as \textit{running of couplings}. While the electromagnetic and weak couplings converge at the electroweak scale, the strong coupling remains separate at accessible energies. A clear manifestation of electroweak unification is seen in neutral-current interactions, where both the photon and the $Z$ boson mediate processes of the same type, and their contributions cannot be separated to make reliable predictions using only one of the mediators.  

This observation motivates the expectation that at even higher energies, the strong force may also unify with the electroweak interaction. Theories that attempt such unification propose frameworks in which all fundamental forces can be integrated, with all coupling constants converging at a single high-energy scale. Typically, these models involve incorporating the gauge group of the SM into a larger, simple gauge group, which breaks to the observed gauge group at lower energies. Unfortunately, the predicted unification scales lie far beyond the reach of current collider experiments. Such a framework is referred to as a \textit{Grand Unified Theory} (GUT), with common examples based on the $SU(5)$ or $SO(10)$ gauge groups~\cite{Georgi:1974sy,Gell-Mann:1979vob,Langacker:1980js,Slansky:1981yr}.

\subsection{Free parameters of the Standard Model}

The SM has at least 18 free parameters\footnote{Other parameter-counting schemes for the Standard Model also appear in the literature. For instance, some authors count in the parameter $\theta'$, related to the strong CP problem, which will not be discussed in the thesis. Furthermore, nine extra parameters appear if the neutrino masses are taken into account, which includes three neutrino masses, three PMNS mixing angles, and three phases.}, whose values must be determined experimentally~\cite{Weinberg:1996kr}. These include six quark masses, three charged lepton masses, three gauge couplings at a given energy scale, the Higgs vacuum expectation value (VEV), the Higgs boson mass, three CKM mixing angles, and one CP-violating phase. Some quantities, such as the masses of the weak bosons, can be derived from the 18 constants, while others, including parameters describing mixing in the lepton sector, require independent measurement. Extensions of the SM that introduce new particles typically bring additional free parameters, such as particle masses or couplings. From this perspective, the SM can be viewed as an effective, low-energy description of nature~\cite{Weinberg:1978kz,Buchmuller:1985jz,Brivio:2017vri}.

Although there is no fundamental theoretical limit on the number of free parameters a \textit{reasonable} theory may have, the relatively large number in the SM is often seen as a conceptual drawback. GUTs offer the possibility of reducing the number of independent parameters by relating gauge couplings or fermion masses, but additional assumptions are needed to specify symmetry-breaking scales. Other approaches invoke new flavour symmetries, extra dimensions, or more fundamental building blocks of nature, as in string theory~\cite{Schwarz:1998ny,Marchesano:2024gul}. However, to date, no framework consistently accounts for the origin of all 18 SM parameters from first principles.

\subsection{Experimental Observations}

The SM provides a very accurate description of collider experiments. However, several effects observed in other types of experiments lie beyond its applicability; they include neutrino oscillations, dark matter density, and the baryon--antibaryon asymmetry in the Universe.

Neutrinos exist in three flavour eigenstates, $\nu_e$, $\nu_\mu$, and $\nu_\tau$, corresponding to the three charged leptons. Due to their extremely weak interactions with matter, detecting neutrinos requires large-scale detectors capable of observing rare events. Already in the 1960s, experiments noted that the number of neutrinos produced in the Sun and detected on Earth was smaller than predicted by solar models~\cite{Davis:1968cp}. Initial attempts to resolve this discrepancy focused on refining the solar models, but at the beginning of the 21st century, the deficit was attributed to neutrino oscillations --- the transformation of neutrinos between flavours during propagation~\cite{SNO:2002tuh}.  

Neutrino oscillations can occur only if neutrinos have non-zero masses, whereas the minimal SM assumes massless neutrinos. To account for neutrino masses, one needs to introduce right-handed neutrinos, which are singlets under the SM gauge group, and a Yukawa term for the neutrinos. The smallness of neutrino masses implies extremely tiny Yukawa couplings; explaining why these couplings are so small may motivate the introduction of a heavy mass scale, as in the seesaw mechanism~\cite{Minkowski:1977sc, Mohapatra:1979ia}.

Another experimental indication for the existence of physics beyond the Standard Model (BSM) comes from the observed density of dark matter. As early as the 1930s, it was noticed that the motion of stellar objects could not be explained if their mass originated solely from visible matter~\cite{Zwicky:1933gu}. Subsequent observations, most notably measurements of galaxy rotation curves by Vera Rubin in the 1970s~\cite{Rubin:1978kmz}, provided strong evidence for the presence of invisible matter affecting the dynamics of galaxies and clusters. Also, other, more modern observations support this theory, most notably microlensing~\cite{Paczynski:1985jf,Clowe:2006eq,Iocco:2011jz}.

All evidence for dark matter appears to arise from gravitational interactions, which are not described within the Standard Model. These observations may instead indicate the need for modifications to the theory of gravitation~\cite{Milgrom:1983ca} or point towards a fundamental quantum theory of gravity. The prevailing interpretation within the scientific community, however, is that dark matter consists of a distinct form of matter composed of new particles. Its name, dark matter, reflects one of its few well-established properties: it does not interact with light~\cite{Garrett:2010hd}. Dark matter may be composed of several particle species with masses spanning multiple orders of magnitude. While the minimal Standard Model does not contain any particle that can account for the observed dark matter density, heavy right-handed neutrinos could already serve as viable candidates. Many other possibilities exist, including axions, WIMPs, primordial black holes, and wimpzillas~\cite{Preskill:1982cy,Jungman:1995df,Chung:1998zb,Kolb:1998ki,Bertone:2004pz,Carr:2016drx}.

Another notable fact is that the Solar System, and the observed Universe more broadly, is composed almost entirely of matter rather than antimatter. Analyses of cosmic radiation reaching Earth indicate that the vast majority of observed objects in the Universe are made of matter, and there is no evidence for large systems of antimatter~\cite{AMS:1999tha,Canetti:2012zc}. If such regions existed, detection of high-energy radiation produced by matter--antimatter annihilation would be expected~\cite{Cohen:1997ac}. Current understanding of the early Universe does not provide any explanation for this imbalance. This phenomenon, known as the baryon--antibaryon asymmetry, cannot be accounted for by mechanisms within the SM and therefore suggests the existence of new physics to explain this cosmological observation.

\chapter{Future colliders}
\label{chap:colliders}
The ideas discussed in the thesis apply to experiments at future colliders, which will achieve higher luminosities and energies than those available today. In this chapter, the context for strategic decisions regarding the next major particle physics facility is outlined, and the most relevant proposals are presented.

\section{Current status of LHC and strategic discussions}
\label{sec:strategic_discussions}

In the early 1980s, the European Organization for Nuclear Research (CERN) decided to build the Large Electron-Positron Collider (LEP) to deepen understanding of fundamental forces~\cite{LEPapproval}. After the discovery of the $Z$ and $W$ bosons in 1983~\cite{UA1:1983crd, UA2:1983mlz}, LEP appeared as a perfect tool to study these particles by colliding electrons and positrons at high energies. With the machine becoming operational in 1989, many aspects of the SM were confirmed. LEP delivered the most precise measurements in several domains, for example, in the electroweak sector. The 27-km circular tunnel hosting LEP was reused in the 2000s to build the Large Hadron Collider (LHC)~\cite{LHCapproval}, which started its operation in 2008. LHC and its four experiments, ALICE~\cite{ALICE:2008ngc}, ATLAS~\cite{ATLAS:2008xda}, CMS~\cite{CMS:2008xjf}, and LHCb~\cite{LHCb:2008vvz}, enabled the confirmation of the Brout-Englert-Higgs mechanism by finding a 125-GeV resonance identifiable with the Higgs boson in 2012~\cite{ATLAS:2012yve,CMS:2012qbp}. Further aspects of the fundamental theory of particle interactions will be studied at the High-Luminosity upgrade of LHC (HL-LHC)~\cite{ZurbanoFernandez:2020cco}, whose operation is expected to conclude in the early 2040s.

Not only did the LHC allow for the discovery of the Higgs boson, but it also enabled precise measurements of its properties, including its mass, decay modes, and couplings to various SM particles~\cite{ATLAS:2013dos,CMS:2013fjq,CMS:2014fzn, ATLAS:2015yey, ATLAS:2016neq,CMS:2018nsn, ATLAS:2020fzp, ATLAS:2020rej,CMS:2020xrn, ATLAS:2020ior,CMS:2022ley,CMS:2022dwd, ATLAS:2022vkf}. Numerous additional measurements have been taken to test the consistency of the SM rigorously and determine its parameters with high precision~\cite{LHCb:2021bjt,CMS:2023ebf,ATLAS:2024erm,CMS:2024lrd,LHCb:2025nob,ATLAS:2025bpp}. Beyond the SM, LHC has placed stringent constraints on a wide range of theoretical extensions, such as supersymmetric models~\cite{ATLAS:2011yrs, CMS:2011bul, CMS:2011xek, CMS:2013cgf, ATLAS:2014zve, ATLAS:2014jxt, CMS:2014ewg, ATLAS:2019lff, ATLAS:2019lng}. It has also provided significant insights into hadronic physics, for instance, through the studies of $B$-meson decays at LHCb~\cite{LHCb:2012skj, LHCb:2014vgu, LHCb:2015gmp, LHCb:2015svh, LHCb:2017avl, LHCb:2019hip, LHCb:2021trn}. Furthermore, several exotic hadrons, including tetraquark and pentaquark states~\cite{LHCb:2015yax, LHCb:2019kea, LHCb:2021auc, LHCb:2021vvq}, have been observed, improving our understanding of the strong interaction. The ALICE experiment has made major contributions to the study of the quark-gluon plasma, offering valuable data for modelling the early Universe~\cite{ALICE:2013osk, ALICE:2015vxz, ALICE:2015mdb, ALICE:2016fzo, ALICE:2016ccg, ALICE:2022wpn}.

As discussed in Section~\ref{sec:openquestions}, several questions within the SM remain unresolved. To address these open issues, the particle physics community explores various experimental paths, with decision-making processes structured regionally. In Europe, the primary process guiding the long-term future of particle physics is the European Strategy for Particle Physics (ESPP)~\cite{EPPSU-website}. Mandated by the CERN Council for the first time in 2005, the ESPP was developed through broad consultation with the European scientific community, with wider participation from physicists and partners around the world, including those from the US and Asia. The first resulting document was adopted by the CERN Council in 2006 and placed LHC at the top of European particle physics’ scientific priorities~\cite{CERN-ESU-001}. Research and Development (R\&D) activities for future accelerators also featured high on the priority list. To accommodate developments in the field, the original Strategy anticipated regular community-driven updates. The first of them was prepared in 2012 and then adopted by the Council in 2013~\cite{ESC-E-106}. Following the discovery of the Higgs boson, LHC confirmed its leading role in particle physics, and the high-luminosity upgrade was recommended to fully exploit the physics potential of the accelerator. 

The second update, initiated in 2018, was expected to deliver an ambitious post-LHC programme for CERN. At that time, two major projects were developed at CERN: the Compact Linear Collider (CLIC)~\cite{Aicheler:2012bya, Linssen:2012hp, Lebrun:2012hj} and the Future Circular Collider (FCC)~\cite{FCC:2018byv, FCC:2018evy, FCC:2018vvp}. Other projects outside Europe got attention: the International Linear Collider (ILC)~\cite{Behnke:2013xla, ILC:2013jhg, Adolphsen:2013jya, Adolphsen:2013kya, Behnke:2013lya} in Japan and the Circular Electron-Positron Collider (CEPC)~\cite{CEPCStudyGroup:2018rmc, CEPCStudyGroup:2018ghi, CEPCStudyGroup:2023quu} in China. The Strategy was approved by the CERN Council in 2020~\cite{CERN-ESU-015}. It emphasised the necessity of the successful completion of HL-LHC over the coming decade, and recommended an \epem \textit{Higgs factory} as the highest priority to follow LHC. The third update was launched in June 2024. One of the goals mandated by the Council for the ongoing update is to recommend a visionary and concrete plan for the next flagship collider at CERN and to explore alternative options if the preferred plan is not feasible or competitive. The community-driven part of the process concluded in December 2025~\cite{CERN-ESU-2025-002}, naming FCC the priority for CERN. The update should be adopted by the Council later in 2026.

The American analogue of the ESPP is the Snowmass Process, organised under the aegis of the Division of Particles and Fields of the American Physical Society. The first meeting took place in 1982 in Snowmass, Colorado, and gave the process its name~\cite{Donaldson:1983mq}, even though subsequent meetings have been held every seven to ten years in various locations. The outcomes of this community-driven process are submitted to the Particle Physics Project Prioritization Panel (P5), a temporary subcommittee of the High Energy Physics Advisory Panel (HEPAP). HEPAP, in turn, was a permanent advisory committee to the United States Department of Energy and the National Science Foundation, the primary funding agencies for particle physics in the USA\footnote{Some changes to the organisation of science funding have been pursued by the current American administration, and it is therefore not clear how the Snowmass process will evolve in the future.}. The most recent Snowmass process~\cite{Butler:2023glv}, culminating in the 2023 P5 Report~\cite{P5:2023wyd}, endorsed the construction of a Higgs factory either in Europe or Japan as the next major collider facility. It also recommended that the particle physics community explore the feasibility of a 10\,TeV parton centre-of-mass (pCM) collider.

Over the past two decades, strategic discussions regarding future facilities have also been active in Japan. In 2004, the International Committee for Future Accelerators (ICFA) endorsed the idea of building a next-generation linear collider using superconducting technology~\cite{Augustin:2004nq}. By 2007, a reference design of ILC was released, comprising a 31-km linear accelerator to collide electrons and positrons with possible upgrades up to 1\,TeV~\cite{ILC:2007oiw}. Japan’s physics community officially expressed interest in hosting ILC in 2011. A full technical design report (TDR) was completed in 2013~\cite{Behnke:2013xla}. Japan has been expected to cover about half of the cost, with international partners sharing the rest~\cite{ILCJapanCostSharing}. However, due to the unclear international situation, Japan's government has never committed to hosting the project, and following the recommendation of the panel formed by the Ministry of Education, Culture, Sports, Science and Technology of Japan (MEXT) in 2022, the pre-lab phase was not initiated. Nevertheless, there are ongoing international and domestic efforts to revive the project in Japan, for example, through the International Development Team (IDT).

The idea of building a large next-generation collider also emerged within the Chinese community. Around 2013, the Institute of High Energy Physics began exploring the feasibility of a circular \epem collider. In 2015, a preliminary Conceptual Design Report for CEPC, followed by plans for a subsequent proton--proton machine called the Super Proton--Proton Collider (SPPC), was published~\cite{CEPC-SPPCStudyGroup:2015csa}. Since then, China has continued to seek international collaboration to realise the project, which was eventually included in the 14th Five-Year Plan for Science and Technology (2021–2025) as a major science initiative. In 2023, the CEPC Technical Design Report was published~\cite {CEPCStudyGroup:2023quu}. In recent years, the preparatory phase has begun, but the project has not been included in China’s 15th Five-Year Plan (2026-2030). As a result, formal government approval and funding are currently on hold.

\section{Collider proposals}
Given the scope of the results presented in this thesis, the findings are relevant to a diverse set of proposals, including both circular and linear \epem colliders, as well as a muon collider. Additionally, further developments based on this research will be valuable for studying hadron--hadron interactions at ultra-high energies. This section introduces the relevant collider and detector proposals currently under discussion. Further details are given in Appendix~\ref{app:technical}. Most of the specifications are presented according to the submissions to the ongoing Update of the ESPP~\cite{ESPPinputs:2024}.

\subsection{Circular \epem colliders}
Two major circular \epem colliders are currently under consideration by the global particle physics community: FCC proposed at CERN and CEPC planned in China. The FCC initiative was launched in 2014 to investigate the feasibility of a next-generation accelerator to succeed LHC at CERN~\cite{Koratzinos:2015fya}. The 2020 Update of the European Strategy for Particle Physics recommended a detailed feasibility study of FCC, which has been actively pursued in recent years. Despite several technical differences between FCC and CEPC, their fundamental operational principles are comparable. In the following, the European FCC design is described for reference~\cite{FCC:2025lpp, FCC:2025uan, FCC:2025jtd}.

\subsubsection{Project overview}
The FCC concept represents the next step in running a flexible experimental infrastructure for the post-LHC era. The proposed strategy follows a staged approach, beginning with the construction of a highest-luminosity electron-positron collider (FCC-ee), envisioned to act as a Higgs, top-quark, and electroweak factory, and eventually succeeded by a highest-energy hadron collider (FCC-hh). FCC-ee could potentially begin its operation in 2048. The staged approach permits control of technical and financial risks. In this way, a long-term scientific programme is secured for decades to come, if resources allow.

FCC-ee offers a platform for electroweak measurements and Higgs boson studies through its operation at four baseline centre-of-mass energies: the $Z$ pole, the $WW$ pair production threshold, the $ZH$ production threshold, and the $t\bar{t}$ production threshold. The collider is designed to remain flexible among the $Z$, $WW$, and $ZH$ runs, allowing for adaptive scheduling of data collection. The FCC-ee luminosity strongly depends on the collision energy. During the $Z$-pole run, which features three energy points at 88, 91, and 94\,GeV to scan the production threshold, the luminosity could reach up to $140 \times 10^{34}$\,cm$^{-2}$s$^{-1}$ per interaction point. The $WW$ run could achieve a luminosity of $20 \times 10^{34}$\,cm$^{-2}$s$^{-1}$, while the $ZH$ run could reach $7.5 \times 10^{34}$\,cm$^{-2}$s$^{-1}$. The $t\bar{t}$ runs are expected to deliver a luminosity of $1.8 \times 10^{34}$\,cm$^{-2}$s$^{-1}$ at 340-350\,GeV, and $1.4 \times 10^{34}$\,cm$^{-2}$s$^{-1}$ at 365\,GeV. It is assumed that the $Z$-pole, $WW$, and $ZH$ datasets would be delivered in less than a decade of operation, and a few additional years would be needed to collect the $t\bar{t}$ data.

\begin{figure}
    \centering
    \includegraphics[width=0.6\textwidth]{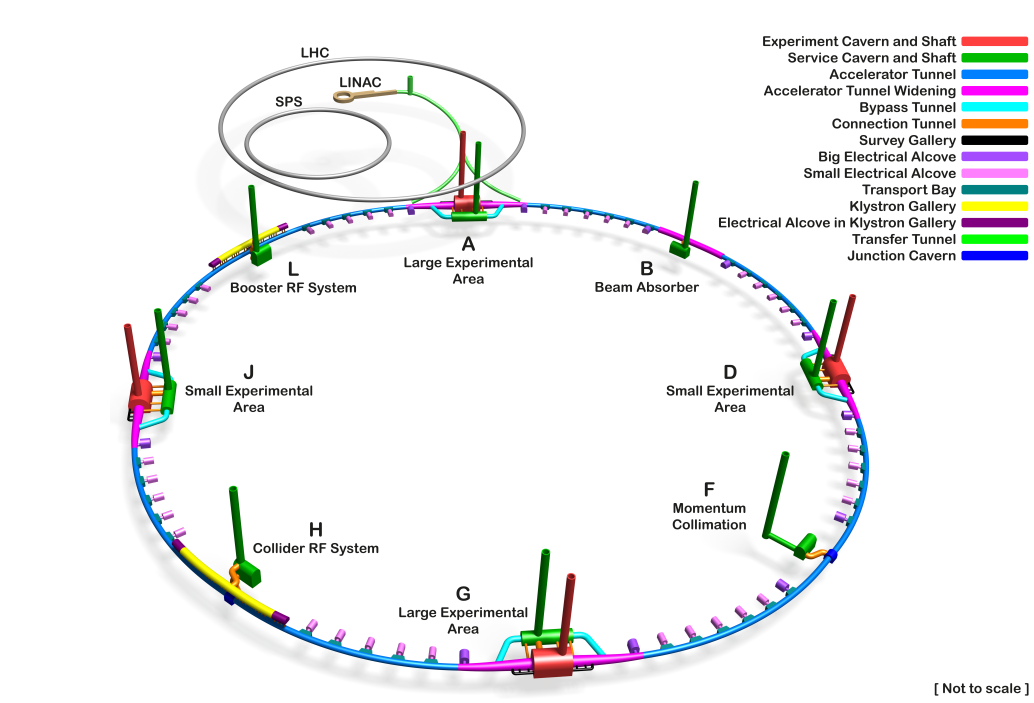}
    
    \caption{Schematic layout of FCC-ee engineering structure~\cite{Benedikt:2025kwi}.}
    \label{fig:FCC_technical}
\end{figure}

To exploit the machine's high luminosity, four multi-purpose detectors would be installed. Over a projected 15-year operation period, FCC-ee is expected to produce more than $6 \times 10^{12}$ $Z$ bosons and $2.7 \times 10^{6}$ Higgs bosons~\cite{Benedikt:2025kwi}. From an infrastructural perspective, the FCC-ee project, comprising tunnel and subsurface structures, surface sites, territorial developments, injector, transfer lines, booster, particle collider and four experiments, as presented in Figure \ref{fig:FCC_technical}, is technically ready for implementation.

The estimated total construction cost for CERN is approximately 14 billion Swiss franc (BCHF), covering the $Z$, $WW$, and $ZH$ stages. Civil engineering represents the highest cost, accounting for 45\% of the total amount (6.2\,BCHF), including all site-related developments. Technical infrastructures, including site-external connections, are estimated at 2.8\,BCHF. The construction of the collider and the full-energy booster accounts for 4.1\,BCHF, while the injector complex and transfer lines contribute an additional 590 million CHF. CERN’s direct investment in the four experiments is estimated at 290 million CHF, while the non-CERN contribution of approximately 1.3\,BCHF is expected.

\subsection{Linear \epem colliders}
Among the proposals currently considered for a linear \epem facility, ILC and CLIC are the most advanced options. As explained in Section \ref{sec:strategic_discussions}, the situation of ILC in Japan remains uncertain due to the lack of commitment from the governmental bodies to advance the project. Similarly, support for CLIC at CERN has become limited, with growing resources spent on the FCC Feasibility Study. Several other proposals are being developed; for example, plasma acceleration technology is now gaining attention from a broader audience~\cite{RevModPhys.81.1229}. The linear collider community remains, however, active and motivated to pursue the idea. In this spirit, a proposal was made in 2024 to consolidate the effort and deliver a single project as Plan A for the entire community. The idea of a Linear Collider Facility (LCF) at CERN was presented for the first time in July 2024 in Tokyo at the Linear Collider Workshop 2024~\cite{LinearColliderVision:2025hlt}. Since it combines several projects into one proposal, this section describes the LCF as a reference linear collider, with a particular emphasis on the ILC concept.

\subsubsection{Project overview}

The LCF presents a fast and cost-effective approach to building a new collider capable of comprehensively probing the Higgs boson’s properties, thanks to its wide range of centre-of-mass energies, from the $Z$ pole up to at least 1\,TeV, and the use of polarised beams~\cite{LinearCollider:2025lya}. With a six-year preparatory phase followed by ten years of construction, the first stage could begin data collection in 2042, a few years before the FCC-ee project. As detailed later in this section, subsequent upgrades would extend the research capabilities for over two decades.

The proposal assumes constructing a 33.5-km tunnel and, similarly to the ILC project, equipping the machine in the first stage with superconducting radiofrequency (RF) cavities suitable for polarised \epem beams (with the polarisation fraction up to 80\% for electrons and 30\% for positrons) at a centre-of-mass energy of 250\,GeV, achieving a luminosity of $2.7 \times  10^{34}$\,cm$^{-2}$s$^{-1}$. The estimated cost of the machine is 8.3\,BCHF. The tunnel length is chosen to accommodate further upgrades; if no energy upgrades are planned, the length can be reduced to about 20\,km, lowering the cost by about 10\%.  In the longer tunnel, the same superconducting technology can reach up to at least 550\,GeV, for an additional cost of about 5.5\,BCHF, by installing more accelerating modules at any time as resources allow. 

The baseline construction scenario assumes first building a 250-GeV machine operating at half of the maximum luminosity. The luminosity can then be doubled by increasing the number of bunches per train, at an additional cost of about 10\% of the initial investment. Furthermore, at 550\,GeV, and building on the experience gained during the earlier phase, it may also be possible to increase the positron polarisation to 60\%. On the other hand, if scientific or strategic considerations arise --- for example, if another project is realised elsewhere in the world --- the machine could instead start operation directly at 550\,GeV. Alternatively, a CLIC-like design could be employed if technically or physically motivated. The CLIC approach uses normal-conducting RF cavities and a two-beam acceleration scheme to achieve accelerating gradients up to 100 MV/m. In this concept, RF power is generated locally by decelerating high-intensity drive beams through specialised power extraction structures, which then transfer energy to the main beams of colliding particles.

Given the two decades of development for the ILC project and more than one decade of operation of the European XFEL~\cite{Altarelli:2006zza}, superconducting radio-frequency cavities (SCRF) are currently the most advanced technology and would be employed in the first stage to minimise the time to first collision. Superconducting cavities with a gradient of 31.5 MV/m are proposed. SCRF technology has also seen important advancements towards higher gradients and quality factors~\cite{LinearColliderVision:2025hlt}. In particular, a factor of two improvement in quality factors over the specification chosen about 15 years ago for ILC is now reliably achievable with a small modification to the cavity production process. Similarly, the klystron efficiency is expected to achieve about 80\%, with 65\% assumed for the original ILC design. Overall, the projected luminosity of the LCF exceeds that of ILC by more than a factor of two for the same power consumption.

The length of 33.5\,km includes a 5-km beam delivery region designed to accommodate collisions up to 3\,TeV. To provide better experimental conditions, the project would feature two interaction points which would share the collider luminosity. Notably, the LCF also offers the opportunity to explore a rich beyond-collider programme, such as beam-dump experiments, which could help, for instance, to search for new particles or probe strong-field QED~\cite{LinearColliderVision:2025hlt}.

\subsection{Muon Collider}
A muon collider could provide a route to achieving the highest centre-of-mass energies in a clean collision environment~\cite{Accettura:2023ked, InternationalMuonCollider:2025sys}. Such a machine would combine the key advantages of \epem and hadron colliders. However, this concept comes with the challenge of the muon’s finite lifetime. Although the idea of a muon collider was already proposed in the 1960s~\cite{budker1969accelerators}, it has remained infeasible for many years. With the muon’s rest-frame lifetime of 2.2\,$\mu$s, acceleration must be extremely rapid, a continuous source of muons is required, and neutrino radiation from muon decays must be mitigated. In recent years, several major advancements have been made, making the Muon Collider a credible proposal on a 20--30-year timescale. This section outlines several key challenges that must be addressed beforehand.

\subsubsection{Project overview}
\begin{figure}
    \centering
    \includegraphics[width=0.9\textwidth]{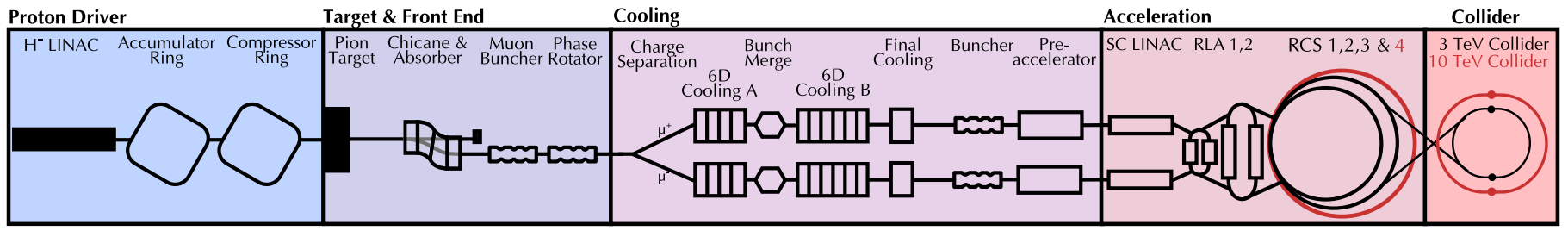}
    
    \caption{Schematic layout of the Muon Collider design~\cite{InternationalMuonCollider:2025sys}.}
    \label{fig:MuC_technical}
\end{figure}

The Muon Collider is a concept of a compact, high-energy machine which would produce, accelerate, and collide two muon beams of opposite charge. The idea represents the first machine to combine the high-energy reach of a hadron collider and the high precision of a lepton collider, reaching a centre-of-mass energy of 10\,TeV, and a luminosity of $21 \times 10^{34}$\,cm$^{-2}$s$^{-1}$. An initial implementation, for example, starting at 3\,TeV, could occur already around 2050. The main components of the collider, as indicated in Figure \ref{fig:MuC_technical}, are the proton driver producing short proton pulses of high intensity, the target on which the proton beam is directed to produce pions, the decay channel guiding the pions and forming muon beams from pion decays, a sophisticated cooling system reducing the beam emittance in a sequence of absorbers and RF cavities in a high magnetic field, a system of linacs and two recirculating linacs accelerating the beams to 64\,GeV followed by a series of high-energy accelerator rings to reach the desired collision energy, and the final collider ring for experimentation.  

One of the main advantages of the Muon Collider is its capability to achieve significant luminosity even at higher energies. In fact, many of the design parameters are optimised to provide the highest collision rates. The luminosity scales approximately with the square of collision energy, which yields a constant rate for very massive particle pair production, as well as a growing rate for vector boson fusion (VBF). This feature is unique to the Muon Collider. In the circular tunnel, the beam can pass through the interaction point multiple times, while the impact of beamstrahlung on the beam focusing is negligible. The luminosity is highest for collider rings having strong dipole fields, and smaller rings are preferred so that muons can collide multiple times. For this reason, a separate collider ring is proposed after the final acceleration stage.

The instantaneous luminosity scales as $\mathcal{L} \sim n_b \cdot (N_\mu)^2$, where $n_b$ is the number of bunches and $N_\mu$ the number of muons per bunch. Effectively, for a fixed total stored muon current, the instantaneous luminosity is inversely proportional to the number of bunches, which favours operation with a single bunch. It also allows for operating the machine with two interaction points. On the other hand, the luminosity decreases with the product of the transverse and longitudinal emittances, so it is crucial to achieve the lowest emittance while preserving high numbers of particles per bunch. If the current target parameters are achieved, the projected integrated luminosity of 10 ab$^{-1}$ could be collected within 5 years. 

The size of the facility depends on future technology and design choices. Currently, the focus of technical studies is on 10\,TeV, but a potential intermediate energy of 3\,TeV is envisaged. The 3-TeV machine would cost about half as much as the full complex and be less demanding in several areas, for example, neutrino flux mitigation. The facility, using 10\,T static dipoles and pulsed dipoles swinging by $\pm$1.8 T, would require a ring of circumference around 10\,km. Achieving the nominal energy of 10\,TeV would require higher-field dipoles of 16\,T and a ring of up to 35\,km. Efforts are being made to fit the facility into the 27-km LHC tunnel, which can be possible if better technology is available or the nominal energy is reduced to about 7.5\,TeV~\cite{InternationalMuonCollider:2025sys}. 

The Muon Collider cost is given as a range for each scenario, together with a most probable value. For a 3.2-TeV machine at CERN, the estimated cost lies between 9 and 18\,BCHF, with a most likely value of about 12\,BCHF. The 7.6\,TeV option at CERN is expected to cost between 12 and 25\,BCHF, with a most probable estimate of 17\,BCHF. A green-field scenario reaching 10\,TeV would require between 14 and 28\,BCHF, with a central estimate of 19\,BCHF. Overall, these cost projections remain highly competitive with those of full-programme \epem colliders or proposed hadron machines.

\subsection{Other collider proposals}
The findings of this thesis are specifically applicable to lepton colliders operating at the $Z$ pole and the multi-TeV scale. The presented proposals, including circular and linear \epem machines and a circular $\mu^{_+}\mu^{_-}$ collider, are the most probable successors of LHC. However, the results can also apply to other facilities, for instance, a $\mu^{_+}\mu^{_+}$ machine or a hadron collider. In the following, two examples of such proposals are briefly described.

\subsubsection{$\mu$Tristan}
The ultra-cold muon technology developed for the muon $g-2$ experiment at J-PARC~\cite{Abe:2019thb} enables the creation of a low-emittance positive muon beam, which can be used for realistic collider experiments. By colliding the positive muon beam of 1\,TeV with a high-intensity electron beam at the Tristan energy (30\,GeV) within a storage ring of the same circumference (3\,km), it becomes possible to realise a collider experiment with a centre-of-mass energy of 346\,GeV~\cite{Hamada:2022mua}. This is sufficient to produce Higgs bosons via VBF, while avoiding $s$-channel annihilation background processes, such as $W^+W^-$ and quark-antiquark production. 

With modern accelerator technologies, the so-called $\mu$Tristan project could potentially achieve a luminosity of the order of $5 \times 10^{33}$\,cm$^{-2}$ s$^{-1}$ per detector, comparable to the design goal of ILC. Assuming ten years of operation with a single detector, the integrated luminosity could reach 1 ab$^{-1}$, resulting in the production of approximately 100,000 Higgs bosons via $W$ boson fusion. If 16-T dipole magnets are available at the time of construction, centre-of-mass energies of 775\,GeV can be achieved with 50-GeV electrons and 3-TeV antimuons. At this energy, due to the enhancement of the $W$ boson fusion process, it is possible to produce 500,000 Higgs bosons with the same luminosity.

A collider using two positive muon beams with centre-of-mass energies up to 2\,TeV is also feasible within the same storage ring. While the production of a low-emittance positive muon beam is already established, a narrow negative muon beam for muon colliders has not yet been achieved. Such a collider would be well-suited for new particle searches, including the pair production of the superpartners of the muon.

Two independent projects in Japan have made significant advancements that enable the planning of such facilities. The first is the SuperKEKB experiment, which has achieved the world’s highest luminosity and aims for an instantaneous luminosity of $8 \times 10^{35}$\,cm$^{-2}$s$^{-1}$~\cite{Belle-II:2010dht}. The second key development is the production of ultra-cold muons, originally developed for precise muon $g-2$ measurements at J-PARC~\cite{Abe:2019thb}. In this method, positive muons from pion decay are stopped on a material surface where they capture electrons to form a bound $\mu^+ e^-$ state (muonium). Laser ionisation is then used to strip the electrons and extract ultra-cold positive muons, which can be accelerated into a low-emittance beam suitable for colliders. 

An upgrade to include polarised beams is under consideration. Muons produced from pion decay are fully polarised due to the vector-axial structure of weak interactions, and given that the muon spin may be preserved by the usage of a longitudinal magnetic field during the formation of muonium, 80\% polarisation can be achieved for positive muon beams. The potential for beam polarisation of the electron beam at SuperKEKB has also been studied~\cite{Roney:2021pwz}, and an electron polarisation of 70\% is targeted.

The cost of the machine has not been estimated yet, but it is expected to be more affordable than other multi-TeV proposals.

\subsubsection{FCC-hh}
FCC-hh, a hadron collider, would be the next major step in high-energy particle physics, building on the technical and operational legacy of LHC~\cite{Benedikt:2928941}. Designed to collide protons at centre-of-mass energies between 80 and 90\,TeV, FCC-hh would deliver an integrated luminosity of approximately 20 ab$^{-1}$ to each of two general-purpose experiments over a 25-year operation period. Two additional specialised experiments are to be hosted. Beyond proton--proton collisions, FCC-hh would also accommodate proton--ion and ion--ion collision modes, and potentially also electron--proton and electron--ion collisions at one of the interaction points. If development proceeds according to schedule, beam operations could begin by the early or mid-2070s. The total construction cost is currently estimated at 18.9\,BCHF, assuming the reuse of the FCC‑ee tunnel, and roughly 10\,BCHF is attributed to the collider magnets.

FCC-hh would be housed in the 90-km tunnel, previously used for FCC-ee. Located in the France-Switzerland region, where electricity is already largely decarbonised, the project could remain environmentally sustainable. Technically, FCC-hh would extend both the centre-of-mass energy and peak luminosity by up to an order of magnitude compared to HL-LHC. The FCC design builds upon the LHC experience, with new challenges arising in magnet technology and cryogenic power consumption due to strong synchrotron radiation at higher energies. Even though the projected luminosity performance of FCC-hh should be attainable, the technical readiness of the project depends critically on the development and availability of the magnets. The principal technology is a high-field dipole magnet system based on Nb$_3$Sn superconductors, targeting a field of 14\,T and operating at 1.9\,K. 

\subsection{Future detectors}
Analogous to the diversity of collider proposals, numerous detector concepts have been developed, often differing substantially in design. However, they typically share similar fundamental operating principles. The International Large Detector (ILD) concept~\cite{Behnke:2013lya, ILDConceptGroup:2020sfq} might be used as a representative example, given its potential suitability for both linear and circular \epem colliders~\cite{ILD:2025yhd}. The ILD is chosen as the most mature design, as it is based on full simulations and prototype beam test results.

The ILD is a detector concept developed for experiments at future high-energy lepton colliders, optimised to perform precision physics across a broad energy range, starting from 90\,GeV up to 1\,TeV. The project incorporates a high-precision, large-volume tracking system that combines silicon and gaseous detectors and a highly granular calorimeter, all situated within a central solenoidal magnetic field. From its beginning, the ILD concept has remained open to new technologies, and no final decisions have been made regarding many sub-detector components. In several areas, multiple technological options remain under investigation.

The main driver of ILD’s design is the goal of achieving the highest possible precision, guided by the ``particle flow'' approach. This concept aims to reconstruct all particles in an event individually. To accomplish this, the detector must effectively distinguish single objects even within dense jets and measure neutral and charged particles separately. This requires exceptional particle separation capabilities, particularly in the calorimeters, which are built with extremely high granularity in both transverse and longitudinal dimensions.

\chapter{Physics at future lepton colliders}

\label{chap:futurephysics}

Based on the technical developments described in the previous chapter, future lepton colliders will provide unprecedented experimental conditions. Their experimental programme will particularly advance Higgs physics, top-quark physics, electroweak studies, and searches for BSM effects. Thanks to the large data sample collected at the $Z$ pole, the FCC can also significantly improve measurements in flavour physics. Although this division is somewhat arbitrary --- for instance, new Higgs states could simultaneously influence top-quark properties and modify the electroweak sector --- it remains a common framework in strategic documents. The same categorisation is therefore adopted in the thesis. Selected research directions, relevant to all major collider proposals, are highlighted in this chapter.

\section{Higgs physics}
For many years, the particle physics community sought to understand how fundamental particles acquire mass. One of the proposed solutions connects the origin of mass to the existence of a new scalar field and to the spontaneous breaking of a gauge symmetry assumed in the massless SM. This framework is commonly referred to as the \textit{Higgs mechanism}. Although the abbreviated term will be used throughout this thesis, it is historically imprecise, as it neglects the prior contribution of Englert and Brout, published shortly before that of Higgs, as well as the independent work of Guralnik, Hagen, and Kibble~\cite{Englert:1964et,Higgs:1964pj,Guralnik:1964eu}. Furthermore, earlier theoretical developments by Nambu~\cite{Nambu:1960tm} and Anderson~\cite{Anderson:1963pc} provided essential conceptual groundwork~\cite{kibble2009englert}. 
Alternative explanations for mass generation have also been proposed. For instance, a yet unidentified spectrum of heavy particles interacting through a new fundamental force could lead to \textit{dynamical symmetry breaking}~\cite{Weinberg:1975gm,Susskind:1978ms}.

In the absence of a mass-generation mechanism, the SM becomes theoretically inconsistent and loses predictive power. Specifically, the amplitudes of certain processes --- such as longitudinal vector-boson scattering --- increase with energy, eventually violating the unitarity bound at sufficiently high energies. With the Higgs mechanism, this growth is cancelled by the inclusion of Higgs-exchange diagrams, whose amplitudes possess the same energy dependence but opposite signs. This cancellation restores unitarity and ensures the self-consistency of the electroweak sector.

The requirement of unitarity places quantitative constraints on the energy scale of the symmetry-breaking mechanism~\cite{Lee:1977yc,Lee:1977eg}. The Higgs boson must have a mass below approximately 1\,TeV, whereas any new heavy states associated with alternative scenarios should appear below the scale of a few TeV. This reasoning gave rise to the so-called \textit{No-Lose Theorem}, according to which the LHC would have necessarily uncovered the phenomenology of electroweak symmetry breaking, irrespectively of the mechanism realised in nature.

The 2012 discovery of a scalar particle with a mass of approximately 125\,GeV at the LHC marked a major success in particle physics~\cite{ATLAS:2012yve,CMS:2012qbp}. Interpreted as the Higgs boson, this particle completed the spectrum of fundamental particles predicted by the SM. As a consistent and predictive framework whose free parameters are now experimentally determined, the SM has so far provided an accurate description of all collider measurements. No significant deviations from its predictions have yet been observed.

Despite this success, the Higgs mechanism leaves several fundamental questions unresolved~\cite{Salam:2022izo}. The observed pattern of Higgs couplings, and consequently the masses and mixings of quarks and leptons, arises from parameters inserted into the theory by hand, without any underlying explanation. Even the key feature of spontaneous symmetry breaking remains unexplained by any deeper, microscopic, theoretical principle.

A genuine dynamical origin for these features would likely require new interactions that couple to the Higgs field. Probing such effects is a central objective of future electron-positron colliders. At centre-of-mass energies around 250\,GeV, the cross section for the Higgs production process $e^+ e^- \to ZH$ reaches its maximum. Running at this energy will enable precise studies of the Higgs sector, making a future collider a true \textit{Higgs factory}~\cite{LinearColliderVision:2025hlt,Blondel:2025kbl}. In this so-called \textit{Higgsstrahlung} process, each Higgs boson is produced in association with a recoil $Z$ boson. This enables straightforward identification of Higgs events from four-momentum conservation and facilitates precise measurements of Higgs decay branching ratios, including searches for decays to invisible final states. Additionally, the recoil $Z$ boson provides a direct way to determine the Higgs boson mass. By analysing the distribution of recoil $Z$ bosons, the absolute production cross section can be measured, which in turn allows for the absolute normalisation of the Higgs boson partial widths.

Additional measurements at higher energies --- such as $WW$ fusion, $t\bar{t}H$ production, and double-Higgs production --- will further constrain possible extensions of the SM. Comparing precision measurements from different processes will provide a critical test of the significance of any deviations from the SM predictions. These probes will also enhance the precision of top-quark measurements, as discussed later.

One of the key parameters of the Higgs boson is its trilinear self-coupling, which is directly related to the shape of the Higgs potential. It is not a free parameter in the SM, but its modifications appear in various extensions of this fundamental theory. At the tree level, two processes are particularly sensitive to this coupling: $e^+ e^- \to ZHH$ and $e^+ e^- \to \nu \bar{\nu} HH$, both of which become accessible at centre-of-mass energies above 500\,GeV. At lower energies, sensitivity to the trilinear coupling can only be obtained through loop-induced effects; however, this approach provides limited discriminatory power between different BSM scenarios. The HL-LHC is expected to measure the Higgs trilinear coupling with a precision of around 25\%~\cite{deBlas:2944678}. A full LCF programme operating at energies up to 1\,TeV could directly improve this precision to about 7\% (for all allowed values of the coupling). In contrast, FCC-ee is projected to reach approximately 17\% assuming the SM value of the coupling.

Higher-energy colliders open new opportunities for probing the Higgs potential. At the TeV scale, Higgs bosons are abundantly produced via 
VBF, potentially improving the precision of Higgs measurements by an order of magnitude or more. The Muon Collider or FCC-hh could probe the Higgs self-coupling with a precision of roughly 4\%. Furthermore, not only could triple Higgs production be observed at these energies, but also the restoration of electroweak symmetry at multi-TeV energies could be investigated through the polarisation of the final-state vector bosons. Compared to high-energy hadron machines, these studies benefit from the clean environment of the Muon Collider, free from QCD backgrounds. These topics will be addressed in the studies presented below.

\section{Top physics}
The top quark is the heaviest particle in the SM, with a mass of about 173\,GeV, exceeding that of the Higgs boson. Consequently, due to its large Yukawa coupling, the top quark provides a sensitive probe of the Higgs sector and the mechanism of EW symmetry breaking. It also has a unique feature in QCD: its lifetime is much shorter than the hadronisation time-scale, so its spin information is retained in the decay products and can be used to study the chirality structure. Recent experimental findings indicating a threshold enhancement in $t\bar{t}$ production consistent with a pseudoscalar state~\cite{CMS:2025kzt,ATLAS:2025woc} further motivate top-quark measurements.
Future electron---urements,- particularly of the top-quark mass and its electroweak and Higgs couplings. The mass of the top quark is a key parameter in testing the consistency of the SM, - it influence-s electroweak loop corrections that determine the predicted $W$~-boson mass.

For collision energies up to a few TeV, top-quark pair production in electron--positron collisions is dominated by $s$-channel $Z/\gamma$ exchange. Each top quark decays predominantly into a $b$ quark and a $W$ boson. Consequently, the topology of the process $e^+ e^- \to t\bar{t} \to b W^+ \, \bar{b} W^- \to 6f$ is determined by the subsequent decays of the $W$ bosons, allowing for the classification of events into hadronic, leptonic, and semi-leptonic final states. Backgrounds to the production process arise, for example, from single-top production and from triple gauge boson production~\cite{ChokoufeNejad:2016qux,CLICdp:2018esa}.

Near the top-quark pair production threshold, the cross section rises sharply over a few GeV in collision energy. The position of this threshold is directly related to the top-quark mass, while its shape provides information about the top-quark width. Background contributions can be efficiently suppressed near the threshold via kinematical cuts. Even for small data sets, well below a single inverse attobarn, the theoretical uncertainties are dominant~\cite{LinearColliderVision:2025hlt, Bach:2017ggt, Defranchis:2025auz}. The FCC-ee targets a precision of 7\,MeV for the top mass measurement, based on fast-simulation analysis, while linear colliders are expected to achieve about 20\,MeV. The results strongly depend on the running scenario; for example, if the same integrated luminosity of 410\,fb$^{-1}$ is assumed, the statistical uncertainty reaches 4.3\,MeV at FCC-ee and 6.3\,MeV at ILC~\cite{Defranchis:2025auz}. The experimental uncertainties remain smaller than the theoretical ones, which are currently estimated at around 35\,MeV~\cite{deBlas:2944678}.

The top-threshold cross section is also affected by the exchange of soft gluons and virtual Higgs bosons, resulting in strong sensitivity to both the strong coupling constant and the top-quark Yukawa coupling. The top-quark Yukawa coupling, which is of order one in the SM, can be extracted from threshold-scan data with a precision of about 4\%~\cite{Horiguchi:2013wra}. However, its contribution across the threshold region is nearly flat. When performing a simultaneous fit with several parameters, the Yukawa coupling is also nearly degenerate with the strong coupling constant, and both are significantly influenced by theoretical uncertainties arising from missing higher-order corrections~\cite{Hoang:2013uda,Beneke:2016cbu,Robson:2920434}. Nevertheless, the Yukawa coupling can be effectively probed at higher energies, above 500\,GeV, where the associated production of a top-quark pair with a Higgs boson becomes relevant, allowing percent-level precision to be achieved at a linear collider operating at 1\,TeV.

Other top-quark measurements can also be performed more effectively --- or even exclusively --- at higher energies, when additional top-related processes might be considered. In the multi-TeV regime, VBF production, $e^+ e^- \to t\bar{t}\nu_e\bar{\nu}_e$, becomes dominant due to the logarithmic enhancement of the cross section. The process provides an independent probe of the top-quark Yukawa coupling, as well as the electroweak top-quark interactions~\cite{Ake:2023xcz,Han:2024gan}. In several BSM models in which the Higgs boson is a composite particle~\cite{Dimopoulos:1981xc,Kaplan:1983fs,Kaplan:1983sm}, the top quark is involved in the composite structure, potentially leading to large deviations in the couplings. The process is also sensitive to potential effects from heavy new particles. Moreover, it serves as a valuable tool for studying theoretical aspects of VBF, owing to the large invariant mass of the produced top-quark pairs regulating potential soft divergences and the sensitivity to different vector-boson polarisations, as will be discussed later in the thesis.

\section{Electroweak sector}
As elaborated earlier, the SM is a remarkably successful theory that precisely describes all collider data from past and ongoing experiments. Nevertheless, it requires setting at least 18 fundamental parameters to specific values, which can only be determined experimentally. Within the SM, other quantities can be derived from these constants. For example, the $W$ boson mass depends on the weak coupling and the Higgs VEV, while the top Yukawa coupling is directly related to the top-quark mass and the Higgs VEV. In this sense, the choice of \textit{fundamental} parameters is somewhat arbitrary --- for instance, one could treat the top Yukawa coupling as a fundamental constant and derive the top mass from it. However, these relations hold only within the SM; if the theory is extended by new forces or particles, they may no longer apply. From this perspective, it is essential to measure as many SM parameters as possible to overconstrain the system and identify potential deviations from the theory’s predictions. At future lepton colliders, the electroweak sector will be probed with unprecedented precision.

The major \epem collider proposals envisage running at 91\,GeV. The so-called $Z$-pole run would produce between $10^9$ and $10^{12}$ $Z$ bosons, depending on the specifics of the facility. These large samples could be used to improve the precision of $Z$-boson coupling measurements, as shown in Chapter~\ref{chap:Zcouplings}. Additional information can be obtained from the $WW$ threshold scan, as considered for FCC-ee, and other higher-energy runs. Currently, the $Z$-boson mass and width are known with a precision of about 2\,MeV; FCC-ee could reduce these uncertainties to 100\,keV and 12\,keV, respectively, while the ILC could achieve 200\,keV and 125\,keV~\cite{deBlas:2944678}. Further improvements are also expected for the $W$ boson, whose mass and width are currently known with uncertainties of about 10\,MeV and 42\,MeV. FCC-ee could reach precisions of roughly 200\,keV for the mass and 300\,keV for the width, while the ILC could achieve 1.6\,MeV and 2\,MeV, respectively. At the same time, these estimates should be treated with caution, owing to subtle differences in the treatment of theoretical uncertainties. The precision attainable at the ILC may be sufficient to saturate the experimental precision for certain other measurements, for example, in the Higgs sector~\cite{FCC:2025lpp}.

Another important feature of the SM is the unification of the electromagnetic and weak interactions. The weak mixing angle describes how the gauge fields of the SM mix to form the physical fields associated with the $Z$ boson and the photon, and it is also related to the ratio of the $W$ and $Z$ boson masses. Measuring this parameter provides one of the most stringent tests of the underlying principles of the SM. At \epem colliders, the weak mixing angle can be determined via measurements of various asymmetries. Linear colliders have a significant advantage due to beam polarisation: the left--right asymmetry, which measures the difference in event rates between left- and right-handed electron beams, directly probes the difference between the electron left- and right-handed couplings to the $Z$ boson~\cite{Fujii:2017vwa}. 

Both linear and circular colliders can also measure forward--backward asymmetries, which quantify the difference in event numbers with fermions emitted in the forward and backward directions. These measurements, combined with auxiliary inputs (e.g., from $\tau$-pair production), allow for a model-dependent determination of the weak mixing angle. In general, linear colliders could directly improve the current precision of the weak mixing angle measurement by a factor of 40--50, while circular colliders could achieve an improvement of up to a factor of 200 under additional assumptions~\cite{deBlas:2944678}. The high luminosity of circular colliders would also enable an order-of-magnitude improvement in the determination of the fine-structure constant.

Another path becomes available if higher, multi-TeV energies are reached, as expected for the Muon Collider or future upgrades of linear \epem facilities. A unique feature of these machines is the simultaneous access to both high energy and precision, allowing detailed studies of the energy-dependent features of the SM~\cite{InternationalMuonCollider:2025sys}. In the multi-TeV regime, new phenomena emerge, with abundant electroweak radiation being a prominent example. Analogous to photon initial-state radiation, weak bosons can be emitted from highly energetic charged particles, effectively making a multi-TeV collider a weak boson collider~\cite{Costantini:2020stv}. VBF processes provide a complementary probe of the electroweak sector compared to standard annihilation processes, giving access to the gauge couplings of the SM and enabling tests of gauge symmetry restoration above the electroweak scale~\cite{Beyer:2006hx,Fleper:2016frz}.

Furthermore, a global interpretation of experimental data can offer new insights into the structure of the SM or its possible extensions. The SM Effective Field Theory (SMEFT) provides a convenient and \textit{quasi}-universal\footnote{The framework assumes that the effects of new physics appear above a particular energy scale, often connected to the collision energy. Ambiguities in comparisons between different colliders can arise if, for example, new particles are within the reach of the highest-energy facility but not others. Additionally, due to the large number of free parameters, extra constraints are typically imposed on the coefficients. Interpretation may also be complicated if, for instance, the Higgs sector is extended. These issues are, however, beyond the scope of this thesis and will not be further discussed.} framework for parametrising deviations from the SM. SMEFT is based on adding higher-dimension operators to the SM Lagrangian, each multiplied by a dimensionless coefficient which is expected to be zero in the SM. Experimentally, these coefficients can be constrained through a variety of measurements at a given collider, providing a global picture of the facility's capabilities and enabling meaningful comparisons between different machines. Future lepton colliders will deliver an impressive volume of data, enabling coefficient fits to be improved by orders of magnitude~\cite{deBlas:2025xhe,deBlas:2944678}.

\section{BSM searches}
Various BSM theories have been proposed to address open questions about the Universe. Future colliders hold enormous potential not only for testing the SM through high-precision measurements, but also for performing efficient direct searches for new particles or interactions. The following sections highlight a few examples of such searches, based on the studies by the Author of this thesis.

\subsection{Heavy Neutral Leptons}
Heavy Neutral Leptons (HNLs) appear in extensions of the SM, which aim to explain the smallness of light neutrino masses through the seesaw mechanism. In many BSM models, introducing additional neutrino states can also account for the baryon asymmetry of the Universe or the observed dark matter density~\cite{Canetti:2012vf, Gninenko:2012anz, Caputo:2018zky}. In realistic scenarios, typically at least two HNLs are needed.

Lepton colliders provide an ideal environment for searching for HNLs. Depending on the HNL’s properties, a variety of experimental signatures can be explored. For light HNLs, one can investigate the invisible decays of the $Z$, $W$, or Higgs bosons. If the HNL has very weak couplings to SM particles, it may be long-lived, leading to the distinctive signature of displaced vertices~\cite{Blondel:2014bra,Bellagamba:2025xpd}. Conversely, for HNL masses above the $W$ mass, decays are generally prompt. These scenarios have been extensively studied in the context of lepton collider research~\cite{Mekala:2022cmm,Mekala:2023diu,Mekala:2023kzo}.

The study was based on an effective extension of the SM, assuming that only one HNL is kinematically accessible at colliders. All heavy-neutrino couplings to SM leptons were taken to be equal, as motivated by the close-to-maximal mixing in the SM neutrino sector. For each of the considered running scenarios, the sensitivity extends nearly up to the centre-of-mass energy, and the expected limits on the heavy-neutrino coupling to SM leptons are significantly stronger than those achievable at high-energy hadron colliders. 

\begin{figure} 
    \centering
    \includegraphics[width=0.7\textwidth]{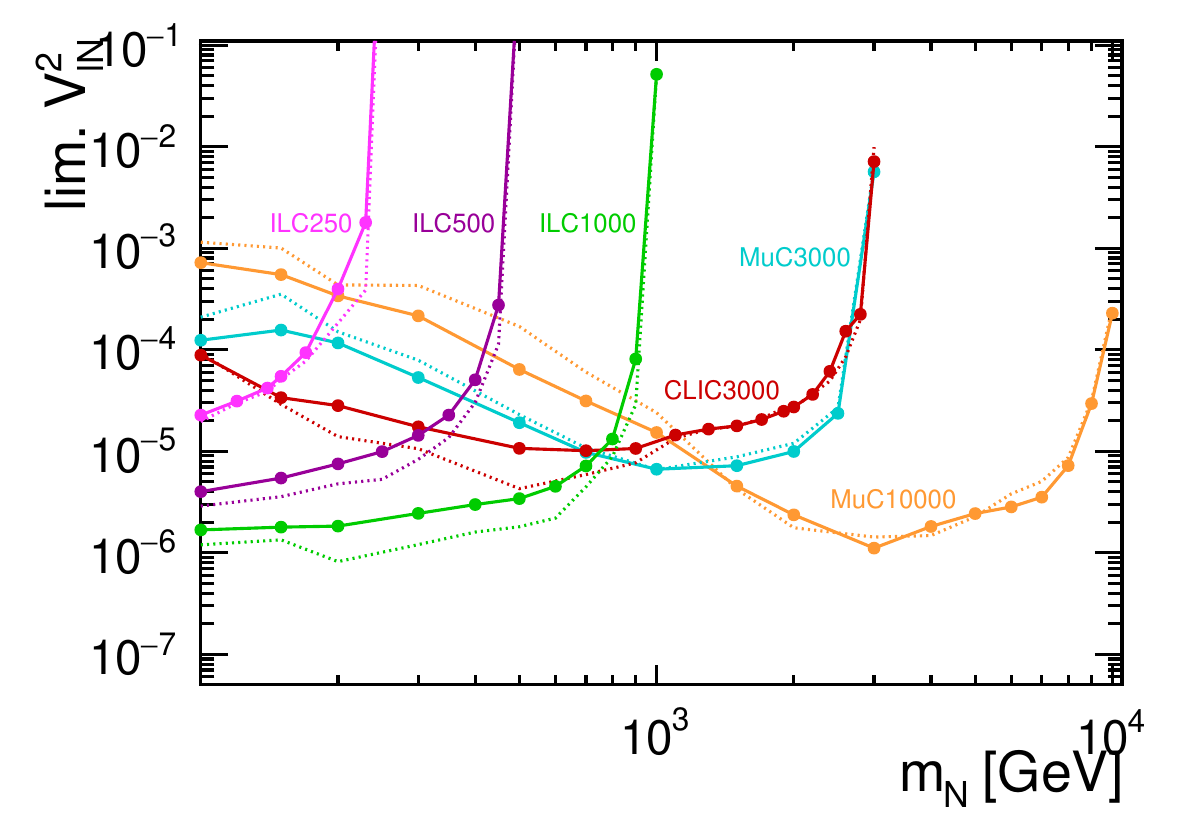}
    \caption{Comparison of the expected 95\% confidence level discrimination limits between Majorana and Dirac neutrinos (solid lines), as a function of the heavy neutrino mass, $m_N$, and the 5$\sigma$ discovery limits (dotted lines) for different collider scenarios~\cite{Mekala:2023kzo}.}
    \label{fig:results_DvM}
\end{figure}

HNLs can be of Dirac nature, exhibiting only lepton-number conserving decays, or of Majorana nature, allowing also lepton-number violating decays. If heavy neutrinos are discovered at a future lepton collider with a significance of $5\sigma$, their nature should also be established at the 95\% confidence level, as illustrated in Figure \ref{fig:results_DvM}.

\subsection{Dark Matter}
As explained in Section~\ref{sec:openquestions}, one of the main motivations for BSM searches is the existence of dark matter. \epem colliders offer a unique possibility for general searches for dark matter particles using the mono-photon signature. Pair production of dark matter particles can be tagged by a photon emitted from the initial state. Its kinematics, including the distributions of energies and angles, can provide information about the nature of the new particles, independently of how the interactions are mediated. Such searches were carried out in~\cite{Kalinowski:2021tyr}.

\begin{figure} 
    \centering
    \includegraphics[width=0.4\textwidth]{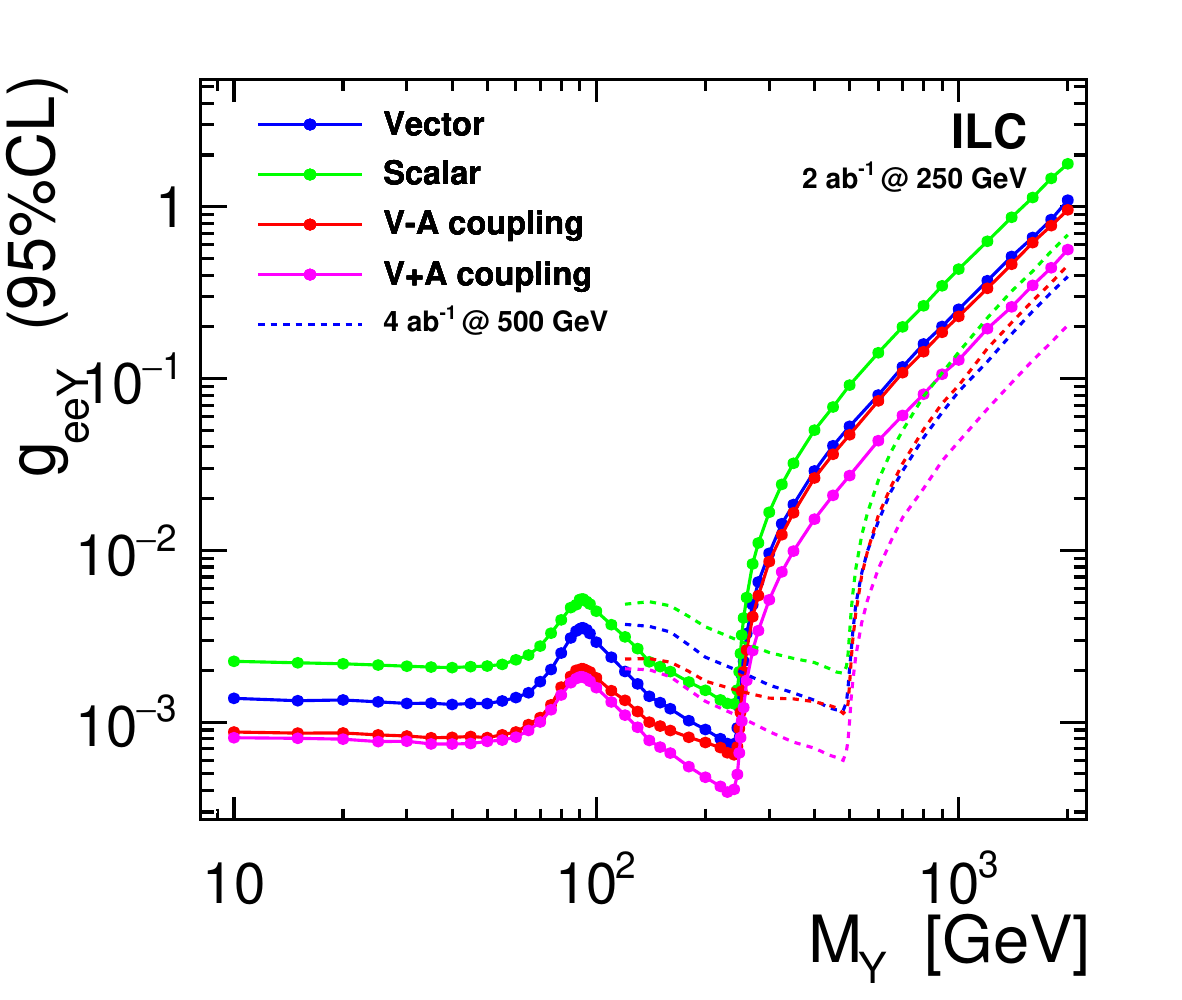}
    \includegraphics[width=0.4\textwidth, trim=0 0 0 1cm]{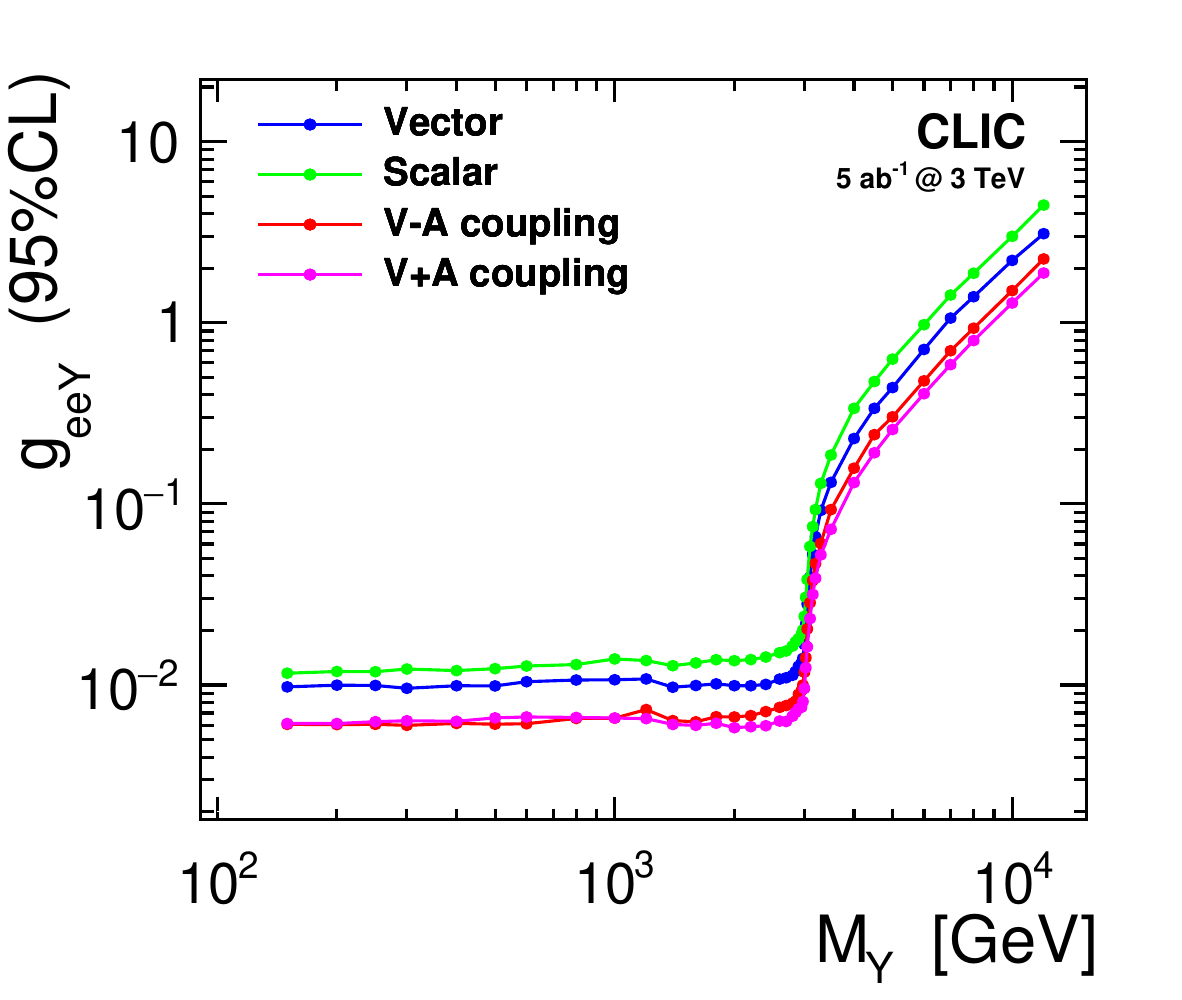}
    \caption{Expected limits on mediator coupling to electrons for the ILC running at 250 and 500\,GeV (left) and CLIC running at 3\,TeV (right), for relative mediator width, $\Gamma$/M = 0.03, and different mediator coupling scenarios, as indicated in the plot. Presented are combined limits corresponding to the baseline running scenarios, with systematic uncertainties taken into account~\cite{Kalinowski:2021tyr}.}
    \label{fig:results_DM}
\end{figure}

Figure \ref{fig:results_DM} shows example limits on the mediator coupling to electrons, $g_{eeY}$, for the ILC running at 250 and 500\,GeV and CLIC running at 3\,TeV. The mass of the DM particles was set to 1\,GeV for ILC and 50\,GeV for CLIC. Different structures of mediator couplings were studied. The mediator coupling to DM was fixed by setting its width to 3\% of its mass. For the considered couplings, the mono-photon analysis gives higher sensitivity to processes with light mediator exchange than the direct resonance searches.

\subsection{Additional gauge bosons}
One of the distinctive features of the SM is its underlying gauge symmetry (see Section~\ref{sec:formSM}). Many BSM models extend this symmetry to provide an explanation for \textit{ad-hoc} effects present in the SM. If any of these scenarios is realised in nature, the existence of additional neutral gauge bosons, denoted here as $Z'$, is its natural consequence. In this context, simplified models containing a single $Z'$ boson can be useful for studying potential deviations from SM predictions in a generic way. Recent measurements at the LHC have excluded the existence of a $Z'$ with a mass of up to about 5\,TeV~\cite{ATLAS:2016gzy, ATLAS:2019fgd, ATLAS:2019erb, CMS:2019gwf, ATLAS:2020fry, CMS:2021ctt, CMS:2021klu,Alvarez:2020yim, Lozano:2021zbu}. \epem colliders could probe masses up to approximately 20\,TeV, depending on the collision energy~\cite{Cvetic:1995zs, Leike:1996pj, Leike:1998wr, Godfrey:2005pm, Osland:2009dp, Andreev:2012cj, Han:2013mra, Godfrey:2013eta, Kapukchyan:2013hfa, Pankov:2017dkv, Gulov:2017uqa}.

\begin{figure}
    \centering
    \includegraphics[width=0.5\textwidth]{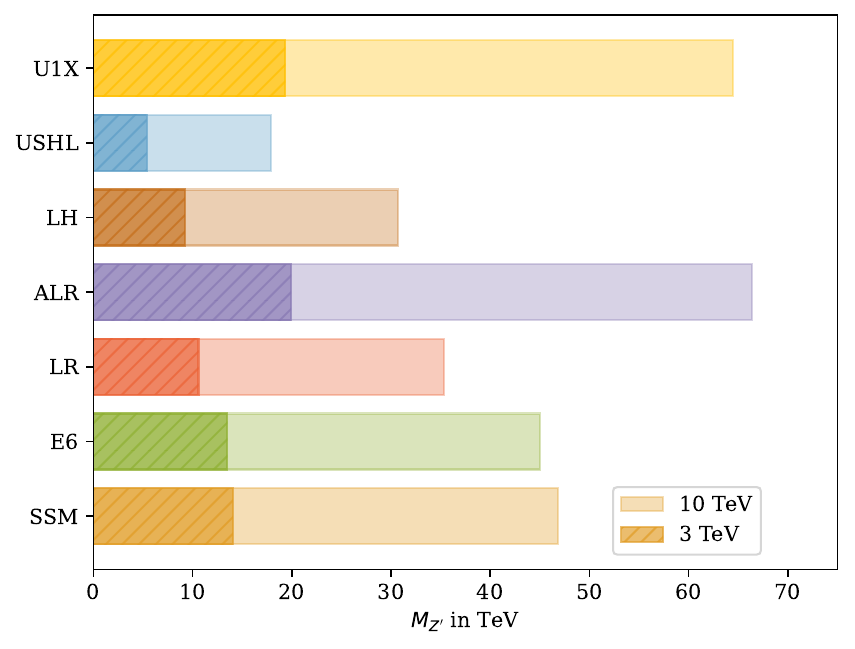}
	\caption{The reach in $M_{Z'}$ for a 3-TeV and 10-TeV muon collider with luminosities of $0.9\,\text{ab}^{-1}$ and $10\,\text{ab}^{-1}$, respectively assuming no beam polarisation. Systematic errors are included. The bars correspond to the exclusion limit of the given $Z'$-model at 95\% confidence level~\cite{Korshynska:2024suh}.}
	\label{fig:reach_MuC}
\end{figure}

The Muon Collider has the potential to significantly improve these limits, reaching scales up to 70\,TeV based on very conservative assumptions, as shown in Figure~\ref{fig:reach_MuC}. Such an analysis was carried out in~\cite{Korshynska:2024suh} within a statistical framework employing leptonic observables in 2-fermion processes, including the total cross section, forward--backward asymmetry, left--right asymmetry, and polarisation asymmetry. The method also allows for efficient discrimination between different scenarios. Employing hadronic observables could further enhance the sensitivity.

\subsection{Non-standard neutrino interactions}
Since experimental results from precision measurements and direct collider searches show no evidence for new particles or interactions, the scale of new physics responsible for generating neutrino masses is likely to lie above the electroweak scale. In such a case, the SM serves as an excellent low-energy description of a more complete theory via SMEFT. At energy scales well below the electroweak scale, the theory can be further simplified by integrating out the heavy SM particles. The resulting framework, the \textit{Weak Effective Field Theory} (WEFT)~\cite{Jenkins:2017jig}, can be used to constrain non-standard interactions in the neutrino sector.

Particle colliders produce large numbers of neutrinos. A particularly intense neutrino flux is expected at the Muon Collider, where neutrinos are copiously produced from muon decays. Another important advantage of the Muon Collider is the well-defined composition and energy spectrum of the flux. This feature can be exploited to host a dedicated experiment probing the neutrino interactions, as proposed in~\cite{Kling:2025zsb}.

\begin{figure}
  \centering
  \includegraphics[trim={0 2.2cm 0 2.5cm},clip, width=0.9\textwidth]{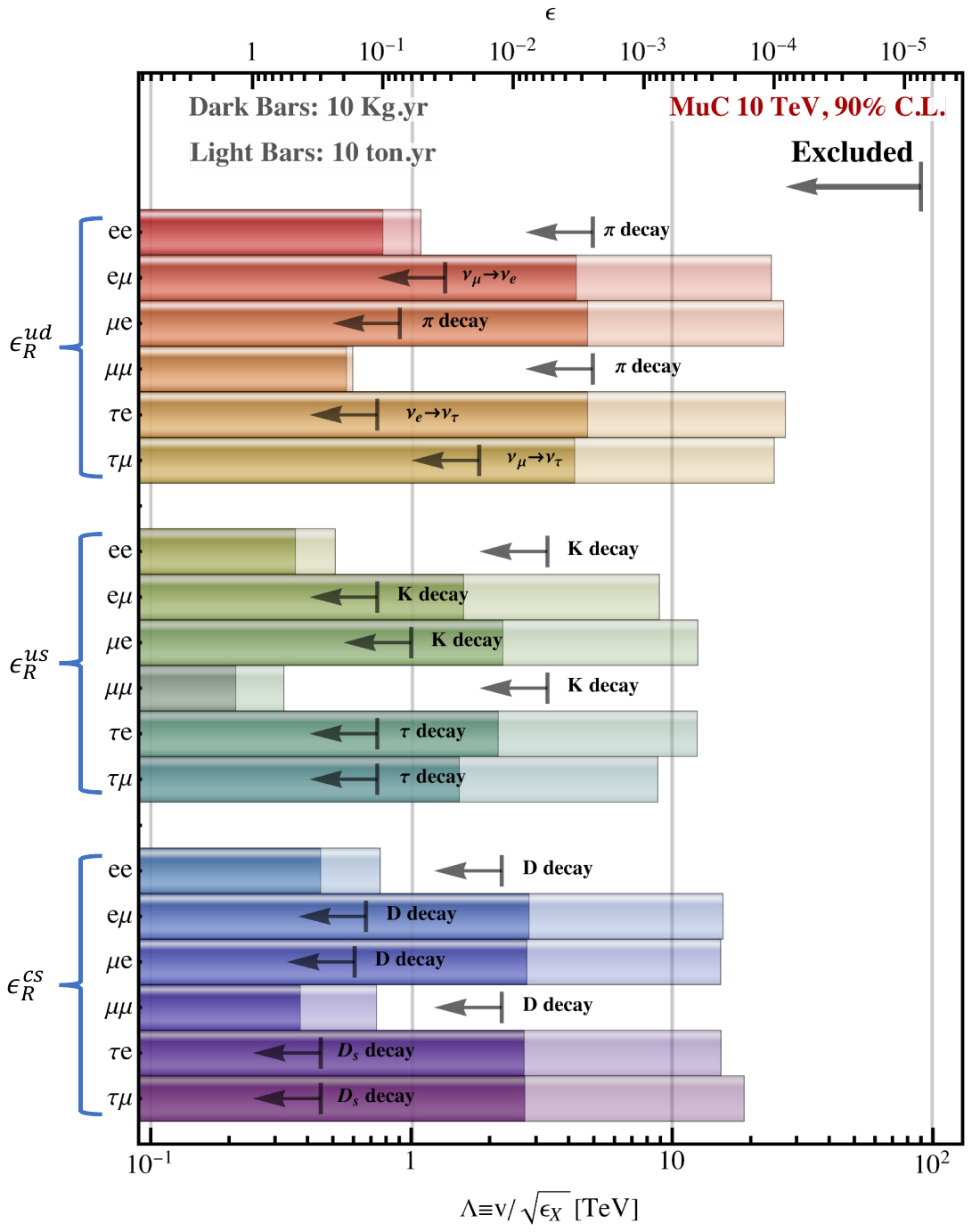}
  \caption{The projected constraints on the Wilson coefficients of new right-handed interactions in the WEFT framework. Projected limits are reported both in terms of the dimensionless couplings $[\epsilon_X^{jk}]_{\alpha\beta}$ (top axis) and in terms of the effective new physics scale $[\Lambda_X^{jk}]_{\alpha\beta} \equiv v / \sqrt{[\epsilon_X^{jk}]_{\alpha\beta}}$ (bottom axis). Each coloured bar indicates the limit on one particular interaction. Darker bars show the projected limits for a 10~kg$\cdot$year detector, while the lighter bars show what a $10~\text{ton}\cdot\text{year}$ neutrino detector can do at a 10-TeV muon collider. The constraints coming from low-energy experiments, such as precision meson decay measurements, are indicated with dark arrows~\cite{Kling:2025zsb}.}
  \label{fig:constraints-summary-R}
\end{figure}

Example constraints for right-handed interactions in the WEFT framework are shown in Figure~\ref{fig:constraints-summary-R}. By considering differential event rates, one can obtain very stringent limits on flavour-changing processes, significantly improving upon data from dedicated flavour experiments. Observation of such processes would, in turn, be a clear sign of new physics.

\chapter{Monte Carlo generators}
\label{chap:MonteCarlo}

Given the statistical nature of quantum interactions, experiments in particle physics require collecting large datasets. These data can only be interpreted efficiently through comparison with the results of computer simulations that encode the current state of knowledge in the field. Monte Carlo techniques are therefore widely used in collider physics to predict experimental outcomes. Typically, the simulation chain can be divided into three stages: particle-level event generation, simulation of particle propagation in matter, and simulation of detector response. In this chapter, the basic principles of Monte Carlo generators used for particle-level event production are introduced, followed by a presentation of the two codes employed in this thesis: \textsc{Pythia} and \textsc{Whizard}.

\section{General concept}
\label{sec:general_concept_MC}
Collider physics typically describes the collision of two beam particles, with momenta denoted as $p_{I,1}$ and $p_{I,2}$ and masses as $m_{I,1}$ and $m_{I,2}$ in the following, resulting in the production of $n$ final-state particles, with momenta $q_{F,i}$. In this case, the differential cross section for such processes is given by
\begin{equation}
  d\sigma_{I\to F} =
  \left(\prod_{i=1}^n \frac{d^3q_{F,i}}{(2\pi)^3 2q_{F,i}^0}
    \right) (2\pi)^4  \delta^4\left(p_{I,1}+p_{I,2} - \sum_{i=1}^n q_{F,i}\right)
    \frac{ \left|\mathcal{M}_{FI}\right|^2}{4 \sqrt{(p_{I,1}\cdot
      p_{I,2})^2 - m^2_{I,1}m^2_{I,2}}}
      \; .
\end{equation}
Three main components can be identified: the matrix element $\mathcal{M}_{FI}$ describing the hard process $I \to F$, the kinematic flux factor in the denominator, and the phase-space integral~\cite{Reuter:2025imz}. The dimension of the integral is $3n - 4$. For many \textit{simple} processes, it may be calculated analytically. However, for multiple particles in the final state, the dimensionality becomes high, and only numerical methods can be applied. Traditional geometrical methods of integration face significant challenges due to the high dimensionality, singularities in the denominator, and experimental cuts that distort the integration space. The Monte Carlo (MC) approach appears to be best suited for both numerical integration of cross sections and generation of events distributed according to the corresponding differential probabilities.

MC methods are computational algorithms which rely on random sampling~\cite{Cowan:1998ji}. One of their most common applications is numerical integration. For example, consider a shape drawn within a unit square (or unit volume in higher dimensions). Its volume can be estimated by randomly sampling points within the unit shape and calculating the ratio of points that fall inside the drawn shape to the total number of points sampled. This simple example illustrates that increasing the number of sampling points improves precision. Additionally, the spatial distribution of these points plays a crucial role. A uniform random distribution is often a natural choice. Still, in cases where the shape is highly irregular, it may be beneficial to concentrate more points in the complex regions to enhance accuracy. Although problems in particle physics typically involve high dimensionality, the same principles apply. Monte Carlo generators are often equipped with high-efficiency pseudo-random number generators and tools that efficiently partition phase space into regions of varying importance, allowing for more effective sampling.

Nevertheless, the efficient calculation of the hard-process matrix element is only one of the many tasks required of Monte Carlo event generators. These codes must also account for several other aspects which significantly alter the overall picture of particle collisions, due to the intrinsically complex nature of the underlying theory of particle physics. For instance, because quarks are only observed within bound states, it becomes necessary to distinguish between the physical particles that are accelerated and the partons --- their constituents --- that participate in the hard-scattering process, which gives rise to the concept of beam structure functions. Likewise, particles produced in the hard interaction must be ``dressed'' into hadrons, a process known as hadronisation. Another challenge comes from the presence of multiple soft and collinear emissions, which cannot be adequately described using fixed-order perturbation theory. To handle these, approximate methods such as parton showers are introduced to complement the hard-scattering picture. The different ways in which these effects are modelled have led to the development of multiple Monte Carlo generators, each with its own strengths and assumptions. Among the most well-known and widely used are \textsc{Herwig}~\cite{Bahr:2008pv, Bellm:2015jjp, Bewick:2023tfi}, MadGraph~\cite{Alwall:2011uj,Alwall:2014hca,Frederix:2018nkq}, \textsc{Pythia}~\cite{Sjostrand:2006za,Sjostrand:2007gs,Bierlich:2022pfr}, \textsc{Sherpa}~\cite{Gleisberg:2003xi,Sherpa:2019gpd,Sherpa:2024mfk}, and \textsc{Whizard}~\cite{Moretti:2001zz,Kilian:2007gr,Reuter:2024dvz}. In the following, further discussion will focus on hard-process matrix elements, beam structure functions, parton showers, and hadronisation models --- which are relevant to the studies in this thesis.

\subsection{Hard scattering}
The core of Monte Carlo generation is the calculation of the matrix element for the hard scattering process. Since multiple, often similar, Feynman diagrams typically contribute to a given process, dedicated techniques are required to simplify the calculations and avoid redundancies. The SM consists of a finite set of well-defined building blocks that can be used to construct matrix elements, which allows for straightforward automation by recursion. However, when exploring BSM scenarios, the variety of possible building blocks becomes extremely large. In the past, BSM theories had to be implemented individually, but automated methods have existed for over a decade.

To match the precision of experimental data, Monte Carlo generators must go beyond leading-order (LO) calculations in perturbation theory. Modern codes include at least next-to-leading order (NLO) corrections, which account for single real-emission diagrams as well as interference between tree-level and one-loop amplitudes. For certain applications, even higher-order corrections, involving multiple real emissions or additional loop diagrams, are required. This will certainly be the case for precision measurements at future colliders~\cite{Blondel:2019qlh,Heinemeyer:2021rgq,deBlas:2944678}.

At hadron colliders, the relatively large value of the QCD coupling compared to the EW or QED couplings means that NLO EW corrections are typically comparable in size to next-to-next-to-leading order (NNLO) QCD corrections. This justifies the primary focus on QCD corrections in theoretical predictions for hadron collider physics. In contrast, at lepton colliders, typical production channels are electroweak in nature, and as a result, EW and QED corrections at NLO or --- in some cases --- NNLO become the dominant sources of theoretical uncertainty.

\subsection{Beam structure functions}
As mentioned in Section \ref{sec:general_concept_MC}, beam particles are not always point-like. It is a well-known fact that protons consist of quarks and gluons, which are the partons that participate in the hard-scattering process. Parton Distribution Functions (PDFs) describe how these partons contribute to the momentum of the beam particles~\cite{Feynman:1969ej}. Similar language can, however, also be applied to leptons, which at high energies resemble a collection of particles arising from QED/electroweak emissions~\cite{Kuraev:1985hb,Skrzypek:1990qs,Cacciari:1992pz,Frixione:2019lga,Bertone:2019hks,Frixione:2012wtz,Han:2020uid}. Beam--beam interactions further smear the energy distribution of the beam particles~\cite{Bonvicini:1989ci}. All these effects can be described within the broader technical concept of beam (structure) functions, which allow for the incorporation of arbitrary distortions to the simple picture of a point-like beam particle with a well-defined energy.

A simple example of a structure function is PDFs, which are widely used in LHC physics, serving as an interface between the beam protons and their constituents participating in hard scattering. They are extracted from the QCD analysis of the data coming from deep-inelastic lepton--nucleon scattering experiments, mainly at the HERA electron--proton collider, but also from fixed target experiments or from proton--proton collisions. There are several different collaborations, each performing independent fits. PDF fits from different groups are available in a standardised manner via the \textsc{LHAPDF} library~\cite{Buckley:2014ana}. The generator typically calls the PDFs as a function of the parton flavour, the energy fraction of the incoming beam particle, the factorisation scale, and some other parameters, for example, the QCD order of the fit. In the so-called \textit{five-flavour scheme}, one hadron includes 11 components: a gluon, five quarks, and five anti-quarks, with several PDF sets additionally including a photon component from coupled QED and QCD DGLAP evolution~\cite{Manohar:2017eqh}.

At high-energy lepton colliders, beam particles also cannot be considered point-like due to multiple photon (and weak boson) emissions. Even though the coupling constant is small, the emission probability is enhanced by the large ratio between the energy scale and the lepton mass. Similar to the QCD case, resummation --- which can be done (semi-) analytically due to the well-defined initial conditions --- is necessary. The resulting object gives the probability of finding a lepton, an anti-lepton, or a photon inside the incoming beam object. A generalised approach, in which more partons are considered, is one of the topics of this thesis and will be described in greater detail in Chapter~\ref{chap:EWPDF}.

Lepton colliders exhibit another phenomenon important for the beam structure: beamstrahlung, which is classical radiation emitted by a beam due to the electromagnetic field of the opposing beam. The effect is especially pronounced at linear colliders, where the beams interact over much longer time scales. In addition, the bunch charge densities at linear colliders are much higher, since the beam bunches are compressed to significantly smaller transverse sizes at the interaction point. In the simplest case, the effect can be approximated by applying a Gaussian spread to the beam energy; however, more sophisticated methods involving advanced parametrisations or dedicated Monte Carlo tools exist.

\subsection{Parton showers}
Typical events at high-energy hadron colliders contain tens of different final-state particles. Even though events at lepton colliders have lower multiplicities due to the simpler initial state, multiple particles are still produced in final-state QCD interactions and particle decays. Generating all of these particles using the standard matrix-element approach would not be feasible --- or at least not efficient --- due to arising singularities, the high dimensionality of the phase-space integrals, and complications in the automated generation of matrix elements themselves~\cite{Reuter:2025imz}. This gives rise to the concept of parton-shower algorithms, which approximate the production of multiple particles as a chain of single splittings.

In QCD, the emission rate for a branching such as $q \to qg$ diverges when the gluon is emitted at a small angle (collinear limit) or when the gluon energy vanishes (soft limit)~\cite{Sjostrand:2009ad}. It is similar to the QED splitting $f \to f \gamma $, with differences in the behaviour of the couplings. Furthermore, the non-Abelian structure of QCD allows for $ g \to gg $ splittings, which are absent in QED. The splittings $g \to q\bar{q}$ and $\gamma \to f\bar{f}$ are less important, as they do not have soft divergences. 

Parton showers allow one to iteratively build a cascade of simple expressions for single splittings to construct a multi-parton final state, as well as to introduce a Sudakov term, maintaining the unitarity of the shower by resumming virtual and unresolved real emissions in perturbative QCD. The core idea is to factorise a complex $2 \to n$ process into a simple hard-scattering process (e.g. $2 \to 2$) convoluted with splitting functions. In the soft and collinear limits, not only do amplitudes factorise, but also the phase space~\cite{Hoche:2014rga}. The subsequent splittings are combined one by one, and a whole shower develops. The incoming and outgoing particles must be on-shell, but closer to the hard interaction --- or at shorter timescales --- the partons may be more off-shell, as follows from the uncertainty principle. Final-state radiation is typically generated in an evolution variable leading to progressively softer and more collinear emissions. On the other hand, initial-state radiation is usually constructed via backward evolution, which corresponds to emissions that become harder as the hard scattering is approached~\cite{Buckley:2011ms}. In general, the hard process must be characterised by the largest scale, for example, in terms of its virtuality $Q^2$, which helps avoid potential double-counting.

Nevertheless, this picture does not eliminate the singularities. They are an inherent feature of QCD, resulting from the fact that at small $Q^2$, the strong coupling constant exceeds values for which perturbative expansion remains valid. To avoid this, a lower cut-off, typically at around 1\,GeV, is needed to regulate divergences. In this way, the singularities are removed, but the branching probabilities can still exceed unity. To mitigate this issue, the Sudakov factor is introduced. It corresponds to the probability of no emissions occurring at a given \textit{(shower) time}. The time might be related to the parton virtuality, but there are also implementations with showers ordered in emission angles or transverse momenta. More deeply, the Sudakov factor approximates the complete virtual corrections from loop diagrams, in the same way that parton showers can be seen as approximations to the full matrix elements for real emissions.

Most current work focuses on QCD showers, which are crucially important for LHC physics. This involves not only improving the precision of showers themselves, for instance, through better tuning to data~\cite{Skands:2014pea,Fieg:2023kld}, but also matching showers to higher-order perturbative corrections~\cite{Nason:2004rx,Frixione:2007vw}. In the latter case, avoiding double-counting is one of the main challenges. At next-generation lepton colliders, QED showers are becoming increasingly important, and work in this direction will be necessary~\cite{Heinemeyer:2021rgq}. For multi-TeV colliders, whether hadron or lepton machines, the effects of weak radiation must also be included~\cite{Kleiss:1986xp,Brooks:2021kji}. These issues are discussed further in the thesis in the context of photon and weak-boson radiation.

\subsection{Hadronisation and decays}
At low energy scales, QCD becomes strongly coupled, causing quarks, except for the top, and gluons to bind together into composite states known as hadrons, which are observed experimentally. Because hadronisation is an inherently non-perturbative process, it cannot be described from first principles, and phenomenological models must be employed. Two such models --- string fragmentation~\cite{Andersson:1983ia,Sjostrand:1984ic} and cluster fragmentation~\cite{Webber:1983if} --- were developed in the 1970s--1980s and, in various forms, have remained essential components of all modern MC event generators. The Lund string model was implemented in \textsc{Jetset}~\cite{Sjostrand:1982fn}, which was later integrated into \textsc{Pythia}. The cluster fragmentation was adopted as part of \textsc{Herwig}.

Hadronisation models are generally motivated heuristically, starting from simple, semi-classical pictures. Despite their empirical basis, they contain multiple free parameters that must be fitted to experimental data. While this might seem unsatisfactory, it is not surprising considering that there are hundreds of known hadrons, each characterised by its mass, width, couplings, and decay patterns~\cite{Sjostrand:2009ad}. The origins of these parameters remain poorly understood; lattice QCD offers some insight~\cite{Wilson:1974sk,FlavourLatticeAveragingGroupFLAG:2024oxs}, but the understanding is still limited.

The string model is based on the assumption of linear confinement. A quark--antiquark pair is connected by a colour flux tube of typical hadronic size~\cite{Sjostrand:2009ad}. As the quark and antiquark move apart, the potential energy stored in the string increases. If the energy becomes sufficient to produce a new quark--antiquark pair, the system may split into two colour singlets, and the process can repeat as long as each resulting subsystem has enough invariant mass to produce new objects. In the Lund string model, this continues until only on-shell hadrons remain. While this mechanism becomes more complex for systems involving gluons or when modelling baryon production, the model achieves strong predictive power once its parameters are fitted to data.

Cluster fragmentation takes a different approach. At the end of the parton shower, all gluons are split into quark--antiquark pairs. By controlling the colour flow in the event, colour singlets can be formed from these quark--antiquark combinations. Typically, the quark and antiquark with complementary colours are created in adjacent shower branches and are close in angle, resulting in small invariant masses. These pairs serve as the building blocks for hadrons. Unlike the string model, no additional internal structure is needed, and the model often allows for a compact description with relatively few parameters, such as masses and flavour content. However, to accurately reproduce event kinematics, more sophisticated techniques must be employed.

Hadronisation models ultimately produce a collection of hadronic resonances. Except for the proton, the hadrons are unstable on macroscopic timescales, so MC generators must also simulate their decays. The decay patterns measured experimentally must be implemented in any MC framework, and given the constant progress in hadron spectroscopy, the databases must be kept up to date.

\section{Pythia}
\textsc{Pythia} is one of the most widely used general-purpose Monte Carlo event generators. It incorporates hard-scattering matrix elements, PDFs, initial- and final-state parton showers, multiparton interactions, hadronisation, and particle decays. While some aspects of its physics are firmly based on theoretical derivations, others rely on phenomenological models with parameters tuned to experimental data.

The software’s origins date back to 1978 with the first release of \textsc{Jetset}~\cite{Sjostrand:1982fn, Sjostrand:1982am, Sjostrand:1985ys, Sjostrand:1986hx}, a program implementing the Lund string model for jet fragmentation. Many components of \textsc{Jetset} were later integrated into \textsc{Pythia}~\cite{Bengtsson:1982jr, Bengtsson:1984yx, Bengtsson:1987kr, Sjostrand:1993yb, Sjostrand:2000wi, Sjostrand:2006za, Sjostrand:2007gs, Sjostrand:2014zea, Bierlich:2022pfr}, making the current version the result of nearly 50 years of development. Although the current major version \textsc{Pythia 8} has been public for over a decade, the legacy version \textsc{Pythia 6} is still partially in use.

Initially, \textsc{Pythia} was written in Fortran 77. The first complete rewrite in C++ was released as \textsc{Pythia 8.100}~\cite{Sjostrand:2007gs}. Although \textsc{Pythia 8} introduced a number of conceptual and technical improvements, some features present in the final Fortran version, \textsc{Pythia 6.4}, were initially absent or less mature. Over time, these aspects have been progressively implemented and refined in \textsc{Pythia 8}. For specific applications --- such as studies involving lepton colliders, which are less central to the \textsc{Pythia} collaboration’s focus --- both versions are often used and cross-checked~\cite{Zhao:2023qaz}.

\textsc{Pythia} is designed to be self-contained and suitable for a wide range of standalone physics analyses. While the number of implemented scattering processes is limited, \textsc{Pythia} can be easily interfaced with other software packages. A prominent example is the Les Houches Accord~\cite{Boos:2001cv} and its associated Les Houches Event Files~\cite{Alwall:2006yp, Andersen:2014efa}, which enable matrix element calculations from external providers to be combined with \textsc{Pythia}’s parton showers and hadronisation.

\textsc{Pythia} supports a wide range of initial-state configurations, including hadron--hadron, lepton--lepton, and photon--photon collisions, as well as mixed combinations. Several PDFs are implemented internally for protons, pions, and leptons. A broader selection of PDFs can be accessed via interfaces to the \textsc{LHAPDF} library~\cite{Whalley:2005nh, Buckley:2014ana}. On the other hand, \textsc{Pythia~8} does not support beam polarisation or beam spectra beyond a simple Gaussian spread, which are important features for linear \epem colliders. Detector simulation is also not included, but thanks to the \textsc{HepMC} interface~\cite{Dobbs:2001ck}, it can be handled by external software tools.

The implementation of hard processes in \textsc{Pythia~8} primarily focuses on $2 \to 1$ and $2 \to 2$ interactions, with some support for $2 \to 3$ processes. For instance, $2 \to 3$ QCD processes are available and useful for validating parton shower algorithms. The generator also includes a wide range of electroweak and top-quark production processes, as well as Higgs production, including scenarios within a generic two-Higgs-doublet model~\cite{Branco:2011iw}. BSM physics is also supported. Users can simulate events involving supersymmetric particles, new gauge bosons, left--right symmetric models, leptoquark production, and extra-dimensional states. For processes unsupported internally, parton-level events can be generated using external tools like \textsc{MadGraph~5} or \textsc{Whizard} and passed to \textsc{Pythia} for showers and hadronisation.

Given its historical origins, parton showers play a central role in \textsc{Pythia}. The current implementations of initial- and final-state radiation are based on dipole-based, $p_\perp$-ordered evolution, introduced in \textsc{Pythia~6.3}~\cite{Sjostrand:2004ef}. \textsc{Pythia~8} also includes photon splitting into quark or lepton pairs, as well as weak boson emissions~\cite{Christiansen:2014kba}. Hadronisation is another distinct feature of \textsc{Pythia}, stemming from its origin in the Lund string fragmentation model. This model remains a core part of the code and is also used by other generators, such as \textsc{Whizard}, when interfaced to \textsc{Pythia}. Additionally, the treatment of particle decays is relatively sophisticated --- for example, $\tau$ decays are simulated with full spin correlations~\cite{Ilten:2012zb}. \textsc{Pythia} also includes a model of multi-parton interactions, which naturally arise in hadron--hadron collisions due to the composite structure of the incoming particles. This feature, however, has limited relevance in lepton--lepton collisions.

\section{Whizard}
\label{sec:Whizard}
\textsc{Whizard}~\cite{Kilian:2007gr} is a comprehensive software package for computing multi-particle scattering cross sections and generating event samples, written using the modern Fortran 2008 standard. It supports NLO QCD and EW corrections in the SM for both lepton and hadron colliders. Tree-level matrix elements for arbitrary partonic processes are automatically generated using the built-in Optimized Matrix Element Generator O'Mega~\cite{Moretti:2001zz}. External matrix element providers, including those incorporating loop corrections, such as Recola~\cite{Actis:2012qn,Actis:2016mpe}, OpenLoops~\cite{Cascioli:2011va,Buccioni:2019sur}, and GoSam~\cite{GoSam:2014iqq,Braun:2025afl}, are also interfaced. The program reliably computes cross sections and can efficiently generate unweighted event samples for processes of various complexity. The events can be exported in standard formats, including ASCII, \textsc{StdHEP}, \textsc{LHEF}, and \textsc{HepMC}, making them usable for subsequent hadronisation and detector simulation in external codes.

A distinct feature of \textsc{Whizard} is its dedicated command language, SINDARIN (Scripting INtegration, Data Analysis, Results display and INterfaces). Although it might appear as a complication, the language offers a wide range of configurable options, giving users substantial flexibility in controlling the simulations. The language is thoroughly documented in the official \textsc{Whizard} manual~\cite{whizard-manual}.

\textsc{Whizard} was originally developed around 1999 for the then-planned linear electron-positron collider TESLA~\cite{ECFADESYLCPhysicsWorkingGroup:2001igx}, with the goal of providing a tool capable of precisely describing electroweak processes involving multiple weak and Higgs bosons using full matrix elements. The acronym ``WHiZard'' stood for W, Higgs, Z, And Respective Decays. Several components of the package had been published before the first official version of the code was released. VAMP, a multi-channel adaptive Monte Carlo integration package, dates back to 1998~\cite{Ohl:1998jn}, and \textsc{CIRCE}, a package for the simulation of lepton collider beamstrahlung, to 1996~\cite{Ohl:1996fi}. The first official version of \textsc{Whizard} appeared in 2000~\cite{Ohl:2000pr}.

\textsc{Whizard 1} was developed for a decade, during which several major steps occurred. Key milestones included full support for a wide range of BSM scenarios, including particles with spin 3/2 and spin 2, for example, the Minimal Supersymmetric SM, the Next-to-Minimal Supersymmetric SM, the Little Higgs, models with anomalous couplings, and extra dimensions. Initially, various methods were used for matrix element generation until the in-house generator O'Mega became available with Version 1.20. Version 1.40 added essential configurations to support a variety of collider setups, including asymmetric beams or intrinsic transverse momentum in structure functions. Support for full-colour amplitudes was introduced in Version 1.50~\cite{Kilian:2012pz}. 

In 2010, Version 2.0 was released as an almost complete rewrite of the original code. This version was built around the internal density-matrix formalism, which enables the factorisation of production and decay processes while fully preserving colour and spin correlations. This native, distinctive feature of \textsc{Whizard} also supports arbitrary beam polarisation states, a crucial aspect for electroweak precision studies. In this release series, \textsc{LHAPDF 6}~\cite{Buckley:2014ana} was also integrated.

The next major version, \textsc{Whizard 3.0}, was released in 2021, introducing full automation of NLO QCD calculations and support for Majorana fermions in UFO models~\cite{Stienemeier:2021cse}. This release also included NLO EW and mixed NLO EW/QCD corrections. Development is ongoing, with the latest version incorporating features such as Equivalent Vector Boson Approximation (EVA) and electroweak PDFs, as discussed in this thesis. As of writing of this thesis, the current version is \textsc{Whizard 3.1.7}.

From the perspective of lepton colliders, several \textsc{Whizard} features related to beam structure are particularly important. One key effect is beamstrahlung, which occurs when charged particles emit radiation as they are accelerated or deflected by the strong electromagnetic fields of the opposing beam during collisions. This phenomenon is especially significant at linear colliders, where extremely dense particle bunches collide, leading to energy loss and beam energy spread. As a result, the initial collision energy becomes less sharply defined. In \textsc{Whizard}, beamstrahlung effects are modelled using the \textsc{CIRCE1} and \textsc{CIRCE2} packages, which fit energy spectra generated by, for example, \textsc{GuineaPig}~\cite{Schulte:1999tx,Rimbault:2007wfy}. Additionally, a parametric Gaussian option is available to model beam energy spread. As mentioned earlier, \textsc{Whizard} allows full flexibility in selecting arbitrary beam polarisation states. It also supports asymmetric beam configurations and crossing angles.

Thanks to these capabilities, \textsc{Whizard}  has been widely used in design studies and physics simulations for future lepton colliders such as the ILC, CLIC, CEPC, FCC-ee, and the Muon Collider~\cite{Linssen:2012hp, ILC:2013jhg, CEPCStudyGroup:2018ghi, FCC:2018evy, InternationalMuonCollider:2025sys}.

\chapter{$Z$ boson couplings to quarks}
\label{chap:Zcouplings}

As explained in Chapter~\ref{chap:futurephysics}, future colliders operating at the $Z$ pole will collect unprecedented statistics of $Z$ bosons in a clean collision environment, allowing for exact measurements of the SM parameters, including the measurement of the electroweak couplings to quarks. The most stringent constraints in this domain still come from the LEP experiments. According to the recent Particle Data Group Review~\cite{ParticleDataGroup:2024cfk}, the partial width of the $Z$ boson to hadrons is known with the precision of 0.11\%. The partial width to $b$~quarks, $\Gamma_{b\bar{b}}$, is measured at the level of 0.34\%, while the SM value is calculated at a precision of 0.05\%. The corresponding value for $c$ quarks, $\Gamma_{c\bar{c}}$, is measured up to 1.8\%, and the theoretical predictions are estimated with a precision of 0.07\%. Remarkably, although the SM calculations estimate $\Gamma_{u\bar{u}}$, $\Gamma_{d\bar{d}}$ and $\Gamma_{s\bar{s}}$ with a sub-permille precision, no experimental results are quoted. It highlights the need to develop methods for measuring these parameters at future experiments. Measurements of these couplings are important tests on both the chirality and the flavour structure of the SM.

Future experiments will benefit from vastly larger data samples than those achieved by LEP: depending on the collider type, $10^9$--$10^{12}$ $Z$ bosons will be available for physics studies. Improvements in detector technologies and heavy-quark tagging algorithms (see e.g. \cite{Bilokin:2017lco, Fuster:2021ekh, Irles:2023ojs,Tagami:2024gtc,Liang:2023wpt,Zhu:2023xpk,Blekman:2024wyf,Albert:2022mpk}) are expected to significantly enhance the experimental sensitivity and deliver more stringent constraints, even with the same analysis methods. Nevertheless, due to the lack of efficient light-quark tagging algorithms, a dedicated approach is required to measure the light-quark couplings to the $Z$ boson. The approach may rely on radiative processes. At high energies, photon emission becomes abundant, and differences in the radiation probabilities of particles with different quantum numbers can be used to distinguish among them. This chapter presents an analysis of a method applicable at the $Z$-pole run of future \epem facilities for measuring the light quark couplings to the $Z$ boson~\cite{Mekala:2025rlk}.

\section{Outline of the $Z$-pole measurement}

The measurement can be taken by simultaneously examining both radiative and inclusive hadronic decays of the $Z$ bosons. The probability of photon emission is related to the electric charge of a particle; as a consequence, up-type quarks emit more photons than down-type quarks, because their electric charge is twice as large. At the same time, down-type quarks dominate the non-radiative sample due to the larger value of their SM electroweak coupling. These facts can be leveraged to fit the cross sections in the two categories of radiative and inclusive decays and therefore disentangle the electroweak couplings of different quarks.

The idea of the measurement originates from LEP~\cite{DELPHI:1991utk, L3:1992kcg, L3:1992ukp, DELPHI:1995okj, OPAL:2003gdu}. The best measurements yielded the precision of 3\% for $d$-type and 6\% for $u$-type quarks. However, given the limited tagging capabilities, the LEP experiments were unable to treat all quark flavours separately and only delivered results for the two isospin states. Prospects for the efficient $s$-quark tagging~\cite{Bedeschi:2022rnj, Liang:2023wpt, Tagami:2024gtc, Ntounis:2025itg} at future facilities make the full flavour decomposition possible. The goal of this study is to propose a method to disentangle the couplings of the five kinematically accessible quark flavours and estimate the level of systematic uncertainties which have to be achieved to benefit from the increased data statistics of future colliders. From this perspective, the analysis might serve as the most precise test of flavour universality in the electroweak sector of quarks.

The coupling strength of the $Z$ boson to a given fermion $f$ is conventionally defined as
\begin{equation*}
    c_f = 4\left((g^f_V)^2 + (g^f_A)^2\right),
\end{equation*}
where $g^f_V$ ($g^f_A$) is the vector (axial) coupling. As mentioned in Chapter~\ref{chap:SM}, the couplings are expressed in terms of the third component of the fermion weak isospin ($T^f_3$) and its charge ($Q_f$):
\begin{equation*}
    g^f_V = T^f_3-2Q_f \sin^2{\theta_W}, \quad \quad g^f_A = T^f_3,
\end{equation*}
where $\theta_W$ is the weak mixing angle.
The total width of the $Z$ boson to hadrons, $\Gamma_{had}$, is given at fixed order in $\alpha_s$ by~\cite{L3:1992kcg,OPAL:2003gdu,ParticleDataGroup:2024cfk}
\begin{equation}
    \Gamma_{had} = N_c \frac{G_\mu M^3_Z}{24\pi \sqrt{2}}\left(1 + \frac{\alpha_s}{\pi}+ \mathcal{O}(\alpha_s^2)\right) \sum_{q = u,d,s,c,b} c_q\;,
    \label{eq:hadronic_Z}
\end{equation}
where $N_c$ is the number of colours, $G_\mu$ is the Fermi constant, $M_Z$ is the mass of the $Z$ boson and $c_q$ is the coupling to a given quark. For simplicity, it will be assumed throughout the text that the sum always runs over the five quark flavours. 

The total width for radiative hadronic decays, $\Gamma_{had+\gamma}$, can be expressed at leading order in QED as
\begin{equation}
    \Gamma_{had+\gamma} = N_c \frac{G_\mu M^3_Z}{24\pi \sqrt{2}} f(y_{cut})\frac{\alpha}{2\pi} \sum_{q} c_q Q_q^2\;,
    \label{eq:radiative_Z}
\end{equation}
where $f(y_{cut})$ is an acceptance factor depending on a parameter $y_{cut}$ incorporating the isolation criteria for photons, $\alpha$ is the electromagnetic coupling constant and $Q_q$ is the electric charge of the given quark, respectively. Radiative cross sections exhibit divergences arising from soft and collinear photon emissions. In practice, this does not pose a problem, as detector resolution is finite. From a theoretical perspective, infrared safety is restored by introducing appropriate photon isolation criteria. The function $f(y_{cut})$ parametrises the acceptance associated with the applied cut. The choice of the parameter $y_{cut}$ will be discussed in Section~\ref{sec:analysis}.

Since the electric charges of $u$-type and $d$-type quarks are different, the expressions for the total hadronic width (Eq. \ref{eq:hadronic_Z}) and the radiative partial width (Eq. \ref{eq:radiative_Z}) include different combinations of the quark couplings $c_q$. In this way, by solving the set of equations for total and radiative events, the dependence on the couplings can potentially be resolved.

Similarly, the hadronic cross section at the $Z$ pole, $e^+e^- \to Z \to q\bar{q}$, $q = u, d, s, c, b$, can be expressed as
\begin{equation}
    \sigma_{had} = \sum_{q} \sigma_q  = \mathcal{C}_1 \cdot \sum_{q} c_q, 
    \label{eq:cross_hadr}
\end{equation}
where $\mathcal{C}_1$ is a constant which can be derived with high accuracy from theoretical calculations at a given order. Here, it is assumed that the contribution from the $\gamma$ exchange is negligible at the $Z$ pole.

The radiative hadronic cross section with exactly one photon identified in the final state (the one-photon cross section) can be parametrised as
\begin{equation}
       \sigma_{had+\gamma} = \mathcal{C}_2 \cdot \sum_{q} c_q Q_q^2,
       \label{eq:cross_rad}
\end{equation}
where $\mathcal{C}_2$ is another constant, which can be derived from theoretical calculations and simulations for a given set of cuts and isolation criteria for photons. This parametrisation assumes that hard initial-state radiation is strongly suppressed at the $Z$ pole.

\section{Statistical framework}
\label{sec:statframe}

In the following, the details of the statistical framework for fitting data for different flavours are presented in two cases: first, when only statistical uncertainties are included, and second, when both statistical and systematic errors are taken into account.

\subsection{Fit with statistical uncertainties only}
Formally, the cross section for each quark flavour can be presented as a sum of processes with and without hard photon radiation  in the final state (a zero-photon exclusive sample and a one-photon inclusive sample):
\begin{equation*}
    \sigma_q  =   \sigma_{0q}  + \sigma_{\gamma q}\,.
\end{equation*}
The separation between the two samples depends on the cuts defined for the photon isolation.
The resulting fraction of hard \textit{radiative} events is given by
\begin{equation*}
    f_{q} = \frac{\sigma_{\gamma q}}{\sigma_q}\,.
\end{equation*}

The definition of a radiative event is directly related to photon reconstruction criteria. Thus, a set of cuts $y$ (including transverse momentum, isolation angle, etc.)  must be introduced to describe hard photons reconstructable in the detector. 
The measured radiative cross section depends on the factors $F_y \equiv F(y)$ and $G_y \equiv G(y)$ (assumed to be
quark-flavour independent for simplicity) which characterise the experimental acceptance of, respectively, radiative and non-radiative events\footnote{In this context, a subtle difference between the two is worth noting: $F_y$ describes how many radiative events were correctly classified into the radiative sample (``acceptance'') and $G_y$ how many non-radiative events were incorrectly classified into this sample (``mis-acceptance'').}; for the latter, measurable photons come mostly from hadronisation and decays of neutral hadrons. The cross section can be written as
\begin{equation*}
    \sigma^{(y)}_{\gamma q} = \sum_{q} 
  F_y \cdot \sigma_{\gamma q} + G_y \cdot \sigma_{0 q}  = \sum_{q} \left( f_q F_y + (1-f_q) G_y \right) \cdot \sigma_q\,.
\end{equation*}
Here, $\sigma^{(y)}_{\gamma q}$ denotes the experimentally accessible cross section, while $\sigma_{\gamma q}$, $\sigma_{0q}$, and $\sigma_{q}$ are theoretically computed cross sections, practically obtained from Monte Carlo generators.

Experimentally, based on dedicated flavour-tagging algorithms, one can classify ha\-dronic 2-jet-like events into $N$ flavour-based categories. In the simplest case, same-flavour quark pair production can be assumed, resulting in, for example, $N=3$ when differentiating between $b\bar{b}$,
$c\bar{c}$ and lighter-quark final states only, or $N=5$ in the case of complete flavour decomposition ($b\bar{b}$, $c\bar{c}$, $s\bar{s}$, $u\bar{u}$, $d\bar{d}$). On the other hand, one can also allow for more freedom and classify events based on single jet-tagging results (which can differ between the two jets in a single event), obtaining more possible event categories (up to $N=15$ for full flavour decomposition, assuming that the difference between jets formed from quarks and antiquarks is neglected). The main advantage of the single-jet classification is that systematic uncertainties can be better constrained from data. In this study, flavour-tagging with four single-jet categories (``light jet'', ``$s$ jet'', ``$c$ jet'', ``$b$ jet'') will be assumed.

Let $j$ be an index referring to a given double-jet category of an event, $j=1,\dots, N$. For example, $j=1$ might correspond to an event with two light jets, $j=2$ to an event with one light jet and one $s$ jet, etc. The total number of hadronic events expected in each category, after the final selection and classification, can be written as:
\begin{equation}
\label{eq:Nj}
    N_j = \sum_q {\cal E}_q \cdot M_{q\,j} \cdot {\cal L}_{int} \cdot \sigma_q\,,
\end{equation}
where ${\cal E}_q$ is the event selection efficiency for flavour $q$, with results obtained from simulation, ${\cal L}_{int}$ is the total integrated luminosity and 
$M_{q\,j}$ gives the probability of a $q\bar{q}$ event being classified as category $j$. The values for the efficiency ${\cal E}_q$ resulting from the simulation studies (as described later in the text) are given in Table \ref{tab:eff}.

A similar equation can be set for radiative events:
\begin{equation}
\label{eq:NAj}
    N^{(y)}_{\gamma j} = \sum_q \Big(
 {\cal E}_{\gamma q} \cdot  M_{\gamma q\,j} \cdot f_q \cdot F_y  + 
 {\cal E}_{0 q} \cdot  M_{0 q\,j} \cdot (1-f_q) \cdot G_y 
 \Big) \cdot {\cal L}_{int} \cdot \sigma_q\,,
\end{equation}
where ${\cal E}_{\gamma q}$ and ${\cal E}_{0 q}$ are the selection efficiencies for radiative events and non-radiative background, i.e. ``fake'' radiative or category migration events (excluding the photon selection cut efficiencies incorporated in $F_y$ and
$G_y$; results from this study are given in Table \ref{tab:eff}). $M_{\gamma q\,j}$ and $M_{0 q\,j}$ are the jet-flavour classification matrices for radiative and non-radiative events. In the following, it is assumed that $M_{q\,j} = M_{\gamma q\,j} = M_{0 q\,j}$. The single-jet tagging-efficiency matrix adapted from \cite{Liang:2023wpt} reads\footnote{The values for the $u$ and $d$ tagging were averaged, as explained later.}:
\begin{equation}
\begin{bmatrix}
0.773 & 0.190 & 0.031 & 0.006\\
0.773 & 0.190 & 0.031 & 0.006\\
0.311 & 0.645 & 0.038 & 0.006\\
0.107 & 0.068 & 0.795 & 0.030\\
0.026 & 0.007 & 0.059 & 0.908
\end{bmatrix}.
\label{eq:matrix}
\end{equation}
The five rows of the matrix correspond to five quark flavours ($d$, $u$, $s$, $c$, $b$), and the four columns to four available single-jet classification labels (``light jet'', ``$s$ jet'', ``$c$ jet'', ``$b$ jet''). In this study, the elements of $M_{q\,j}$ are obtained from this matrix as the product of the corresponding single-jet classification efficiencies. Differences between tagging for quarks and antiquarks are neglected, as well as contributions from possible gluon jets. Since all tagging probabilities are taken into account in the analysis, i.e. contributions in each row sum up to one, the exact values of the entries are not critically important. However, their associated uncertainties must be carefully estimated, as they significantly influence the migration between classification categories. This issue is discussed in more detail in Section~\ref{sec:sys_unc}.

\begin{table}[]
\centering
\begin{tabular}{c|c|c|c|c|c}
$q$           & $d$     & $u$     & $s$     & $c$     & $b$     \\ \hline
${\cal E}_{q}$        & 0.955 & 0.955 & 0.947 & 0.949 & 0.936 \\ \hline
${\cal E}_{0 q}$     & 0.902 & 0.903 & 0.852 & 0.892 & 0.870 \\ \hline
${\cal E}_{\gamma q}$ & 0.917 & 0.922 & 0.876 & 0.914 & 0.888
\end{tabular}
\caption{Event selection efficiencies obtained from the simulation. See text for details.}
\label{tab:eff}
\end{table}

To simplify the formulae, two matrices are defined:
\begin{equation*}
    A_{q\,j} \equiv  {\cal E}_q \cdot M_{q\,j} \cdot {\cal L}_{int}\,,
\end{equation*}
\begin{equation*}
    B_{q\,j} \equiv  {\cal E}_{\gamma q} \cdot  M_{q\,j} \cdot {\cal L}_{int}
  \cdot f_q \cdot F_y   + 
        {\cal E}_{0 q} \cdot  M_{q\,j} \cdot {\cal L}_{int}
        \cdot (1-f_q) \cdot G_y
\end{equation*}
and Eqs. \ref{eq:Nj} and \ref{eq:NAj} are rewritten:
\begin{equation}
    N_{j} = \sum_q A_{q\,j}\cdot \sigma_q\,, 
    \label{eq:nj}
\end{equation}
\begin{equation}
    N_{\gamma j} =  \sum_q B_{q\,j}\cdot \sigma_q\,.  \label{eq:ngj}
\end{equation}
Now, the dependence on $y$ is implicitly assumed and hence the superscript in $N^{(y)}_{\gamma j}$ has been dropped.

Future experiments will measure event numbers in $2N$ categories (total hadronic and radiative), and the goal is to extract $M=5$ quark-level cross
sections which are directly related to the quark couplings (Eqs. \ref{eq:cross_hadr} and \ref{eq:cross_rad}). Assuming the number of events is large for each category $j$, one can consider the $\chi^2$ function: 
\begin{equation*}
    \chi^2 = \sum_j \frac{(n_j - N_j)^2}{N_j} + 
   \sum_j \frac{(n_{\gamma j} - N_{\gamma j})^2}{N_{\gamma j}}\,,
\end{equation*}
where $n_j$ and $n_{\gamma j}$ are the numbers of hadronic and radiative events, respectively, observed in given categories. 

Following the maximum likelihood principle, partial derivatives with respect to $\sigma_q$ are calculated:
\begin{equation*}
\frac{\partial \chi^2}{\partial \sigma_i}
\approx -2 \sum_j \frac{(n_j - N_j)}{N_j}\, A_{ij} -2 \sum_j \frac{(n_{\gamma j} - N_{\gamma j})}{N_{\gamma j}}\, B_{ij}\,,
\end{equation*}
where higher powers of $N_j$ and $N_{\gamma j}$ have been neglected.
For the extremum, this expression evaluates to zero, giving
\begin{equation*}
\sum_j A_{ij}\left( \frac{n_j}{N_j} - \frac{N_j}{N_j} \right) + \sum_j B_{ij}\left( \frac{n_{\gamma j}}{N_{\gamma j}} - \frac{N_{\gamma j}}{N_{\gamma j}} \right) = 0\,.
\end{equation*}

To linearise the equation, $N_j$ and $N_{\gamma j}$ in the numerators can be expanded using Eqs.~\ref{eq:nj} and \ref{eq:ngj} while keeping the occurrences in the denominators unchanged. In this way, one obtains a system of $M$ equations for cross-section values, which can be represented in a matrix form:
\begin{equation*}
     \mathbb{H} \cdot \vec{\sigma}  =   \mathbb{V}\,,
\end{equation*}
or alternatively ($i = 1, \ldots, M$):
\begin{equation*}
     \sum_q \mathbb{H}_{iq} \cdot \sigma_q  =   \mathbb{V}_{i}\,,
\end{equation*}
where $\mathbb{H}$ is a Hessian matrix and $\mathbb{V}$ is a Hessian vector, defined as:
\begin{eqnarray*}
     \mathbb{H}_{iq} &=&  \frac{1}{2}
  \frac{\partial^2\chi^2}{\partial \sigma_i \partial \sigma_q} 
   =   \sum_j \frac{A_{ij} A_{qj}}{N_j} + 
  \sum_j \frac{B_{ij} B_{qj}}{N_{\gamma j}}\,, \\
  \mathbb{V}_i &=&  \sum_j \frac{n_j}{N_j} \, A_{ij} + \sum_j  \frac{n_{\gamma j}}{N_{\gamma j}} \, B_{ij} \,.
\end{eqnarray*}

The solution can be found by inverting the matrix $\mathbb{H}$:
\begin{equation*}
     \vec{\sigma} = \mathbb{H}^{-1} \cdot \mathbb{V}\,.
\end{equation*}
While more numerically stable methods exist for solving such equations, the inverse of the Hessian matrix directly provides the covariance matrix of the extracted cross sections. In particular, cross-section uncertainties are given by the square root of the diagonal elements of the inverse matrix:
\begin{equation*}
     \Delta \sigma_i = \sqrt{ \left(\mathbb{H}^{-1}\right)_{ii}}\,.
\end{equation*}
Since the cross section depends on the quark coupling linearly, and the proportionality constant can be precisely determined from theoretical calculations, the uncertainty on the coupling is assumed to be the same as the uncertainty of the cross-section measurement, $\Delta \sigma_i \equiv \sigma_C$.

\subsection{Fit with systematic uncertainties}
\label{s:systematic_uncertainties}
The matrices $A_{q\,j}$ and $B_{q\,j}$ can be obtained from simulations of physics processes and detector performance with negligible statistical uncertainty. However, they are influenced by multiple systematic uncertainties which must be included in the analysis for reliable fitting.
Assuming $K$ independent systematic uncertainties described by variations $\delta_k$, the $\chi^2$ formula can be extended to include the dependence on the systematic variations: 
\begin{equation*}
     \chi^2(\vec{\delta}) =
  \sum_j \frac{(n_j - N_j(\vec{\delta}))^2}{N_j}  +  \sum_j \frac{(n_{\gamma j} -N_{\gamma j}(\vec{\delta}))^2}{N_{\gamma j}} + \sum_k \delta_k^2
     + \sum_q 2 \lambda_q \; w_q(\vec{\delta})\,,
     \label{eq:chi2full}
\end{equation*}
where the third term corresponds to the assumed normal distributions for the systematic variations $\delta_k$ and the last term includes Lagrange multipliers $\lambda_q$ introduced to enforce proper normalisation of the tagging probabilities, as the sum of their variations, $w_q$, has to be zero for each quark flavour $q$.  The dependence of the numbers of expected hadronic and radiative events, $N_j$ and $N_{\gamma j}$, on the considered systematic effects is described by the variation of the matrices $A_{qj}$ and $B_{qj}$ (see Eqs. ~\ref{eq:nj} and \ref{eq:ngj}). To allow for a semi-analytical solution, the linear dependence of $A_{q\, j}$ and $B_{q\,j}$ is imposed on the systematic variations, $\delta_k$, and the matrices are rewritten:
\begin{eqnarray*}
     A_{q j} = A_{q j}^{MC} + \sum_k a^k_{q j} \cdot \delta_k\,,\\
  B_{q j} = B_{q j}^{MC} + \sum_k b^k_{q j} \cdot \delta_k\,,
\end{eqnarray*}
where $a^k_{q j}$ and $b^k_{q j}$ correspond to $1 \sigma$ variations of $A_{q j}^{MC}$ and $B_{qj}^{MC}$, which are the nominal values obtained from the Monte Carlo simulation, due to systematic uncertainty $k$.
Following the linear approximation, one can rewrite the experimentally measured cross sections by defining the expected cross section (theoretical value calculated within the SM), $\sigma^{th}_q$, and the cross-section deviation parameters, $\Delta_q$:
\begin{equation*}
     \sigma_q = \sigma^{th}_q + \Delta_q\,.
\end{equation*}
When solving the maximum likelihood problem with systematic uncertainties included, one tries to find a solution for $\Delta_q$, $\delta_k$, and $\lambda_q$, assuming small deviations from the nominal
predictions, i.e. $|\Delta_q| \ll \sigma_q^{th}$, $|a_{qj}^k| \ll A_{qj}$ and $|b_{qj}^k| \ll B_{qj}$, 
so that the terms including products
of deviations  ($\Delta_q \Delta_{q'}$, $\delta_k \delta_{k'}$ and $\Delta_q \delta_k$) do not play a major role and can be neglected.

By calculating respective derivatives of \ref{eq:chi2full} and assuming they are equal to zero at the extremum, the problem can be reduced to solving the system of $2M+K$ linear equations, similar to the one previously discussed:
\begin{eqnarray*}
     \tilde{\mathbb{H}} \cdot
  \left(\! \begin{array}{c} \vec{\Delta}\\ \vec{\delta} \\ \vec{\lambda} \end{array} \!\right)
  =  \tilde{\mathbb{V}}\,,
\end{eqnarray*}
where $\tilde{\mathbb{H}}$ and $\tilde{\mathbb{V}}$ are the extended Hessian matrix and vector, respectively.

\section{Event simulation}

Reliable measurements require comparison of data with accurate and realistic Monte Carlo simulations. The process $e^+e^- \to q\bar{q}$ is one of the simplest electroweak processes and is thus often perceived as a ``standard candle'' for event generators (see e.g.~\cite{Robson:2920434}); however, the reconstruction of isolated photons poses several challenges. A proper description necessitates the full matrix-element generation of measurable photons. On the other hand, due to soft and collinear divergences, a theoretically infrared-safe pseudo-observable must be constructed. Using ISR structure functions, resummation techniques, and parton showers, one can reproduce the experimental measurement as accurately as possible. Additionally, hadronisation effects must also be included, as neutral hadrons decay into photons. In the following, we will use this modelling prescription: data samples shall be generated using fixed-order ME calculations, with exclusive emissions of hard photons, and matched with initial-state radiation (ISR) structure functions and final-state radiation (FSR) showers, accounting for collinear and soft emissions, and hadronisation models.

\begin{figure}
    \centering
    \includegraphics[width=0.6\textwidth]{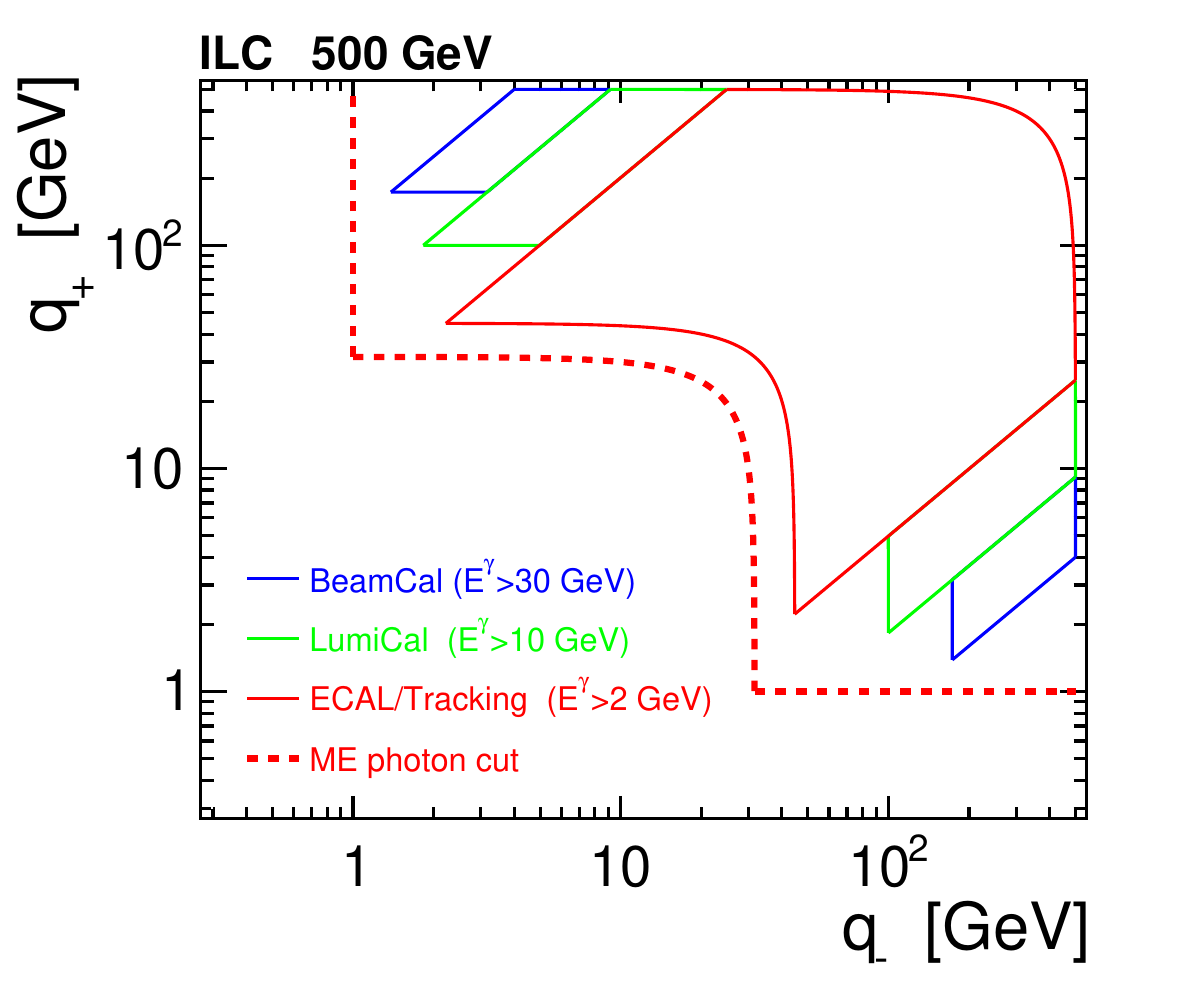}
    \caption{Detector acceptance in the $(q_+, q_-)$ plane expected for the future experiments at 500\,GeV ILC. Red dashed lines indicate the cut used to restrict the phase space for ME photon generation. The plot is taken from~\cite{Kalinowski:2021tyr}.}
    \label{fig:DM_q_acceptance}
\end{figure}

A similar matching was studied in the literature in the context of dark-matter production detected in the mono-photon signature at lepton colliders~\cite{Kalinowski:2020lhp, Kalinowski:2021tyr}. A matching procedure for simulating photons using both ME calculations (for hard emissions) and ISR structure function (for soft emissions) was developed for \textsc{Whizard}. It is based on two variables, $q_-$ and $q_+$, defined for each photon as:
\begin{align*}
    q_- = \sqrt{4E_0 E_\gamma}\sin \frac{\theta_\gamma}{2},\\
    q_+ = \sqrt{4E_0 E_\gamma}\cos \frac{\theta_\gamma}{2},
\end{align*}
where $E_0$ is the nominal beam energy, $E_\gamma$ is the energy of the emitted photon and $\theta_\gamma$ is its emission angle with respect to the electron beam direction. The variable $q_-$ ($q_+$) represents the virtuality of an electron (positron) in the case of a single photon emission. As a rule, photons with $q_\pm > q_{min}$  and $E_\gamma > E_{min}$ are generated in the full ME picture, while all the \textit{softer} emissions are modelled via the built-in ISR structure function handler for the ISR photons in \textsc{Whizard}. Figure~\ref{fig:DM_q_acceptance} shows how the variables $q_\pm$ separated the phase space in the case of~\cite{Kalinowski:2021tyr}, where ILC running at 500\,GeV was considered.
As indicated in the plot, only photons in the ME domain can be detected; all photons from the ISR structure function are outside the detector acceptance.

However, as dark matter is typically assumed to be electrically neutral and colourless, the procedure of~\cite{Kalinowski:2020lhp} does not account for photons originating from the final-state showers or hadronisation. To include these effects, the procedure was extended in the following way: parton-level events are generated with \textsc{Whizard} and include ISR, while \textsc{Pythia\,6}~\cite{Sjostrand:2006za} is employed to simulate parton showers and hadronisation. A standard cut is applied, requiring the invariant mass of the parton-level hadronic system to be greater than 10~GeV. The results were found to be qualitatively consistent when using \textsc{Pythia\,8}~\cite{Sjostrand:2014zea}.

To separate hard and soft final-state radiation regimes, an additional criterion based on the invariant mass of the photon-quark pairs, $M_{\gamma\, q_{1,2}}$, is used. A \textit{hard} photon is defined as one fulfilling all the following criteria:
\begin{itemize}\setlength\itemsep{-0.2em}
    \item $q_\pm > 0.5$\,GeV,
    \item $E_\gamma > 0.5$\,GeV,
    \item $M_{\gamma\, q_{1,2}} > 1$\,GeV,
\end{itemize}
and is generated using fixed-order ME calculations. All remaining \textit{soft} photons, not meeting at least one of these criteria, are simulated via the internal structure function handler for ISR photons in \textsc{Whizard} and the \textsc{Pythia} parton shower for FSR. Events with at least one ISR or FSR photon passing the \textit{hard} photon selection are rejected to avoid double counting. 

The above matching criteria are somewhat technical. They are applied at the event-generation level to ensure that all photons observable in the detector are simulated using the full ME approach, accounting, for instance, for proper interference between initial- and final-state radiation. Meanwhile, soft effects are incorporated through corresponding modelling methods. The matching parameters are set well below the experimental acceptance; for example, the energy cut of 0.5\,GeV is below the threshold of 2\,GeV required for the photon to be reconstructed in the detector. In this context, variations of the matching parameters below the experimental sensitivity do not alter the final results but may influence computational performance.

The authors of \cite{Kalinowski:2020lhp} validated the matching procedure applied for CLIC running at 380\,GeV and at 3\,TeV against \textsc{KKMC}~\cite{Jadach:1999vf,Jadach:2013aha,Jadach:2022mbe}, a Monte Carlo code designed to simulate two-fermion production processes in the electron–positron colliders and based on the coherent exclusive exponentiation of the initial and final state bremsstrahlung~\cite{Jadach:2000ir}. Similar cross-checks for $Z$-pole running were performed, resulting in no significant deviations between \textsc{KKMC} and the matching method, confirming the method's applicability in this regime. However, prior to the construction of the experiments, a more comprehensive study and comparison of various Monte Carlo generators will be required to evaluate systematic uncertainties associated with this aspect of the analysis.

For this study, the following samples were generated: 10 million events with 0 ME photons, 10 million events with 1 ME photon and 1 million events with 2 ME photons. Higher photon multiplicities were neglected, given that the cross section for the 2-ME-photon sample is already approximately 30 times smaller than that of the 1-ME-photon sample. The samples were subsequently matched using the procedure outlined above.
Table \ref{tab:cs} summarises the cross sections for each flavour for all the \textit{unmatched} samples (with generator-level cuts applied only to ME photons, but not those modelled in showers) and the final matched sample. As expected, the cross section for the matched samples is consistent with the cross sections of the unmatched 0-ME-photon samples, which include higher-photon multiplicities via modelling methods. The contribution of radiative events is larger for up-type quarks than for down-type quarks; however, the ratio deviates from the naïvely expected value of four due to the specific values of the quark couplings themselves and ISR.
\begin{table}[h!]
\begin{tabular}{c|c|c|c|c|c|c}
             $\sigma_q$ [fb] & $d$ & $u$ & $s$ & $c$ & $b$ & \textbf{total} \\ \hline
0$\gamma$ unmat.   & 5.46e+06   & 4.28e+06   & 5.46e+06   & 4.28e+06   & 5.47e+06   & 2.49e+07     \\ \hline
1$\gamma$ unmat.   & 3.63e+05   & 4.63e+05   & 3.63e+05   & 4.65e+05   & 3.64e+05   & 2.02e+06     \\ \hline
2$\gamma$ unmat.   & 8.48e+03   & 2.19e+04   & 8.51e+03   & 2.19e+04   & 8.50e+03   & 6.92e+04     \\ \hline
0+1+2$\gamma$ mat. & 5.44e+06   & 4.34e+06   & 5.45e+06   & 4.33e+06   & 5.45e+06   & 2.50e+07     \\
\end{tabular}
\caption{Cross sections for each quark flavour for unmatched samples (generator-level cuts applied only to ME photons) and the final matched sample. Integration uncertainties are below $10^{-3}$.}
\label{tab:cs}
\end{table}

\begin{figure}
    \centering
    \includegraphics[width=0.6\textwidth]{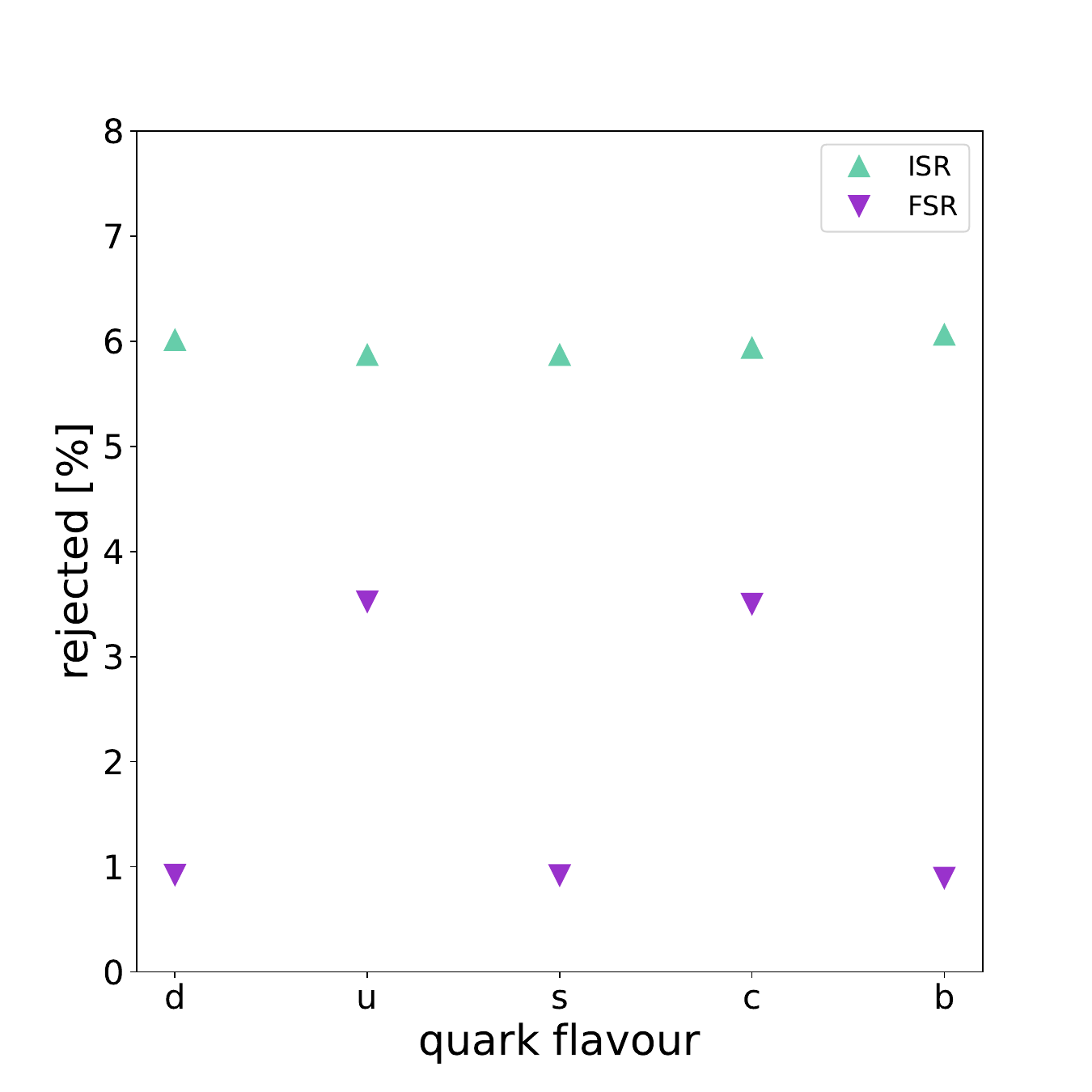}
    \caption{The fraction of events from the sample with no photons generated at the ME level rejected by the matching procedure for different quark flavours for ISR (aquamarine upwards) and FSR (purple downwards).}
    \label{fig:rejection}
\end{figure}

Figure~\ref{fig:rejection} shows the fraction of events rejected by the matching procedure in the 0-ME-photon sample, categorised by quark flavour, for both ISR and FSR contributions. As expected, the ISR rejection efficiency is independent of the quark flavour. In contrast, the FSR rejection efficiency differentiates between up- and down-type quarks, with a factor of approximately four between the two, reflecting the difference in the squared quark charges.

To account for experimental effects, the detector response was simulated in \textsc{Delphes 3.5}~\cite{deFavereau:2013fsa}, a fast simulation framework designed to model general-purpose collider experiments. \textsc{Delphes} provides a simplified yet realistic description of detector components, a tracking system, embedded into a magnetic field, calorimeters and a muon system. Standard event formats from Monte Carlo tools are supported. \textsc{Delphes}  outputs reconstructed objects like jets, leptons, and missing transverse energy.

The ILCgen cards, based on the performance of the ILD and SiD concepts~\cite{ILDConceptGroup:2020sfq, White:2015dxj}, were used. Although originating from linear-collider studies, ILD is currently also considered for circular colliders~\cite{Einhaus:2023npl,Robson:2920434}. The ILCgen cards assume the photon tagging efficiency of 95\% for photons with energies above 2\,GeV and the absolute value of pseudorapidity below 3, and 90\% for photons with energies above 2\,GeV and pseudorapidities between 3 and 4. Other photons are considered undetectable.

\section{Analysis procedure and results}
\label{sec:analysis}

In this section, different steps in the analysis procedure are explained, and results for various assumptions are given. 

\subsection{Event reconstruction}
An important advantage of performing measurements at the $Z$ pole is the reduced contribution from ISR, due to the limited phase space available for photon emission. However, photons produced in hadronic decays still present a significant background and must be accounted for in the analysis. To maximise the sensitivity to true hard-photon emissions, photon reconstruction criteria should be carefully optimised at the experimental level. In the analysis, other background channels to the inclusive sample of hadronic $Z$ decays (e.g. $Z\to \tau\tau$) are neglected, as their impact is expected to be negligible and can be easily mitigated by standard cuts, for instance, on the invariant mass of the di-jet system.

At the ME level, the photon isolation parameter is defined as
\begin{equation}
    q^T = E_\gamma \sin \theta^{min}_{\gamma q_i},
\end{equation}
where $E_\gamma$ is the photon energy and $\theta^{min}_{\gamma q_i}$ is the angle between the photon and the closest quark. This variable corresponds to the photon transverse momentum with respect to the quark direction. The redefinition of the variable to the detector-level analysis is straightforward: the angle $\theta^{min}_{\gamma q_i}$ is replaced by $\theta^{min}_{\gamma j_i}$, the angle between the photon and the closest jet.

The Durham jet algorithm in the exclusive mode was used in \textsc{Delphes} to cluster all final-state objects into two jets. The photon isolation criteria were not applied in \textsc{Delphes}, so all photons were included in the jet clustering. In this way, radiative event tagging could be separated from the standard reconstruction procedure, focusing on the analysis-specific approach. It is worth mentioning that in this context, the physical interpretation of the $q^T$ variable becomes somewhat vague, as including all photons in jets changes jet kinematics. However, the definition remains comprehensive at the analysis level and will be used in the following.

\begin{figure}
    \centering
    \includegraphics[width=0.43\textwidth]{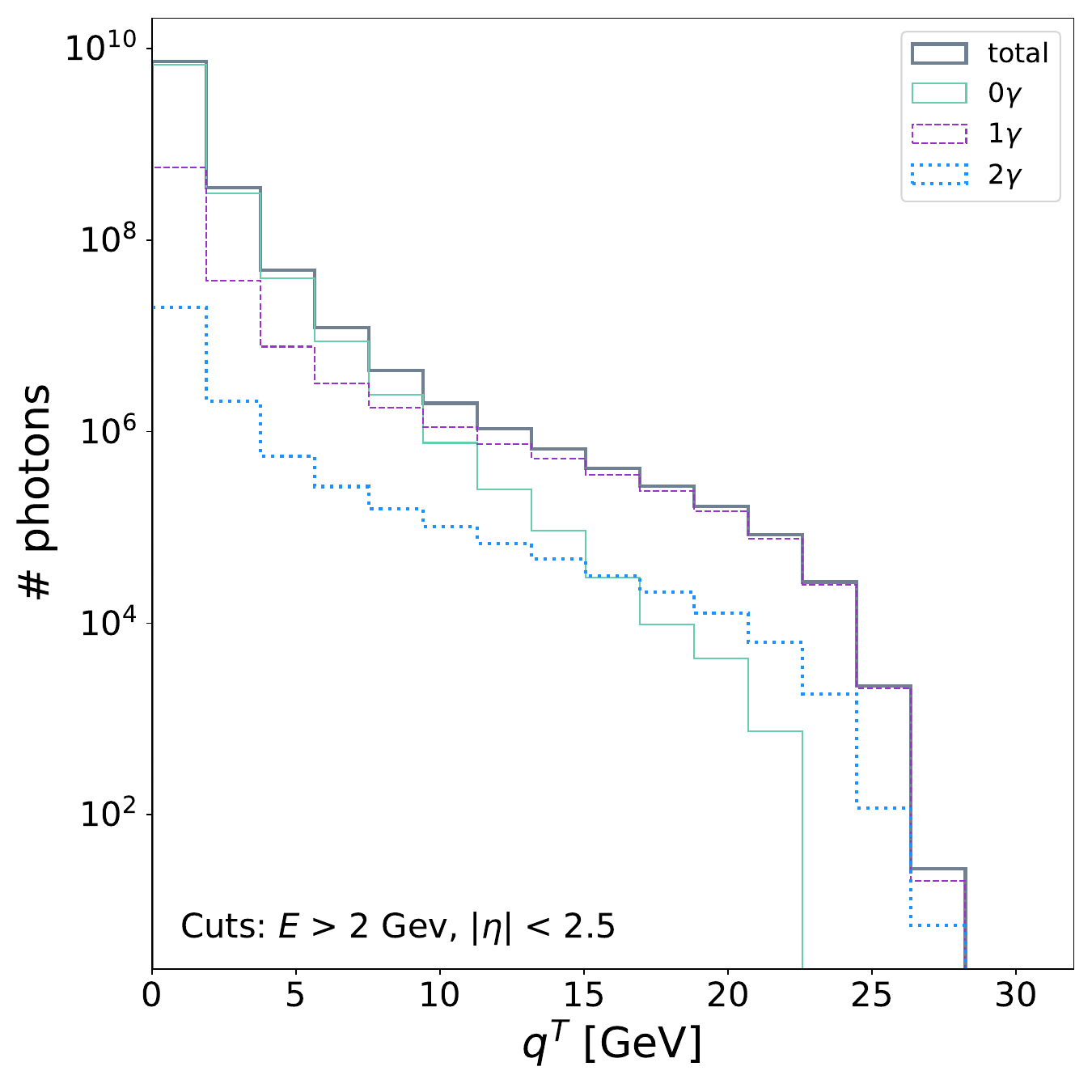}
    \includegraphics[width=0.43\textwidth]{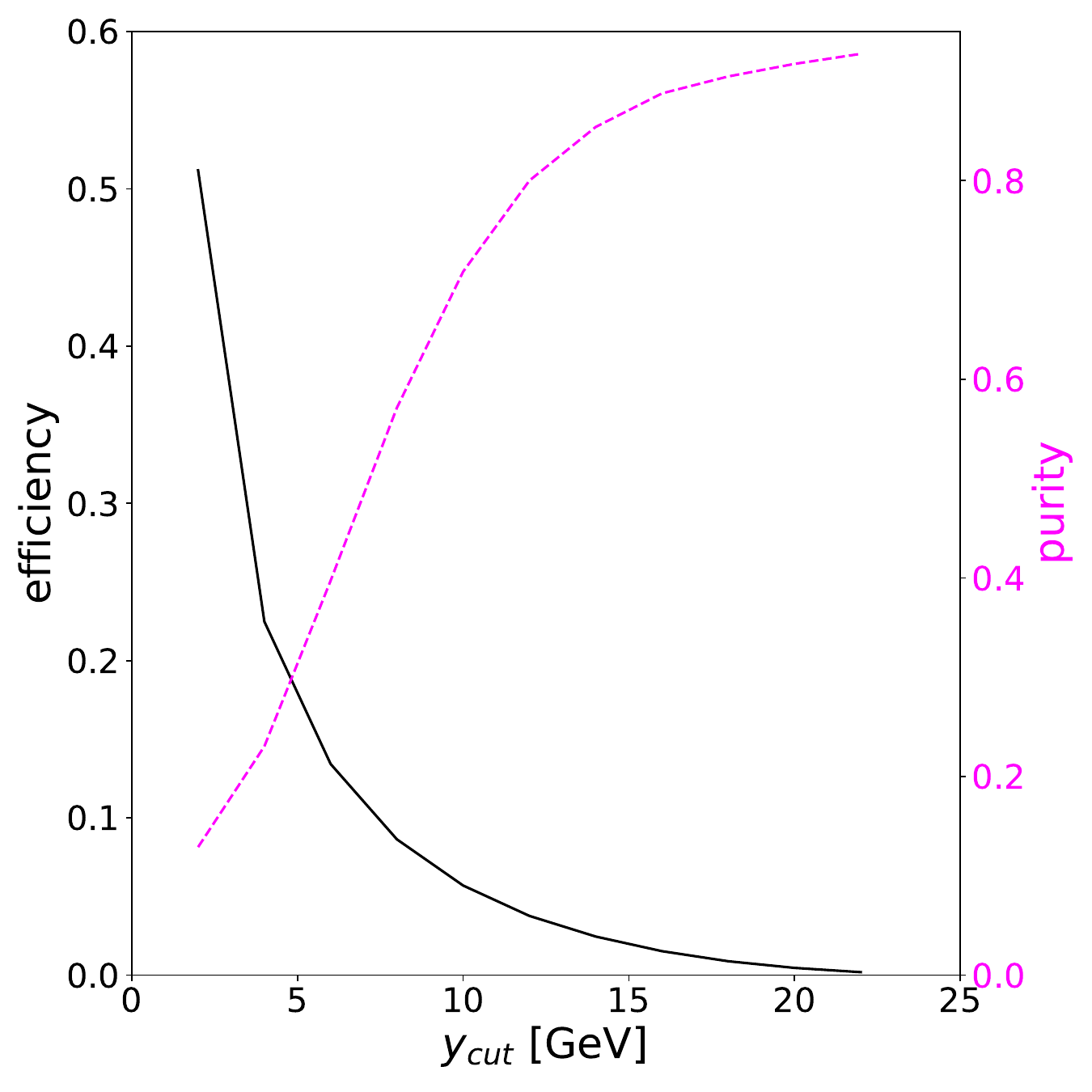}
    
    \caption{\textbf{Left:} the distribution of selected photons (normalised to an integrated luminosity of 100\,fb$^{-1}$) as a function of $q^T$. The thick grey line stands for the sum of all the samples, the solid aquamarine line for the 0-ME-photon sample, the purple dashed one for the 1-ME-photon sample and the blue dotted one for the 2-ME-photon sample. Fast detector simulation is included in the results. \textbf{Right:} the signal efficiency (left axis, black solid line) and the signal purity (right axis, pink dashed line) for different values of $y_{cut}$, starting at a minimal value of 2\,GeV considered in the study. Note that these distributions are defined at the event level, i.e. they relate to the number of those events in which exactly one photon was tagged.}
    \label{fig:y}
\end{figure}

Figure \ref{fig:y} (left) shows the distribution of $q^T$ for all reconstructed photons at the detector level. The number of events corresponds to 100\,fb$^{-1}$ of unpolarised data, the lowest integrated luminosity considered in the study, and can be easily scaled to other integrated luminosities. At the detector level, only photons with an energy above 2\,GeV and an absolute value of pseudorapidity below 2.5 were selected, which corresponds to the central region in the detector. For low values of $q^T$, the total distribution is dominated by the 0-ME-photon sample --- for which the reconstructed photons originate from hadronic decays, with 90\% of the photons coming from $\pi^0$ decays, 6.4\% from $\eta$ decays and 0.9\% from $\omega(782)$ decays --- while for larger values the 1-ME-photon sample becomes dominant. The transition between the two regions occurs at about 10\,GeV. Above this value, the photon sample is completely dominated by hard photons, i.e. those generated at the ME level and not those modelled in showers. Figure \ref{fig:y} (right plot) shows the \textit{signal} efficiency (ratio of the accepted radiative events over all the radiative events) and purity (ratio of the accepted radiative events over all the accepted events) as a function of $y_{cut}$. For the purpose of this study, only events with exactly one hard-tagged photon, i.e. for which $q^T > y_{cut}$, are selected. Events with one proper photon generated at the ME level become dominant for a cut value of about 7\,GeV, for which about 10\% of events are preserved.

In this context, a brief remark is in order. For the statistical analysis, a small conceptual change was introduced compared to the procedure described in Section~\ref{sec:statframe}. The one-photon sample was treated exclusively by requiring exactly one tagged photon, while the full hadronic sample was taken inclusively. This choice simplifies the technical implementation of the tagging algorithm by avoiding ambiguities arising from events with multiple photons. With this approach, radiative events were not removed from the hadronic sample, and the two samples were treated as independent. This is justified by the fact that the fraction of radiative events in the full hadronic sample is small, at the percent level. An alternative strategy would be to explicitly remove the radiative events from the hadronic sample and modify the fit procedure accordingly.

\subsection{2-flavour fit}
In the next step, the statistical framework described in Section~\ref{sec:statframe} was adopted. The quark flavour tagging efficiencies were taken from \cite{Liang:2023wpt}, assuming that both jets are tagged independently and that light jets have the same tagging probabilities, which follows from the fact that the goal of this analysis is to separate the light quark couplings using photon radiation and no other methods. To avoid any bias coming from the small difference present in \cite{Liang:2023wpt}, the values for the $u$ and $d$ tagging of \cite{Liang:2023wpt} were averaged.

First, a sample of events with two tagged light jets was considered, i.e. all events in which at least one jet was not classified into the first category were rejected. A transverse momentum cut $p^T_{j} > 10$\,GeV and a pseudorapidity cut $|\eta_{j}| < 4$ were applied on both jets. In this sample, a subsample consisting of events with one tagged photon ($E_\gamma > 2$\,GeV, $|\eta|<2.5$) was identified. The numbers of events in the main inclusive sample (``hadronic'') and its subsample (``radiative'') were used to fit the data and extract $c_d$ and $c_u$ using $\chi^2$ minimisation. 

Subsequently, the impact of systematic uncertainties was studied. The measurement is assumed to be affected by the following factors:
\begin{itemize}\setlength\itemsep{-0.2em}
    \item relative uncertainty on the integrated luminosity,
    \item relative uncertainty on the radiative event selection efficiency,
    \item relative uncertainty on the background to the radiative sample due to hadronisation photons,
    \item tagging uncertainties (introduced for all tagging and mis-tagging probabilities used in the fit procedure).
\end{itemize}
By setting the tagging uncertainties as an absolute shift to the elements of the tagging matrix (\ref{eq:matrix}), it is ensured that the relative tagging uncertainty for the light quarks is larger than for the heavy quarks, as the corresponding elements of the tagging matrix are smaller, and an absolute shift of the same magnitude is more significant.

\begin{figure}
    \centering
    \includegraphics[width=0.46\textwidth]{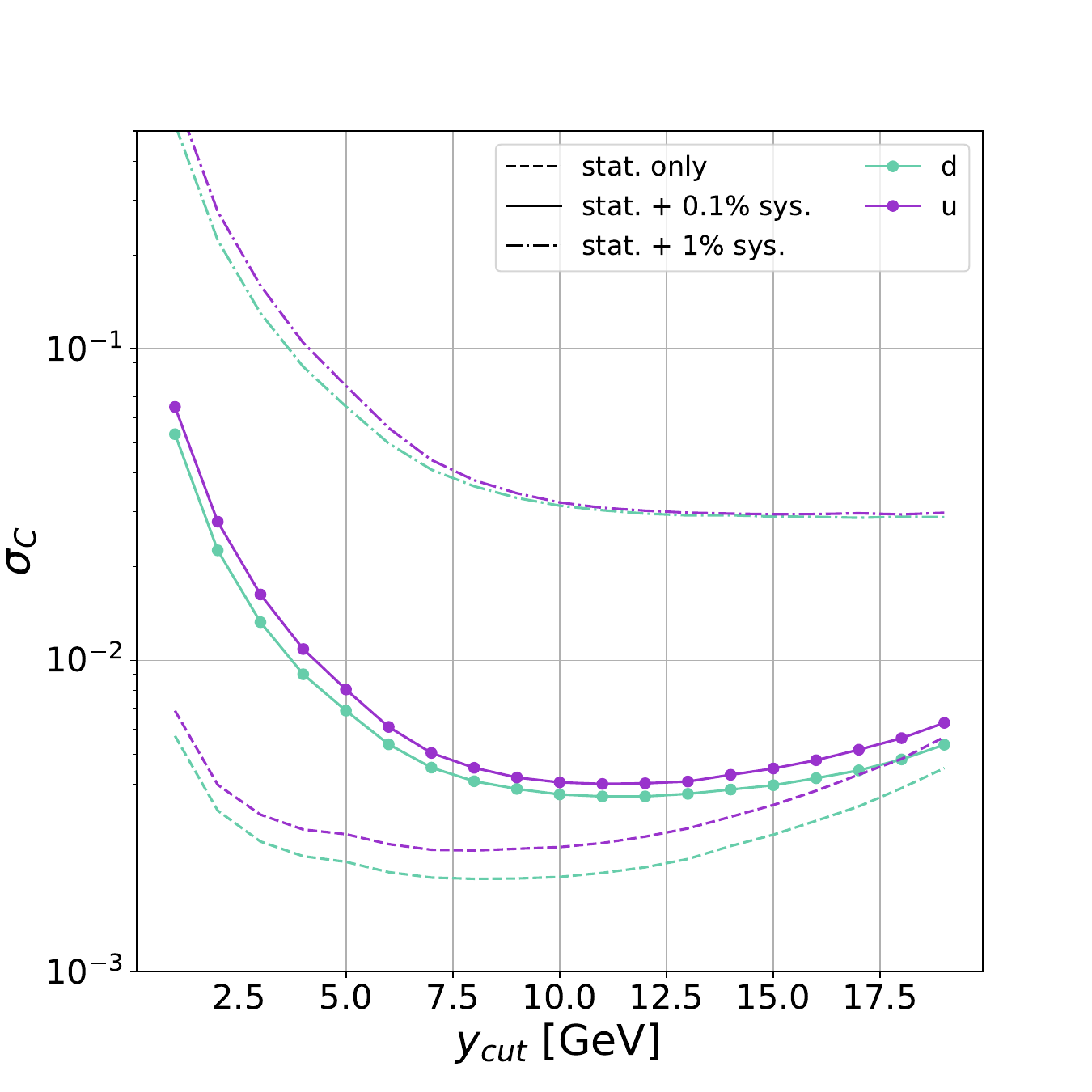}
    \includegraphics[width=0.46\textwidth]{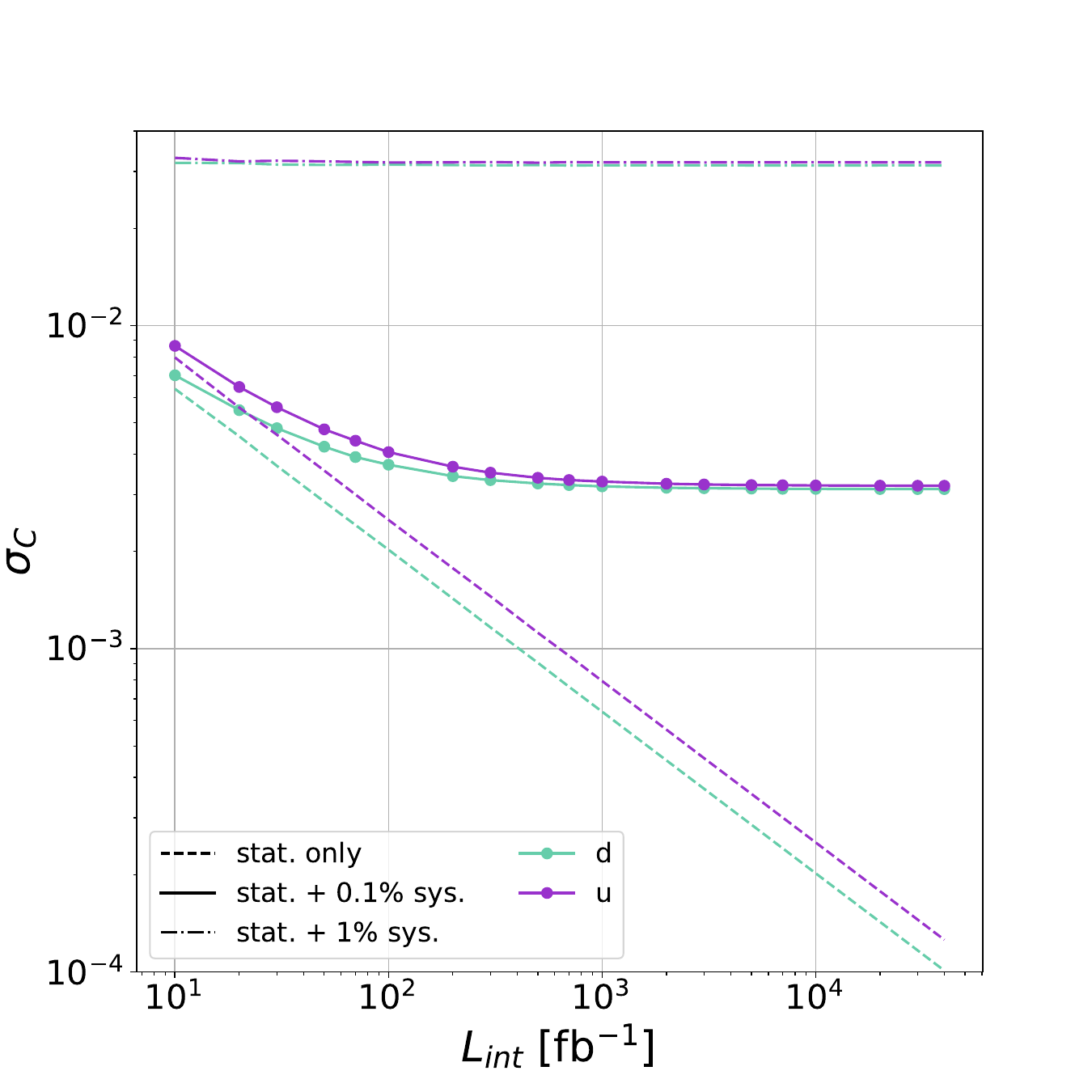}
    
    \caption{The uncertainty of the $d$ (aquamarine) and $u$ (purple) coupling measurements as a function of $y_{cut}$ (left) and the integrated luminosity (right). Dashed lines indicate statistical uncertainty only, solid lines with points --- statistical and systematic uncertainties of 0.1\% combined, dash-dotted lines --- statistical and systematic uncertainties of 1\% combined. The luminosity uncertainty was fixed at 10$^{-4}$. For the left plot, 100\,fb$^{-1}$ of data is assumed and for the right plot, the value of $y_{cut}$ is set to 10\,GeV.}
    \label{fig:2flav}
\end{figure}

For simplicity, uncertainties were initially set to the same value for all the contributions, except for the luminosity uncertainty which was always kept at $10^{-4}$ (see e.g. \cite{BozovicJelisavcic:2013lni, Jadach:2018jjo, Dam:2021sdj,Gong:2026qnu}). The results are shown in Figure \ref{fig:2flav}, where the relative statistical uncertainties on the $u$ and $d$ quark couplings are compared with the total uncertainties calculated assuming the systematic uncertainties to be either 0.1\% or 1\%. The left plot presents the dependence of the results on the $y_{cut}$ parameter. As naively estimated from Figure \ref{fig:y}, the optimal value of the parameter corresponds to about 10\,GeV, where the FSR-initiated photons start to dominate over photons coming from hadronic decays. The right plot illustrates the dependence on integrated luminosity. The measurement is dominated by systematic uncertainties, as increasing the integrated luminosity beyond $10^2$\,fb$^{-1}$ yields no significant improvement. In the following, possible enhancements to the measurement, together with the impact of specific systematic effects, are discussed.

\begin{figure}
    \centering
    \includegraphics[width=0.46\textwidth]{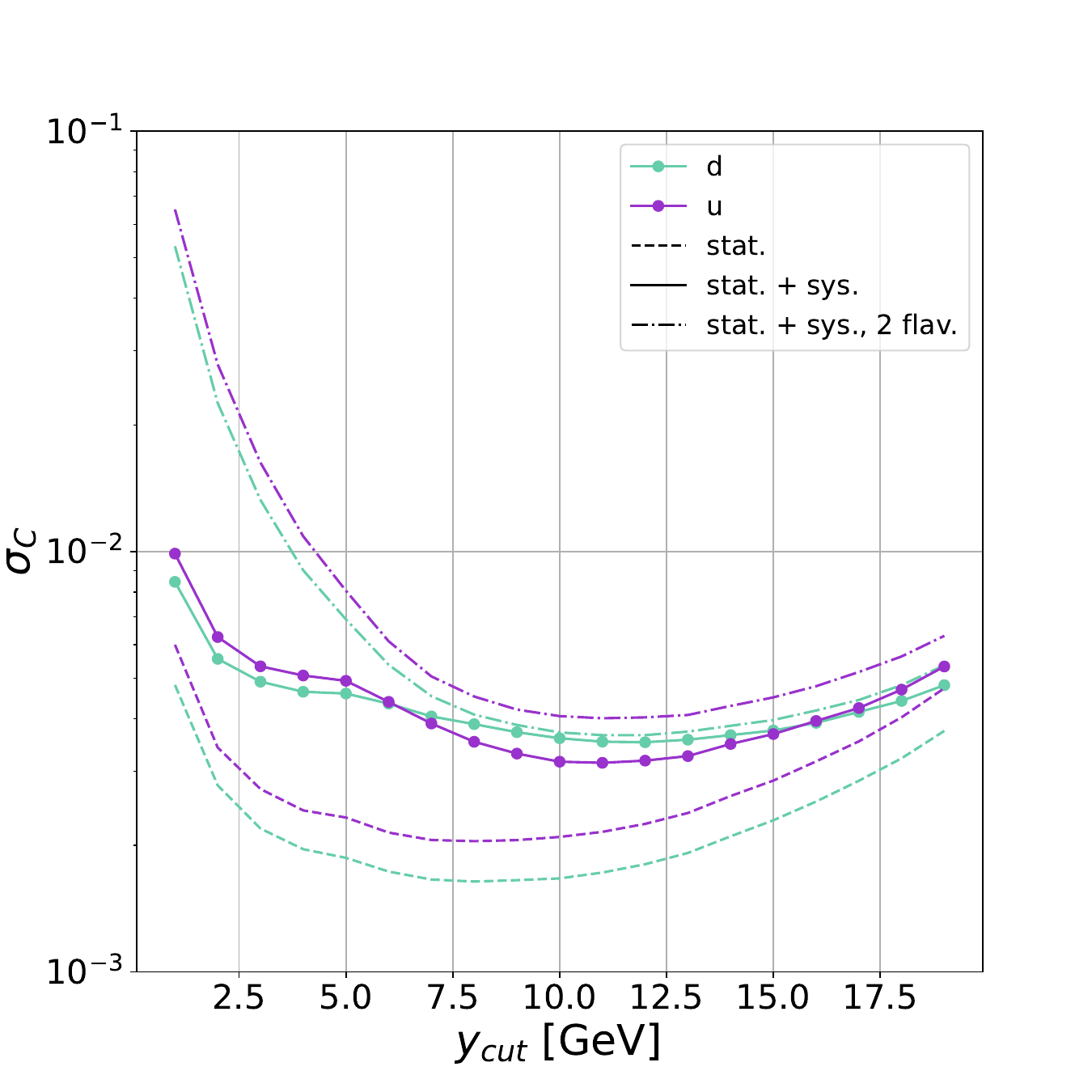}
    \includegraphics[width=0.46\textwidth]{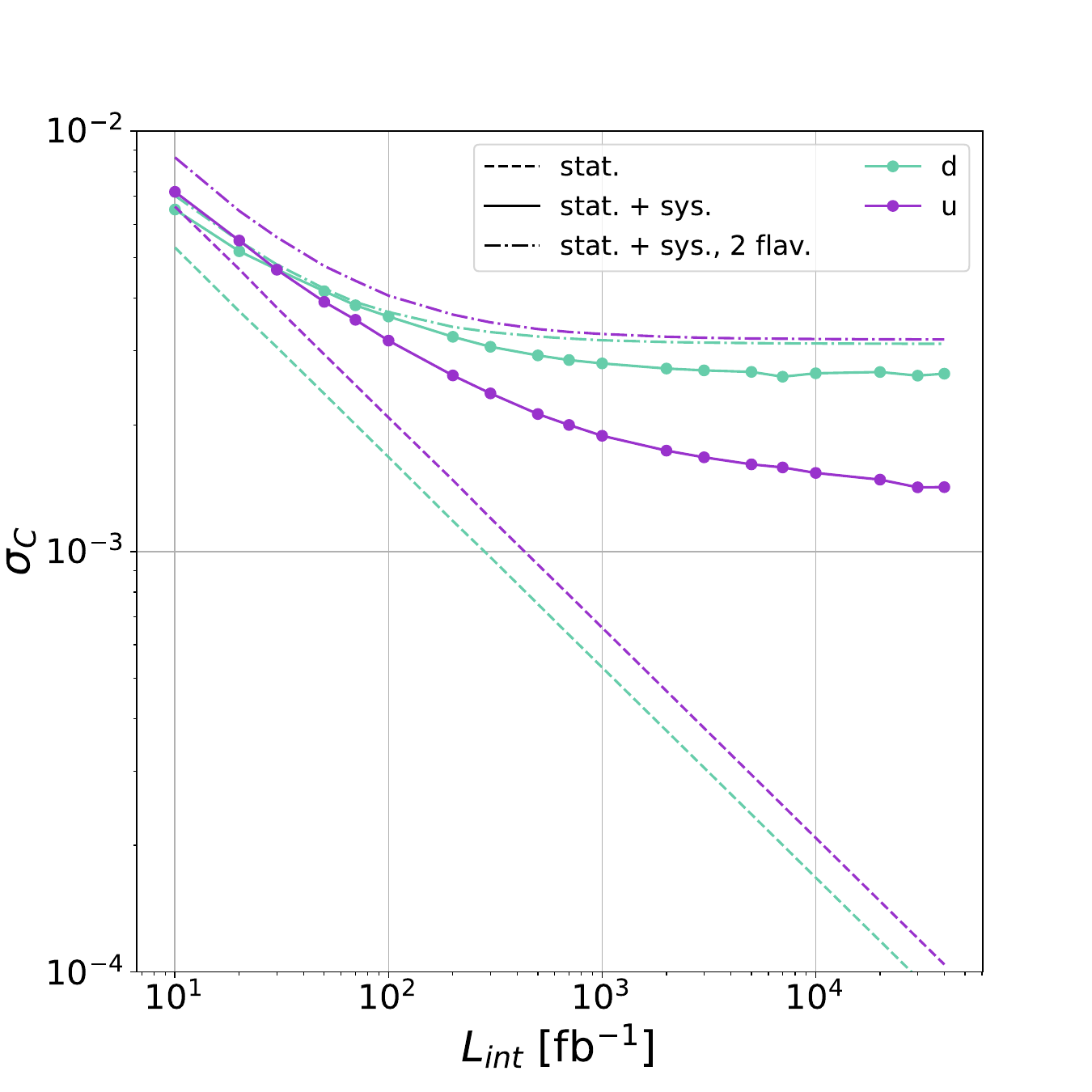}
    
    \caption{The uncertainty of the $d$ (aquamarine) and $u$ (purple) coupling measurements as a function of $y_{cut}$ (left) and the integrated luminosity (right). Dashed lines indicate statistical uncertainty only, solid lines with points --- statistical and systematic uncertainties for the 5-flavour fit, and dash-dotted lines --- statistical and systematic uncertainties for the 2-flavour fit. 0.1\% uncertainty is assumed for all the contributions, except for the luminosity uncertainty (fixed at 10$^{-4}$). For the left plot, the integrated luminosity was fixed at 100\,fb$^{-1}$ and for the right plot, the value of $y_{cut}$ was set to 10\,GeV.}
    \label{fig:2vs5flav}
\end{figure}

\subsection{5-flavour fit}

To enhance the fit precision, an inclusive approach incorporating all jet flavours was considered. This strategy encompasses events involving heavy-flavour jets (e.g., $b$-jet pairs) as well as mixed-flavour combinations (e.g., one light jet and one $s$ jet). By simultaneously utilising all available information, the fitting procedure was expected to benefit from improved constraints on certain uncertainties, derived directly from the data.

For this purpose, the statistical framework was extended to include all possible jet combinations, and the fitting procedure was repeated. A comparison between the 2- and 5-flavour fits is given in Figure \ref{fig:2vs5flav}. The improvement is especially remarkable at lower values of $y_{cut}$ and larger integrated luminosities. As expected, the inclusion of other channels helps constrain systematic uncertainties from the data, which results in better scaling of the measurement uncertainty with growing luminosity. It is also interesting to note that the precision for the $d$ quark becomes worse than for the $u$ quark at larger statistics, which can be explained by the signal contamination originating from the $s$-quark channel.

\begin{figure}
    \centering
    \includegraphics[width=0.46\textwidth]{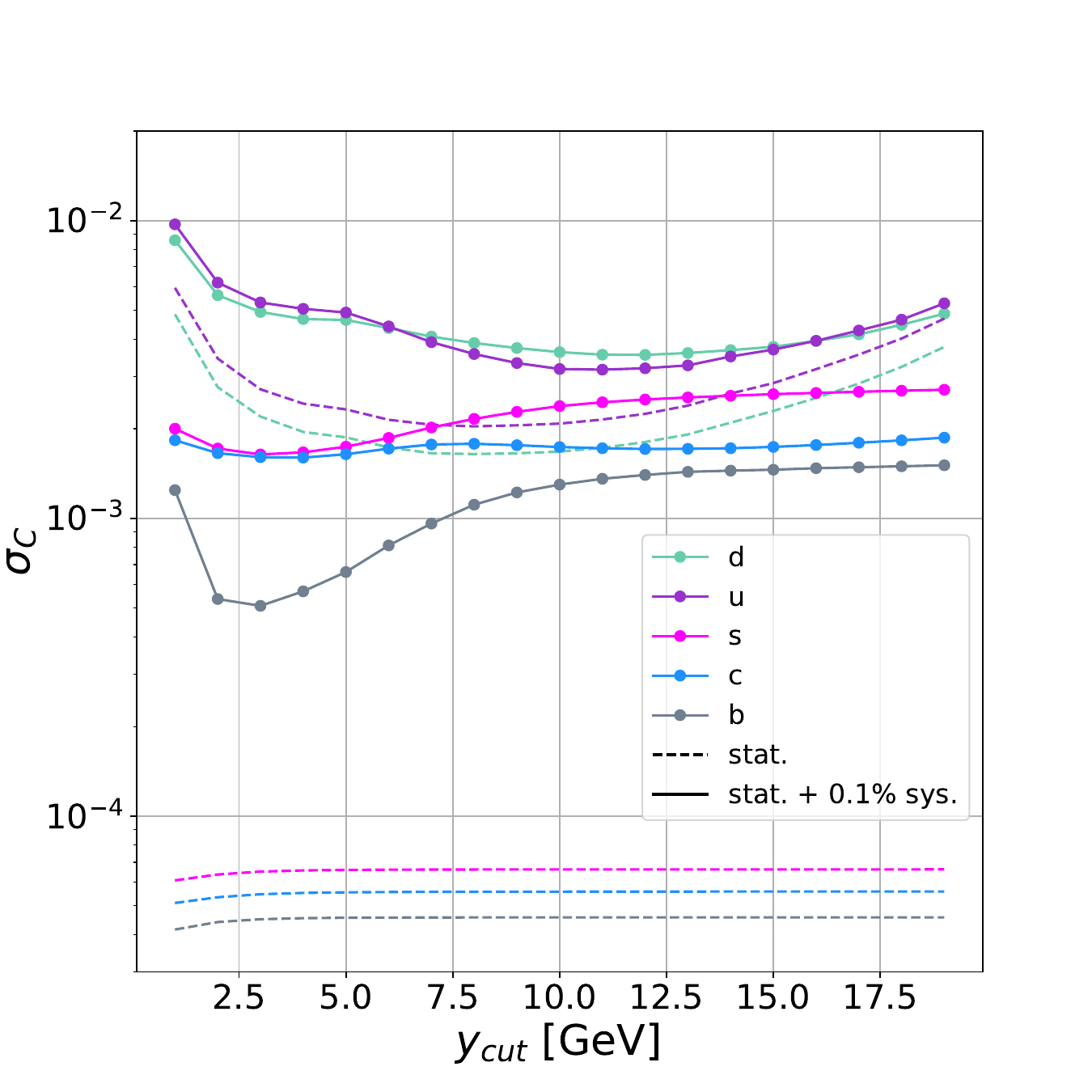}
    \includegraphics[width=0.46\textwidth]{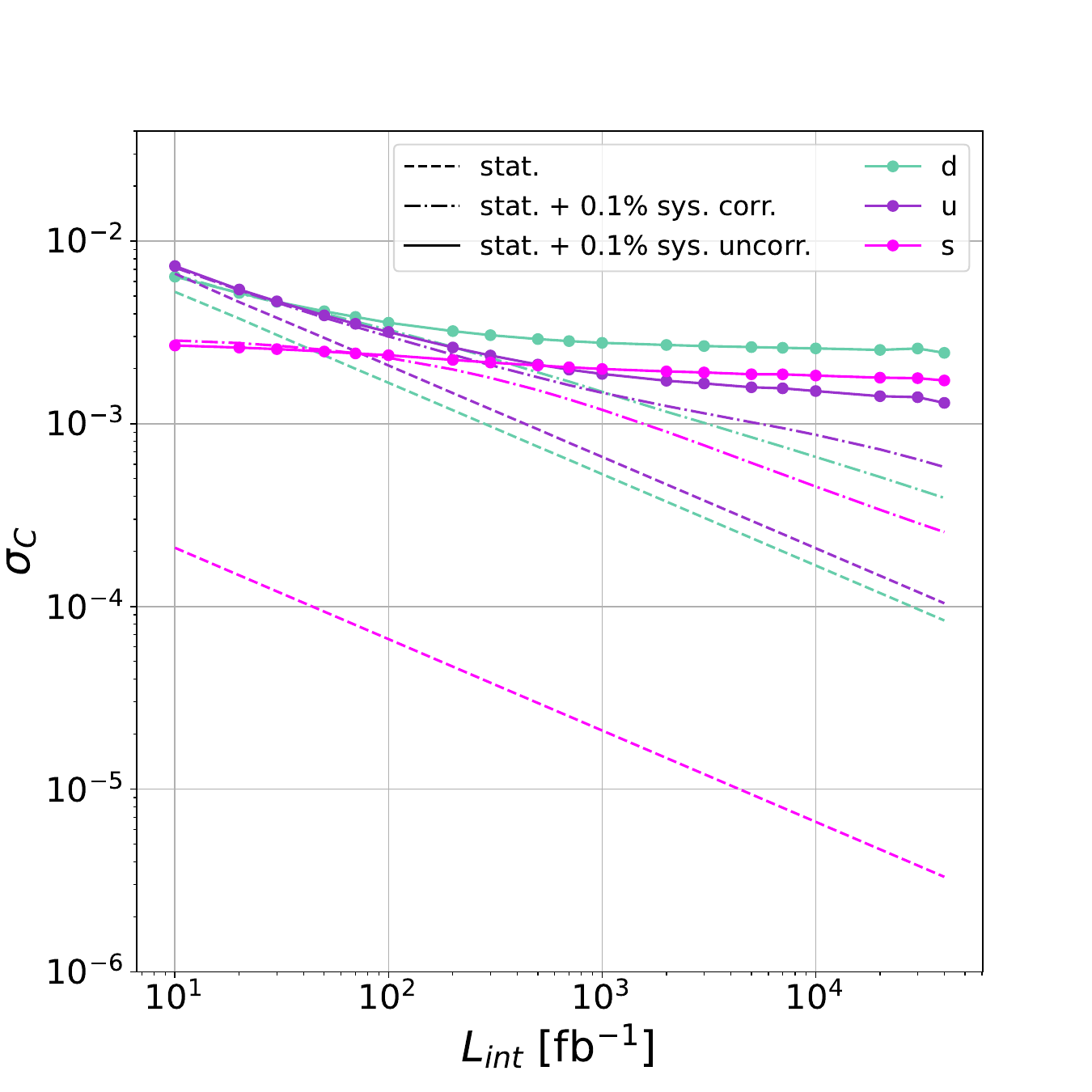}
    
    \caption{The uncertainty of the $d$ (aquamarine), $u$ (purple), $s$ (pink), $c$ (blue) and $b$ (grey) coupling measurement as a function of $y_{cut}$ (left) and the integrated luminosity (right). Dashed lines indicate statistical uncertainty only; the two other lines --- statistical and systematic uncertainties combined for two cases: solid lines with points for $u$- and $d$-tagging uncertainties uncorrelated, dash-dotted lines for $u$- and $d$-tagging uncertainties fully correlated. 0.1\% uncertainty is assumed for all the contributions, except for the luminosity (fixed at 10$^{-4}$). For the left plot, the integrated luminosity was fixed at 100\,fb$^{-1}$ and for the right plot, the value of $y_{cut}$ was set to 10\,GeV. Data for heavy quarks (correlated $u$- and $d$-tagging uncertainties) are skipped in the right (left) plot to improve readability.}
    \label{fig:5flav}
\end{figure}

Figure \ref{fig:5flav} gives more insight into these results. In the left plot, the dependence of the results on $y_{cut}$ for all five flavours is shown. The developed statistical method results in stringent limits for all quark flavours. In the most favourable case of the $b$ quark, this analysis, optimised for the light quark case, can still provide results better than the current measurement by a factor of five. The light quark couplings can be constrained at the sub-percent level, yielding an essential prospect for electroweak precision measurements. It is also worth noting that the results for light quarks are significantly worse than for heavy quarks already at the statistical level, which is connected to the lack of $u$/$d$ discrimination in the flavour tagging.

In the right plot, the results are presented as a function of the integrated luminosity for the three lightest quarks. In addition, scenarios assuming identical tagging and mis-tagging probabilities for the two light quark flavours are considered. Under this assumption, the corresponding systematic variations are treated as fully correlated, thereby reducing the number of systematic uncertainties included in the fit. For all other results presented in this work, a more conservative approach is adopted, in which all systematic variations are assumed to be independent and are thus treated as fully uncorrelated.
The figure shows that this assumption has a crucial impact on the results. Even though the $u$ and $d$ jets are classified into the same category of ``light jet'', the uncertainties for the two physical particles are, in principle, independent. For this reason, the conservative approach was followed in this work.

\subsection{Discussion of systematic uncertainties}
\label{sec:sys_unc}

\begin{figure}
    \centering
    \includegraphics[width=0.32\textwidth]{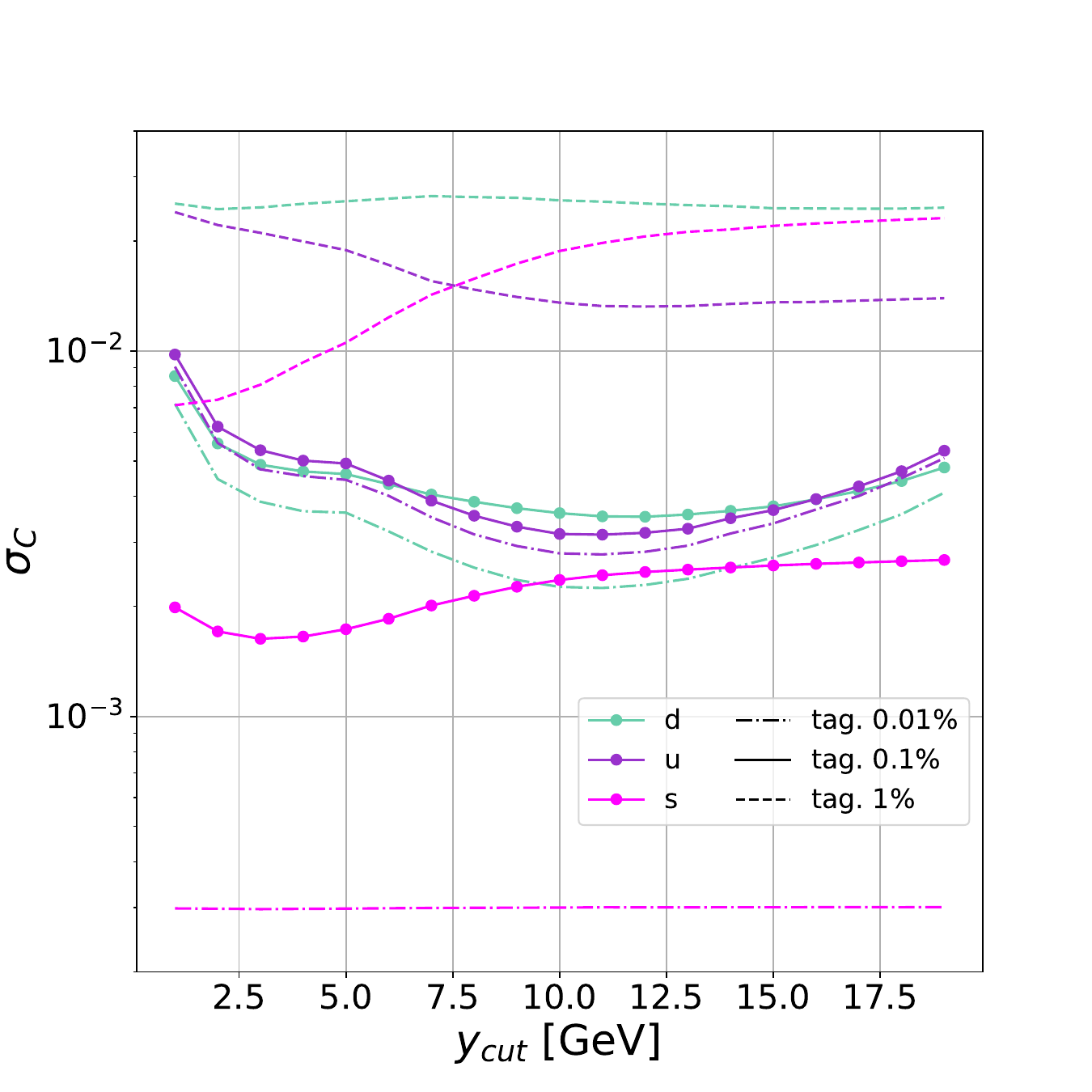}
    \includegraphics[width=0.32\textwidth]{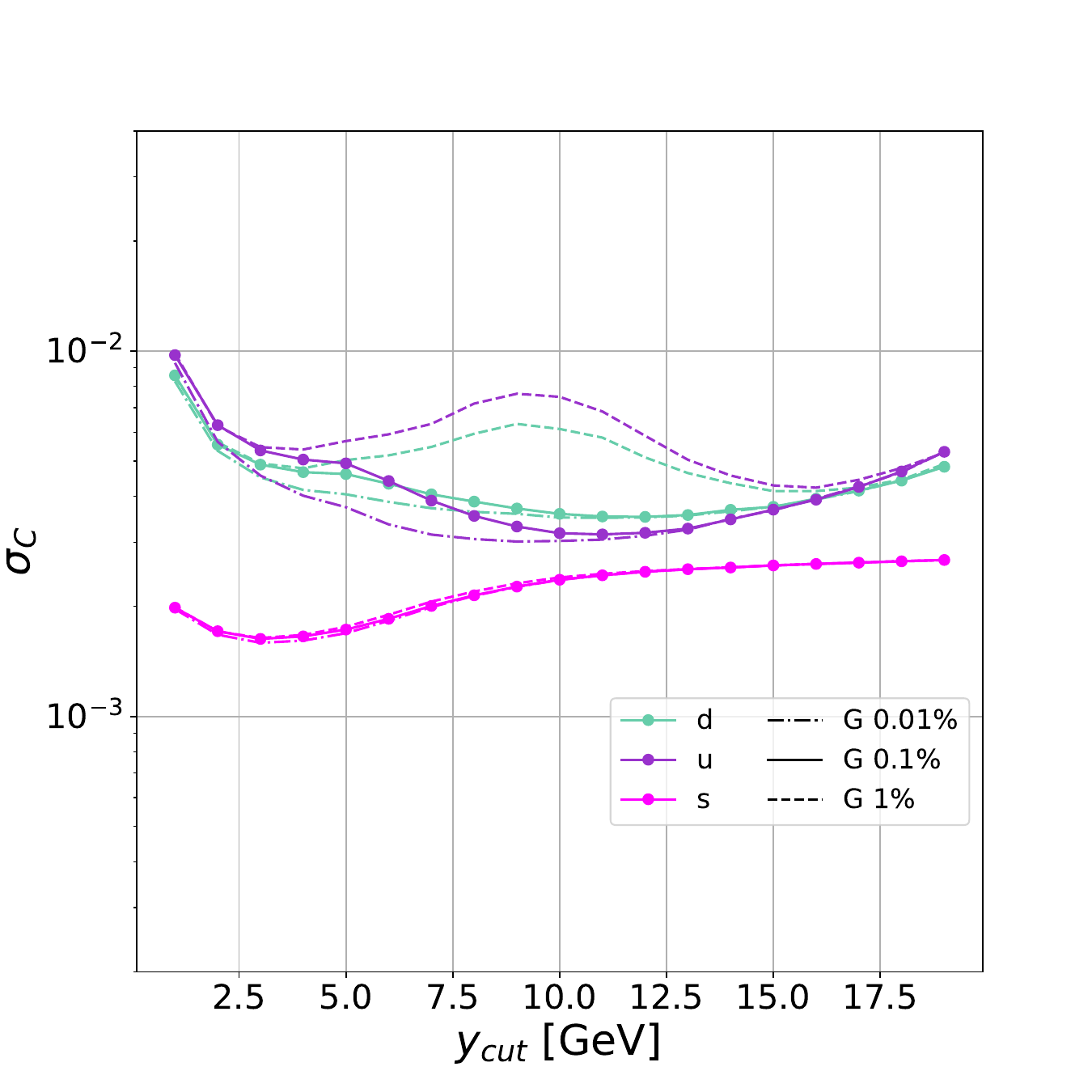}
    \includegraphics[width=0.32\textwidth]{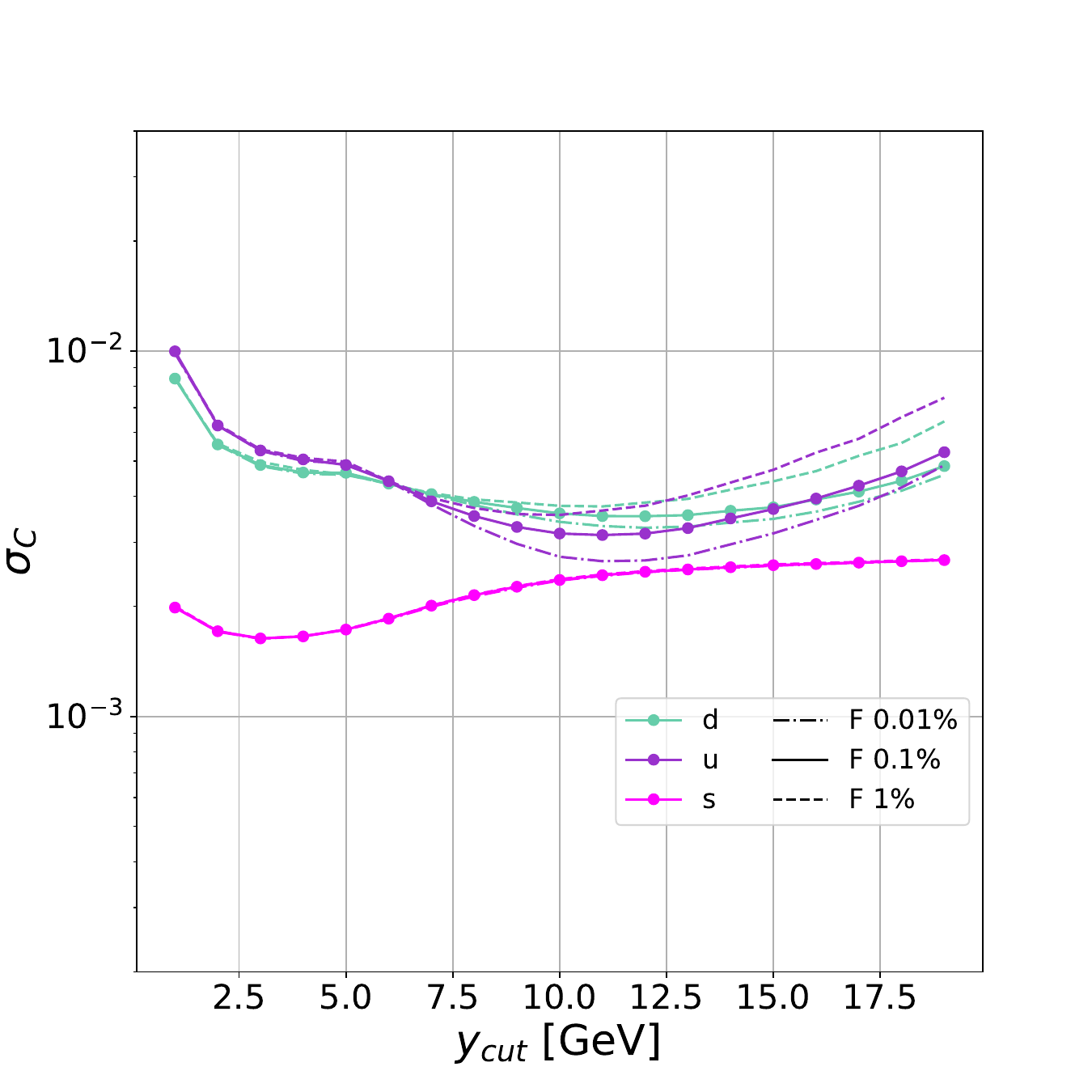}
    
    \caption{The uncertainty of the $d$ (aquamarine), $u$ (purple) and $s$ (pink) coupling measurement as a function of $y_{cut}$. The left plot shows the tagging uncertainty, the central one --- the uncertainty of the acceptance of hadronic events, and the right one the variation of the uncertainty of the acceptance of radiative events. The solid lines indicate the uncertainty of 0.1\%, the dashed lines of 1\%, and the dash-dotted ones of 0.01\%. 0.1\% uncertainty is assumed for all the unvaried contributions, except for the luminosity (fixed at 10$^{-4}$). The integrated luminosity was set to 100\,fb$^{-1}$.}
    \label{fig:5flav_unc}
\end{figure}

In the preceding considerations, all systematic uncertainties, except for the luminosity uncertainty, were assigned a common value. However, it is instructive to relax this assumption and evaluate the individual impact of specific sources of uncertainty. Figure \ref{fig:5flav_unc} shows this comparison by varying the uncertainties of tagging, of the selection efficiency of hadronisation background and of radiative events, respectively. Three values are considered: 0.1\% (treated elsewhere as the ``reference'' value, and used also for non-varied sources of systematic uncertainties in each plot), 0.01\%, and 1\%. 

The tagging uncertainty has a crucial impact on the results, though its estimation remains a non-trivial task, as this topic is rarely addressed in the context of future facilities. Nonetheless, reasonable estimates can be inferred from the current performance of $b$-tagging algorithms. For instance, at ATLAS, the present uncertainty is at the level of a few percent, arising from both statistical limitations and hadronisation uncertainties, including experimental effects, such as jet energy resolution, and Monte Carlo simulation~\cite{ATLAS:2019bwq}. Given the anticipated large amount of hadronic data at future colliders, along with significant advancements in Monte Carlo generators and detector technologies, a 1\% uncertainty is a conservative estimate. An uncertainty of 0.1\% appears optimistic yet achievable, while a 0.01\% uncertainty may be potentially attainable with major improvements in simulation and reconstruction techniques. Nevertheless, to fully explore the physics potential of future $Z$-pole runs at lepton colliders, a deep understanding of jet tagging is mandatory.

The dependence on the uncertainty of the acceptance of hadronic events is rather small and appears around the transition value of $y_{cut}$ (see Figure \ref{fig:y}). The impact of the uncertainty on the acceptance of radiative events is marginal in the considered range. Both sources of uncertainty receive contributions from experimental effects (e.g. photon tagging efficiency, energy resolution) and theoretical predictions (e.g. radiative corrections, hadronisation models). Given the substantial theoretical work undertaken for the corresponding LEP measurements~\cite{Gorishnii:1990vf, Surguladze:1990tg, Kramer:1991yc, Kirkby:1992xm, Kunszt:1992np, Mattig:1993jv}, it might be anticipated that theoretical uncertainties will remain well controlled by the time such measurements are conducted at a future facility. Furthermore, it is assumed that collecting a large dataset will enable effective mitigation of uncertainties associated with hadronisation models. Advances in detection techniques required for employing the particle flow concept are expected to enable high precision in constraining the experimental uncertainty. Therefore, sub-percent levels appear achievable. In any case, the results are not strongly sensitive to the exact values assumed and can be rescaled once more precise information on experimental performance becomes available.

\begin{figure}
    \centering
    \includegraphics[width=0.52\textwidth]{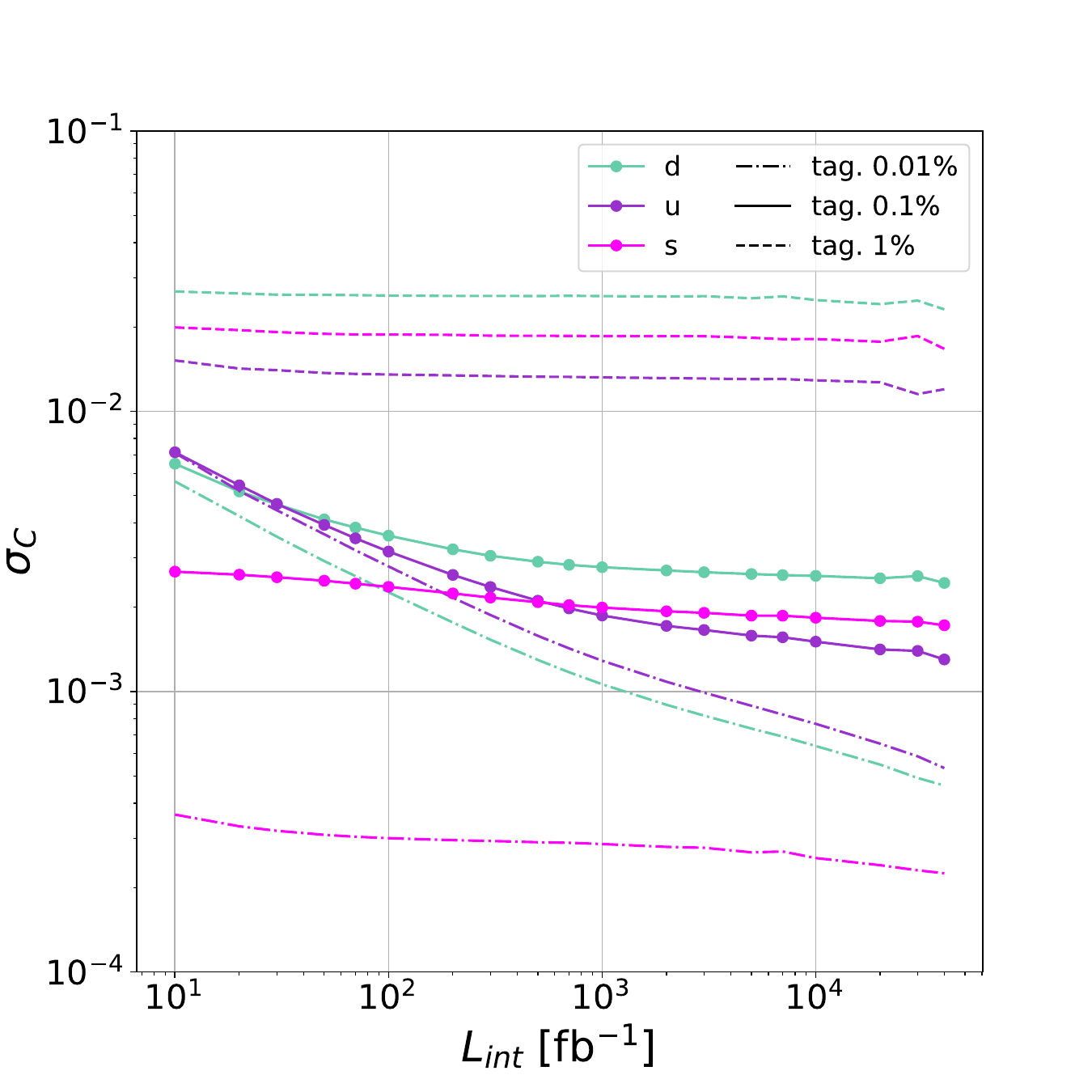}
    
    \caption{The uncertainty of the $d$ (aquamarine), $u$ (purple) and $s$ (pink) coupling measurement as a function of the integrated luminosity. The plot shows the variation of the tagging uncertainty; the solid lines indicate the uncertainty of 0.1\%, the dashed lines of 1\%, and the dash-dotted one of 0.01\%. 0.1\% uncertainty is assumed for all the unvaried contributions, except for the luminosity (fixed at 10$^{-4}$). The $y_{cut}$ parameter was set to 10\,GeV.}
    \label{fig:5flav_unc_lumi}
\end{figure}

In Figure \ref{fig:5flav_unc_lumi}, the results are shown as a function of the luminosity for different values of the tagging uncertainty. The plot clearly shows that to benefit from higher data statistics, sub-permille tagging uncertainty must be achieved. According to this study, sub-percent level precision for the light quark couplings to the $Z$ boson can be achieved at a future experiment at a linear collider collecting 100\,fb$^{-1}$ of data, provided that the tagging uncertainty is reduced below 0.3\%, and the relative uncertainty on the background to the radiative sample from hadronisation photons remains at 1\% or below. The effect of the uncertainty on the radiative event selection efficiency is found to be marginal. Higher integrated luminosities yield an improvement in the precision; for instance, at 40 ab$^{-1}$, corresponding to a single experiment at a circular collider, with the tagging uncertainty of 0.3\%, the results improve by 25-50\%, depending on the quark flavour. By achieving sub-permille precision on the tagging efficiency, the results can improve further, up to a factor of 5-6 for the $u$ and $d$ quarks.

\subsection{Discussion of correlations in the measurement}
\begin{figure}
    \centering
    \includegraphics[width=0.52\textwidth]{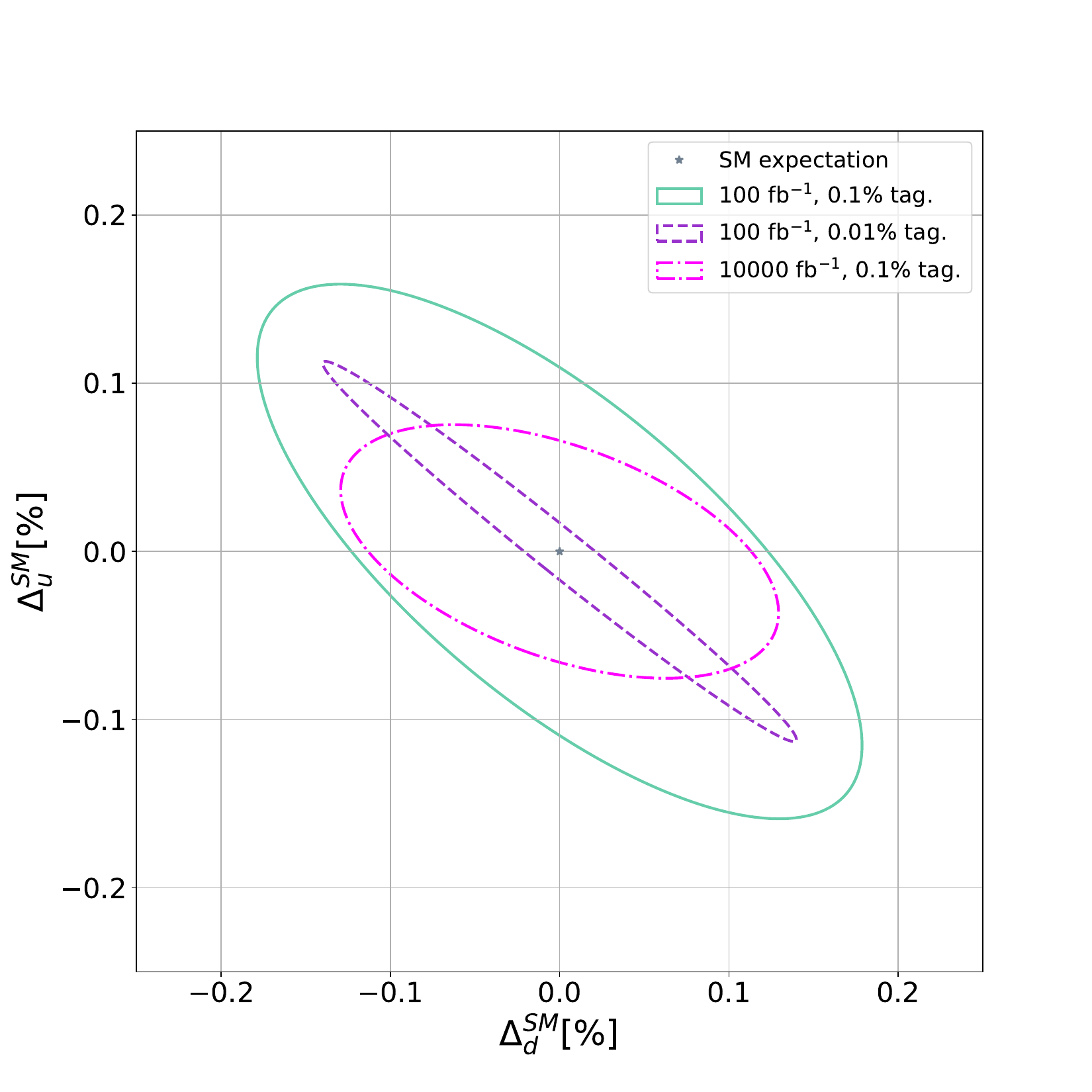}
    
    \caption{The correlation ellipses for the measurement of the $d$ and $u$ couplings for three different cases: 100\,fb$^{-1}$ of collected data with tagging uncertainty of 0.1\% (solid aquamarine), 100\,fb$^{-1}$ of collected data with tagging uncertainty of 0.01\% (dashed purple), 10,000\,fb$^{-1}$ of collected data with tagging uncertainty of 0.1\% (dash-dotted pink), respectively. The deviation from the SM value is shown. 0.1\% uncertainty is assumed for all the other contributions, except for the luminosity (fixed at 10$^{-4}$). The $y_{cut}$ parameter was set to 10\,GeV.}
    \label{fig:eli}
\end{figure}

The measurement aims to disentangle the dependence of the cross sections on the quark couplings, but the measured parameters of the SM are all correlated. By collecting more data of better quality, one can minimise this effect. In Figure \ref{fig:eli}, the correlation ellipses are shown for the $d$ and $u$ quark couplings as their deviation from the SM values. The improvements yielded by increasing luminosity or improving tagging efficiency are compared. Both an improvement in the amount of data collected and a higher tagging efficiency change the sensitivity of this measurement and can significantly enhance the precision.

It is interesting to note that improving the tagging uncertainty affects the uncertainties of both couplings in a similar way, while increasing the integrated luminosity leads to a noticeable improvement mainly for the $u$ quark. The same conclusion can already be drawn from Figure~\ref{fig:5flav_unc_lumi} by inspecting the behaviour of the $u$- and $d$-quark coupling measurements as a function of the assumed tagging uncertainty. The worse results for the $d$ quark at the default tagging uncertainty of 0.1\% most likely stem from confusion with the $s$ quark, which is significantly reduced once the tagging uncertainty is improved.

\section{Summary of the measurement}
Future \epem colliders operating at the $Z$ pole will further constrain the Standard Model parameters and possibly pave the path beyond. Precision measurements of the $Z$ boson couplings to fermions will be possible thanks to very high data statistics. The method to extract the values of the $Z$ couplings proposed in this thesis builds upon the LEP analyses, far exceeding their scope and capabilities, based on up- and down-quarks' different contributions to radiative and non-radiative $Z$ boson decays. However, thanks to the inclusion of the $s$-tagging, the full flavour decomposition is now possible.

For uncertainties of 0.1\%, one can expect a precision of about 0.3\% for the measurement of the light quark couplings to the $Z$ boson, improving upon the statistically limited LEP data by at least an order of magnitude. In general, for integrated luminosities considered for future experiments at \epem colliders, the tagging uncertainty should achieve a precision of about 0.3\% to measure the light quark couplings to the $Z$ bosons at the sub-percent precision. At the same time, the same method allows for constraining the precision of the $b$ ($c$) coupling up to 0.05\% (0.1\%), which yields an improvement of a factor of 5--15 and can potentially be further optimised. Possible extensions of the analysis are highlighted in Chapter~\ref{chap:prospects}.
\chapter{Collinear factorisation and 
Equivalent Vector Boson Approximation}
\label{chap:EVA}

In the previous chapter, photon radiation at low-energy lepton colliders was used as a tool to improve tagging capabilities in the light-quark sector. However, the radiation itself is also of theoretical interest. At sufficiently high energies, well above the electroweak scale, photon emissions from the initial and final states represent only one manifestation of a more general phenomenon of boson radiation; other electroweak bosons can also be emitted and may subsequently participate in the hard-scattering process.

An important feature of physics predictions at the LHC is the ability to disentangle different components of an interaction by applying --- in the relevant energy regimes --- the concept of \textit{factorisation}. Since protons are composite particles and the underlying interaction occurs at the partonic level, the factorisation leads to the definition of PDFs, which describe the probability of finding a given parton with specified kinematic properties inside the proton, before it participates in the hard interaction described by a proper matrix element. The concept is inherently connected to the separation of scales in the underlying subprocesses — in this case, the confinement scale and the scale of the hard process. A similar concept can be applied to high-energy lepton colliders. Although leptons are elementary, point-like particles, they can be effectively described as states composed of various constituents arising from initial-state radiation of electroweak bosons. For example, if a process initiated by a photon emitted collinearly from a charged lepton is considered, the interaction can be described by convoluting a universal structure function encoding the probability of finding a photon \textit{inside} the lepton with the matrix element of the corresponding photon-initiated hard process.

This chapter provides an overview of such processes at leading order, which must be carefully studied to obtain reliable predictions for the discussed Muon Collider. The framework of collinear factorisation for lepton colliders is introduced in the context of the Equivalent Vector Boson Approximation, its Monte Carlo implementation is presented, and the range of applicability of the approach is examined~\cite{Dahlen:2025udl}.

\section{Historical remarks}

The concept of collinear factorisation discussed in this chapter dates back to the first half of the 20th century. This approach was first proposed for photons and is known as the \textit{Equivalent Photon Approximation} (EPA) or the \textit{Weizs\"acker-Williams Approximation}~\cite{Fermi:1924tc, vonWeizsacker:1934nji, Williams:1935dka}. Over the years, the EPA has been broadly discussed in the literature to assess its applicability to collider experiments and establish a set of conditions for its validity~\cite{Brodsky:1971ud,Budnev:1975poe,Frixione:1993yw,Engel:1995yda,Vysotskii:2018eic}. 

In the 1980s, the approximation was extended to the weak-boson case. The so-called \textit{Equivalent Vector Boson Approximation} (EVA)~\cite{Cahn:1983ip,Dawson:1984gx,Kane:1984bb} is based on the fact that the masses of weak bosons can be viewed as negligible for collision energies well above the electroweak scale, even though a truly collinear kinematics is never realised due to the finite masses. The EVA is much more complex than the EPA; the structure functions depend on the isospin of the radiating fermions and differ for transverse and longitudinal polarisations. For the last four decades, the approximation has been studied from different perspectives, often yielding indefinite conclusions regarding its validity~\cite{Lindfors:1985yp,Gunion:1986gm,Kleiss:1986xp,Kunszt:1987tk,Johnson:1987tj,Abbasabadi:1988ja,Kuss:1995yv,Kuss:1996ww,Accomando:2006mc,Borel:2012by,Costantini:2020stv,Ruiz:2021tdt,Bigaran:2025rvb,Frixione:2025guf,Frixione:2025wsv,Dittmaier:2025htf}. 

As a remark, it should be noted that the fundamental difference between the EPA and EVA lies in the massiveness of the weak bosons. This distinction changes the motivation for introducing a structure-function formalism. In the case of EPA, the numerical stability of the full matrix-element description becomes challenging at low momentum transfer. The approximation is therefore introduced to capture the problematic region of phase space where the photon is nearly on shell. Since there is a large (infinite) hierarchy of scales between the vanishing photon mass and the collision scale, EPA is naturally restricted to the low-momentum-transfer regime, while full matrix-element calculations are employed at higher momentum transfer, and matching between the two is necessary.

In contrast, EVA was originally developed to describe VBF processes at a time when full $2\to 4$ matrix-element calculations were computationally prohibitive. It therefore served as an approximate alternative rather than as a component to be matched. The approximation was expected to provide a reasonable description over the full phase space, noting that the hierarchy between the electroweak scale and the collider energy is much less pronounced than in the photon case. With modern Monte Carlo tools, however, VBF topologies can now be computed straightforwardly using full matrix-element calculations, and the original motivation for EVA has largely disappeared. Nevertheless, EVA remains in use because of its simplicity and computational efficiency; in practical applications, an order-of-magnitude speed-up compared to full matrix-element calculations can be observed. In the following, a wide set of processes will be discussed to assess in which cases the approximation is justified.

Both approaches rely on identifying regions of phase space that produce logarithmic enhancements from collinear splittings in the initial state. In this regime, processes initiated by electroweak bosons emitted from the incoming leptons become non-negligible and can eventually dominate over standard annihilation channels. The transition typically occurs at energies of a few TeV, depending on the specific process. This makes such studies particularly relevant today, as future facilities discussed in the current update of the ESPP are designed to reach the multi-TeV scale. This is the primary reason for revisiting the EVA framework and reconsidering its applicability.

\section{Derivation of the EVA structure functions}
In this section, the derivation of the EVA structure functions is presented in a formalism that resembles its implementation and usage in \textsc{Whizard}. The derivation follows to some extent the ideas of \cite{Kilian:2003pc,Ruiz:2021tdt}, but several different choices are made.

\subsection{Matrix-element picture}

\begin{figure}
	\centering
	\includegraphics[width=0.5\linewidth]{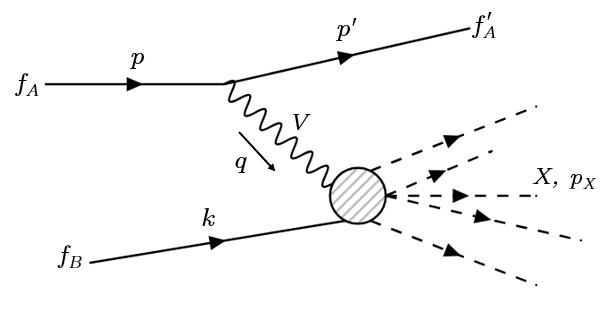}
	\caption{Prototype diagram in the EW deep inelastic scattering (DIS) picture as the starting point for the derivation of the EVA.}
	\label{fig:derivdiag}
\end{figure}

To derive the EVA structure functions, one may consider a simple process which starts with the emission of a single weak boson\footnote{The derivation for photon emission follows a similar approach but is simpler, because the photon is massless, and can be easily found in standard textbooks~\cite{Peskin:1995ev}.}. Suppose a weak boson $V$ is emitted from an incoming fermion $f_A$ with momentum $p$, as depicted in Figure \ref{fig:derivdiag}. After emitting the boson, the fermion becomes $f^{\prime}_A$ and carries momentum $p^{\prime}$. The emitted boson $V$, which has momentum $q$, interacts with another incoming particle $f_B$ with momentum $k$. This interaction produces a final state $X$, which carries momentum $p_X$. The exact form of the interaction $f_B V \to X$ is irrelevant.

In general, the matrix element of this process can be written as

\begin{equation*}
\mathcal{M}_\textnormal{full} = \mathcal{M}^\mu (f_A \to V f^{\prime}_A) \Delta^V_{\mu\nu}(q) \mathcal{M}^\nu (f_B V \to X)\,,
\end{equation*}
where $\mathcal{M}^\mu (f_A \to V f^{\prime}_A)$ is associated with the splitting $f_A \to V f^{\prime}_A$, $\Delta^V_{\mu\nu}(q)$ is the vector-boson propagator, and $\mathcal{M}^\nu (f_B V \to X)$ corresponds to the hard scattering $f_B V \to X$. As a side note, care should be taken to treat the polarisation vectors of the intermediate vector boson correctly, i.e. either by including them as external legs in the emission and hard-scattering subprocesses, or by treating them as part of the propagator. More specifically, in the latter case, the propagator of the vector boson $V$ reads 
\begin{equation*}
     \Delta_{\mu\nu}^V(q) = \frac{d^G_{\mu \nu} (q)}{q^2-m_V^2}\,,
\end{equation*}
where $d^G_{\mu \nu} (k)$ is the propagator numerator in a given gauge $G$ (e.g. $R_\xi$-gauge or axial gauge) and $m_V$ is the vector-boson mass~\cite{Peskin:1995ev}. At the same time, the numerator might be rewritten as a polarisation sum:
\begin{equation*}
	 d^G_{\mu \nu} (q) =  \sum_{\lambda^{\prime}, \lambda^{\prime\prime}} \varepsilon^G_\mu(q,\lambda^{\prime})^* \varepsilon^G_\nu(q,\lambda^{\prime\prime})\,.
\end{equation*}
In this way, $\varepsilon^G_\mu(q,\lambda^{\prime})^*$ is associated with the splitting process $f_A \to V f^{\prime}_A$ and $\varepsilon^G_\nu(q,\lambda^{\prime\prime})$ with the hard scattering $f_B V \to X$.

The first step of the approximation is to discard the off-diagonal spin-correlations between the vector bosons in the internal line, i.e. it is assumed that $\lambda^{\prime} = \lambda^{\prime\prime} \equiv \lambda$.
Depending on the gauge, the sum goes over transverse ($\lambda = \pm1$), longitudinal ($\lambda = 0$), and, possibly, also unphysical (scalar) polarisations ($\lambda = S$).

Assuming that only one polarisation state, $\lambda = \lambda_V$, contributes to the hard process, one can write
\begin{equation}
 \mathcal{M}^\mu (f_A \to V f^{\prime}_A) \Delta^V_{\mu\nu}(q) = \bar{u}(p^{\prime}, \lambda_{f^{\prime}_A}) \gamma^\mu (g_V - g_A \gamma^5) u(p, \lambda_{f_A}) \frac{\varepsilon_G^\mu(q,\lambda_V)^* \varepsilon_G^\nu(q,\lambda_V)}{q^2-m_V^2}\,, 
 \label{eq:general_form}
\end{equation}
where the standard spinor formalism, outlined in Section~\ref{sec:calculationsSM}, is used, and $g_V$ and $g_A$ are, respectively, the vector and axial couplings of a given boson $V$. Due to the commutativity, the polarisation vector $\varepsilon_G^\nu(q,\lambda_V)$ might be associated with the expression for the hard process $f_B V \to X$. To construct structure functions for the vector-boson emission accounting for particle polarisation, one needs to find explicit expressions for the helicity-dependent amplitudes from Eq.~\ref{eq:general_form}.

\subsection{EVA kinematics}
For simplicity, suppose the massless particles $f_A$ and $f_B$ collide collinearly in the $z$-direction, and their energies are equal. Then, the kinematics is given by:
\begin{align*}
p^\mu &= (E,\, 0,\, 0,\, E)\,, \\
k^\mu &= (E,\, 0,\, 0,\, -E)\,, \\
p^{\prime\mu} &= \left(E^{\prime},\, E^{\prime}\sin\theta \cos\phi,\, E^{\prime} \sin\theta \sin\phi,\, E^{\prime} \cos\theta\right)\,, \\
q^\mu &= p^\mu - p^{\prime\mu} = \left(E - E^{\prime},\, -E^{\prime} \sin\theta \cos\phi,\, -E^{\prime} \sin\theta \sin\phi,\, E - E' \cos\theta\right)\,.
\end{align*}
$\theta$ is the angle between the collision axis and the outgoing fermion $f_A^{\prime}$, and $\phi$ is its azimuthal angle. Furthermore, let $x$ be the fraction of the initial energy carried by the vector boson, i.e. $x = \frac{E-E^{\prime}}{E}$ (the \textit{Bjorken x}).

Suppose that $f_A^{\prime}$ and V are emitted at very small angles in the splitting, i.e. their transverse momenta are small as compared to the collision energy. In the collinear limit, assuming $\phi = 0$, one can expand 
\begin{equation*}
E'\sin\theta = E'\theta + \mathcal{O}(\theta^2)\,.
\end{equation*}
Let $p_\perp \equiv E'\theta = E(1-x)\theta$ (the transverse momentum of $f_A^{\prime}$ at leading order in $\theta$). Then, by expanding all the functions of $\theta$ to leading order in $\theta$ and replacing $\theta \to \frac{p_\perp}{E(1-x)}$, the momenta read:
\begin{align*}
p^{\prime\mu} &= \left((1-x)E,\, p_\perp,\, 0,\, (1-x)E - \frac{p_\perp^2}{2(1-x)E}\right)\,, \\
q^\mu &= \left(xE,\, -p_\perp,\, 0,\, Ex + \frac{p_\perp^2}{2(1-x)E}\right)\,.
\end{align*}
Following the same procedure for the polarisation vectors of Eqs.~\eqref{eq:polpl}--\eqref{eq:pollong}, one obtains:
\begin{align*}
\varepsilon^\mu(q, +1) &= -\frac{1}{\sqrt{2}} \left(0,\, 1,\, i,\, \frac{p_\perp}{xE}  \right)\,, \\
\varepsilon^\mu(q, -1) &= -\frac{1}{\sqrt{2}} \left(0,\, 1,\, -i,\, \frac{p_\perp}{xE}  \right)\,, \\
\varepsilon^\mu(q, 0) &= \frac{m_V}{\sqrt{2xE}} \left(-1,\, 0,\, 0,\, 1  \right)\,.
\end{align*}
An additional step is carried out beforehand for the longitudinal polarisation, for which the vector can be decomposed as
\begin{equation*}
    \varepsilon^\mu (q,0) = \frac{q^\mu}{m_V} + \tilde{\varepsilon}^\mu (q)\,.
\end{equation*}
The first term above vanishes via the Dirac equation and hence is neglected in this derivation~\cite{Dawson:1984gx}.

In this approximation, the amplitudes are given by 

\begin{center}
\begin{tabular}{c|c|c}
 & $\lambda_{f_A}=-1,\,\lambda_{f^{\prime}_A}=-1$ & $\lambda_{f_A}=+1,\,\lambda_{f^{\prime}_A}=+1$ \\[3pt] \hline
$\lambda_V = +1$ &
$\dfrac{g_L\,p_\perp\,\sqrt{2-2x}}{x}$ &
$-\dfrac{\sqrt{2}g_R\,\,p_\perp}{\sqrt{1-x}\,x}$ 
\\[15pt]
$\lambda_V = -1$ &
$\dfrac{\sqrt{2}\,g_L\,p_\perp}{\sqrt{1-x}\,x}$ &
$-\dfrac{g_R\,p_\perp\,\sqrt{2-2x}}{x}$
\\[15pt]
$\lambda_V = 0$ &
$-\dfrac{2\,g_L\,m_V\,\sqrt{1-x}}{x}$ &
$\dfrac{2\,g_R\,m_V\,\sqrt{1-x}}{x}$
\end{tabular}
\end{center}
Here, the left- and right-handed coupling constants, $g_L = g_V + g_A$ and $g_R = g_V - g_A$, were introduced. It is important to mention that no helicity flip is possible for massless fermions at leading order in EW coupling, i.e. $\lambda_{f_A} = \lambda_{f^{\prime}_A}$ or the amplitude vanishes.

\subsection{Phase-space integration}
The differential cross section for the process depicted in Fig.~\ref{fig:derivdiag} might be written as

\begin{equation}
\label{eq:diffxs}
\dd\sigma = \frac{x}{2s} \sum_{\lambda} \overline{\left|\mathcal{M}_\lambda(f_A \to f_A^\prime V_\lambda)\right|^2} \frac{1}{\left( q^2 - m_V^2\right)^2}  \overline{\left|\mathcal{M}_\lambda(f_B V_\lambda \to X)\right|^2} \dd PS\, ,
\end{equation} 
where $\dd PS$ is the phase space measure given by 
\begin{equation*}
\dd PS = \frac{\dd^3 p^\prime}{(2\pi)^3 2E^\prime}\frac{\dd^3 p_X}{(2\pi)^3 2E_X} (2\pi)^4 \delta (p+k-p^\prime - p_X).
\end{equation*} 
The factor $x/2s$ is the flux factor ($s = 4E^2$) multiplied by $x$, which comes from the boost of the splitting frame along the $z$-axis to the frame of the hard process. $\mathcal{M}_\lambda$ denotes a matrix element contracted with its corresponding polarisation vector for the emitted/absorbed vector boson, and the overline indicates averaging over initial-state spins and summing over final-state quantum numbers (while keeping the vector-boson polarisation explicit).

The following factor can be expressed in terms of $\bar{x} = 1-x$, $p_\perp$ and the azimuthal angle $\phi$:
\begin{equation*}
    \frac{\dd^3 p^\prime}{(2\pi)^3 2E^\prime} \frac{1}{(q^2 - m_V^2)^2}
= \frac{\dd x\, \dd p_\perp^2\, d\phi}{4(2\pi)^3} \frac{\bar{x}}{(p_\perp^2 - \bar{x} m_V^2)^2} \, .
\end{equation*}
Then, integrating Eq.~\ref{eq:diffxs}, one obtains:
\begin{align*}
\sigma(f_A f_B &\to f_A' X) = \\ 
&= \sum_{\lambda} \int_{\xmin}^1 dx \int_0^{\ptmax^2} \frac{dp_\perp^2}{16\pi^2} \cdot \frac{x \bar{x}}{(p_\perp^2 - \bar{x} m_V^2)^2}
\overline{\left|\mathcal{M}_\lambda(f_A \to f_A^\prime V_\lambda)\right|^2} \hat{\sigma}(f_B V_{\lambda} \to X)\,.
\end{align*}
Here, the trivial azimuthal-angle integration has been performed, as well as the integration over $p_X$, yielding the hard-scattering cross section for $f_B V_{\lambda} \to X$. This factorisation of phase-space integrations is analogous to electroweak factorisation in QCD or QED. 

The minimal sensible value of $\xmin$ is $m_V/E$, which corresponds to the minimal energy fraction to produce the boson $V$ on-shell. The maximal possible value of $\ptmax$ is $E$. At the same time, the allowed emission angle is restricted by on-shellness and four-momentum conservation within the collinear approximation.

From the above equation, one can identify the EVA structure functions as:
\begin{equation*}
    F_{\lambda_V}(x,\ptmax^2) =  \int_0^{p^2_{\perp,max}} \frac{dp_\perp^2}{16\pi^2} \cdot \frac{x \bar{x}}{(p_\perp^2 - \bar{x} m_V^2)^2}
\overline{\left|\mathcal{M}_\lambda(f_A \to f_A^\prime V_{\lambda_V})\right|^2}\, .
\end{equation*}
They can be directly computed, yielding:
\begin{equation}
	\label{eq:strucfunc}
	\begin{aligned}
            F_+(x, \ptmax^2) &= \frac{g_R^2 + g_L^2 \xbar^2}{4\pi^2 x}
		\left[
		\ln\left( \frac{\ptmax^2 + \xbar m_V^2}{\xbar m_V^2} \right)
		- \frac{\ptmax^2}{\ptmax^2 + \xbar m_V^2}
		\right], \\
		F_-(x, \ptmax^2) &= \frac{g_L^2 + g_R^2 \xbar^2}{4\pi^2 x} \left[
		\ln\left( \frac{\ptmax^2 + \xbar m_V^2}{\xbar m_V^2} \right)
		- \frac{\ptmax^2}{\ptmax^2 + \xbar m_V^2}
		\right], \\
		F_0(x, \ptmax^2) &= \frac{(g_L^2 + g_R^2) \xbar}{2\pi^2 x} \frac{\ptmax^2}{\ptmax^2 + \xbar m_V^2}.
	\end{aligned}
\end{equation}

\begin{figure}
	\centering
		\includegraphics[width=0.6\textwidth]{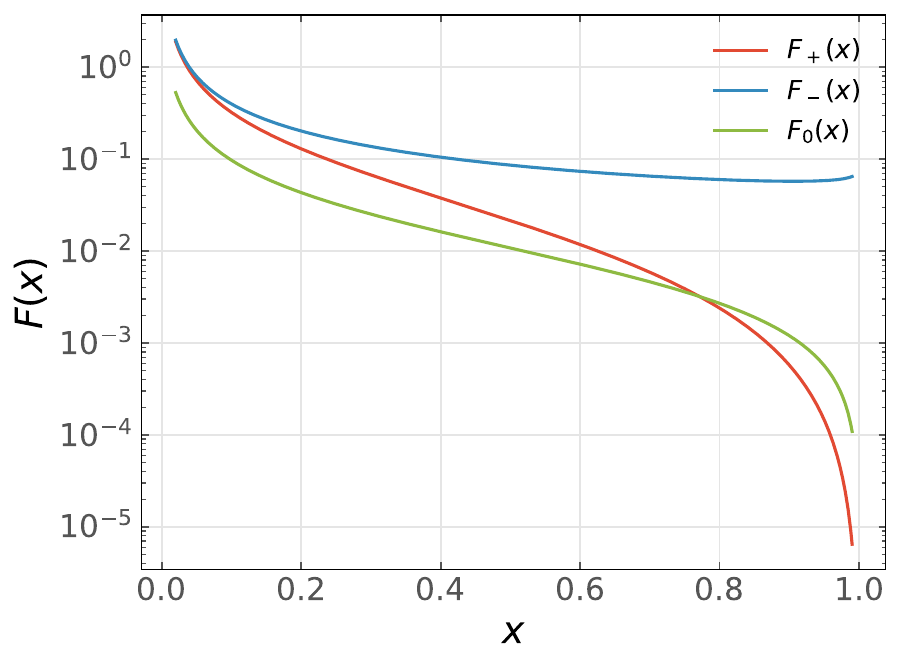}

	\caption{The EVA structure functions for the $W$ bosons. The SM values of the parameters are used, and the \ptmax parameter is set to 5 TeV.}
	\label{fig:sf}
\end{figure}

The values of the structure functions $F$ for the $W$ boson are shown in Figure \ref{fig:sf}, assuming \ptmax of 5 TeV.
All three structure functions depend on the scale choice \ptmax, but only the transverse ones entail a logarithmic enhancement. Overall, the approximation also depends on the minimum energy fraction $\xmin$ carried by the vector boson, appearing in the integration of the structure function convolved with the hard-process cross section. One of the goals of the analysis is to study the influence of these parameters on the quality of the EVA in comparison to full matrix-element results. From physical considerations, $\xmin = m_V/E$ is needed to have enough energy for the production of an on-shell vector boson, and $\ptmax \ll E_A$ is required for the collinear (or at least \textit{small-angle}) approximation to remain valid.
Nevertheless, in the following, different choices will be discussed.

\section{EVA in Whizard}
As mentioned in Section~\ref{sec:Whizard}, \textsc{Whizard} originates from precision electroweak studies; interestingly, difficulties with the EVA description in~\cite{Boos:1997gw,Boos:1999kj} triggered the first version of the code allowing for the direct computation of the $2\to 4$ processes. In 2008, the EVA framework was implemented in Version 1.91. However, in 2010, \textsc{Whizard 2} was released with a major overhaul, primarily to support the LHC physics. The EVA framework was reimplemented in 2014, but the treatment of polarisation was simplified by averaging over the helicity states of the incoming partons. As part of this thesis, the implementation was corrected and revalidated, with proper inclusion of beam polarisation. The updated implementation was released in Version 3.1.7.

In the current \whizard\ implementation, the EVA mode \verb|default| (set by the variable \verb|$ewa_mode|) incorporates the full leading power structure functions from Eq.~\eqref{eq:strucfunc}. Additionally, two other modes with different choices of approximation are included to compare with previous studies:
\begin{itemize}
\item EVA mode \verb|log_pt| for which:
\begin{equation*}
	\begin{aligned}
		G_+(x, \ptmax^2) &= \frac{g_R^2 + g_L^2 \xbar^2}{16\pi^2 x}
		\ln\left( \frac{\ptmax^2}{\xbar m_V^2} \right), \\
		G_-(x, \ptmax^2) &= \frac{g_L^2 + g_R^2 \xbar^2}{16\pi^2 x} 
		\ln\left( \frac{\ptmax^2}{\xbar m_V^2} \right), \\
		G_0(x, \ptmax^2) &= \frac{(g_L^2 + g_R^2) \xbar}{8\pi^2 x},
	\end{aligned}
\end{equation*}
\item EVA mode \verb|log| for which:
\begin{equation*}
	\begin{aligned}
		H_+(x, \ptmax^2) &= \frac{g_R^2 + g_L^2 \xbar^2}{16\pi^2 x}
		\ln\left( \frac{\ptmax^2}{m_V^2} \right), \\
		H_-(x, \ptmax^2) &= \frac{g_L^2 + g_R^2 \xbar^2}{16\pi^2 x} 
		\ln\left( \frac{\ptmax^2}{m_V^2} \right), \\
		H_0(x, \ptmax^2) &= \frac{(g_L^2 + g_R^2) \xbar}{8\pi^2 x}.
	\end{aligned}
\end{equation*}
\end{itemize}

In \textsc{Whizard}, the EVA is a single-beam structure function, which means it can also be applied to asymmetric beam setups. The code allows users to specify the minimal value of $x$ and the maximal value of the scale $\ptmax^2$ (either as fixed constants or as dynamical expressions, e.g., the hard-process scale, $\hat{s}$). The \verb|legacy| mode was introduced to facilitate comparisons with Version 2. An example SINDARIN input file for running EVA in \textsc{Whizard} is provided in Appendix~\ref{app:sinEVA}.

\section{EVA and full matrix-element calculations}
\label{sec:EVAfullME}
In this section, the application of the EVA to a set of processes is discussed to assess under which conditions the description provided by the EVA can reliably emulate the full calculation. The processes are organised according to complexity, understood in terms of both the contributing polarisations --- where the first processes are dominated by a single vector-boson polarisation --- and the increasing multiplicity of topologies constituting the full process. The default parameter settings are $\xmin = 2 m_V /\sqrt{s}$ for the lower Bjorken cutoff, while the central value of the momentum scale is $\ptmax = \sqrt{\hat{s}}/4$. The latter is heuristically chosen, as it yields good agreement for most processes and will be discussed in more detail in the following. If not stated otherwise, the results are presented for a fictitious electron--muon collider operating at 10~TeV to facilitate the technical comparison between full ME calculations and the EVA by removing $s$-channel contributions. 

As a general remark, it is worth emphasising that the EVA can only describe well phase space regions that are dominated by VBF-like topologies. For most processes, additional diagrams, such as $s$-channel annihilation or bremsstrahlung, give significant contributions to the full results. Unfortunately, a separation into different sets of topologies and a diagram-by-diagram comparison between the full calculation and the approximation is not possible due to large gauge cancellations between the different topologies in the full matrix-element case~\cite{Boos:1997gw,Boos:1999kj,Accomando:2006mc}. While these gauge cancellations can be partially separated on a diagram-by-diagram basis, for instance, by working in an axial gauge, they cannot be ignored; a diagram-based selection within the full process is never physically meaningful. Therefore, non-diagonal beam flavour combinations are considered, and technical cuts are applied in order to extract the VBF-like contributions for which the EVA should yield reasonable results, which can then be relaxed for more realistic beam setups and viable physical cuts.

\subsection{Di-Higgs}
\begin{figure}
	\centering
	\begin{subfigure}{0.49\textwidth}
		\centering
		\includegraphics[width=\textwidth]{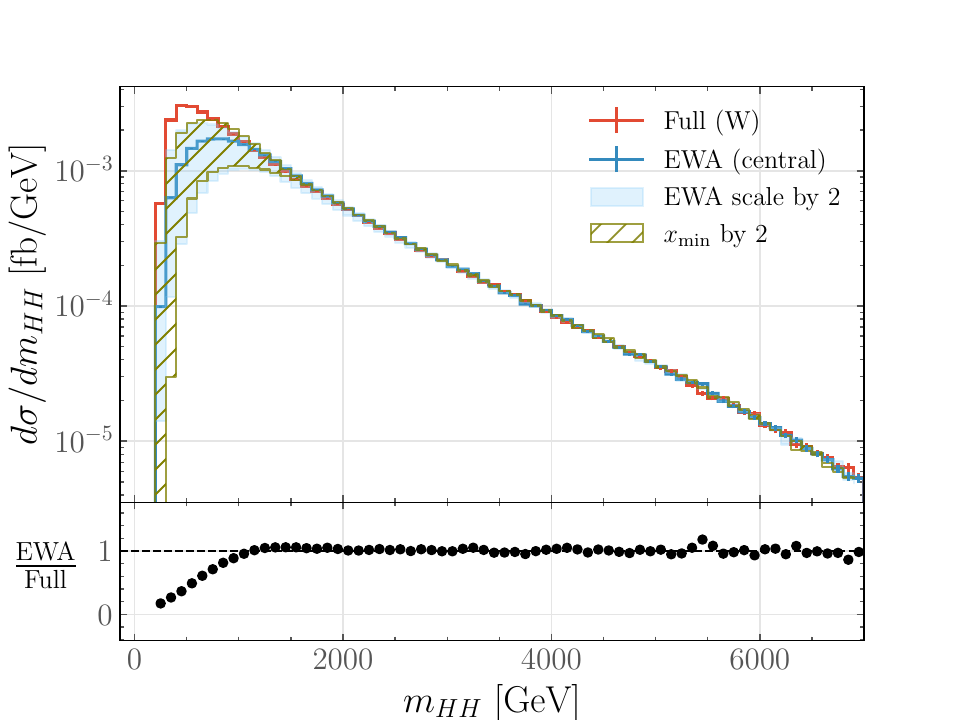}
		\caption{EWA}
	\end{subfigure}
	\hfill
	\begin{subfigure}{0.49\textwidth}
		\centering
		\includegraphics[width=\textwidth]{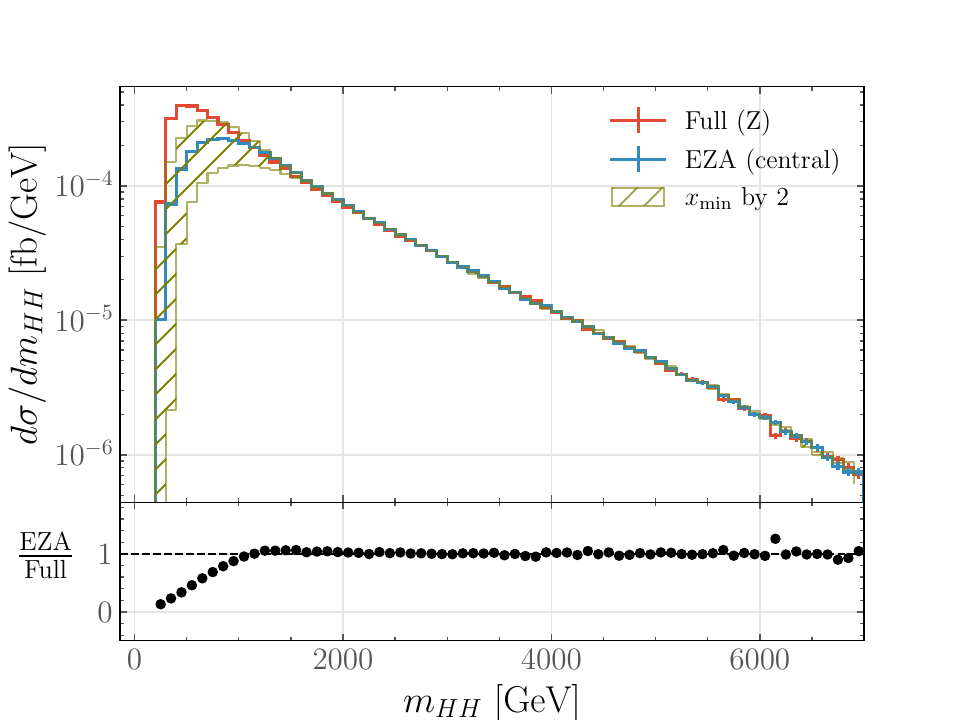}
		\caption{EZA}
	\end{subfigure}
	\caption{Invariant mass distributions of the di-Higgs system in the process $e^+ \mu^- \to HH + X$ (left: charged current, right: neutral current) at $\sqrt{s} = 10$\,TeV for the full matrix element evaluation (\textit{red}) and the EVA for different values of $\xmin$ and \ptmax, respectively. The \textit{green} band represents a two-sided variation of $x_{min} = 2m_V/\sqrt{s}$ by a factor of two. The \textit{blue} band in the left panel represents a scale variation by a factor of two around the central scale  $\ptmax = \sqrt{\hat{s}}/4$. To facilitate the comparison between the two effects, the scale variation is not shown in the right plot.}
	\label{fig:HH}
\end{figure}

The first process of interest is $e^+ \mu^- \to HH + (\bar{\nu}_e \nu_\mu \text{ or } e^+ \mu^-)$, depending on which vector bosons mediate the process. Figure~\ref{fig:HH} shows the invariant mass of the di-Higgs system. Here, the EVA is considered in two incarnations: as the Equivalent \textit{W-boson} Approximation (EWA) and as the Equivalent \textit{Z-boson} Approximation (EZA). Here, no generator-level cuts are applied. In this case, by using beam leptons of different flavours and ignoring light-lepton Yukawa couplings, all topologies contributing to the process are of the VBF type, and one can therefore expect them to be well within the realm of the EVA. Nevertheless, the approximation always assumes the vector bosons to be on shell, which suggests that it cannot accurately describe regions of phase space in which one of the vector bosons is far off shell, for example in configurations where two Higgs bosons can be interpreted as emitted from a single vector-boson line. Such emissions are expected to be predominantly collinear, thereby driving the invariant mass of the Higgs pair towards lower values.

Indeed, the EVA agrees very well with full ME calculations for large values of $m_{HH} \gtrsim 1.5\, \text{TeV}$, independently of the parameter choice. On the other hand, the threshold region --- where the major part of the total cross section comes from  --- heavily depends on the choice of both \xmin and \ptmax. The weak dependence on \ptmax in the tails follows from the fact that the main contribution ($\sim 99\%$) for this process occurs through the longitudinal structure function $F_0$ of Eq.~\ref{eq:strucfunc}, which is not logarithmically enhanced and for which the scale factor becomes approximately constant for $\bar{x} m_V^2 \lesssim \ptmax^2$, i.e. in the high-energy regime where $\bar{x} \to 0$.

The dependence on $x_\text{min}$ is obviously the strongest in the low invariant mass regime, where the Bjorken-$x$ values are small for each beam. As explained earlier, a natural choice for this variable is $x_\text{min} = 2 m_V/\sqrt{s}$, because this represents the minimum energy fraction to produce a vector boson $V$ off a single beam. In Figure \ref{fig:HH}, a variation around this value by a factor of two up and down is shown. It becomes apparent that the naïve choice is not optimal. Instead, one can artificially enhance the population of the EVA phase space in the low-$x$ region by choosing smaller values for $x_\text{min}$. Nevertheless, there does not seem to be a unique choice for $x_\text{min}$ which works for all processes. As the approximation performs well for $m_{HH} > 1.5~\text{TeV}$, an invariant mass cut could be applied to remove the threshold region. Below this value, one must rely on full matrix-element calculations.

\begin{figure}
	\centering
	\includegraphics[width=0.69\textwidth]{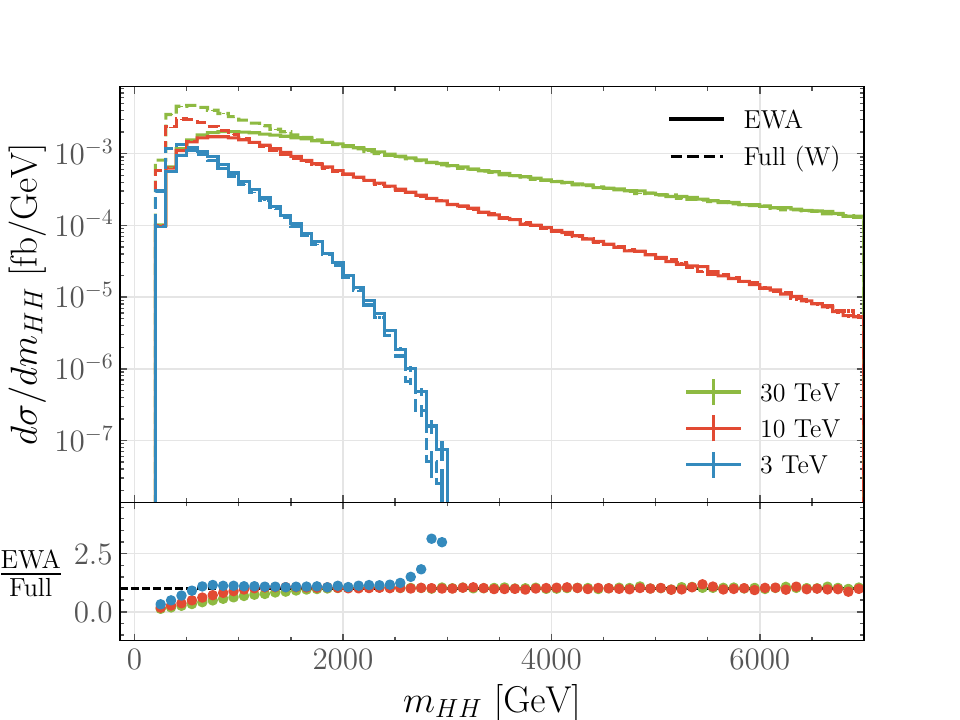}
	\caption{Invariant mass distributions of the di-Higgs system in $e^+ \mu^- \to HH + X$ for different collision energies.}
	\label{fig:HH_energies}
\end{figure}

In Figure \ref{fig:HH_energies}, the di-Higgs invariant mass distributions for three different centre-of-mass energies are shown. Two notable effects emerge: first, the threshold cut-off --- beyond which the EVA and full results begin to align --- shifts to higher values as the centre-of-mass energy increases; the ratio between the two results slowly approaches one and then stabilises. The higher the $\sqrt{s}$, the higher this point moves up in the invariant mass spectrum. Second, a sharp rise in the ratio is visible at the tail end of the 3-TeV distribution. This points to an unphysical divergence of the transverse structure functions when $x\to 1$, an artifact of the collinear approximation in the EVA derivation. The effect can be mitigated by introducing an upper cut-off, e.g. $x_\text{max} \sim 1 - \frac{2 m_V}{\sqrt{s}}$. In any case, this sharp discrepancy from the full matrix element is not concerning for the total cross section, because it only appears at the far end of the distribution, where the differential cross section is tiny.

\begin{figure}
	\centering
	\begin{subfigure}{0.49\textwidth}
		\centering
		\includegraphics[width=\textwidth]{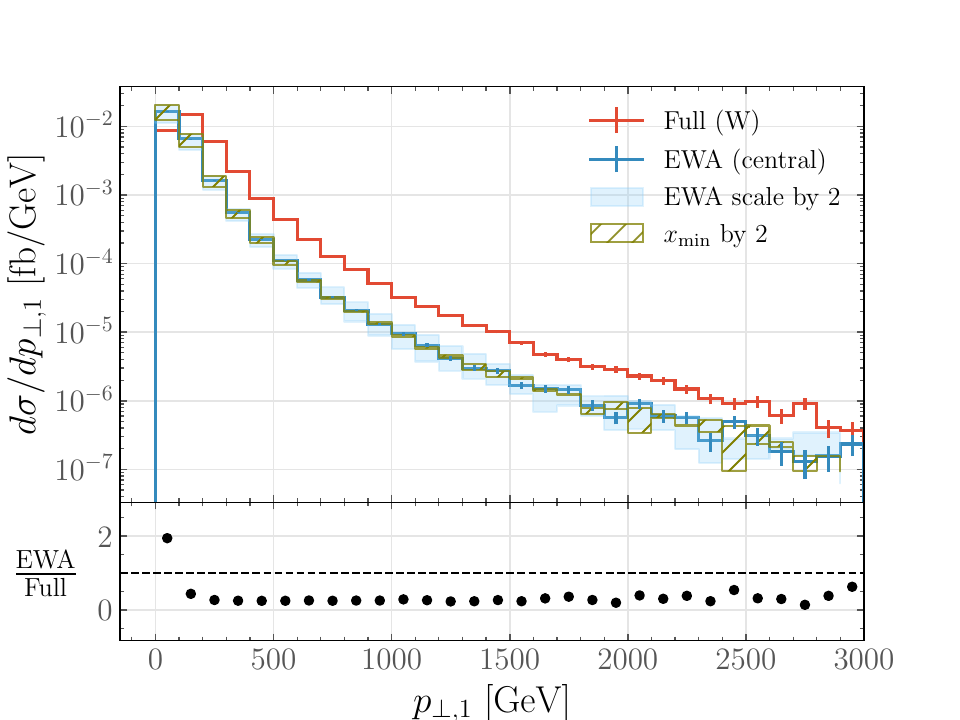}
		\caption{leading $p_\perp$}
	\end{subfigure}
	\hfill
	\begin{subfigure}{0.49\textwidth}
		\centering
		\includegraphics[width=\textwidth]{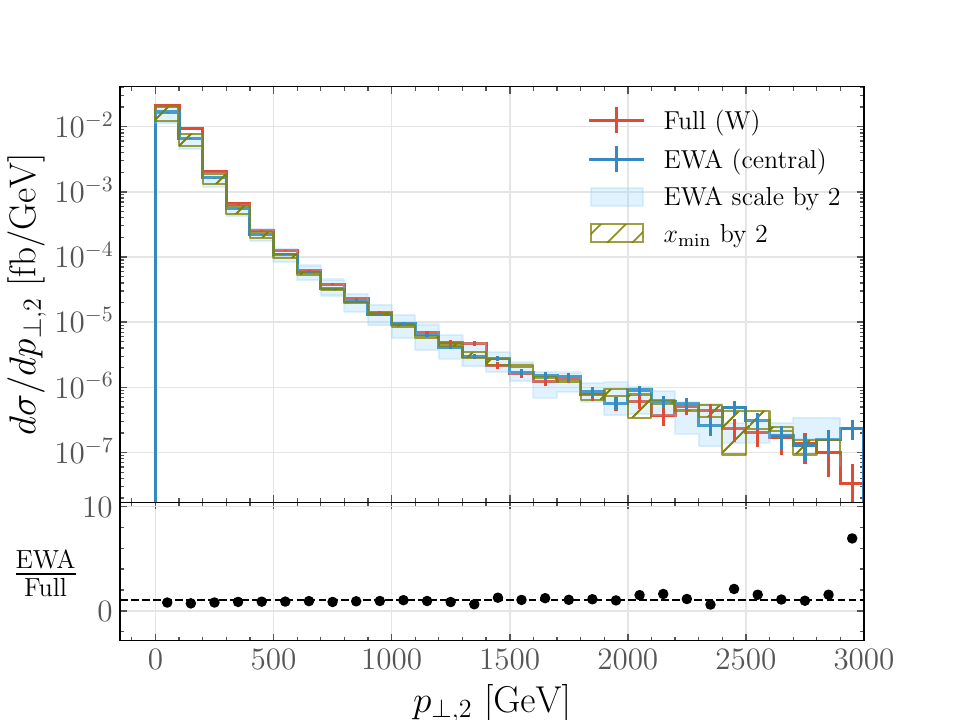}
		\caption{sub-leading $p_\perp$}
	\end{subfigure}
	\caption{Transverse momentum distributions for the leading-Higgs $p_\perp$ (left) and the sub-leading  $p_\perp$ (right) in $e^+ \mu^- \to HH + X$ at $\sqrt{s} = 10$\,TeV for the full matrix element evaluation (\textit{red}) and the EWA (\textit{blue}) for different values of $\xmin$ and \ptmax. The \textit{blue} band represents a scale variation by a factor of two around the central scale  $\ptmax = \sqrt{\hat{s}}/4$ and the \textit{green} one around  $x_\text{min} = 2 m_V/\sqrt{s}$ at this central scale.}
	\label{fig:HH-pT}
\end{figure}

\begin{figure}
	\centering
    \begin{subfigure}{0.49\textwidth}
		\centering
		\includegraphics[width=\textwidth]{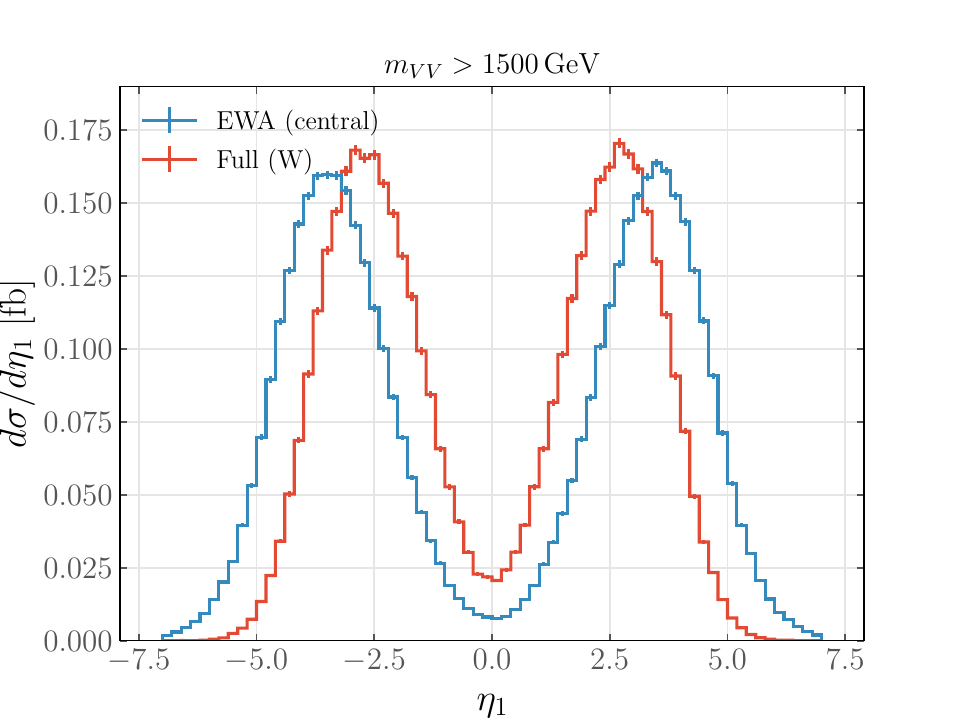}
		\caption{leading $p_\perp$}
	\end{subfigure}
	\hfill
	\begin{subfigure}{0.49\textwidth}
		\centering
		\includegraphics[width=\textwidth]{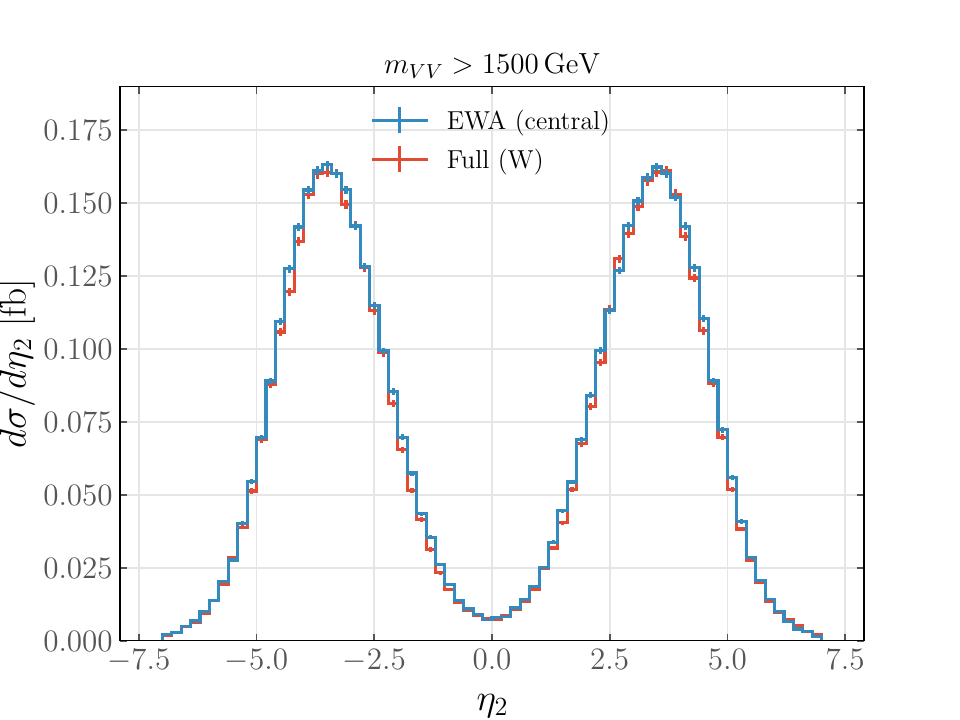}
		\caption{sub-leading $p_\perp$}
	\end{subfigure}
	\caption{Pseudorapidity distributions for the leading (left) and sub-leading Higgs in $p_\perp$ (right) for $e^+ \mu^- \to HH + X$ for $\sqrt{s} = 10$\,TeV evaluated with the EWA (\textit{blue}) and with the full matrix element. The cut $m_{VV} > 1500\,\text{GeV}$ is imposed to show events from the region where the invariant mass distributions for the two approaches agree.}
	\label{fig:HH-eta}
\end{figure}

The distributions of transverse momenta and rapidities of the two Higgs bosons are shown in Figures \ref{fig:HH-pT} and \ref{fig:HH-eta}, respectively. Interestingly, the kinematics of the parton subleading in $p_\perp$ is much better described by the EVA than the leading one, and this behaviour persists when cutting off the threshold region in $m_{HH}$, as shown in Figure \ref{fig:HH-eta}. This is expected, since EVA assumes the vector bosons to be on shell and collinear with the incoming beam particles, suppressing hard recoil effects that significantly affect the particle leading in $p_\perp$. This shifts the peak in the $p_{\perp,1}$-distribution to slightly larger values as compared to the EVA. Nevertheless, after a cut of $m_{HH}>1.5\,\text{TeV}$, the EVA perfectly describes the kinematics of the subleading parton, which, being softer, is less sensitive to these approximations\footnote{For consistency, $m_{VV}$ is used in all plots to denote the invariant-mass cut. In this way, a single notation is employed for all final states, irrespective of whether they are produced via $WW$ or $ZZ$ fusion.}.

It is worth noting that this behaviour of the leading parton occurs despite the relative simplicity of the process, with VBF topologies only and negligible contributions from transverse vector bosons. Thus, the mismatch is likely more severe in more complex processes, and any kinematic cuts should be treated carefully. \whizard allows users to generate a $p_{\perp,1}$ kick in the EVA in the so-called \textit{recoil scheme}, but this option remains to be validated.

\subsection{Tau neutrinos}
\begin{figure}
	\centering
	\begin{subfigure}{0.49\textwidth}
		\centering
		\includegraphics[width=\textwidth]{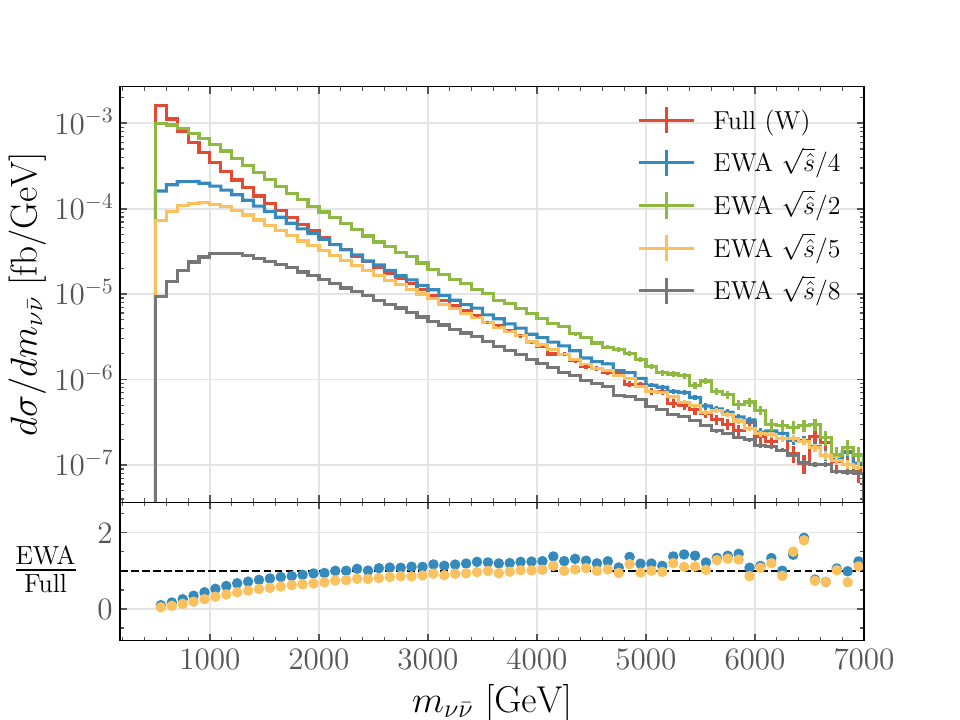}
		\caption{EWA}
	\end{subfigure}
	\hfill
	\begin{subfigure}{0.49\textwidth}
		\centering
		\includegraphics[width=\textwidth]{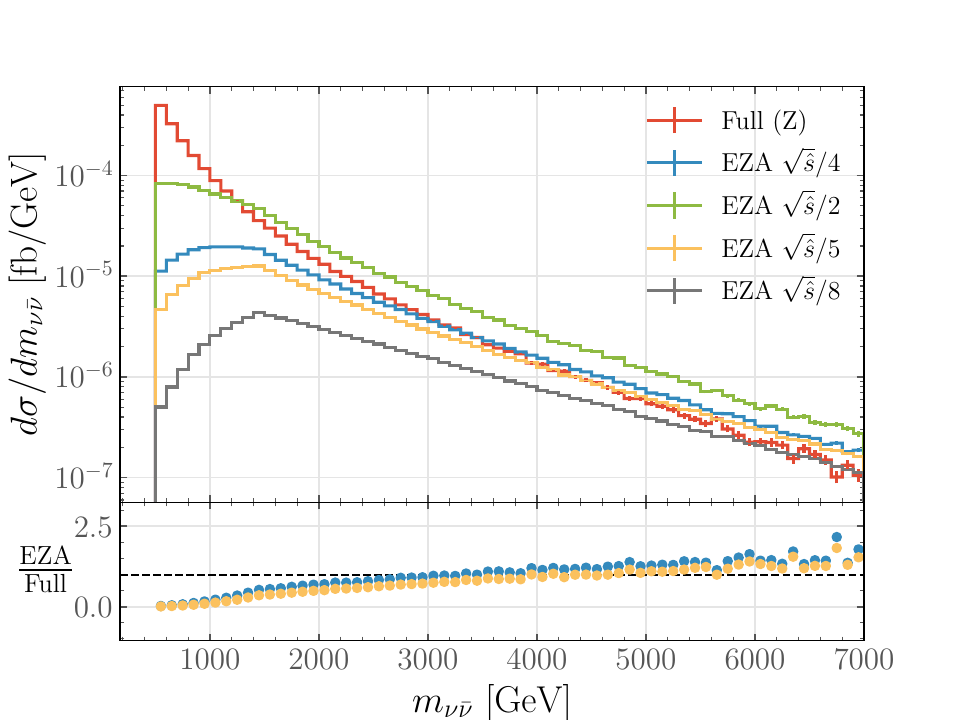}
		\caption{EZA}
	\end{subfigure}
	\caption{Invariant mass distributions of the $\nu_\tau \bar \nu_\tau$-system  in $e^+ \mu^- \to \nu_\tau \bar \nu_\tau + X$ (left: charged current, right: neutral current) at $\sqrt{s} = 10$\,TeV for the full matrix element evaluation (\textit{red}) and EVA results for different values of \ptmax.}
	\label{fig:nunu}
\end{figure}

Next, the process $e^+ \mu^- \to \nu_\tau \bar \nu_\tau + X$ is studied, where the neutrinos of the third family are chosen to avoid non-VBF topologies. Nevertheless, there are weak bremsstrahlung diagrams from $Z$ boson radiation with subsequent $Z\to \nu_\tau\bar{\nu}_\tau$ decay contributing to the process. These, in turn, can be addressed by cuts on $m_{\nu \bar{\nu}}$. It is worth underlining that this process is not directly measurable experimentally, but in contrast to the di-Higgs case, its hard matrix element, $VV \to \nu_\tau \bar \nu_\tau$, is dominated by transversely polarised vector bosons, which makes it a good testing ground for the transverse structure functions $F_\pm$ and their scale dependence.

The results in Figure \ref{fig:nunu} are, indeed, much more scale-dependent, even with a large generator-level invariant mass cut of 500 GeV applied. Four different scale choices are probed: $\sqrt{\hat{s}}/2$, $\sqrt{\hat{s}}/4$, $\sqrt{\hat{s}}/5$ and $\sqrt{\hat{s}}/8$. The largest scale choice clearly overshoots the correct result, while the smallest one undershoots it. The two scales in the middle yield a similarly reasonable performance; among them, $\sqrt{\hat{s}}/4$ provides better agreement for larger cross sections and is therefore heuristically selected as the default. The strong sensitivity to the scale choice makes the EVA less reliable for processes dominated by transverse polarisations.

Even for the scale choices that bring the EVA results close to the full calculation, the slopes of the tails in the distributions differ. This clearly shows a breakdown of the EVA whenever non-VBF diagrams contribute significantly to a process. Moreover, the performance of the EVA depends strongly on the mass scale of the final state. For light final states, where large transverse momenta and off-shell effects play an important role, the approximation loses predictive power, because the condition $\sqrt{\hat{s}} \gg m_V$ is not guaranteed.

\begin{figure}
	\centering
	\includegraphics[width=0.69\textwidth]{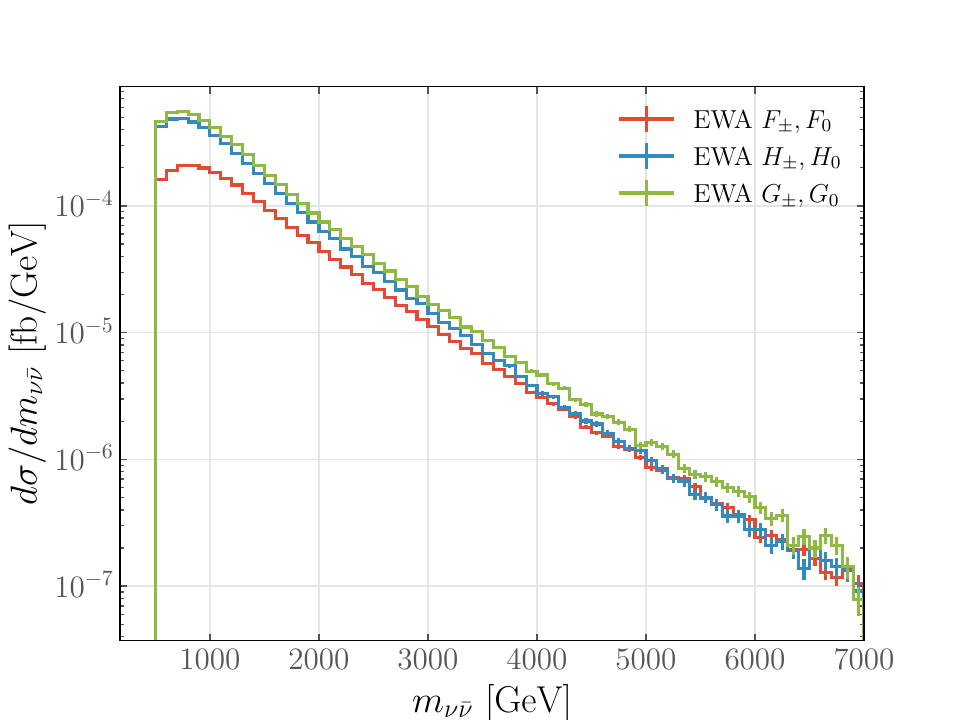}
	\caption{Invariant mass distributions of the process $\nu_\tau \bar \nu_\tau$ system in $e^+ \mu^- \to \nu_\tau \bar \nu_\tau + X$ at $\sqrt{s} = 10$\,TeV with the three different major EWA modes implemented in \whizard. The same scale choice of $\sqrt{\hat{s}}/4$ is set for all the cases.}
	\label{fig:nunu_modes}
\end{figure}

In Figure \ref{fig:nunu_modes}, the invariant mass distributions arising from the three different EVA implementations of $F_\lambda$, $G_\lambda$, and $H_\lambda$ with the same scale choice of $\sqrt{\hat{s}}/4$ are shown. $F_\lambda$ and $G_\lambda$ agree for large invariant masses and deviate significantly only in the low-$x$ regime, because the structure functions $G_\lambda$ only contain the logarithmic terms of $F_\lambda$, so they should be equal where these terms are largest. The structure functions $H_\lambda$ agree with $G_\lambda$ in the low energy regime and deviate at larger invariant masses, because $H_\lambda$ contain the same logarithmic terms as $G_\lambda$, but with a different dynamical scaling, i.e. they are related by the replacement $p_{\perp,\text{max}} \to \bar{x} p_{\perp,\text{max}}$. Therefore, differences between the two are mainly expected when $\bar{x} \to 0$, which is the case for the high-energy tails.

\subsection{Di-photon}

Photon production has also been studied in the literature to assess the applicability of the EWA for the processes dominated by transverse polarisation~\cite{Ruiz:2021tdt}. As only electrically charged particles couple to photons, there is no EZA equivalent in this case. The di-photon production in the VBF picture, $e^+\mu^- \to \gamma\gamma + \bar\nu_e  \nu_\mu$, is indeed dominated by transverse $W$ bosons; nevertheless, the full process receives large contributions from bremsstrahlung-like topologies, in which photons are emitted from the initial-state leptons. Additionally, the final-state particles are massless. Hence, no agreement between the EVA and the full process should be expected, contrary to some claims in the literature.

\begin{figure}
	\centering
	\begin{subfigure}{0.49\textwidth}
    \centering
    \includegraphics[width=\textwidth]{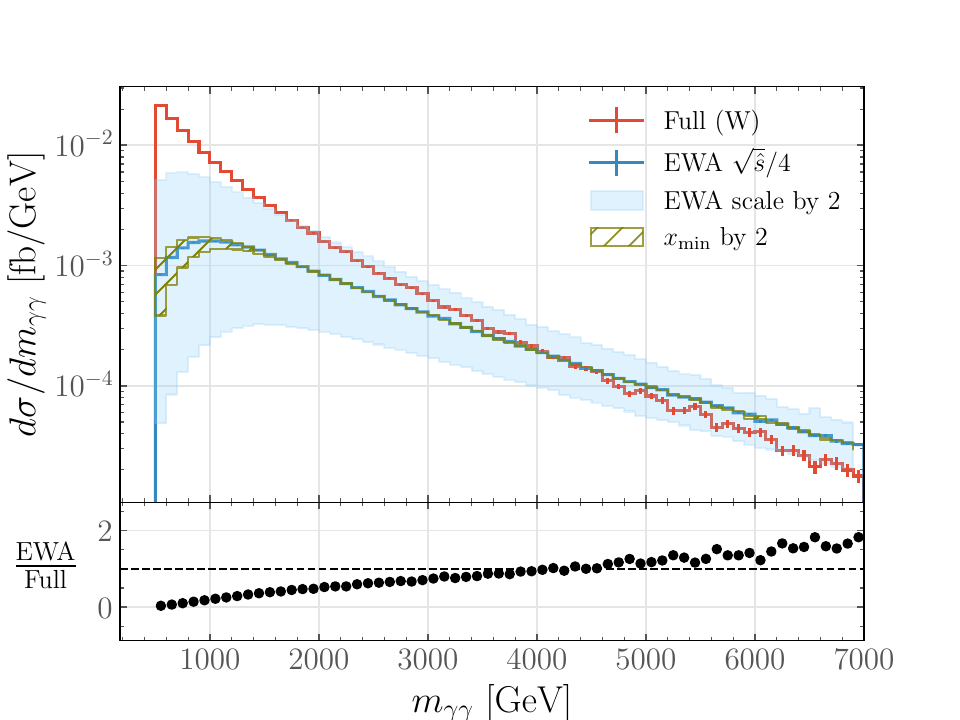}
    \caption{$m_{\gamma\gamma}$}
    \label{fig:di-gamma-a}
    \end{subfigure}
    \hfill
    \begin{subfigure}{0.49\textwidth}
    \centering
    \includegraphics[width=\textwidth]{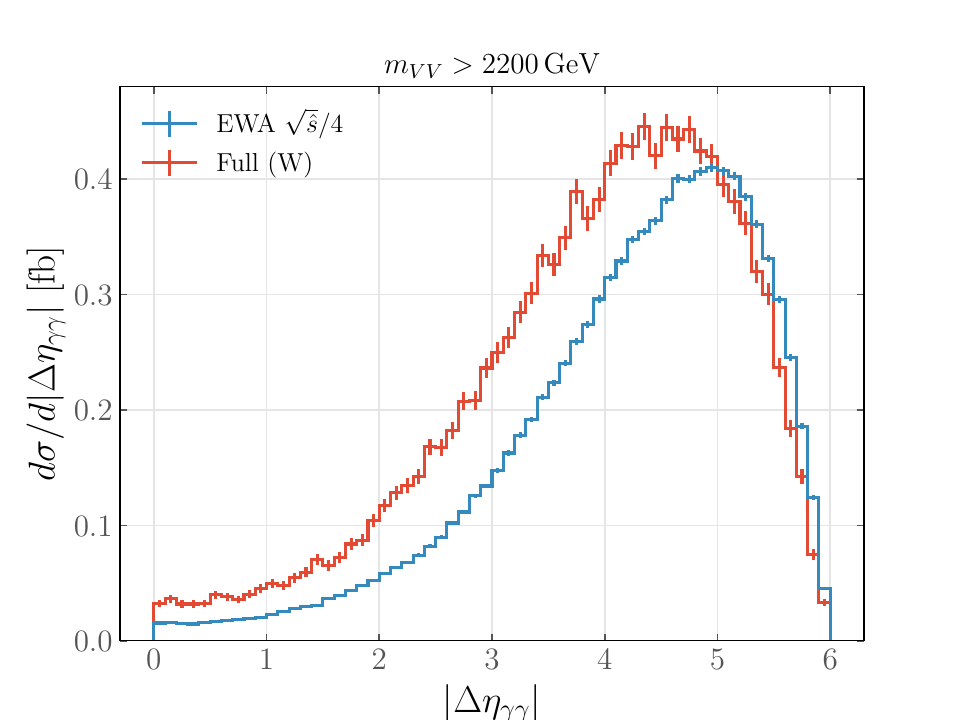}
    \caption{$\lvert \Delta \eta_{\gamma \gamma} \rvert$}
    \label{fig:di-gamma-b}
    \end{subfigure}
	\caption{Left: invariant mass distributions of the $\gamma \gamma$-system  in $e^+ \mu^- \to \gamma \gamma + X$ at $\sqrt{s} = 10$\,TeV for the full matrix element evaluation (\textit{red}) and EVA results for different values of \xmin and \ptmax. The \textit{blue} and \textit{green} bands show the scale and $\xmin$ variation by a factor of two, respectively. Right: the $\lvert \Delta \eta_{\gamma \gamma} \rvert$ distribution with an additional cut of $m_{VV} > 2200\,\text{GeV}$.}
	\label{fig:di-gamma}
\end{figure}

\begin{figure}
	\centering
	\begin{subfigure}{0.49\textwidth}
		\centering
		\includegraphics[width=\textwidth]{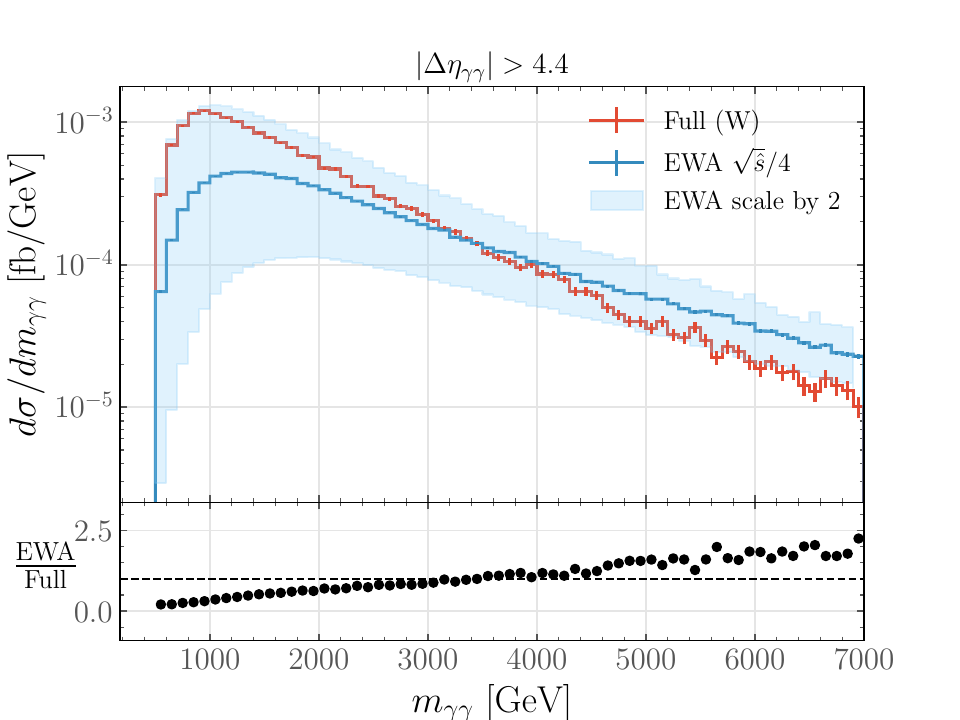}
		\caption{$\Delta \eta_{\gamma\gamma}$ cut only}
	\end{subfigure}
	\hfill
	\begin{subfigure}{0.49\textwidth}
		\centering
		\includegraphics[width=\textwidth]{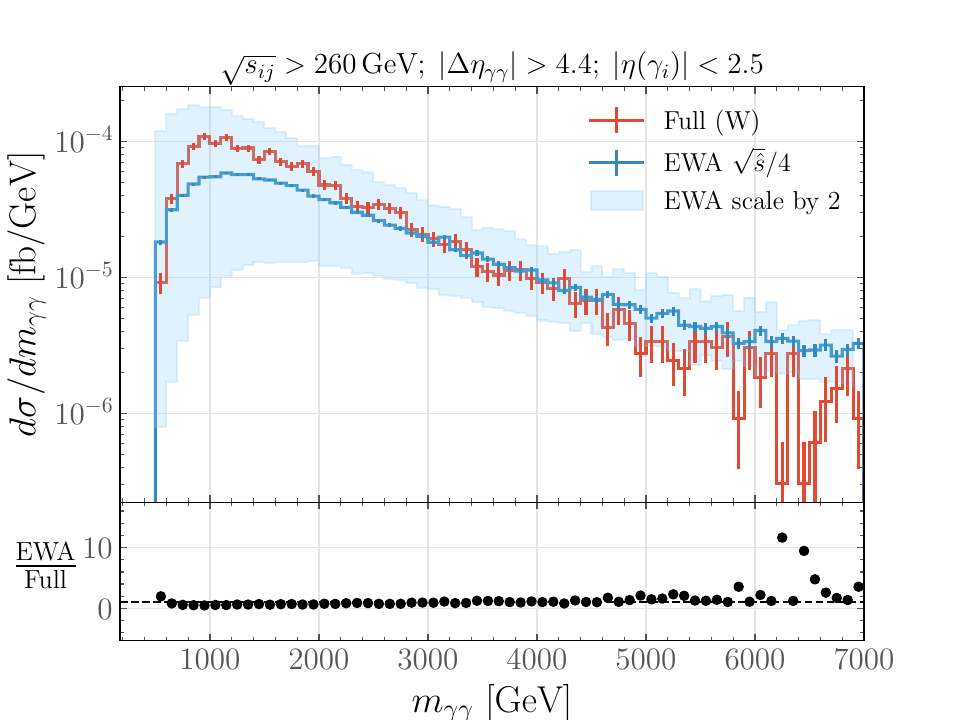}
		\caption{All cuts}
	\end{subfigure}
	\caption{Invariant mass distributions of the $\gamma \gamma$-system  in $e^+ \mu^- \to \gamma \gamma + X$ at $\sqrt{s} = 10$\,TeV for the full matrix element evaluation (\textit{red}) and EVA results for different cut setups and a scale variation by a factor of two in the \textit{blue} band.}
	\label{fig:di-gamma-cut}
\end{figure}

In Figure \ref{fig:di-gamma-a}, the invariant-mass distribution for $e^+ \mu^- \to \gamma \gamma + X$ is shown.
Similar to the $\nu_\tau \bar{\nu}_\tau$ case, this process is heavily scale-dependent due to the large transverse contributions. It also exhibits additional non-VBF-like topologies that undermine the accuracy of the EVA. Their impact is minimised by imposing generator-level cuts on the photons, namely an invariant mass cut $m_{\gamma\gamma} > 500\,\text{GeV}$, a transverse-momentum cut $p_{\perp,\gamma_i} > 50\,\text{GeV}$, an angular separation cut $\Delta R > 0.4$ and a pseudorapidity cut $|\eta_{\gamma_i}| < 3$. These cuts are expected to suppress soft contributions, which are likely to originate from the initial-state radiation. Nonetheless, as discussed earlier, any cuts must be applied with care due to the kinematic mismatch between the EVA and the full calculation. Even with these cuts in place, the impact of the non-VBF-like topologies remains visible, as reflected in the differing slopes of the ratios observed in the distributions.

In Figure \ref{fig:di-gamma-a}, the variation of $\xmin$ is again shown in the green hatched region. This variation solely affects the low $m_{\gamma\gamma}$ region due to the lower invariant mass cut-off. However, due to the sizeable transverse contributions, the variation of the scale $\ptmax$ affects the whole range, including the tail of the distribution. The agreement between EVA and the full matrix evaluation is poor. Since the ratio of the two shows a constant slope over the whole range of the $m_{\gamma\gamma}$ distribution, the accuracy cannot be improved even by a large invariant mass cut, in contrast to the di-Higgs case.

To find a region in the phase space in which the two approaches agree, a set of technical cuts may be applied. The pseudorapidity difference between the photons $\Delta \eta_{\gamma\gamma}$ can be considered, as indicated in Figure \ref{fig:di-gamma-b}.
Expecting that the EVA would agree better for large invariant di-photon masses, the distribution is examined for events satisfying $m_{\gamma\gamma} > 2200\,\text{GeV}$. In the EVA, events typically exhibit a large forward--backward separation $\Delta \eta_{\gamma\gamma}$, in contrast to the full matrix-element calculation, where additional non-VBF-like topologies populate the region of smaller $\Delta \eta_{\gamma\gamma}$.
A lower cut on $\lvert \Delta\eta_{\gamma\gamma} \rvert$ thus mainly affects the results of the full matrix element evaluation. The agreement of the EVA with the full matrix elements in the $m_{\gamma\gamma}$ tail region can also be improved with an additional cut removing events with large pseudorapidity $\lvert \eta(\gamma_i) \rvert < 2.5$. 
Finally, a cut on $s_{ij} = (p_i + p_j)^2$, for all pairs $i\neq j$ of the final state four-momenta, reduces the full matrix element mostly in the low $m_{\gamma\gamma}$ region. This technical cut is expected to improve agreement because it introduces a large scale to the process, enhancing the validity range of the small-angle expansion. Nevertheless, it cannot be applied in a real experimental setup, when the neutrinos cannot be reconstructed. The combination of all cuts yields better agreement, as indicated by Figure \ref{fig:di-gamma-cut}; however, the full process is still not reproduced reliably, and differences in the slopes remain visible.

Furthermore, the modest agreement is achieved only for the specific cut values displayed in the plot. For example, increasing the cut on $s_{ij}$ does not result in any further improvement. Additionally, even when relaxing the generator-level cuts mentioned earlier, it was not possible to identify universally applicable kinematic cuts that yield satisfactory results within the EVA for this process. The conclusion is that the EVA fails to accurately describe the $\gamma\gamma$ production due to missing diagrams from non-VBF topologies and the absence of an inherent large scale. It is only achievable to approximate the full matrix element calculation within a limited region of phase space, defined by carefully tuned technical cuts.

\subsection{Top pairs}
\begin{figure}
	\centering
	\begin{subfigure}{0.49\textwidth}
		\centering
		\includegraphics[width=\textwidth]{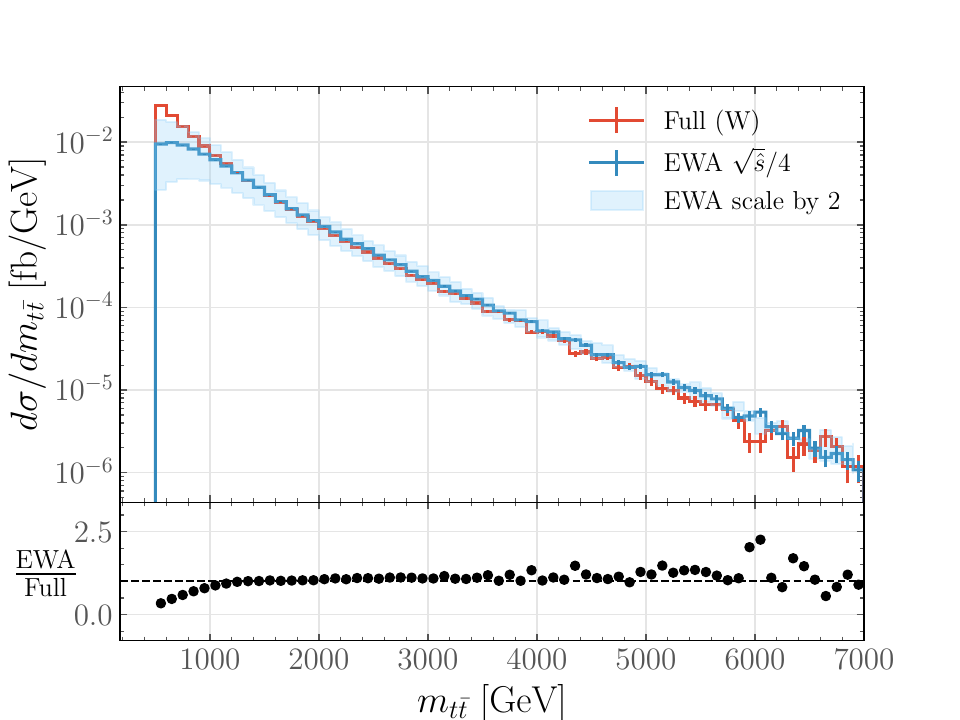}
		\caption{$e^+ \mu^-$ collider}
	\end{subfigure}
	\hfill
	\begin{subfigure}{0.49\textwidth}
		\centering
		\includegraphics[width=\textwidth]{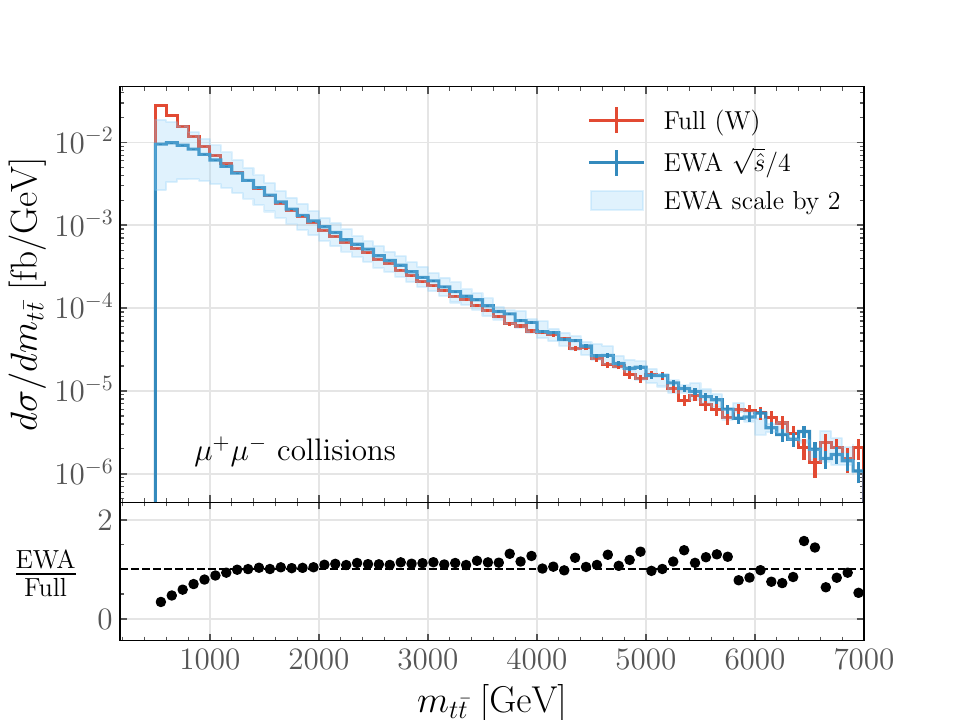}
		\caption{$\mu^+ \mu^-$ collider}
	\end{subfigure}
	\caption{Invariant mass distributions of the $t \bar{t}$-system at an $e^+ \mu^-$ (left) and a $\mu^+ \mu^-$ collider (right) for the full matrix element evaluation (\textit{red}) and EVA results for different values of \ptmax, respectively. The collider energy is $\sqrt{s} = 10$\,TeV.}
	\label{fig:ttbar}
\end{figure}

The next process to consider is top-pair production. In Figure \ref{fig:ttbar}, the results for $e^+ \mu^- \to t \bar{t} + X$ and $\mu^+ \mu^- \to t \bar{t} + X$ are shown. In the latter case, the full matrix element calculation receives a small, few-percent contribution from additional $s$-channel diagrams. 
Top-pair production proceeds both through purely longitudinal and mixed transverse--longitudinal polarisations of the initial-state vector bosons.
Interestingly, the EVA yields accurate results for $m_{t\bar{t}} \gtrsim 1\,\text{TeV}$ in both cases, despite the additional non-VBF topologies and scale-dependent contributions from transversely polarised bosons; similarly to di-Higgs production, the low-$m_{t\bar{t}}$ is not well described due to the missing contributions from off-shell $W$ bosons.

This indicates that the presence of a heavy state, the top quark pair, enhances the reliability of the EVA, as the heavy scale helps suppress effects not captured by the small-angle approximation, as previously mentioned. It suggests that the EVA may also apply to more realistic collider scenarios involving same-flavour leptons, once the energy scale allows for neglecting the $s$-channel contributions. Furthermore, the observed agreement suggests that off-diagonal transverse--longitudinal interference between the amplitude and its complex conjugate plays only a minor role, contrary to the expectation of~\cite{Dawson:1984gx}.
The scale variation is naturally more pronounced than in the di-Higgs case because of the transverse--longitudinal mixed contributions. Nevertheless, the width of the scale variation bands is significantly reduced as compared to the $\gamma \gamma$ and $\nu_\tau \bar{\nu}_\tau$ cases.

\subsection{Associated ZH production}
\begin{figure}[h]
	\centering
	\begin{subfigure}{0.49\textwidth}
		\centering
		\includegraphics[width=\textwidth]{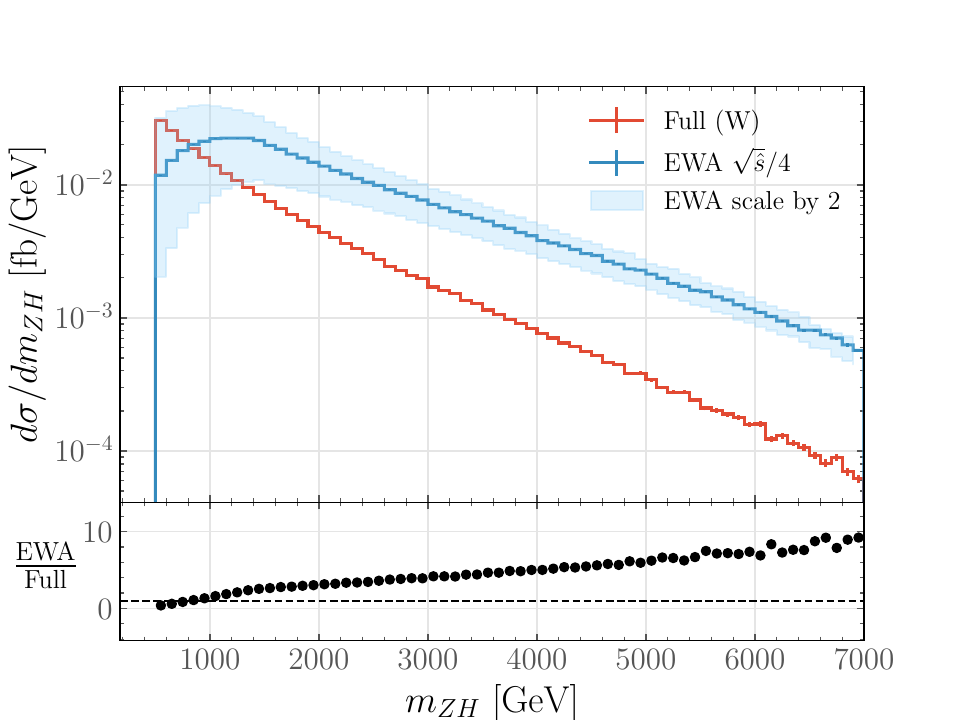}
		\caption{No cut}
		\label{fig:ZHa}
	\end{subfigure}
	\hfill
	\begin{subfigure}{0.49\textwidth}
		\centering
		\includegraphics[width=\textwidth]{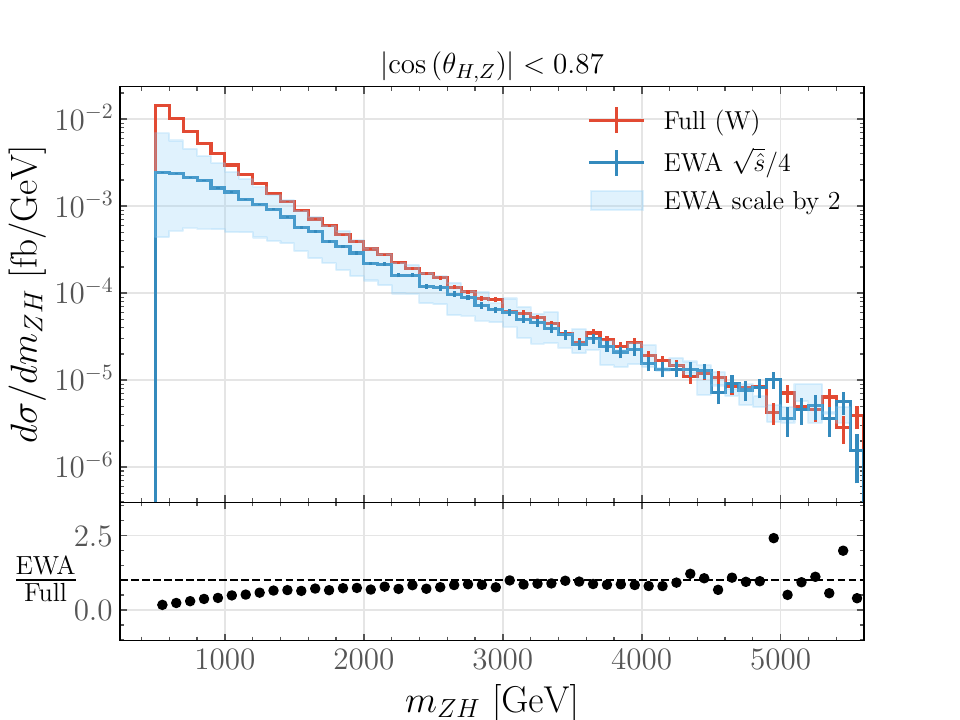}
		\caption{$\cos(\theta_{Z,H})$ cut}
		\label{fig:ZHb}
	\end{subfigure}\\
	\begin{subfigure}{0.49\textwidth}
		\centering
		\includegraphics[width=\textwidth]{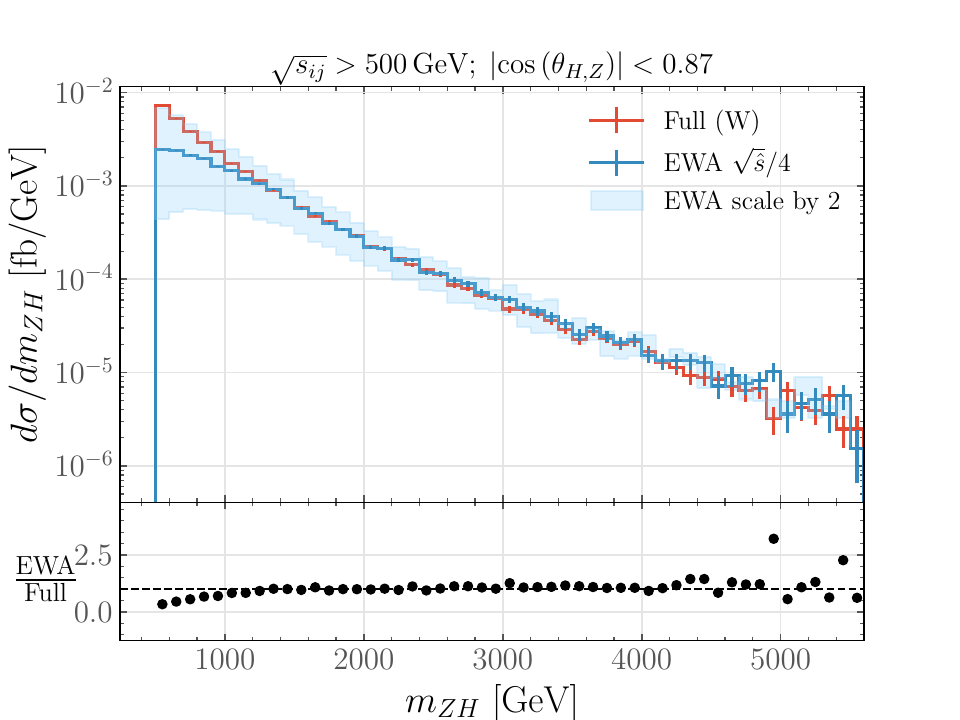}
		\caption{$\cos(\theta_{Z,H})$ and $s_{ij}$ cuts}
		\label{fig:ZHc}
	\end{subfigure}
    \hfill
 	\begin{subfigure}{0.49\textwidth}
		\centering
		\includegraphics[width=\textwidth]{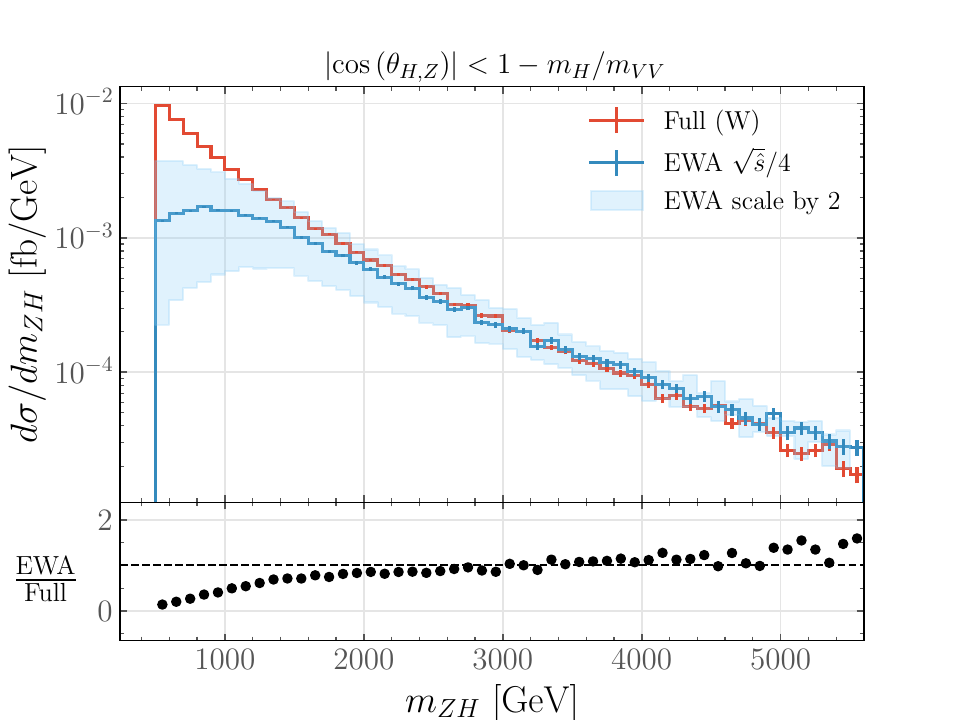}
		\caption{Dynamic $\cos(\theta_{Z,H})$ cut}
		\label{fig:ZHd}
	\end{subfigure}   
	\caption{
		Invariant mass distributions of the $ZH$-system  in the process $e^+ \mu^- \to ZH + X$ for the full matrix element evaluation (\textit{red}) and EVA results for different values of \ptmax with the \textit{blue} band indicating the scale variation by a factor of two. The collider energy is $\sqrt{s} = 10$\,TeV.
	}
	\label{fig:ZH}
\end{figure}

In the case of associated $ZH$ production, the EVA result differs significantly from the full calculation over the whole range of the invariant mass spectrum, as shown in Figure~\ref{fig:ZHa}, where no generator-level cuts are applied.
The discrepancy can be alleviated by a carefully tuned cut on $\theta_{Z,H}$, the angle between the final-state particles and the beam axis in their rest frame, see Figure~\ref{fig:ZHb}. Rejecting values of $\lvert{\cos\theta_{Z,H}}\rvert$ close to unity corresponds to removing the peaking region of low Mandelstam variable $\hat{t}$, which is artificially enhanced in the EVA results, following the collinear approximation. The agreement can be further improved by applying a technical cut on the invariant masses of all final-state parton pairs, which suppresses contributions from non-VBF-like topologies (as can be seen from the fact that only the full result is affected), as shown in Figure~\ref{fig:ZHc}.
Alternatively, a dynamical cut that depends on the invariant mass $m_{ZH}$ of each event can also work similarly (Figure~\ref{fig:ZHd}).

\begin{figure}[h]
	\centering
	\begin{subfigure}{0.49\textwidth}
		\centering
		\includegraphics[width=\textwidth]{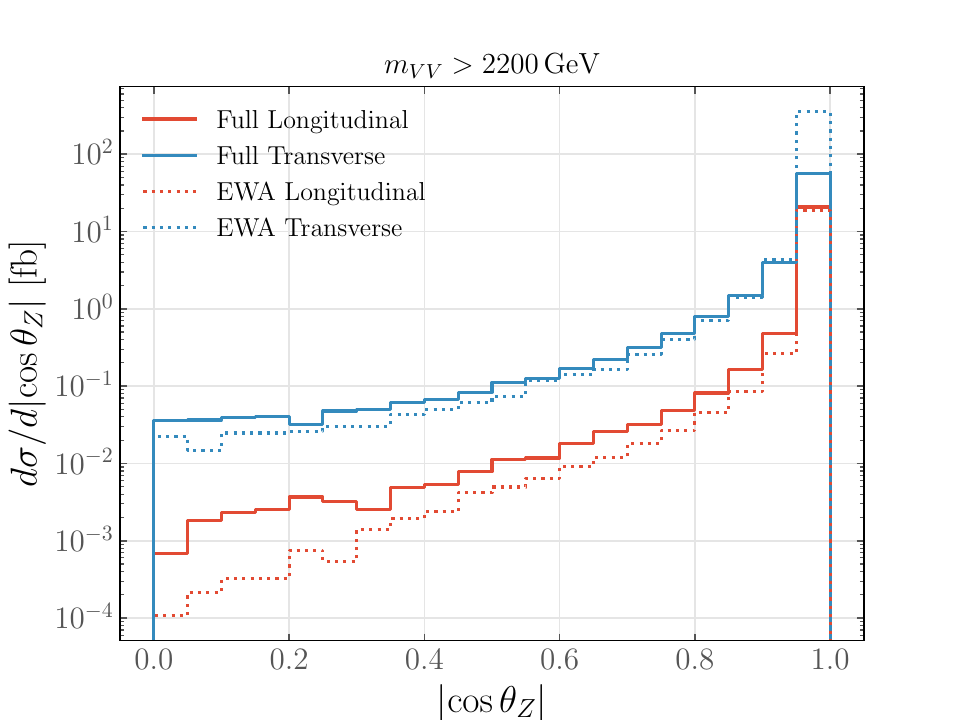}
		\caption{No $\cos\theta_{Z,H}$ cut}
	\end{subfigure}
	\hfill
	\begin{subfigure}{0.49\textwidth}
		\centering
		\includegraphics[width=\textwidth]{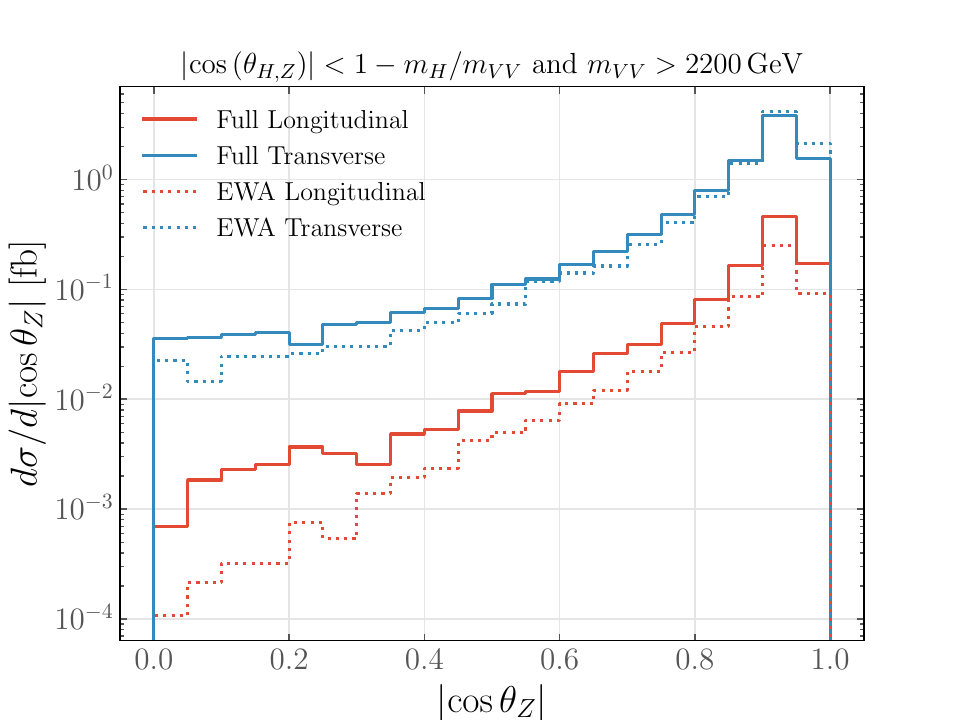}
		\caption{dynamical $\cos\theta_{Z,H}$ cut}
	\end{subfigure}
	\caption{
Distributions of the cosine of the $Z$-boson emission angle in the rest frame at $\sqrt{s} = 10$\,TeV. Both plots correspond to a high invariant-mass region, $m_{VV} > 2200\,\text{GeV}$, while in the right plot, an additional dynamical cut on $\cos\theta_{Z,H}$ is applied.
    \label{fig:costhetaZH}
}
\end{figure}

In order to better understand the impact of the angular cut, the differential distribution in $\lvert\cos\theta_{Z}\rvert$ is presented in Figure~\ref{fig:costhetaZH}, where transverse and longitudinal contributions of the final-state $Z$ boson are shown separately. The process is dominated by transverse contributions, and the agreement with the full result increases significantly for values of $\lvert\cos{\theta_{Z,H}}\rvert \lesssim 0.9$.
This is not true for the longitudinal contributions.
Therefore, a cut on $\theta_{Z,H}$ cannot improve the agreement in this case.
This means that in general, the cut can only be useful for processes where the longitudinal contributions are small, so one should not impose it, for example, on the $HH$ production.
Lowering the cut even further does not increase the agreement because in this regime, the discrepancy for the longitudinal modes cannot be neglected.

Furthermore, Figure~\ref{fig:costhetaZH} shows how a dynamical cut on $\cos\theta_{Z,H}$ can somewhat improve the results by reducing the excess transverse contributions of the EVA at large values of $\lvert \cos\theta_{Z,H} \rvert$.
Nonetheless, a comparison of Figures~\ref{fig:ZHc} and \ref{fig:ZHd} proves that a constant, fine-tuned cut on $\lvert\cos \theta_{Z,H}\rvert$ performs better in achieving a flat ratio between the EVA and the full result. The EVA does not provide a viable alternative to the full matrix-element approach.

\subsection{Vector boson scattering}
\begin{figure}
	\centering
	\begin{subfigure}{0.49\textwidth}
		\centering
		\includegraphics[width=\textwidth]{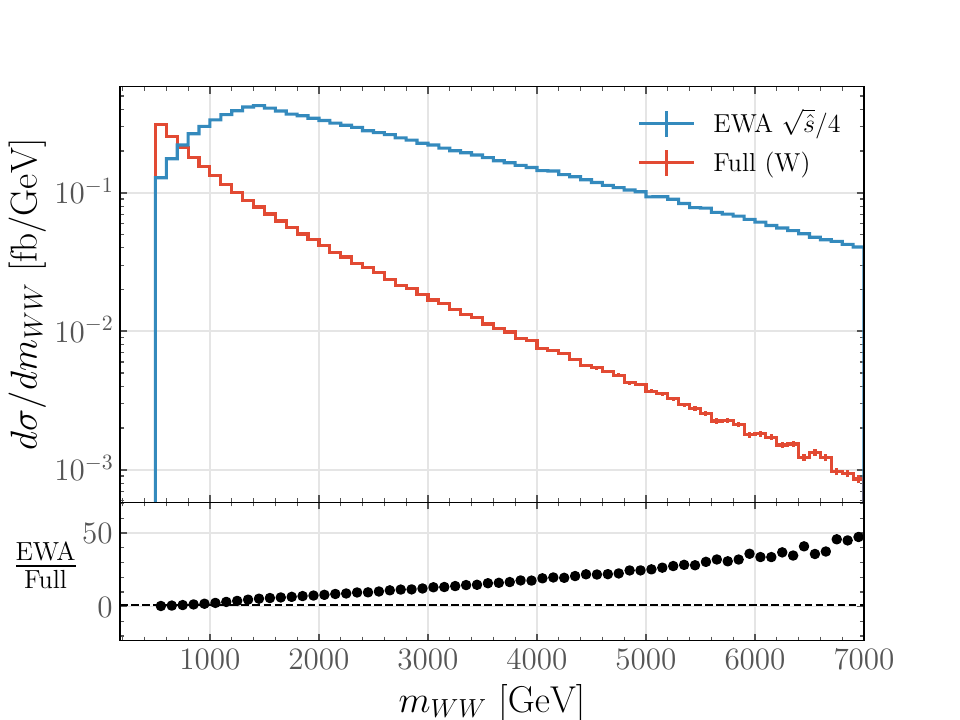}
		\caption{No cut}
		\label{fig:WWa}
	\end{subfigure}
	\hfill
	\begin{subfigure}{0.49\textwidth}
		\centering
		\includegraphics[width=\textwidth]{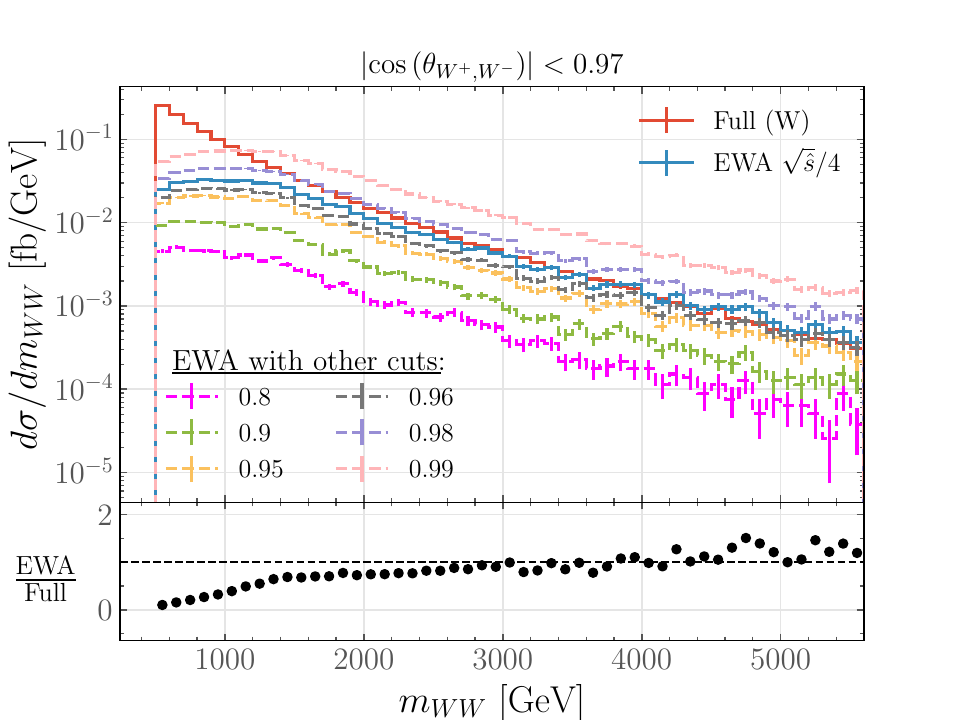}
		\caption{$\cos(\theta_{W^+,W^-})$ cut}
		\label{fig:WWb}
	\end{subfigure}
	\caption{Invariant mass distributions of the $W^+W^-$ final state in the vector boson scattering $e^+ \mu^- \to W^+ W^- + X$ at $\sqrt{s} = 10$\,TeV for the full matrix element evaluation (\textit{red}) and EVA (\textit{blue}). Left: the distribution with no cut. Right: the distribution with different cuts on $\cos(\theta_{W^+,W^-})$. For the latter, the $\lvert \cos (\theta_{W^+, W^-})\rvert < 0.97$ cut for the full matrix element and EWA as solid lines are shown, while EWA with other cut values is indicated in dashed lines.}
	\label{fig:WW}
\end{figure}

The last investigated process is vector boson scattering, $e^+\mu^- \rightarrow W^+ W^- + X$, as shown in Figure \ref{fig:WWa}. A 20-GeV generator-level cut on $p_\perp$ is applied to the final-state $W$ bosons for better convergence. The EVA deviates significantly from the full matrix element calculation, differing by two orders of magnitude in the high $m_{WW}$ region.
The situation resembles the associated $ZH$ production, where large contributions in the EVA case arise when the final-state $W$ bosons are produced at low angle. Therefore, applying a specific cut on $\lvert \cos(\theta_{W^+,W^-}) \rvert$, the angle of the $W$-boson emission with respect to the beam axis in the rest frame of $W^+ W^-$, can significantly improve the agreement between the EVA and the full matrix element calculation, as illustrated in Figure \ref{fig:WWb}. The cut value of 0.97 was chosen to optimise overall agreement. Nevertheless, comparison with other possible values shows that this choice is merely fine-tuned, and either more or less restrictive cuts deteriorate the accuracy, assuming they barely affect the full calculation.
A slight flattening of the ratio in Figure~\ref{fig:WWb} is visible above $\sim 1.5$\,TeV for the chosen cut, but in general, the EVA fails to describe this process.

\section{Implications for EVA phenomenology}
A broad range of physics processes was analysed to revalidate the EVA and its implementation within the Monte Carlo event generator \whizard. While precision measurements at the highest energies will ultimately rely on fixed-order calculations of the full processes, the EVA remains an interesting concept. It might enhance the understanding of phase-space regions dominated by electroweak splitting kinematics and enable the development of efficient tools for probing this regime with reduced computational requirements. These studies confirmed that the EVA offers a reasonably accurate description for processes primarily driven by the fusion of longitudinal $W$ or $Z$ gauge bosons, such as di-Higgs production. In contrast, for processes dominated by final-state transverse electroweak gauge bosons, additional, often fine-tuned, selection cuts might be necessary to isolate the VBF-enhanced kinematic regions. 

The key findings can be summarised as follows:

\begin{itemize} 
    \item The processes mediated by longitudinal vector bosons are described better within the realm of the EVA than transverse and mixed transverse--longitudinal modes due to the strong dependence of the transverse structure functions on the scale $\ptmax$, which modifies the shape of differential distributions over the full spectrum. The best agreement can typically be achieved for dynamical scales around $\sqrt{\hat{s}}/4$.

    \item The dependence on lower Bjorken cutoff $x_\text{min}$ is most pronounced in the low-invariant mass regime, where the $x_i$ are small for each beam. The naïve choice of this variable corresponding to the minimum energy fraction needed to radiate a massive vector boson $V$ off a single beam, $x_\text{min} = 2 m_V/\sqrt{s}$, might not bring the best agreement. 

    \item A universal approach to enhancing the reliability of the EVA is to impose a lower bound on the invariant mass of the final state to remove the region dominated by off-shell vector-boson interactions. A sensible cut-off should consider not only the masses of the produced particles but also the collision energy, e.g. the ratio of these quantities. Similarly, the presence of a heavy final state tends to improve the reliability of the EVA.

    \item The kinematics of the final states produced via vector boson fusion in the EVA does not fully match that from the complete matrix element calculation, so all kinematic cuts must be applied with care. Notably, the EVA tends to describe the kinematics of the particle subleading in $p_\perp$ more accurately than that of the leading one.

    \item For instance, by the excellent agreement for top pair processes, it is apparent that interferences between amplitudes for hard processes initiated by differently polarised vector bosons at the same leg (which are included in the full process but excluded by definition in the factorised EVA) only play a minor role.
\end{itemize}
\chapter{Electroweak Parton Distribution Functions}
\label{chap:EWPDF}

As presented in the previous chapter, the framework of collinear factorisation can be extended to high-energy lepton colliders. It provides an approximate and efficient method for obtaining reliable predictions in selected cases by separating the beam-structure function and the hard interaction, analogously to the concept used at the LHC for proton--proton collisions. Another parallel emerges at very high energies: the relevant logarithmic enhancements become large, making higher-order corrections non-negligible and motivating resummation, as in the context of QCD PDFs.

In this chapter, an overview of Electroweak Parton Distribution Functions (EWPDFs) for leptons is presented, together with their implementation in the Monte Carlo generator \textsc{Whizard} and their application to high-energy lepton collisions. While the EVA is based simply on a kinematic approximation, EWPDFs are deeply rooted in quantum field theory and entail systematic improvements, e.g. by means of resummation.

\section{The concept of Electroweak Parton Distribution Functions}

In high-energy lepton collisions above the electroweak (EW) scale, collinear splitting of electroweak gauge bosons becomes the dominant process, mostly through initial-state radiation. Under these conditions, VBF-like topologies tend to prevail over $s$-channel annihilation processes, allowing for the use of approximate frameworks. As an example, consider a simple EPA formula:
\begin{equation*}
    f_{\gamma}^{\rm EPA}(x) = \frac{\alpha}{2\pi} P_{\gamma \ell}^{QED}(x) \log \frac{E^2}{m_\ell^2}\,,
\end{equation*}
where $\alpha$ is the fine-structure constant, $P_{\gamma \ell}^{QED}(x) = (1+(1-x)^2)/x$ is the splitting function for $\ell \to \ell\gamma$, $E$ is the lepton energy and $m_\ell$ is its mass. The logarithmic factor grows with energy. Given that higher orders contribute with increasing powers of $\alpha^n \log^k (E^2/m_\ell^2)$, accounting only for the leading contribution might not be enough already for energy scales available at multi-TeV colliders.

A need for a new conceptual framework arises: logarithms can be systematically resummed using the Dokshitzer-Gribov-Lipatov-Altarelli-Parisi (DGLAP) equations~\cite{Gribov:1972ri,Altarelli:1977zs,Dokshitzer:1977sg} in a similar way to the standard QCD PDF. Multiple emissions --- also representing different interaction types --- appear in the evolution from subsequent splittings of comparable probabilities. In this way, the leading-order structure functions evolve into EWPDFs~\cite{Han:2020uid,Han:2021kes}, which account for various partons representing the full spectrum of the SM particles. For example, a gluon content of the lepton might be considered via a chain $\ell \to \ell\gamma \to \ell q\bar{q} \to \ell q\bar{q}g$.

The EWPDF framework is considerably more complex than the well-established QCD formalism broadly used for hadron collisions. Above the EW scale, the full SM gauge symmetry is restored, requiring the inclusion of unbroken-phase electroweak gauge bosons in the evolution, including their mixing and interference~\cite{Chen:2016wkt,Bauer:2017isx,Bauer:2018arx}. Additionally, since the SM is a chiral theory, parton polarisation effects arise even when the initial lepton beams are unpolarised. The residual effects of EW symmetry breaking also appear through the inclusion of longitudinally polarised gauge bosons.

For leptons, gauge-boson radiation is predominantly electromagnetic below the EW scale, as emissions of weak bosons are suppressed by their masses. However, above this scale, all electroweak states become dynamically activated. Denote an EWPDF as $f_i(x,Q^2)$ with $i$ labelling a specific particle carrying a fraction $x$ of the beam energy at factorisation scale $Q^2$. Then, the evolution is given by the full EW DGLAP equations~\cite{Ciafaloni:2005fm}:
\begin{equation*}
    \frac{\dd f_i}{\dd \ln Q^2} = \sum_I \frac{\alpha_I}{2\pi} \sum_j P^I_{i,j} \otimes f_j\, ,
\end{equation*}
where $I$ specifies the gauge group, and the $P^I_{i,j}$ are the splitting functions for $j \to i$. The convolution $\otimes$ is given by:
\begin{equation*}
\left[f\otimes g\right](x_0)
=\int_{x_0}^1\frac{\dd\xi}{\xi}f(\xi)g\left(\frac{x_0}{\xi}\right).
\end{equation*}
The complete list of the splitting functions $P^I_{i,j}$ can be found in the literature~\cite{Chen:2016wkt,Bauer:2017isx}. 

In contrast to hadron PDFs, for which the initial conditions must be determined from fits to experimental data, the initial conditions for lepton PDFs can be deduced. At leading order, for a beam lepton $\ell$, $f_\ell (x, m_\ell ^2) \approx \delta(1-x) + \mathcal{O}(\alpha)$. All other PDFs vanish for $Q^2 = m_\ell ^2$ at this order~\cite{Garosi:2023bvq}. At the QCD scale, the QCD interaction becomes available with the initial condition $f_g(x,\mu^2_{QCD}) = 0$. At the EW scale, the matching conditions are $f_\gamma (x,m^2_Z) \neq 0$, $f_Z (x,m^2_Z) = 0$ and $f_{\gamma Z} (x,m^2_Z) = 0$. The last PDF represents interference between the photon and the $Z$ boson, and should be included to properly account for the constructive or destructive interference in neutral-current diagrams.

For illustration, consider a muon beam with energy below the EW scale, such that the five light quarks, the three charged leptons, the photon, and the gluon are kinematically accessible. Since QED and QCD interactions play a dominant role in this regime, no additional particles need to be included\footnote{Neutrinos, despite their negligible masses, can participate only through $W$/$Z$ interactions and therefore become relevant only above the EW scale. The small neutrino component in the muon beam originating from muon decay is neglected in this derivation.}. Adopting the standard QCD PDF notation --- where singlet and non-singlet combinations are introduced to separate closed-loop emissions from final-state quark contributions --- the singlet distributions may be written as~\cite{Han:2021kes}:
\begin{equation*} f_L=\sum_{i=e,\mu,\tau}(f_{\ell_i}+f_{\bar{\ell}_i}), \quad f_U=\sum_{i=u,c}(f_{u_i}+f_{\bar{u}_i}), \quad f_D=\sum_{i=d,s,b}(f_{d_i}+f_{\bar{d}_i}).
\end{equation*}
The corresponding non-singlet PDFs are
\begin{equation*} 
f_{\ell_i}^{\rm NS}=f_{\ell_i}-f_{\bar{\ell}_i}, ~f_{\ell,12}=f_{\bar{e}}-f_{\bar{\mu}}, ~f_{\ell,23}=f_{\bar{\mu}}-f_{\bar{\tau}}\,, 
\end{equation*} 
\begin{equation*} f_{u_i}^{\rm NS}=f_{u_i}-f_{\bar{u}_i}, f_{u,12}=f_u-f_c\,, 
\end{equation*} 
\begin{equation*} f_{d_i}^{\rm NS}=f_{d_i}-f_{\bar{d}_i}, ~ f_{d,12}=f_{d}-f_{s}, ~ f_{d,13}=f_{d}-f_{b} 
\end{equation*}

The DGLAP equations for the singlet contributions are given by
\begin{equation*}
\frac{\dd}{\dd\log Q^2}
\begin{pmatrix}
f_L\\
f_U\\
f_D\\
f_\gamma\\
f_g
\end{pmatrix}
=
\begin{pmatrix}
P_{\ell\ell} & 0 & 0 & 2N_\ell P_{\ell\gamma} & 0\\
0 & P_{uu} & 0 & 2N_uP_{u\gamma} & 2N_uP_{ug}\\
0 & 0 & P_{dd} & 2N_dP_{d\gamma} & 2N_dP_{dg}\\
P_{\gamma\ell} & P_{\gamma u} & P_{\gamma d} & P_{\gamma\gamma} & 0\\
0 & P_{g u} & P_{gd} & 0 & P_{gg}
\end{pmatrix}
\otimes 
\begin{pmatrix}
f_L\\
f_U\\
f_D\\
f_\gamma\\
f_g
\end{pmatrix}\,,
\end{equation*}
where $N_\ell$, $N_u$, and $N_d$ denote the numbers of active lepton, up-type quark, and down-type quark flavours, respectively. A characteristic feature of these equations is the mixing that occurs among the different states. On the other hand, the DGLAP equation for any non-singlet PDF reads
\begin{equation*}
\frac{\dd}{\dd\log Q^2}f^{\rm NS}=P_{ff}\otimes f^{\rm NS}.
\end{equation*}
It is remarkable that the non-singlet contributions do not mix; therefore, any PDF that vanishes in the initial condition remains zero at all higher scales.

In this basis, the flavour-state PDFs are given as
\begin{eqnarray*}
f_{\mu}=\frac{f_{L}+(2N_{\ell}-1)f_{\mu}^{\rm NS}}{2N_{\ell}}\,,\\
f_{e}=f_{\bar{e}}=f_{\bar{\mu}}=f_{\tau}=f_{\bar{\tau}}=\frac{f_L-f_{\mu}^{\rm NS}}{2N_{\ell}}\,,\\
f_{u}=f_{\bar{u}}=f_{c}=f_{\bar{c}}=\frac{f_{U}}{2N_u}\,, \\
f_{d}=f_{\bar{d}}=f_{s}=f_{\bar{s}}=f_{b}=f_{\bar{b}}=\frac{f_{D}}{2N_d}\,.
\end{eqnarray*}
The above relations for the sea-flavour PDFs hold strictly in the massless limit. When fermion masses are included, corrections proportional to $\ln\left( m^2_f / m^2_\ell \right)$ must be taken into account. Moreover, for scales near or below the QCD scale, the effects of confinement become relevant. Conversely, for scales near or above the EW scale, the full unbroken SM gauge interactions must be incorporated, including interactions of the $W$ and $Z$ bosons, similar to the EVA. Once the massive electroweak bosons become kinematically accessible, the PDFs naturally acquire polarisation due to the chiral structure of the electroweak interactions, while the DGLAP evolution due to QCD and QED is purely vector-like.

The EWPDF approach can potentially help improve the precision of predictions for multi-TeV interactions. In this context, an analogy to the standard EPA treatment becomes apparent. EWPDFs are expected to provide the most accurate description in the kinematic regime where resummation effects dominate and should be matched to fixed-order calculations covering the remaining phase space. Developing a consistent matching prescription for this purpose is a natural next step and constitutes the ultimate goal of the studies initiated within this thesis.

\section{EWPDF in Whizard}

The technical implementation of EWPDFs in \textsc{Whizard}, developed as part of this thesis, draws inspiration from two major features of the generator: the EVA formalism discussed in the previous chapter and the \textsc{LHAPDF}-based~\cite{Whalley:2005nh,Buckley:2014ana} treatment used for hadron collisions. However, two key considerations motivated the development of a new framework rather than a direct reuse of these pre-existing components.

First, the EVA formalism in \whizard relies on analytic expressions for electroweak boson distributions. In contrast, EWPDFs typically require numerical computation due to the intrinsic complexity arising from resummation effects, running couplings, and the interplay among different parton types. Numerical solutions of the DGLAP equations yield grids similar in spirit to those used in the \textsc{LHAPDF} framework. Yet \textsc{LHAPDF}, as implemented for the proton, supports as partons only six quarks and their antiquarks, gluons, and photons. In contrast, the EWPDF case requires treating all SM particles --- along with their distinct helicity states --- as independent degrees of freedom. For these reasons, an entirely new framework had to be established.

In \textsc{Whizard}, the EWPDF is a single-beam structure function, similarly to the EVA case; thus, it can also be applied to beam setups asymmetric in momentum or flavour. The implementation supports 56 partons\footnote{Several PDF components are related to each other if the masses of the corresponding particles are neglected, e.g. $f_{u_L} = f_{c_L}$, $f_{u_R} = f_{c_R}$, etc.~\cite{Garosi:2023bvq}. To allow for maximal flexibility in defining the grids, the correlations are not imposed in the code.}:
\begin{itemize}
    \item 24 entries for quarks (6 flavours $\times$ 2 helicities for particles and antiparticles),
    \item 12 entries for charged leptons (3 flavours $\times$ 2~helicities for particles and antiparticles), 
    \item 6 entries for neutrinos (3 flavours for left-handed neutrinos and right-handed antineutrinos),
    \item 4 entries for massless gauge bosons (2 helicities for the photon and the gluon),
    \item 9 entries for massive gauge bosons (3 helicities for the $Z$, $W^+$ and $W^-$ bosons),
    \item 1 entry for the Higgs boson.
\end{itemize}  
The current version neglects the three mixed parton states, $Z/\gamma_\pm$ and $h/Z_L$~\cite{Han:2020uid,Marzocca:2024ica}. The inclusion of these states, while contributing moderately to the final calculations, poses significant technical challenges and will require major changes in the interface routines of the internal Matrix Element generator, which has been left for further work within the \textsc{Whizard} collaboration. 

In the derivation of the EWPDFs, particle masses at the GeV scale are consistently incorporated into the calculations. The \textsc{Whizard} implementation distinguishes five mass states: the top mass, the Higgs mass, the $Z$ boson mass, the $W$ boson mass, and the massless case. However, the current version of the EWPDF grids provided by the Authors of~\cite{Han:2020uid} and used for this analysis includes only a single scale, associated with the $Z$ boson mass.

Similarly to the EVA case, the code allows users to specify a minimal value of $x$. However, since partons are treated as massive, only $x > m_\text{parton}/\sqrt{s}$ is sensible. Users can additionally set the process scale, either as fixed constants (e.g. related to the nominal collision energy) or as dynamical expressions (e.g. the hard-process scale). The concept will be discussed further in the thesis.

Even though the implementation was developed using a specific set of EWPDFs, it can be used with any future grids. Only minimal adjustments may be required --- for example, adapting the parton numbering scheme, which in the current version combines the PDG code with the helicity code into a three-digit identifier (e.g. 301 denotes the left-handed $s$ quark, while $-240$ denotes the longitudinal $W^{-}$ boson).

An example SINDARIN input file for running EWPDFs in \textsc{Whizard} is provided in Appendix~\ref{app:sinEWPDF}.

\section{Comparison of EWPDF with EVA}

As discussed earlier, EWPDFs are constructed within the framework of collinear splittings. Unlike the EVA, they include the resummation of large logarithms, which can improve the precision of collider predictions, even under the collinear approximation. To separate the impact of higher-order corrections from that of the small-angle expansion, it is useful to compare results obtained with the EWPDF framework to those from the EVA, alongside full matrix-element calculations. 

In the following, a few representative cases are examined, drawing on a reduced selection from Section~\ref{sec:EVAfullME} to eliminate redundancy. Building on previous findings, the default factorisation scale is chosen as $\sqrt{\hat{s}}/4$. Unless stated otherwise, results are shown for an electron--muon collider operating at 10\,TeV, which suppresses the $s$-channel contributions that could otherwise undermine the VBF approximation.

\subsection{Di-Higgs}
\begin{figure}
	\centering
	\begin{subfigure}{0.49\textwidth}
		\centering
		\includegraphics[width=\textwidth]{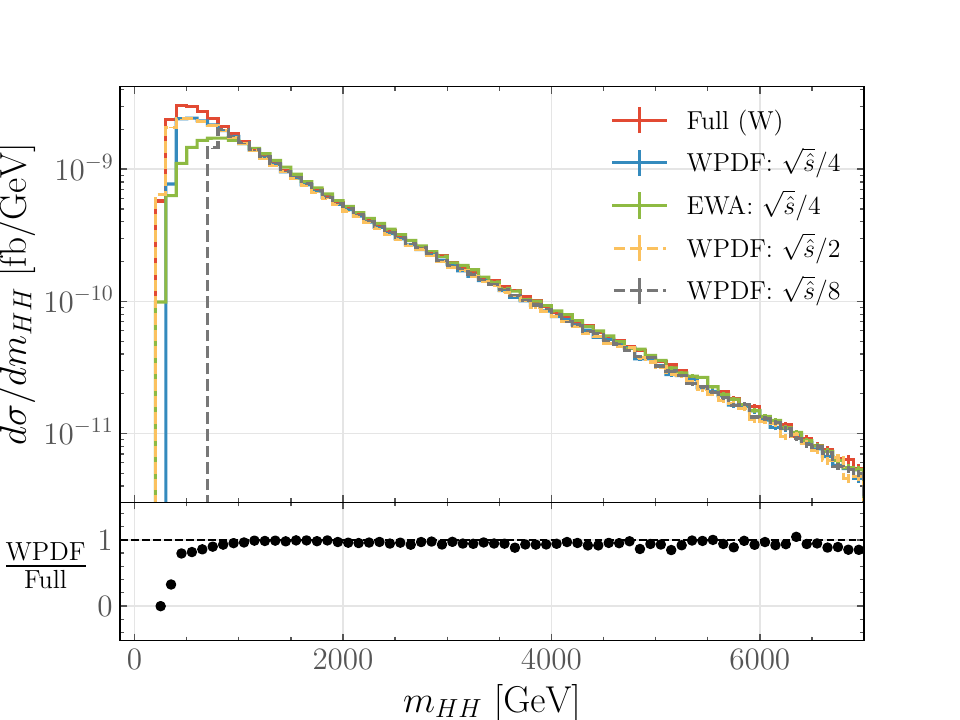}
		\caption{$WW$-initiated}
	\end{subfigure}
	\hfill
	\begin{subfigure}{0.49\textwidth}
		\centering
		\includegraphics[width=\textwidth]{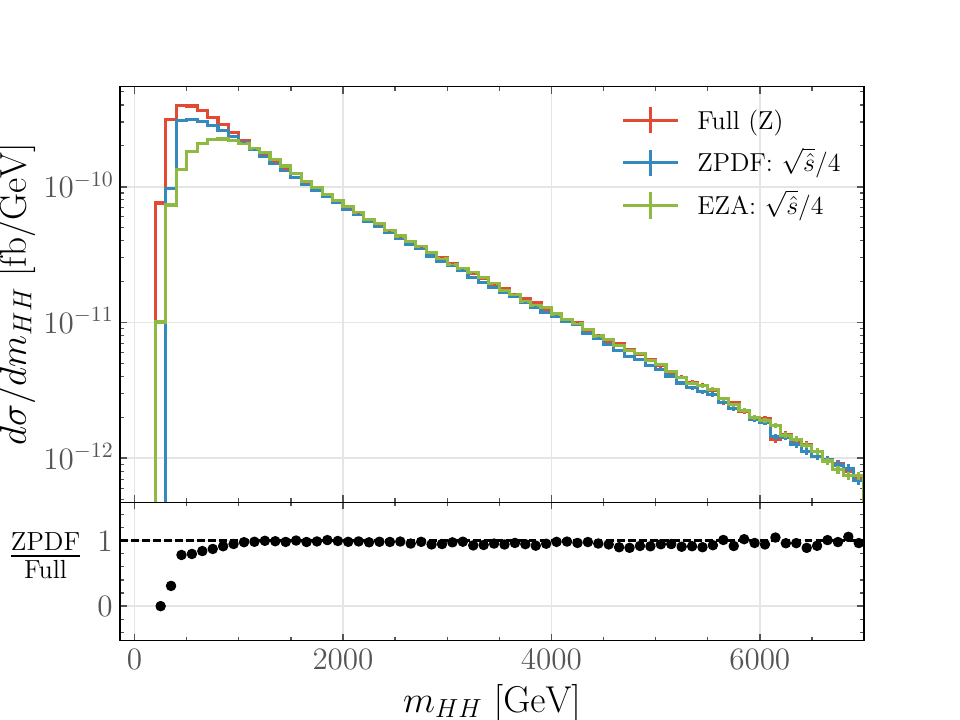}
		\caption{$ZZ$-initiated}
	\end{subfigure}
	\caption{Invariant mass distributions of the di-Higgs system in the process $e^+ \mu^- \to HH + X$ at $\sqrt{s} = 10$\,TeV for the full matrix element calculation (\textit{red}), the EWPDF (\textit{blue}) and the EVA (\textit{green}). 
    The \textit{yellow} and \textit{grey} dashed lines represent a variation of the scale by a factor of two. To facilitate the comparison, the scale variation is not shown in the right plot.}
	\label{fig:EWPDF_HH}
\end{figure}

The first process to consider is again $e^+ \mu^- \to HH + X$, where contributions mediated by the $W$ and $Z$ bosons were included. It should be emphasised that, within the EWPDF formalism, fully inclusive predictions generally require summing over all possible initial-state partons that can contribute to a given final state. In the present analysis, however, the selection is restricted to specific sub-processes in order to enable a direct and transparent comparison between the EWPDF and EVA frameworks. In any case, these parts are considered to yield the dominant contributions.

Figure~\ref{fig:EWPDF_HH} shows the invariant mass distribution of the di-Higgs system, comparing the EVA and EWPDF predictions with the full matrix-element calculations. In the left plot, the scale variation by a factor of two is additionally presented. The conclusions drawn in the EVA case also hold here: the collinear approximation provides a good description of the full process except near threshold, where discrepancies can be mitigated by imposing a lower invariant-mass cut. The tail of the distribution is essentially scale-independent, reflecting the dominance of the longitudinal $W$ boson contribution.

Two noteworthy effects appear in the threshold region. First, the EWPDFs describe this regime significantly better than the EVA, which might stem from either resummation effects or differences in the numerical treatment of PDFs. Second, the EWPDF distributions start at different values depending on the scale choice. This is, however, a technical artifact: in the current EWPDF grid implementation, weak bosons appear as partons only for energies above the $Z$ boson mass, and their contributions vanish identically below this threshold. Consequently, when the scale $\sqrt{\hat{s}}/8$ is used, non-zero entries arise only for $\sqrt{\hat{s}} > 8m_Z$. This issue can be addressed in future releases of EWPDF grids through a more refined treatment of threshold effects.

\subsection{Tau neutrinos}
\begin{figure}
	\centering
	\begin{subfigure}{0.49\textwidth}
		\centering
		\includegraphics[width=\textwidth]{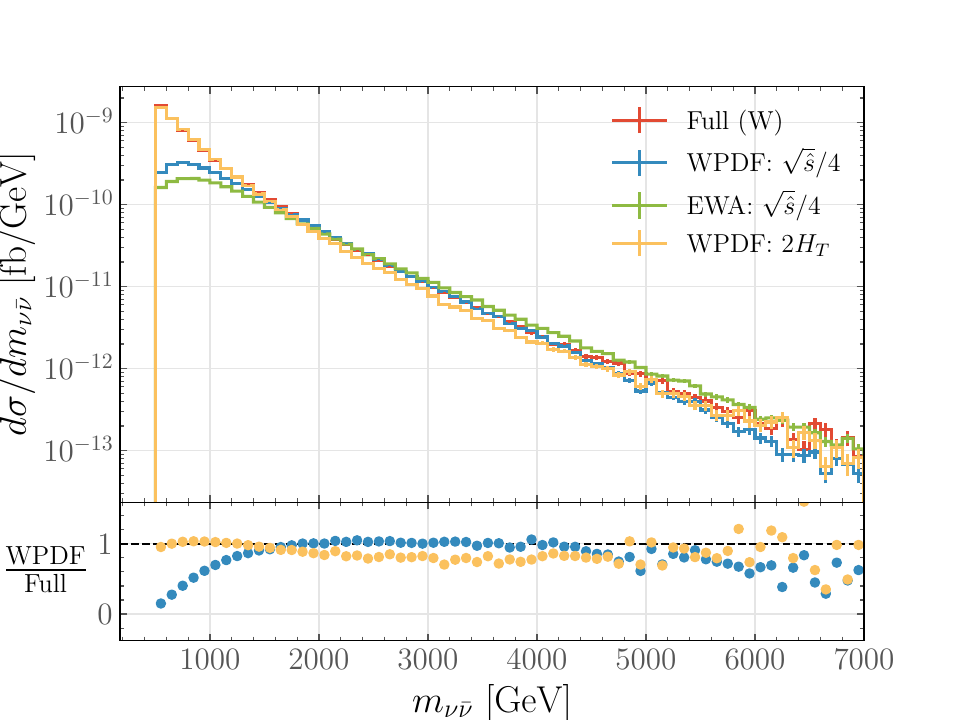}
		\caption{$WW$-initiated}
	\end{subfigure}
	\hfill
	\begin{subfigure}{0.49\textwidth}
		\centering
		\includegraphics[width=\textwidth]{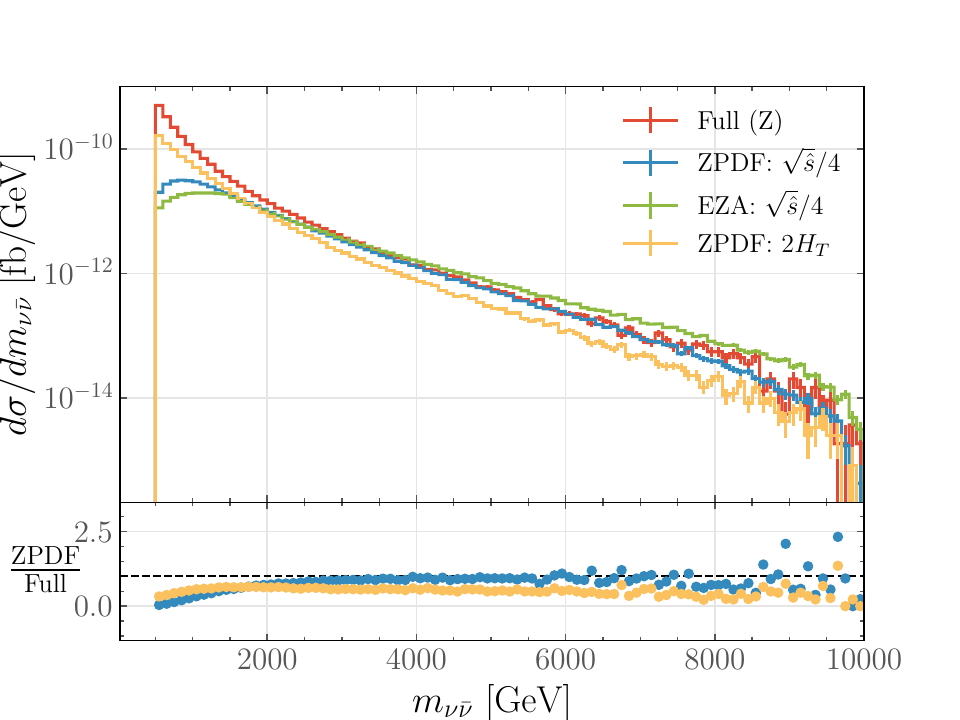}
		\caption{$ZZ$-initiated}
	\end{subfigure}
	\caption{Invariant mass distributions of the $\nu_\tau \bar \nu_\tau$-system  in $e^+ \mu^- \to \nu_\tau \bar \nu_\tau + X$ at $\sqrt{s} = 10$\,TeV for the full matrix element calculation (\textit{red}), the EWPDF (\textit{blue}) and the EVA (\textit{green}). The \textit{yellow} lines indicate the EWPDF result with another scale choice of $2H_T$.}
	\label{fig:EWPDF_nunu}
\end{figure}

\begin{figure}
	\centering
	\begin{subfigure}{0.49\textwidth}
		\centering
		\includegraphics[width=\textwidth]{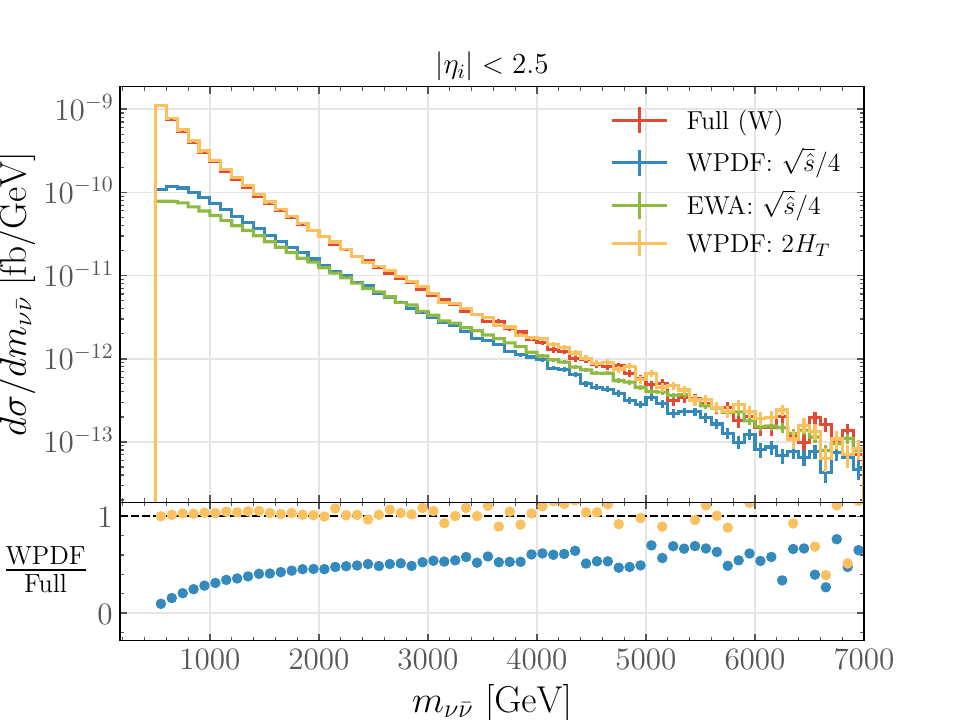}
		\caption{$WW$-initiated}
	\end{subfigure}
	\hfill
	\begin{subfigure}{0.49\textwidth}
		\centering
		\includegraphics[width=\textwidth]{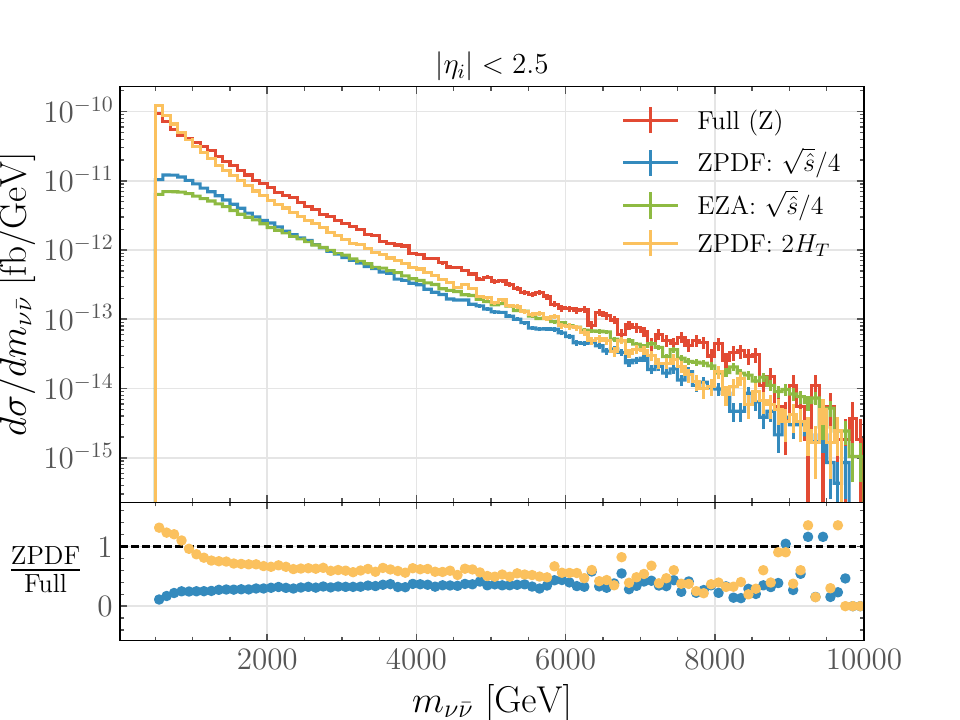}
		\caption{$ZZ$-initiated}
	\end{subfigure}
	\caption{Invariant mass distributions of the $\nu_\tau \bar \nu_\tau$-system  in $e^+ \mu^- \to \nu_\tau \bar \nu_\tau + X$ at $\sqrt{s} = 10$\,TeV for the full matrix element calculation (\textit{red}), the EWPDF (\textit{blue}) and the EVA (\textit{green}). The \textit{yellow} lines indicate the EWPDF result with another scale choice of $2H_T$. With respect to Figure~\ref{fig:EWPDF_nunu}, a cut on the pseudorapidity of the final-state tops, $|\eta_i| < 2.5$, is applied.}
	\label{fig:EWPDF_nunu_cut}
\end{figure}

In the next step, the process $e^+ \mu^- \to \nu_\tau \bar \nu_\tau + X$ is studied. As before, tau neutrinos are selected to suppress non-VBF topologies. The process is dominated by transversely polarised vector bosons and is therefore theoretically interesting, although not experimentally observable. It is important to emphasise that neutrinos produced via $Z$ boson emissions are still present and cannot be removed without violating gauge invariance; their impact is controlled instead through an invariant-mass cut applied to the tau-neutrino pair. 

Figure~\ref{fig:EWPDF_nunu} shows the invariant-mass distribution of the tau-neutrino system. As found in the previous chapter, the agreement between the collinear framework with the standard scale choice and the full calculation remains rather poor in the threshold region. This motivated the search for an alternative scale that performs better in this regime. 

In the literature, a typical scale choice for EVA or its extensions is associated with either the collision energy or the invariant mass of the initial state. On the other hand, in LHC physics, a scale related to the transverse mass is often chosen for various processes involving massive final states with significant transverse momentum. Transverse-mass observables are employed at hadron colliders because the longitudinal momentum of the initial state is unknown on an event-by-event basis. Similarly, in processes involving unobserved neutrinos at lepton colliders, the longitudinal momentum of the final state cannot be fully reconstructed. In such cases, transverse observables appear to be an appropriate choice.

For nearly massless neutrinos, the transverse mass, typically defined as $H_T\! = \!\sqrt{m^2 + p^2_\perp}$, effectively reduces to the transverse momentum of the neutrino, $p_\perp$. This leads to an unphysically small scale in the low-\(p_\perp\) region. To regulate this behaviour, an alternative scale is tested, in which the neutrino mass is replaced by the weak-boson mass, $2H_T = 2\sqrt{m_{W/Z}^2 + p^2_\perp}$. In the collinear framework, the two neutrinos are always produced back-to-back, which motivates the overall factor of two. This choice reflects the presence of a virtual electroweak boson exchange that mediates the interaction. The weak-boson mass is thus used as a proxy for the intrinsic electroweak scale of the process. 

A very good agreement is obtained for the $WW$-initiated process, while the agreement for the $ZZ$-initiated process remains acceptable. This behaviour is expected: the full process $e^+ \mu^- \to \nu_\tau \bar \nu_\tau + e^+ \mu^-$ receives sizeable contributions from $t$-channel photon-exchange diagrams, which do not appear in the former case. Such diagrams are absent in the EWPDF framework, which includes only VBF-like topologies.

Figure~\ref{fig:EWPDF_nunu_cut} demonstrates that the scale choice of $2H_T$ indeed performs better than the invariant-mass scale. Here, an additional cut on the pseudorapidity of the $\tau$ neutrinos, $|\eta_i| < 2.5$, is applied. This cut corresponds to the central detector region (an acceptance of about 10 degrees). For the $WW$-initiated process, it becomes clear that this scale accurately reproduces the full results, with the kinematics implicitly better captured, as the kinematic cut yields nearly perfect agreement in this case and spoils accuracy for the other scales. In contrast, for the $ZZ$ case, the agreement remains moderate due to the missing $\gamma$-exchange diagrams, although the cut still leads to a visible improvement, especially in the peak region.

\subsection{Top pairs}
\begin{figure}
	\centering
	\begin{subfigure}{0.49\textwidth}
		\centering
		\includegraphics[width=\textwidth]{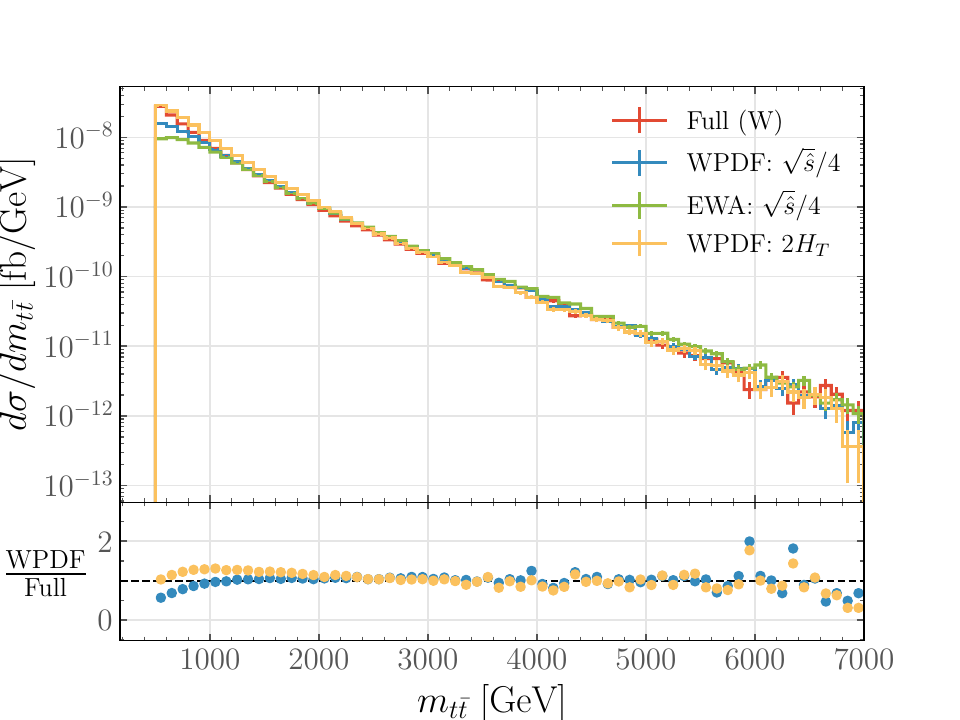}
		\caption{$e^+ \mu^-$ collider}
	\end{subfigure}
	\hfill
	\begin{subfigure}{0.49\textwidth}
		\centering
		\includegraphics[width=\textwidth]{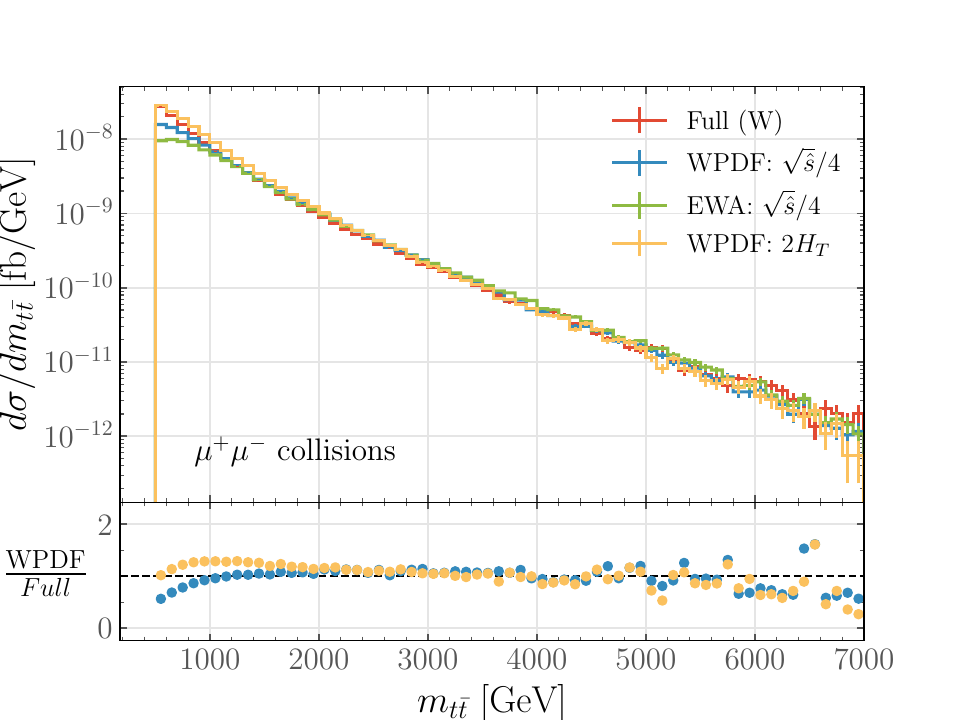}
		\caption{$\mu^+ \mu^-$ collider}
	\end{subfigure}
	\caption{Invariant mass distributions of the $t \bar{t}$-system at an $e^+ \mu^-$ (left) and a $\mu^+ \mu^-$ collider (right) for the full matrix element calculation (\textit{red}), the EWPDF (\textit{blue}) and the EVA (\textit{green}). The \textit{yellow} lines indicate the EWPDF result with another scale choice of $2H_T$.}
	\label{fig:EWPDF_ttbar}
\end{figure}

\begin{figure}
	\centering
	\begin{subfigure}{0.49\textwidth}
		\centering
		\includegraphics[width=\textwidth]{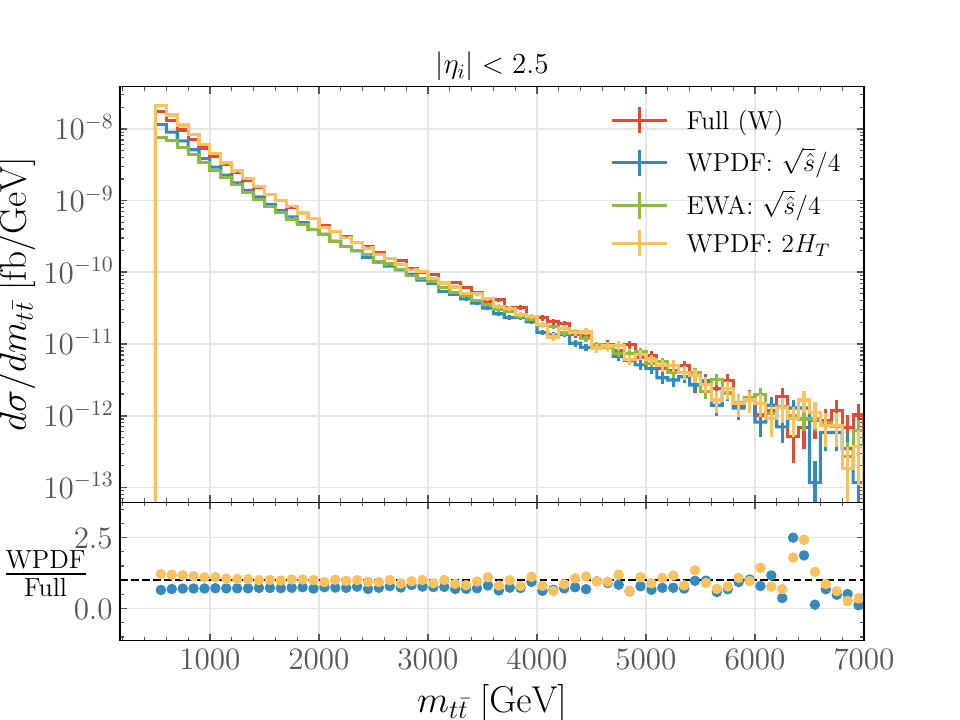}
		\caption{$e^+ \mu^-$ collider}
	\end{subfigure}
	\hfill
	\begin{subfigure}{0.49\textwidth}
		\centering
		\includegraphics[width=\textwidth]{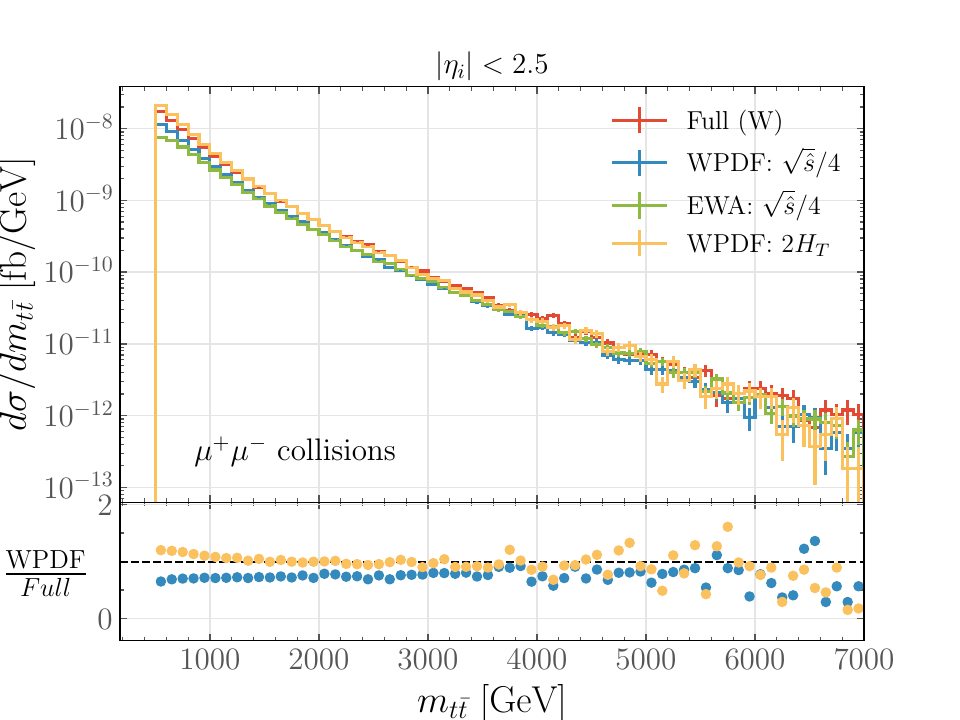}
		\caption{$\mu^+ \mu^-$ collider}
	\end{subfigure}
	\caption{Invariant mass distributions of the $t \bar{t}$-system at an $e^+ \mu^-$ (left) and a $\mu^+ \mu^-$ collider (right) for the full matrix element calculation (\textit{red}), the EWPDF (\textit{blue}) and the EVA (\textit{green}). The \textit{yellow} lines indicate the EWPDF result with another scale choice of $2H_T$. With respect to Figure~\ref{fig:EWPDF_ttbar}, a cut on the pseudorapidity of the final-state tops, $|\eta_i| < 2.5$, is applied.}
	\label{fig:EWPDF_ttbar_cut}
\end{figure}

The next process is the top-pair production. Figure \ref{fig:EWPDF_ttbar} shows the results for $e^+ \mu^- \to t \bar{t} + \nu_\mu\bar{\nu}_e$ and $\mu^+ \mu^- \to t \bar{t} + \nu_\mu\bar{\nu}_\mu$, where the latter includes additional small contributions from $s$-channel diagrams in the full matrix-element calculation. In both channels, the factorised approach provides a reasonably accurate description of the processes. The EWPDF predictions outperform those of the EVA, and the transverse-mass scale (now unmodified) gives a slightly better agreement with the full result than the invariant-mass scale, although the difference remains modest. 

Figure \ref{fig:EWPDF_ttbar_cut} shows the same distributions, now with an additional cut applied to the final-state pseudorapidity. In this case, the transverse-mass scale almost perfectly reproduces the full matrix-element results, even for the $\mu^+\mu^-$ collider. This demonstrates that the collinear framework can be applied reliably to realistic collider setups and can successfully reproduce the full results in specific cases.

\subsection{Vector boson scattering}
\begin{figure}
	\centering
	\begin{subfigure}{0.49\textwidth}
		\centering
		\includegraphics[width=\textwidth]{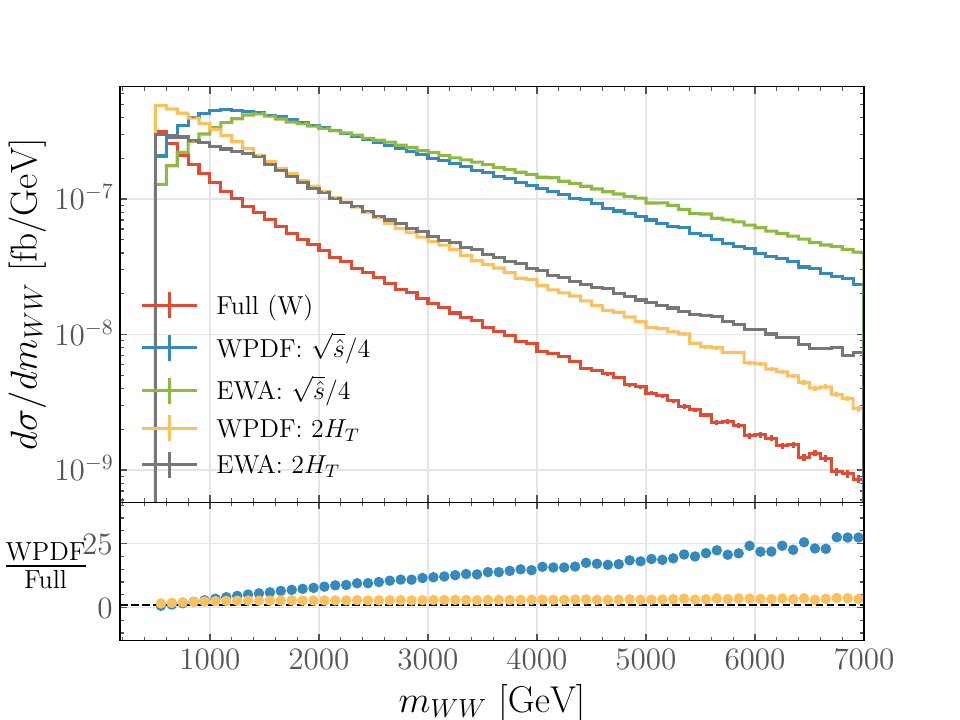}
        \caption{no cut}
	\end{subfigure}
	\hfill
	\begin{subfigure}{0.49\textwidth}
		\centering
		\includegraphics[width=\textwidth]{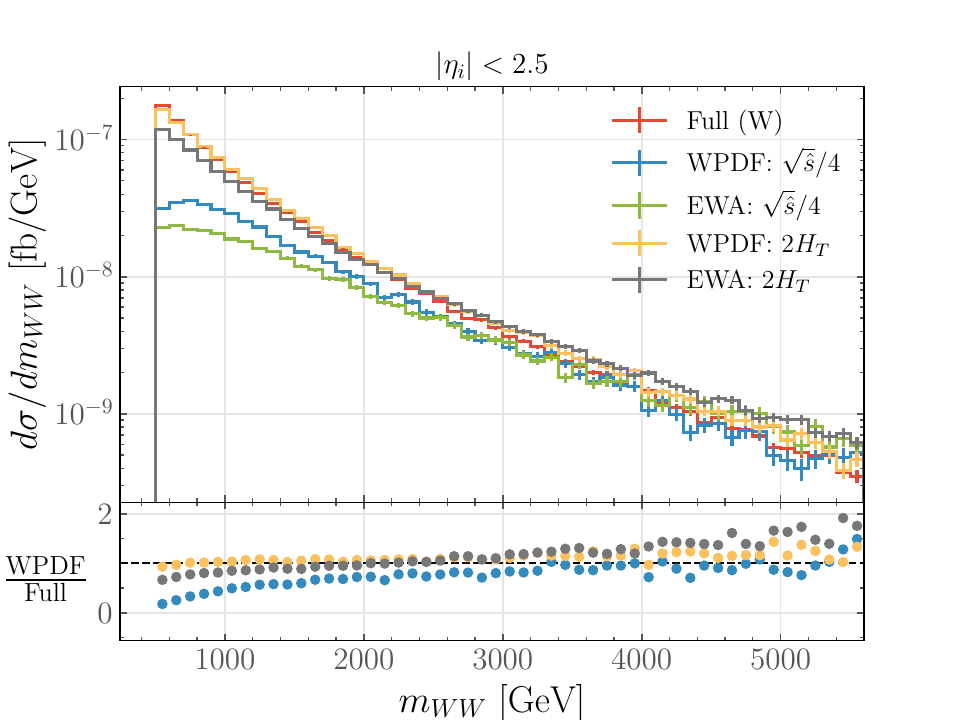}
        \caption{$|\eta_i|< 2.5$}
	\end{subfigure}
    \hfill
	\begin{subfigure}{0.49\textwidth}
		\centering
		\includegraphics[width=\textwidth]{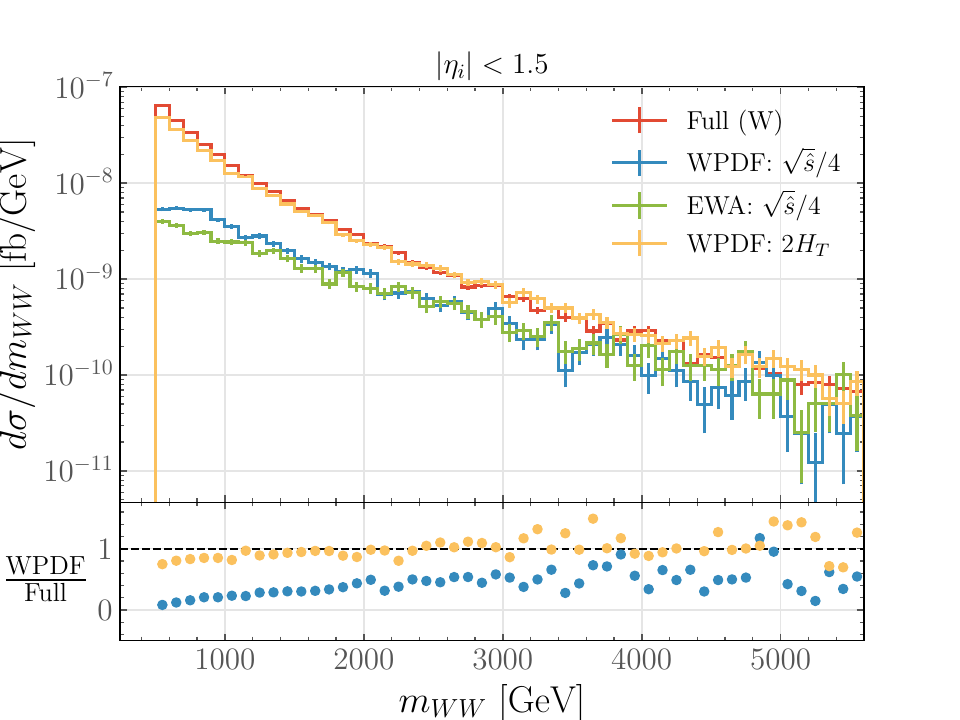}
        \caption{$|\eta_i|< 1.5$}
	\end{subfigure}
	\caption{Invariant mass distributions of the $W^+W^-$ system in the vector boson scattering $e^+ \mu^- \to W^+ W^- + X$ at $\sqrt{s} = 10$\,TeV for the full matrix element calculation (\textit{red}), the EWPDF (\textit{blue}) and the EVA (\textit{green}). The \textit{yellow} (grey) lines indicate the EWPDF (EVA) result with another scale choice of $2H_T$. The other scale for the EVA is not shown in the last plot for clarity.}
	\label{fig:EWPDF_WW}
\end{figure}

Finally, the last process considered is vector boson scattering, $e^+\mu^- \rightarrow W^+ W^- + \nu_\mu \bar \nu_e$.
As in the previous chapter, a generator-level cut of 20\,GeV on the $p_\perp$ of the final-state $W$ bosons is applied to improve convergence. The invariant-mass distribution is shown in Figure~\ref{fig:EWPDF_WW}. Without any additional cuts, the collinear framework differs significantly from the full matrix-element prediction. However, the $2H_T$ scale --- shown here also for the EVA case --- performs noticeably better. Moreover, the EWPDF result follows a similar slope to the full calculation, whereas the EVA result deviates more strongly.

When the pseudorapidity cut $|\eta_i| < 2.5$ is applied, the agreement improves substantially, with the EWPDF reproducing the full calculation almost perfectly. Tightening the cut further to $|\eta_i| < 1.5$ shows that the good performance of the transverse-mass scale persists, in contrast to the findings of the previous chapter with the typical, invariant-mass scale choice. This strongly suggests that the novel scale choice captures the relevant kinematics much more effectively and should be adopted as the default choice for subsequent analyses.

\subsection{EWPDF vs. EVA --- general remarks}

The comparison above shows that good agreement between the EWPDF and the full matrix-element calculations --- and, at the same time, between the EWPDF and EVA frameworks --- can be achieved for the most relevant processes once an appropriate scale is chosen. The transverse-mass scale clearly outperforms the invariant-mass scale used in Chapter~\ref{chap:EVA} and, at the same time, captures the kinematics more accurately. Overall, the differences between EVA and EWPDF appear subtle, and the description provided by the EWPDF framework is, in general, more accurate in the relevant collinear regime, as it is not merely a kinematic approximation but a proper QFT-motivated framework. As such, it offers means to systematically improve the accuracy of the results presented in this thesis.

However, several caveats should be considered before concluding differences and similarities between the two formalisms. First, the results in this section were compared to full leading-order calculations, while the EWPDF framework, by construction, incorporates higher-order effects through resummation. The deviations between the two collinear approaches may therefore originate from these additional contributions. Investigating this lies beyond the scope of the present thesis but will be pursued in future work.

Furthermore, the EWPDF approach introduces a qualitatively new feature: the presence of multiple other particles as partons. This aspect will be explored in the next section, where processes involving colour-charged particles will serve as an illustrative example.

\section{Colour-charged particles in the EWPDF framework}
The Authors of~\cite{Han:2021kes} showed that VBF processes involving colour-charged particles can contribute significantly to the phenomenology of the Muon Collider. In particular, processes such as dijet production can reach cross sections of tens of picobarns, forming an important background at future multi-TeV colliders. Naturally, a fraction of these events originates from purely QED interactions, for example through $\gamma\gamma \to q\bar{q}$. However, the dominant contributions stem from QCD interactions, such as $gg \to jj$ or $gq \to gq$.

The coloured partons involved in these processes originate from a sequence of mixed electroweak and QCD splittings and are well described within the framework of EWPDFs. In this section, a few example processes involving coloured particles at a $\mu^{_+} \mu^{_-}$ collider are presented to complement the discussion given above.

\subsection{Photon-initiated processes}
\begin{figure}
	\centering
	\includegraphics[width=0.69\textwidth]{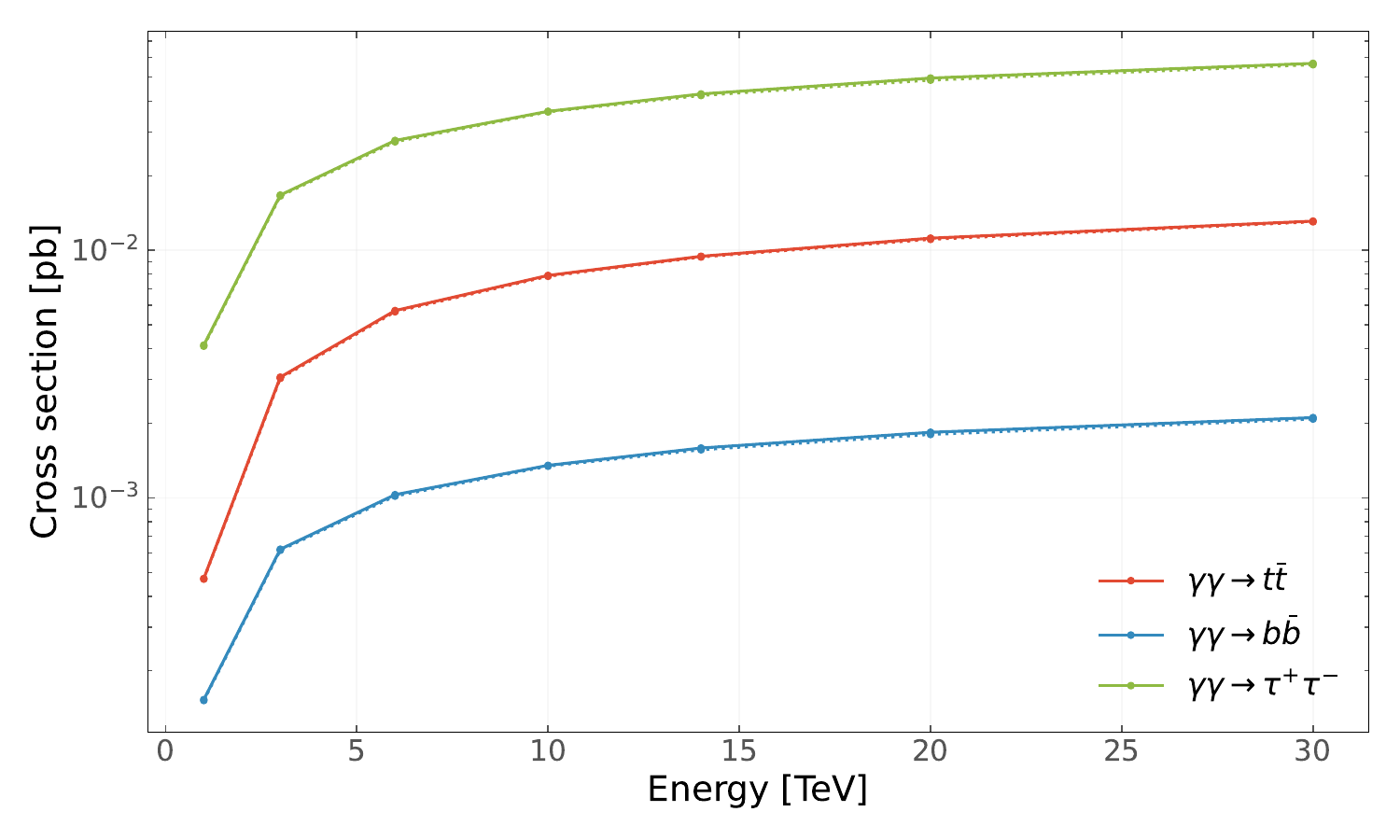}
	\caption{Cross sections for various processes initiated by photons at a muon collider as a function of the collision energy (red: $t\bar{t}$ production, blue: $b\bar{b}$ production, green: $\tau^+\tau^-$ production). Solid lines are calculated within the new EWPDF implementation in \whizard, while the dotted lines were obtained by the Authors of~\cite{Han:2021kes}. A 100-GeV cut is applied on the transverse momentum of outgoing particles and a 300-GeV cut on their invariant mass (trivially satisfied for the $t\bar{t}$ production).}
	\label{fig:aa_EWPDF}
\end{figure}

First, the photon content of the muon is considered. Figure~\ref{fig:aa_EWPDF} shows the cross sections for three processes initiated by photons treated as partons: $t\bar{t}$ production, $b\bar{b}$ production, and $\tau^+\tau^-$ production. Collision energies in the range of 1--30\,TeV are scanned. The factorisation scale is set to $\sqrt{\hat{s}}/2$ to compare with previous studies~\cite{Han:2021kes}. The results shown as solid lines, obtained using the new \whizard implementation, are compared with those privately reported by the authors of~\cite{Han:2021kes}. The two sets of results are in excellent agreement, and the tiny differences visible upon closer inspection can be safely attributed to numerical fluctuations. This agreement further validates the technical implementation of the EWPDF framework in \whizard.

Depending on the collision energy, the cross sections typically lie in the range from single to tens of femtobarns. By applying a cut on the transverse momentum of the final-state particles of $p_T > 100\,\text{GeV}$ and an invariant-mass cut of $m_{jj} > 300\,\text{GeV}$ (the latter being trivially satisfied for $t\bar{t}$ production), it is expected that these events will be experimentally accessible at a future muon collider. The hierarchy among the processes originates from QED and kinematic effects. 
At high energies, the cross section for $\gamma \gamma \to f \bar f$ is expected to scale as $N_{\rm col}\, \cdot Q_f^4$, where $Q_f$ is the fermion electric charge and $N_{\rm col}$ is the number of colours. 
Under this approximation, one would therefore expect the ratios
\[
\sigma(\gamma\gamma \to \tau^+\tau^-) : \sigma(\gamma\gamma \to t\bar t) : \sigma(\gamma\gamma \to b\bar b) 
\;\simeq\; 1 : \frac{16}{27} : \frac{1}{27}.
\]
However, the large mass of the top quark significantly reduces its cross section due to the limited 
$\sqrt{\hat{s}}$ available from the photon spectrum. 
As a result, the top-quark cross section is lowered to approximately one quarter of the tau-lepton cross section.

\subsection{Quark-initiated processes}
\begin{figure}
	\centering
	\includegraphics[width=0.69\textwidth]{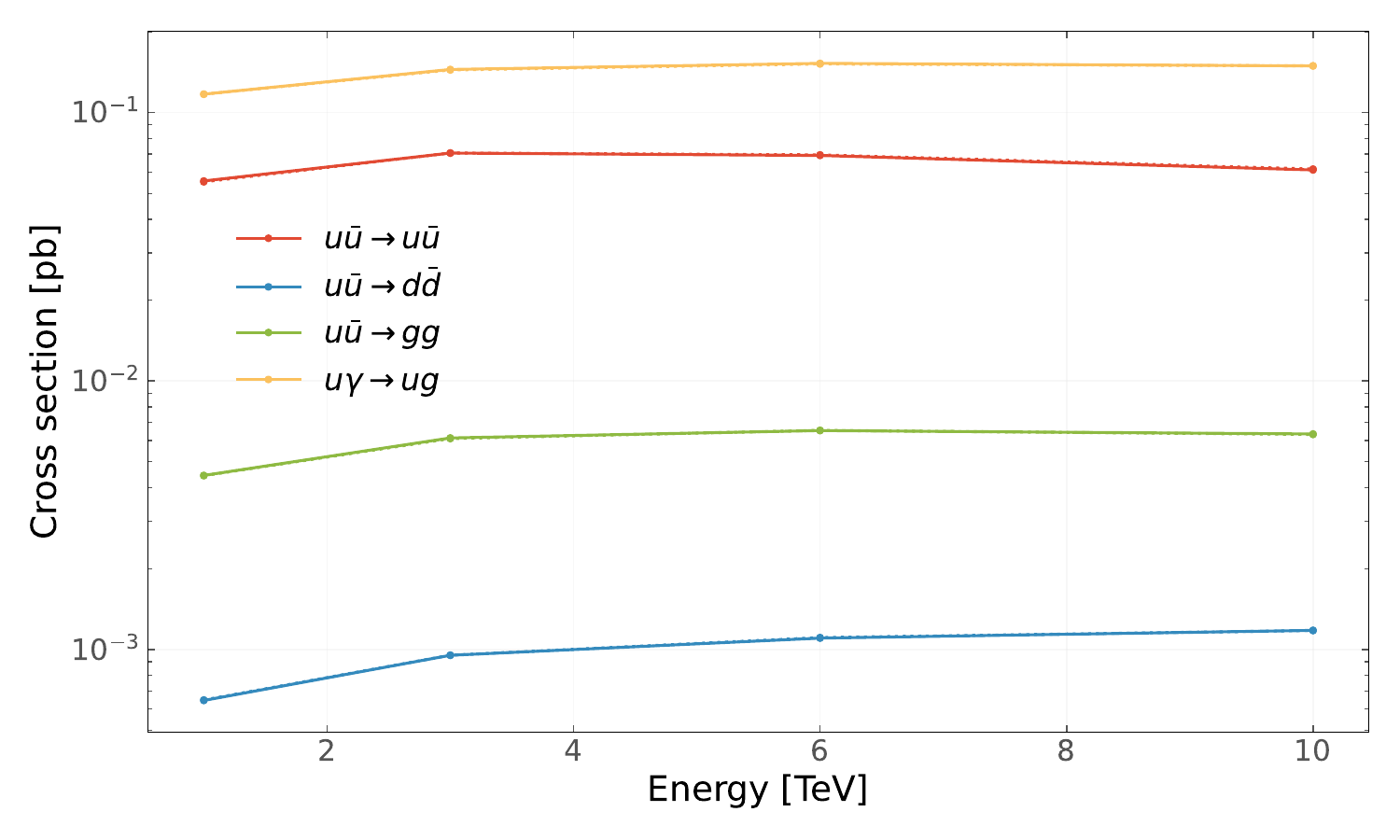}
	\caption{Cross sections for various processes initiated by $u$ quarks and photons at a muon collider as a function of the collision energy (red: $u\bar{u} \to u\bar{u}$, blue: $u\bar{u} \to d\bar{d}$, green: $u\bar{u} \to gg$, yellow: $u\gamma \to ug$). Solid lines are calculated within the new EWPDF implementation in \whizard, while the dotted lines were obtained by the Authors of~\cite{Han:2021kes}. The same set of cuts is adapted (see text for details).}
	\label{fig:uu_EWPDF}
\end{figure}

The $u$-quark content of the muon can arise from secondary splittings of electroweak gauge bosons. In this way, mixed EW--QCD interactions can take place. Figure~\ref{fig:uu_EWPDF} shows the cross sections for processes initiated by $u$ quarks and photons: $u\bar{u} \to u\bar{u}$, $u\bar{u} \to d\bar{d}$, $u\bar{u} \to gg$, and $u\gamma \to ug$. Collision energies in the range of 1--10\,TeV are considered, the factorisation scale is set to $\sqrt{\hat{s}}/2$, and the running of $\alpha_s$ is included. For technical simplicity, in the case of asymmetric initial states, the quark is taken to originate from the muon beam, while the photon comes from the anti-muon beam; an identical contribution is expected from the opposite configuration.

Following~\cite{Han:2021kes}, a set of cuts is imposed to separate the hadronic activity:
\begin{equation}
\label{eq:cuts_EWPDF}
p_T^j > \left(4 + \frac{\sqrt{s}}{3~\text{TeV}}\right)\,\text{GeV}, \quad 
|\eta_j| < 2.44, \quad 
m_{jj} > 20\,\text{GeV}, 
\end{equation}
where $p_T^j$ denotes the transverse momentum of each outgoing particle, $\eta_j$ the pseudorapidity in the laboratory frame, and $m_{jj}$ their invariant mass. The energy-dependent cut on $p_T^j$ is chosen to control collinear logarithms of the form $(\alpha_s/\pi)\log(p_T^j/\sqrt{s})$~\cite{Barklow:1443518}. The pseudorapidity cut corresponds to a minimum angle with respect to the beam axis of approximately $\theta_j \sim 10^\circ$ in the laboratory frame.

The two sets of results (from \whizard and from the previous work) are again found to be in perfect agreement. In this case, the dominant contribution arises from the mixed photon--quark interaction, closely followed by $u\bar{u}$ scattering. For a realistic phenomenological study, all accessible quark flavours should be included; here, they are shown separately to allow a clear comparison with the previous work, while in a full analysis, they would be summed to estimate the total hadronic contribution.

A qualitative comparison between up-type and down-type quarks showed that up-type quarks generally give larger contributions. At the 10-TeV Muon Collider, the cross sections for processes initiated by $u\bar{u}$ pairs can exceed those initiated by $d\bar{d}$ pairs by roughly an order of magnitude. This hierarchy arises because QED-induced production of coloured partons scales with the quark electric charge (squared per beam); in contrast, production via weak bosons is governed by the electroweak couplings, which are larger for down-type quarks. The relative importance of the two contributions can, however, be understood from the logarithmic enhancements in the splitting: for QED, the relevant logarithm involves the muon mass, $\ln(Q^2/m_\mu^2)$, while for weak boson emission it is controlled by the mass of the heavy bosons, $\ln(Q^2/m_{W/Z}^2)$. As a result, photon-induced splittings dominate at high energies, leading to a strong preference for up-type quarks.

\subsection{Gluon-initiated processes}
\begin{figure}
	\centering
	\includegraphics[width=0.69\textwidth]{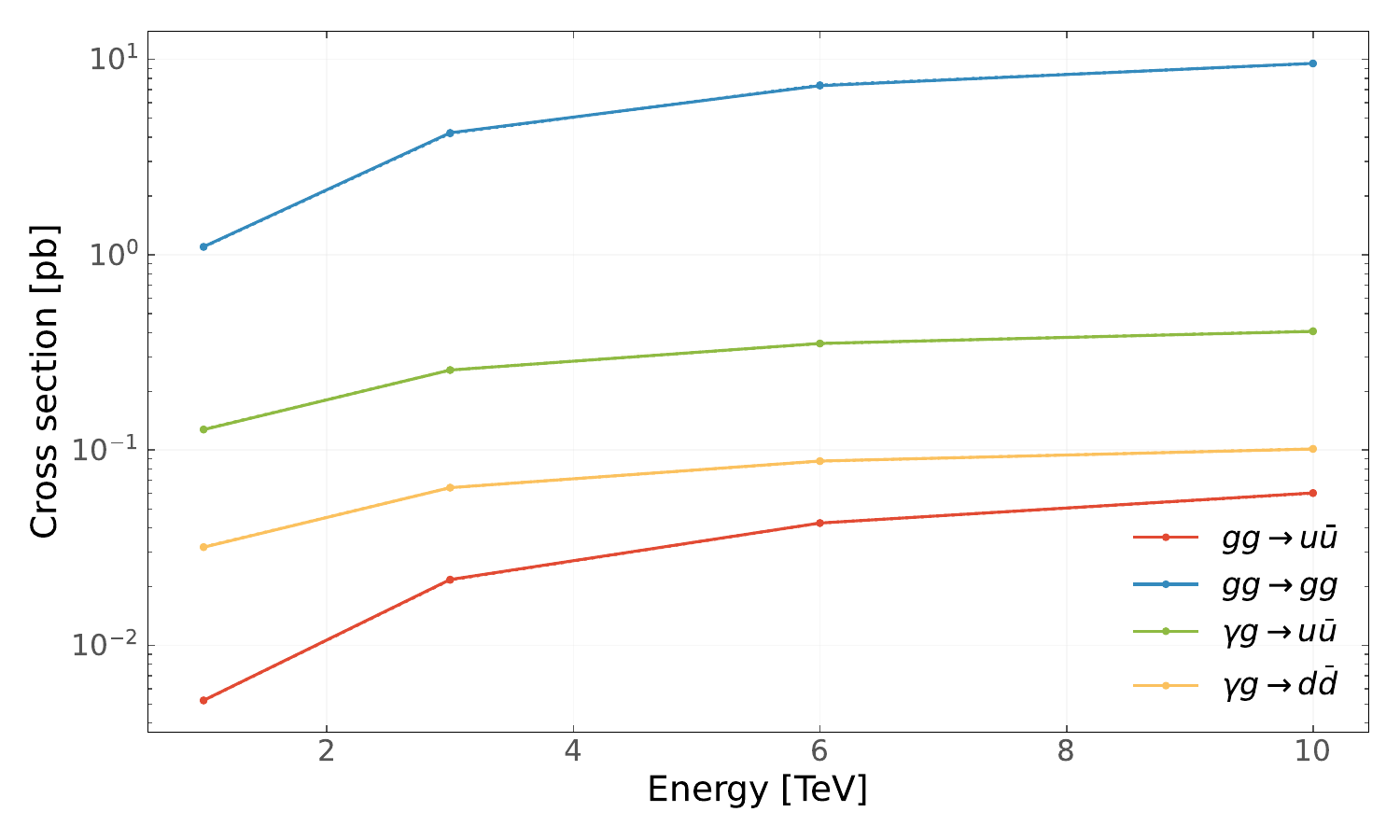}
	\caption{Cross sections for various processes initiated by gluons and photons at a muon collider as a function of the collision energy (red: $gg \to u\bar{u}$, blue: $gg \to gg$, green: $\gamma g \to u\bar{u}$, yellow: $\gamma g \to d\bar{d}$). Solid lines are calculated within the new EWPDF implementation in \whizard, while the dotted lines were obtained by the Authors of~\cite{Han:2021kes}. The same set of cuts is adapted (see text for details).}
	\label{fig:gg_EWPDF}
\end{figure}

The gluon content of the muon arises from a sequence of splittings, beginning with the emission of an electroweak boson that subsequently produces a coloured state. Figure~\ref{fig:gg_EWPDF} shows the cross sections for processes initiated by gluons, or by gluons and photons treated as muon partons: $gg \to u\bar{u}$, $gg \to gg$, $\gamma g \to u\bar{u}$, and $\gamma g \to d\bar{d}$. The same setup as in the previous case is used, including the energy range, scale and cuts of Eq.~\ref{eq:cuts_EWPDF}. For gluon--photon processes, the gluon is taken to originate from the muon beam, while the photon comes from the anti-muon beam. As in the previous examples, the \whizard results agree perfectly with those of~\cite{Han:2021kes}. 

In this case, the dominant contribution arises from gluon--gluon scattering, with cross sections reaching the level of single picobarns. The remaining processes also contribute to the background, although typically at the level of one to two orders of magnitude lower per channel. The large enhancement of gluon--gluon scattering relative to quark--antiquark production originates from the non-Abelian structure of QCD. In particular, triple-gluon vertices lead to a stronger singular behaviour in the forward region, i.e. at low transverse momentum. The impact of this enhancement can be partially mitigated by imposing more stringent transverse-momentum cuts on the final state. Naturally, for final states involving quarks, a sum over all accessible flavours should be performed to obtain the total cross section.

\section{Implications for EWPDF phenomenology}

In this chapter, the EWPDF implementation in \whizard was presented, and a representative set of processes was studied to validate the implementation and to outline the physics potential of the formalism. The EWPDF approach provides a promising path towards improving theoretical precision in the collinear regime by systematically including resummation effects. However, further developments are required to establish a consistent physics prescription for its use, in particular when complemented by analogous treatments of final-state particles as partons in subsequent electroweak showers.

The main conclusions of this chapter can be summarised as follows:
\begin{itemize}
\item The \whizard implementation of the EWPDF framework has been validated and is technically correct, reproducing full matrix-element calculations in the appropriate kinematical regime.

\item The EWPDF formalism for longitudinal functions shows a very mild dependence on the choice of factorisation scale. 

\item In general, the factorisation scale based on the transverse mass provides a better description than the commonly used invariant-mass-driven scale. For light partons, a small modification of the scale by including the weak-boson mass effectively shifts it towards the physically relevant regime and better captures the underlying dynamics of the splittings that produce the parton.

\item Higher-order QCD processes, effectively accounted for within the EWPDF framework, should be included in phenomenological studies at future multi-TeV colliders, as they yield sizeable cross sections in kinematically accessible regions of phase space.
\end{itemize}
\chapter{Future prospects}
\label{chap:prospects}

The two major analyses presented in this thesis --- the measurement of light quark couplings to the $Z$ boson and the implementation of EWPDFs --- can be further developed. This chapter outlines the main ideas that will guide the work carried out in these directions.

\section{Quark couplings at lepton colliders}
Chapter~\ref{chap:Zcouplings} presented a method that can be used to measure the quark couplings to the $Z$ boson at the $Z$-pole run of future \epem colliders. Although the study was motivated by the goal of obtaining constraints for light quarks, the method was applied to measure the couplings of all five accessible flavours. The study can be further extended to separate the left- and right-handed components of the couplings. The key idea is highlighted in this section.

In Chapter~\ref{chap:SM}, the electroweak couplings of the $Z$ boson were introduced in terms of their axial and vector components, $g_V^f$ and $g_A^f$. They can be expressed as the sum and difference of left- and right-handed fermion couplings, 
\begin{equation*}
g_V^f = \frac{g_L^f + g_R^f}{2}, \quad g_A^f = \frac{g_L^f - g_R^f}{2}\,,
\end{equation*}
respectively.
The left- and right-handed couplings in the Standard Model are, in general, given by
\begin{equation*}
g^f_{L/R} = 2T_3^{f_{L/R}} - 2Q_f \sin^2 \theta_W\,,
\end{equation*}
where $T_{3}^{f_{L/R}} $ is the fermion weak isospin, $Q_f$ is its electric charge and $\theta_W$ is the weak mixing angle. The difference between the left- and right-handed couplings arises from the weak isospin, which is zero for the right-handed states. For example, for the left-handed electron:
\begin{equation*}
T_3^{e_L} = -\tfrac{1}{2}, \quad Q_{e_L} = -1 \quad \Rightarrow \quad g^e_L = -1 + 2\sin^2 \theta_W \approx -0.538\,,
\end{equation*}
while for the right-handed one:
\begin{equation*}
T_3^{e_R} = 0, \quad Q_{e_R} = -1 \quad \Rightarrow \quad g^e_R = 2\sin^2 \theta_W \approx 0.462\,.
\end{equation*}
Analogously, the couplings for down-type quarks are given by 
\begin{equation*}
\quad g^d_L = -1 + \tfrac{2}{3} \sin^2 \theta_W, \quad g^d_R = \tfrac{2}{3} \sin^2 \theta_W
\end{equation*}
and for the up-type one by
\begin{equation*}
g^u_L = 1 - \tfrac{4}{3} \sin^2 \theta_W, \quad g^u_R = -\tfrac{4}{3} \sin^2 \theta_W\,.
\end{equation*}

Future linear colliders are expected to provide beams highly polarised in the longitudinal direction. Efforts are also being made to achieve beam polarisation at circular colliders~\cite{Duan:2023cgu}, at which the beams become naturally polarised in the transverse direction through synchrotron radiation, as described by the Sokolov--Ternov effect~\cite{Sokolov:1963zn}. In principle, the use of spin rotators could transform this transverse polarisation into longitudinal polarisation at the interaction point, as demonstrated at HERA~\cite{Beckmann:2000cg}, thereby allowing one to probe the chiral structure of the Standard Model, although this is experimentally challenging and often constrained by luminosity considerations. 

Separating left- and right-handed contributions facilitates drawing meaningful conclusions for such studies. In the massless limit, four contributions can be considered for the fermion pair production:

\begin{equation*}
\begin{array}{ll}
\text{LL:} & e_L^- e_R^+ \rightarrow f_L \bar{f}_R\,, \\
\text{LR:} & e_L^- e_R^+ \rightarrow f_R \bar{f}_L\,, \\
\text{RL:} & e_R^- e_L^+ \rightarrow f_L \bar{f}_R\,, \\
\text{RR:} & e_R^- e_L^+ \rightarrow f_R \bar{f}_L\,. \\
\end{array}
\end{equation*}
In this naming convention, the first letter denotes the incoming electron helicity, while the second letter denotes the outgoing fermion helicity.

\subsection{$Z$ boson exchange}

First, assume that the $Z$ boson is the only mediator of the interaction, i.e. the photon contribution and the $Z/\gamma$ interference are neglected. Such an approach might be suitable, for instance, for the $Z$-pole run of \epem colliders. In this framework, the emission-angle differential cross sections for different polarisation combinations scale as:
\begin{equation*}                    
\frac{d\sigma}{d\cos\theta}\bigg|_{LL} \sim \frac{(g_L^e)^2 (g_L^f)^2}{(s - m_Z^2)^2 + m_Z^2 \Gamma^2_Z} (1 + \cos\theta)^2\,,
\end{equation*}

\begin{equation*}
\frac{d\sigma}{d\cos\theta}\bigg|_{RL} \sim \frac{(g_R^e)^2 (g_L^f)^2}{(s - m_Z^2)^2+ m_Z^2 \Gamma^2_Z} (1 - \cos\theta)^2\,,
\end{equation*}

\begin{equation*}
\frac{d\sigma}{d\cos\theta}\bigg|_{LR} \sim \frac{(g_L^e)^2 (g_R^f)^2}{(s - m_Z^2)^2 + m_Z^2 \Gamma^2_Z} (1 - \cos\theta)^2\,,
\end{equation*}

\begin{equation*}
\frac{d\sigma}{d\cos\theta}\bigg|_{RR} \sim \frac{(g_R^e)^2 (g_R^f)^2}{(s - m_Z^2)^2 + m_Z^2 \Gamma^2_Z} (1 + \cos\theta)^2\,,
\end{equation*}
where $\Gamma_Z$ is the $Z$ boson width, and the angular dependence can be heuristically understood in terms of helicity conservation. Experimentally, the polarisation of the final-state fermion can only be determined in specific cases, such as those of $\tau$ leptons or top quarks. For this reason, it is convenient to sum over the final-state polarisations. Expanding the angular terms, grouping them into even and odd contributions and splitting off the common propagator factor, one arrives at:
\[
\renewcommand{\arraystretch}{1.6}
\begin{array}{|c|c|c|}
\hline
& 1 + \cos^2\theta & \cos\theta \\
\hline
L & (g_L^e)^2 \left[(g_L^f)^2 + (g_R^f)^2\right] & 2 (g_L^e)^2 \left[ (g_L^f)^2 - (g_R^f)^2 \right] \\
\hline
R & (g_R^e)^2 \left[(g_L^f)^2 + (g_R^f)^2\right] & -2 (g_R^e)^2 \left[ (g_L^f)^2 - (g_R^f)^2 \right] \\
\hline
\end{array}
\]
where $L$ stands for the $LL$ and $LR$ contributions summed over, and $R$ stands for the $RL$ and $RR$ contributions. The separation into the two groups follows from the fact that the total cross section is only related to the symmetric $1+\cos^2\theta$ term, while the forward-backward asymmetry can be calculated using only the asymmetric $\cos\theta$ term.

An important remark is in place: the couplings of the final-state fermions factorise from those of the initial electrons. Consequently, the same combinations of the final-state fermion couplings hold in differential cross sections with either a left-handed or right-handed electron beam. From this perspective, the cross-section measurements with different beam polarisations --- while helping to control systematic uncertainties --- do not allow for disentangling left- and right-handed contributions. The degeneracy can be lifted if the forward--backward asymmetry is measured. This, however, introduces experimental challenges related to charge identification, which is needed to distinguish fermions from anti-fermions. This example illustrates that the $Z$-pole run will have limited power to constrain the left- and right-handed couplings.

\subsection{$\gamma/Z$ exchange}
Now, assume that both photon and $Z$ boson contributions are present, and the interference cannot be neglected. This is the case for higher-energy runs of future \epem facilities. The differential cross sections scale then as:

\begin{equation*}
\frac{d\sigma}{d\cos\theta}\bigg|_{LL} \sim  \left[
\frac{Q_f^2}{s^2} + \frac{(g_L^e)^2 (g_L^f)^2}{(s - m_Z^2)^2} - \frac{2Q_f g_L^e g_L^f}{s (s - m_Z^2)}
\right] (1 + \cos\theta)^2\,,
\end{equation*}

\begin{equation*}
\frac{d\sigma}{d\cos\theta}\bigg|_{LR} \sim  \left[
\frac{Q_f^2}{s^2} + \frac{(g_L^e)^2 (g_R^f)^2}{(s - m_Z^2)^2} - \frac{2Q_f g_L^e g_R^f}{s (s - m_Z^2)}
\right] (1 - \cos\theta)^2\,,
\end{equation*}

\begin{equation*}
\frac{d\sigma}{d\cos\theta}\bigg|_{RR} \sim  \left[
\frac{Q_f^2}{s^2} + \frac{(g_R^e)^2 (g_R^f)^2}{(s - m_Z^2)^2} - \frac{2Q_f g_R^e g_R^f}{s (s - m_Z^2)}
\right] (1 + \cos\theta)^2\,,
\end{equation*}

\begin{equation*}
\frac{d\sigma}{d\cos\theta}\bigg|_{RL} \sim  \left[
\frac{Q_f^2}{s^2} + \frac{(g_R^e)^2 (g_L^f)^2}{(s - m_Z^2)^2} - \frac{2Q_f g_R^e g_L^f}{s (s - m_Z^2)}
\right] (1 - \cos\theta)^2\,,
\end{equation*}
The first term in each expression corresponds to the photon-mediated interaction, the second one to the $Z$-boson-mediated one, and the third one to the interference between the two. The $Z$ boson width does not play a major role for energies well above the $Z$-pole and has been neglected for simplicity.

Expanding the angular terms and grouping them in the same way as previously, one obtains:
\[
\renewcommand{\arraystretch}{1.6}
\begin{array}{|c|c|}
\hline
& 1 + \cos^2\theta \\
\hline
L & \frac{2 Q_f^2}{s^2} + \frac{(g_L^e)^2}{(s - m_Z^2)^2}\left[(g_L^f)^2 + (g_R^f)^2\right] - \frac{2 Q_f g_L^e}{s (s - m_Z^2)}(g_L^f + g_R^f) \\
\hline
R & \frac{2 Q_f^2}{s^2} + \frac{(g_R^e)^2}{(s - m_Z^2)^2}\left[(g_L^f)^2 + (g_R^f)^2\right] - \frac{2 Q_f g_R^e}{s (s - m_Z^2)}(g_L^f + g_R^f) \\ \hline

& \cos\theta \\
\hline
L & 
\frac{2 (g_L^e)^2}{(s - m_Z^2)^2}\left[(g_L^f)^2 - (g_R^f)^2\right] - \frac{4 Q_f g_L^e}{s (s - m_Z^2)}(g_L^f - g_R^f) \\
\hline
R & 
-\frac{2 (g_R^e)^2}{(s - m_Z^2)^2}\left[(g_L^f)^2 - (g_R^f)^2\right] + \frac{4 Q_f g_R^e}{s (s - m_Z^2)}(g_L^f - g_R^f) \\ \hline
\end{array}
\]
It is evident that the couplings of the electron and the final-state fermion no longer factorise. Different beam polarisations result in independent combinations of $g_L^f$ and $g_R^f$, even for the symmetric term alone. It implies that if a single flavour can be tagged in the final state (for instance, a cleanly tagged $b$ quark), $g_L^f$ and $g_R^f$ can be disentangled by measuring total cross sections at two different beam polarisation setups\footnote{One caveat is that in this approach, the couplings are measured as a pair, \textit{i.e.} one cannot distinguish which one is $g_L^f$ or $g_R^f$. This can be either mitigated by measuring --- even with moderate precision --- the forward--backward asymmetry or heuristically treated as a freedom in naming parameters of the theory.}, without the necessity of taking the experimentally challenging measurement of the forward--backward asymmetry.

Experimentally, it is not feasible to achieve pure beam polarisations. The beam polarisation fraction for electrons ($P^-$) and positrons ($P^+$) can be defined as
\begin{equation*}
P^\pm = \frac{N_R^\pm - N_L^\pm}{N_L^\pm + N_R^\pm}\,,
\end{equation*}
where $N^\pm_{L/R}$ denotes the number of left-/right-handed (lower index) electrons/positrons (upper index) in a bunch. The total cross section is then given as
\begin{align*}
\sigma &= \frac{1 - P^-}{2} \frac{1 - P^+}{2} \sigma_L +
\frac{1 + P^-}{2} \frac{1 + P^+}{2} \sigma_R\\
&= \frac{1}{4} \left[ (1-P^- P^+)(\sigma_L + \sigma_R) + (P^- - P^+)(\sigma_R - \sigma_L)\right]\,,
\end{align*}
where $\sigma_L$ ($\sigma_R$) denotes the pure left-handed (right-handed) electron contribution. Then, the differential cross section, $\frac{d\sigma}{d\cos\theta}$, can be written as
\begin{align*}
&\frac{1}{4}(1+\cos^2\theta) \left[(1-P^-P^+)\left(\frac{4Q_f^2}{s^2} + \frac{\left[(g_L^e)^2 + (g_R^e)^2\right]}{(s-m^2_Z)^2}\left[(g_L^f)^2 + (g_R^f)^2\right] - \frac{2Q_f (g^e_L + g^e_R)}{s(s-m_Z^2)}(g^f_L + g^f_R)\right)\right.\\
&+ (P^- - P^+) \left.\left( \frac{\left[(g_R^e)^2 - (g_L^e)^2\right]}{(s-m^2_Z)^2}\left[(g_L^f)^2 + (g_R^f)^2\right] - \frac{2Q_f (g^e_R - g^e_L)}{s(s-m_Z^2)}(g^f_L + g^f_R)\right)\right]\\
&+ \frac{1}{2}\cos\theta \left[(1-P^-P^+)\left( \frac{\left[(g_L^e)^2 - (g_R^e)^2\right]}{(s-m^2_Z)^2}\left[(g_L^f)^2 - (g_R^f)^2\right] - \frac{2Q_f (g^e_L - g^e_R)}{s(s-m_Z^2)}(g^f_L - g^f_R)\right)\right.\\
&+ (P^- - P^+) \left.\left( \frac{\left[(g_L^e)^2 + (g_R^e)^2\right]}{(s-m^2_Z)^2}\left[(g_R^f)^2 - (g_L^f)^2\right] - \frac{2Q_f (g^e_L + g^e_R)}{s(s-m_Z^2)}(g^f_R - g^f_L)\right)\right]\,,
\end{align*}
where the symmetric and asymmetric parts have been explicitly separated again. 

As mentioned earlier, the expression can be used to disentangle the left- and right-handed couplings of a single fermion based on cross-section measurements with two different beam polarisations, provided that a clean data sample is collected. In the case of light quarks, when tagging algorithms cannot efficiently distinguish the $u$ and $d$ quarks, the forward--backward asymmetry measured at two different polarisations can be employed to gain additional information. Finally, the method based on the photon radiation presented in Chapter~\ref{chap:Zcouplings} can be used to separate the contribution. A dedicated analysis should be carried out to determine which approach provides the best sensitivity once systematic uncertainties are taken into account. This, however, lies beyond the scope of the thesis and will only be investigated in the future.

\section{Electroweak Parton Distribution Functions}

As elaborated in Chapter~\ref{chap:EWPDF}, the EWPDF framework provides an interesting realisation of the concept of electroweak factorisation at the highest achievable energies and can potentially improve precision thanks to resummation of higher-order effects. The \textsc{Whizard} implementation, being the first of its kind on the market, has certain limitations. Addressing those will be one of the main challenges for future work in this direction.

The EWPDF framework, as developed in~\cite{Han:2020uid}, incorporates the effects of mixed partons in the initial state. These correspond to $\gamma/Z$ interference for transverse polarisations and to $H/Z$ interference for longitudinal polarisations. These effects originate from the structure of the electroweak sector of the SM, unifying the weak and electromagnetic interactions, as well as the electroweak symmetry breaking mechanism giving rise to particle masses. The impact of mixed states is expected to be moderate; nevertheless, they must be included to achieve the highest possible precision~\cite{Chen:2016wkt}. 

From a technical perspective, their implementation would require modifications to the interface with O'Mega, the matrix-element provider of \textsc{Whizard}. By default, it provides squared matrix elements for cross-section calculations, whereas in the presence of mixed states, the cross section is obtained from the product of matrix elements corresponding to the two mixing components. An additional complication arises from the different masses of the mixed states. While the associated effects can be neglected as higher-order corrections, they must be treated consistently within the Monte Carlo framework. For these reasons, the inclusion of mixed initial states is deferred to future work and will be pursued in close collaboration with the \textsc{Whizard} team.

Another issue concerns the treatment of beam remnants. The current technical implementation of the EWPDF framework in \textsc{Whizard} assumes that beam remnants are always massless and electrically neutral; by contrast, colour quantum numbers are preserved in order to allow a potential interface with parton-shower and hadronisation models in \textsc{Pythia}. The implementation will need to be extended to properly account for the remnants' quantum numbers.

As an example, consider the $W$-boson content of the muon. In this case, the leading-order contribution arises from the $\mu\to W\nu_\mu$ splitting. The corresponding beam remnant, the muon neutrino, is indeed massless and electrically neutral, and colour conservation is trivially satisfied. Within the collinear framework, the remnant is emitted collinearly with the incoming beam, so that the only interaction requiring explicit treatment is the hard-scattering process.

A similar situation holds when considering the electron content of the muon, for which the leading contribution involves a sequence of two splittings. The dominant process is $\mu \to \mu \gamma \to \mu e^+ e^-$, although the chain may also be initiated by $W$-boson emission. In this case, the beam remnant consists of two electrically charged particles. While, in principle, the energy sharing between the remnant particles should be modelled, within the collinear approximation, the remnant leptons are expected to escape detection.

Significantly more complications are expected for the gluon content of the muon, which is most likely generated through a chain such as $\mu \to \mu \gamma \to \mu q \bar{q} \to \mu q \bar{q} g$. The resulting beam remnant is composed of multiple particles in various possible configurations and may carry electric charge, colour charge, and, in the case of heavy quarks, non-negligible mass. Since quarks inevitably hadronise (or decay, as in the case of the top quark), some of the hadronisation products may no longer be soft or collinear and could, in principle, enter the detector acceptance, effectively forming an \textit{underlying-event} contribution~\cite{CDF:2001onq} in lepton collisions. These considerations point to the need for a dedicated, parton-shower-like treatment of beam remnants within the EWPDF framework.

Similarly, careful treatment of final-state showers, including electroweak contributions, is required. Although several algorithms are available, a dedicated study is necessary to assess their applicability in this context~\cite{Bauer:2018xag,Kleiss:2020rcg,Brooks:2021kji}. Contributions arising from initial-state splittings, which underpin the formulation of the EWPDF framework, from final-state splittings implemented in parton-shower algorithms, and from the hard-scattering process itself must be clearly identified and conceptually separated in order to avoid potential double counting. The Author has initiated the work on electroweak showers at high energies, and the first results are expected in 2026. 

The discussion in this thesis focused solely on the lepton-collider case. Nevertheless, further developments of the EWPDF framework can be envisaged for hadron-collider studies. Conceptual work in this direction has already begun~\cite{Fornal:2018znf}. In particular, the inclusion of the full spectrum of Standard Model particles inside the proton could become important for precision studies at future colliders such as the FCC-hh. A key conceptual difference is that, for hadrons, the PDFs must ultimately be extracted from experimental data~\cite{MoralFigueroa:2025cev}. There remains considerable scope for further research in this direction.

Finally, as outlined earlier, the EWPDF framework offers the potential to improve the precision of theoretical predictions. This can be achieved by matching the resummed results obtained within the collinear EWPDF formalism to higher-order calculations performed in the full phase space. The development of a consistent matching prescription, together with a demonstration of the accuracy and reliability of the resulting predictions, may be regarded as the ultimate objective of studies based on the EWPDF framework. This clearly lies beyond the scope of the present thesis.

\chapter{Summary}
\label{chap:summary}
In the absence of a deeper theoretical framework explaining the origin of the Standard Model’s structure and given its inability to account for several observed phenomena, the particle physics community broadly agrees on the need for a next-generation facility to explore fundamental physics. Proposed future colliders differ in their technical designs and experimental conditions, including energy staging and achievable data rates. Their energy reach spans from the $Z$ pole up to the multi-TeV regime, enabling a broad programme of precision measurements to test the validity of the Standard Model.

This doctoral thesis demonstrated how several proposed future colliders could probe the electroweak sector of the Standard Model with high precision through radiative processes. It introduced the theoretical structure of the Standard Model, reviewed the main collider proposals and their physics potential, and described methods for simulating their performance using Monte Carlo techniques. A novel method to measure the couplings of the $Z$ boson to quarks at the $Z$-pole run of future \epem facilities was proposed, achieving an order-of-magnitude improvement over current experimental precision. Furthermore, the collinear factorisation framework was presented, alongside a revalidation of the Equivalent Vector Boson Approximation within the Monte Carlo generator \textsc{Whizard}. A new and unique implementation of Electroweak Parton Distribution Functions in the same generator was also developed, and first physics studies were initiated, providing an important tool for the high-energy physics community. Finally, possible future directions for research building on this work were outlined.

The main findings of this thesis can be summarised as follows:

\begin{itemize}
    \item Future \epem colliders operating at the $Z$-pole will enable precision measurements of the $Z$ boson couplings to fermions, driven by the extremely large data samples and anticipated advances in flavour-tagging techniques. Using the proposed method, the electroweak couplings of quarks can be extracted with a precision of about 0.3\% for light quarks, assuming systematic uncertainties of 0.1\%. More generally, achieving sub-percent precision requires flavour-tagging uncertainties to be controlled at or below the sub-percent level.

    \item The collinear approximation provides an efficient and accurate description of several processes mediated by vector bosons. However, the transverse structure functions exhibit a strong dependence on the choice of factorisation scale, which has a significant impact on the shapes of differential distributions. The scale commonly used in the literature, based on the invariant mass of the final state, does not provide the most accurate description. Instead, choosing the transverse mass as the characteristic scale better reproduces both the overall normalisation and the kinematics of processes dominated by vector-boson-fusion topologies. This choice should ensure that the scale sets a minimal reference, allowing the framework to correctly reflect the electroweak nature of interactions mediated by massive gauge bosons. A general strategy to improve the reliability of collinear frameworks is to impose a lower bound on the invariant mass of the final state to suppress contributions from partons being far off shell.
    
    \item An entirely new and self-contained implementation of the EWPDF formalism was introduced in \whizard and validated across a variety of processes. The studies demonstrated that QCD-induced contributions can play a major role at high-energy lepton colliders and should be included for reliable phenomenological predictions. This framework provides a foundation for further work, encompassing both technical improvements and theoretical developments, and offers a novel tool for the broader high-energy collider community.

\end{itemize}

Electroweak radiative processes will be important at future collider facilities, both at low and high energy scales. Their proper modelling can be essential for many measurements. The results presented in this thesis contributed to the development of advanced simulation tools, opening new prospects for precision studies.

\appendix
\chapter{Technical design of future machines}
\label{app:technical}

In this appendix, further details on major collider and detector proposals are given. If not stated otherwise, the specifications are presented according to the submissions to the recent Update of the ESPP~\cite{ESPPinputs:2024}.

\section{Future Circular Collider}
The FCC-ee design is based on technologies validated in previous high-energy collider facilities~\cite{FCC:2025lpp, FCC:2025uan, FCC:2025jtd}. It adopts a double-ring configuration with separate beam pipes for electrons and positrons, which allows for storing a large number of bunches and for tapering of magnet strengths to match the local energy of each beam. Regular top-up injection from a full-energy booster synchrotron, located in the same tunnel, ensures high luminosity, as proven successful in several past colliders, including KEKB~\cite{KEKTsukuba:1995urg}, BEPC II~\cite{Zhang:2009ze}, and SuperKEKB~\cite{Akai:2018mbz}. All energy stages of FCC-ee utilise the ``crab-waist'' collision scheme, implemented at DA$\Phi$NE~\cite{Zobov:2010zza} and SuperKEKB~\cite{Zhou:2023ygh}. Additionally, precision energy calibration at low-energy modes ($Z$ and $WW$) is assured through resonant depolarisation~\cite{Assmann:1998qb}.

The FCC-ee injector complex consists of two separate linacs for electrons and positrons, each accelerating beams to 2.86 GeV. The positron linac features a conventional positron source using electrons directed onto a tungsten target. This system is augmented by a novel non-insulated high-temperature superconductor (HTS) solenoid, operating at a peak field of 12.5 T. Subsequently, the beams are transferred to a damping ring, followed by acceleration in a high-energy linac up to 20 GeV. At the $Z$ pole, each beam consists of 11200 bunches. Such a high number is possible due to the relatively mild synchrotron radiation and beam--beam effects at low energy. As the collision energy increases, these effects become more significant, and fewer bunches are achievable: 1780 at the $WW$ threshold, 440 during the $ZH$ run, and 60 at the $t\bar{t}$ threshold. The damping ring and linacs are connected to the main ring via 5.6-km transfer lines.

The RF system is based on two-cell 400\,MHz superconducting Nb/Cu cavities operating at 4.5\,K and offering an accelerating gradient of 10.6\,MV/m for the $Z$, $WW$, and $ZH$ runs.  For the $t\bar{t}$ run, additional six-cell 800\,MHz bulk Nb cavities operating at 2\,K are employed with an accelerating gradient of 20.1\,MV/m. Two distinct cryomodule types have been developed to accommodate these cavity systems: one housing four 400\,MHz cavities, with a total length of 11.24\,m, and the other containing the 800\,MHz cavities (10.25\,m total length). As an RF power source, a tristron is proposed as a viable compact and cost-effective alternative to traditional two-stage multi-beam klystrons.

The vacuum chamber for the collider arcs is constructed from silver-bearing copper. The collider ring spans approximately 100 km, of which the arcs constitute 77 km. Dipole magnets occupy nearly the entire available length, ensuring a high dipole filling factor and efficient bending over the vast ring circumference.

\section{Linear Collider Facility}
Given the current status of the LCF, the original ILC design is first presented, followed by a discussion including details of anticipated future upgrades.

\subsection{Original ILC technical design}

\begin{figure}
    \centering
    \includegraphics[width=0.7\textwidth]{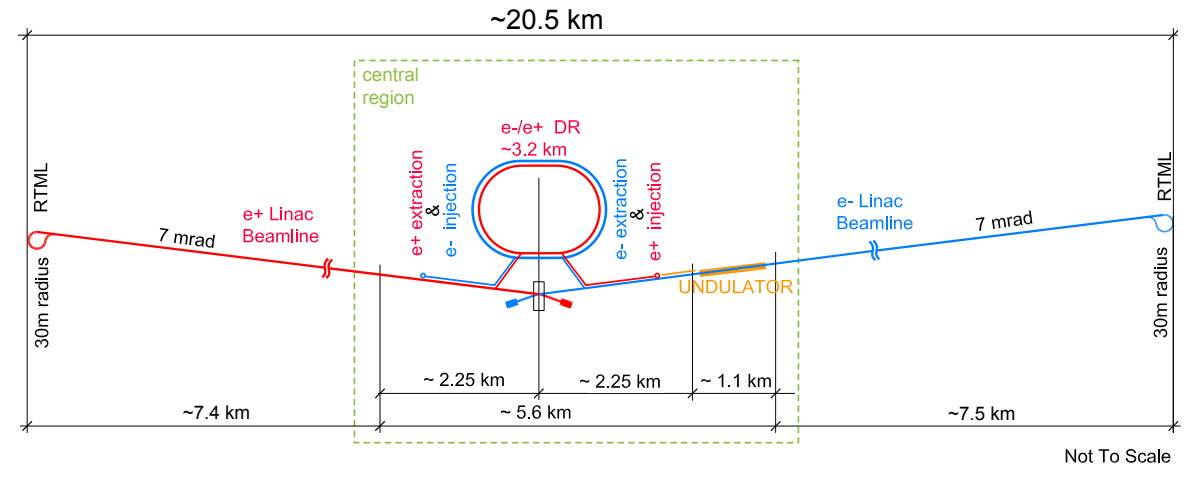}
    
    \caption{Schematic layout of ILC design~\cite{Bambade:2019fyw}.}
    \label{fig:ILC_technical}
\end{figure}

The original ILC concept was described in the Technical Design Report in 2013~\cite{Behnke:2013xla, ILC:2007bjz, Adolphsen:2013jya, Adolphsen:2013kya, Behnke:2013lya}. The machine consists of a few major components. They include a polarised electron source based on a photocathode DC gun, a polarised positron source, in which positrons are produced from electron--positron pairs originating from converting high-energy photons produced by the main electron beam passing through an undulator, a 3.2-km damping rings for electrons and positrons in a common tunnel used to reduce the beam emittance, beam transport systems, followed by a two-stage bunch compressors before injection into the main linac, two 11-km main linacs, two beam-delivery systems of 2.2\,km, bringing the beams into collisions at a 14-mrad crossing angle. The schematic design is shown in Figure \ref{fig:ILC_technical}. The main linacs accelerate beams from 15\,GeV to the assumed collision energy. Each linac consists of over seven thousand 1-m-long superconducting nine-cell niobium cavities operating at 2\,K, assembled into about 850 cryomodules of two types: Type A with nine 1.3\,GHz cavities and Type B with eight cavities and one superconducting quadrupole. To operate at 2\,K, the cavities are immersed in a saturated helium-4 bath. The RF power is provided by 10\,MW multi-beam klystrons, which have achieved the ILC specifications and are well industrialised. The average accelerating gradient in the complex is 31.5\,MV/m. Over ten years of optimisation preceded the determination of these key parameters, culminating in the publication of the TDR.

The electron beam is produced by a laser illuminating a strained GaAs photocathode in a DC gun, providing the necessary bunch train with 90\% polarisation. Two independent laser and gun systems are installed to ensure continuous operation without interruption. The beam is pre-accelerated to 75\,MeV using normal-conducting structures and then accelerated to 5\,GeV in a superconducting linac. Before injection into the damping ring, superconducting solenoids are employed to rotate the spin vector into the vertical and energy is compressed in a cryomodule. The electron beam is also used to produce positrons by transporting it through a superconducting helical undulator, generating photons. These photons are then directed onto a rotating 0.4 radiation-length titanium-alloy target, producing a beam of electron-positron pairs. The positrons are extracted and, after pre-acceleration to 400\,MeV in a normal-conducting system, accelerated to 5\,GeV in a superconducting linac. Similarly to the electron case, the spin is rotated and the energy spread compressed before injection into the damping ring. 

The role of the damping rings is to reduce large transverse and longitudinal beam emittances to achieve the required luminosity. One electron and one positron ring, one above the other, are envisioned in the design. The complex is located in the central region, about 100\,m from the interaction region, to avoid the experimentation hall. Damping is accomplished by approximately 113\,m of superferric wigglers, operating at 4.5\,K, in each damping ring. The electron and positron beams are then transported through the longest continuous beamlines, identical for both particle types, to the main linac. Besides serving as a transport line, this system performs several additional functions, including energy collimation, beam stabilisation, spin rotation, and bunch-length compression, each handled by dedicated subsystems. The final BDS is designed to transport the beams exiting the main linacs, focus them to the required sizes, bring them into collision, and extract the spent beams into dumps. It must also fulfil other critical roles, such as minimising background from the beam halo and monitoring the beam energy and polarisation before the collisions. Due to the presence of two interaction points in the LCF concept, unlike the original ILC design which featured two detectors in a ``push-pull'' configuration, this part of the machine layout requires re-optimisation.

\subsection{Potential upgrades}
Although the primary focus is building a 250-GeV machine to study the properties of the Higgs boson, the $Z$-pole run could also offer a valuable physics programme~\cite{LinearCollider:2025lya,LinearColliderVision:2025hlt}. In its initial configuration, the collider could operate at the $Z$ pole with an instantaneous luminosity of $2.05 \times 10^{33}$\,cm$^{-2}$s$^{-1}$ and polarised beams. If needed, it could also efficiently explore the $WW$ threshold. Two aspects are relevant in this context: the potential for higher luminosity at lower centre-of-mass energies, and the ability to achieve high polarisation for both beams. At lower energies, beamstrahlung is naturally suppressed, which allows for adjustments to beam conditions and achieving higher luminosity per bunch crossing. These adjustments could be made with only modest changes to the luminosity spectrum. Possible improvements --- such as increasing the repetition rate, the number of bunches, the number of particles per bunch, or reducing the beam sizes at the interaction point --- could lead to up to an order-of-magnitude increase in luminosity. As for beam polarisation, the baseline design assumes 80\% polarisation for electrons and 30\% for positrons. However, ongoing efforts aim to reach 90\% and 60\% respectively, which would increase the effective polarisation from 88.7\% to 97.4\%~\cite{Moortgat-Pick:2005jsx}.

One advantage of the circular $e^+e^-$ colliders is the ability to reuse the circular tunnel to accommodate a next-generation hadron facility. Nevertheless, a linear tunnel could also host another collider based on modernised or novel technologies. The SCRF technology used for the first stage of the machine is currently being developed in different directions. Those developments could not only exceed previous record quality factors or avoid degradation with increasing gradients, but also potentially reduce cost. For example, if an accelerating gradient of 60\,MV/m and a quality factor of $1.4 \times 10^{10}$ are achievable, like in the HELEN concept~\cite{Belomestnykh:2023uon, Belomestnykh:2023naf}, the baseline tunnel of 33.5 km could be equipped with a 1-TeV machine. Similarly, the use of CLIC high-gradient RF cavities, widely studied in recent years, could realistically enable reaching 1.5\,TeV within the same footprint~\cite{Adli:2025swq, LinearColliderVision:2025hlt}.

Among other ideas, $C^3$ has been proposed as a compact, high-gradient facility with a normal-conducting distributed-coupling accelerator operated at cryogenic temperatures~\cite{Vernieri:2022fae, Andorf:2025dnn}. Originally prepared for 250- and 550-GeV collisions, the project could be adapted to run in the LCF tunnel, reaching up to 2\,TeV in a 15-km facility and up to 3\,TeV in the longer tunnel. It is also interesting that the same tunnel could potentially host a facility reaching the 10-TeV scale. Such a machine must be based on unprecedented technology with GeV-per-meter accelerating gradients. It could be possibly realised, for instance, with the future wakefield accelerator concept~\cite{Cros:2019tns, HALHF:2025hqj, Foster:2025gsj, Caldwell:2025hsa}.

\section{Muon Collider}

In the Muon Collider, the beam production starts at the proton complex~\cite{Accettura:2023ked, InternationalMuonCollider:2025sys}. The first part of the system, the proton source, has not been determined, but it is expected to share many characteristics with proton sources used for neutrino experiments~\cite{Neuffer:2024dka}. The first storage ring, the accumulator, accumulates the produced protons. The incoming beam is chopped to allow a clean injection into the ring. The second ring, the compressor, rotates bunches from the accumulator in longitudinal phase space to shorten them. The intense short bunches are then passed onto the target, while ensuring that neither the transverse nor longitudinal properties of the beam are compromised in the process. Since the final collider is assumed to work in a single-bunch mode, the bunches must be recombined and hit the target simultaneously. The target is immersed in a 20-T solenoid field. After removing beam impurities in the beryllium absorber, the beam passes through a longitudinal drift where pions decay. The muons are captured into bunches and split by charge.

The bunches are cooled in a 6-dimensional ionisation cooling system, which helps ensure high luminosity. Ionisation cooling appears to be the only solution compatible with the short muon's lifetime, which, for many years, remained prohibitive for the concept of colliding muons. The system consists of a series of solenoids manipulating the beam to pass through energy absorbers. The reduced beam's momentum is restored longitudinally using RF cavities. An initial rectilinear cooling system reduces the beam's 6-dimensional emittance enough to allow for merging of multiple initial bunches into a single bunch. Further emittance reduction is achieved through another stage of rectilinear cooling, followed by final transverse cooling using a series of high-field solenoids operating with a low-momentum beam. The beam is then re-accelerated using a pre-accelerator to reach semi-relativistic speeds, preparing it for injection into the main acceleration system. The design of this stage is crucial for achieving high luminosity, and it is currently thoroughly optimised, although the feasibility of the concept has already been proven~\cite{MICE:2019jkl}.

To mitigate muon decays, the highest possible acceleration gradient should be achieved. The initial acceleration phase (``low-energy acceleration'') is composed of a set of linacs and recirculating linacs. The beams after final cooling pre-acceleration at 255\,MeV pass first through a linac accelerating them to 1.25\,GeV and then by a pair of multi-pass recirculating linacs accelerating the beams to 5\,GeV and 63\,GeV, respectively. Then, a series of rapid cycling synchrotrons is employed to accelerate bunches to the desired collision energy. The current green-field design assumes intermediate stages of 0.31\,TeV, 0.75\,TeV, and 1.5\,TeV before the final 5\,TeV ring, but this acceleration scheme might differ for other options, for instance, if the LHC tunnel is to be reused.

The collider ring consists of two straight interaction regions positioned opposite each other, where the counter-rotating muon bunches collide. These interaction points provide two distinct locations for separate experiments. Additional systems, such as those designed to mitigate neutrino radiation, are also located in these areas. A significant flux of neutrinos would be produced along the direction of the long straight sections housing the experiments, leading to high radiation doses where these neutrinos reach the Earth's surface. As a result, the affected area must be controlled by the host laboratory and appropriately fenced off. Although neutrinos interact only weakly, their number presents a challenge in meeting radiation safety standards. Key mitigation strategies include minimising the length of the straight collider sections and placing the machine deep underground. Further reduction of neutrino flux density can be achieved by intentionally altering the muon beam trajectory to increase the angular spread of emitted neutrinos. At 3\,TeV, periodic orbit variations within the beam pipe are sufficient. However, for 5-TeV beams, a more active solution is necessary: arc beamline components are mounted on movable supports, allowing small, periodic deformations of the ring. This dynamic adjustment gradually changes the muon beam direction over time, thereby spreading the neutrino flux and reducing exposure at any fixed surface location. The final design of the system remains under study.

\section{International Large Detector}

Several performance targets define the ILD’s design requirements~\cite{Behnke:2013lya, ILDConceptGroup:2020sfq}. First, the detector must achieve an excellent impact-parameter resolution to identify displaced vertices for heavy-flavour tagging. Second, a good momentum resolution is needed for the Higgs recoil mass measurement, which can be provided by the combined silicon-TPC tracker at high momenta and by minimising the detector's material budget at lower momenta. Third, the goal for jet energy resolution of 3\% for light-flavour jets is set to enable statistical separation of hadronic decays of the $W$, $Z$, and Higgs bosons. Additional requirements include near-complete hermeticity, allowing detection of particles at angles as low as possible by the presence of the beam pipes, which is essential for measuring missing energy. Finally, the readout system is expected to operate without a hardware trigger, ensuring maximal efficiency for all possible event topologies.

To optimise the precision reconstruction of both charged and neutral particles, all major detector components are placed within a strong 3.5-T solenoidal magnetic field. This configuration enables separation of charged and neutral particles in the calorimeter and helps eliminate low-energy background from the primary detection region. Furthermore, to preserve resolution, the material in front of the calorimeters must be minimised, requiring that both the tracking and calorimetry systems be located inside the magnet coil.

\begin{figure}
    \centering
    \includegraphics[width=0.5\textwidth]{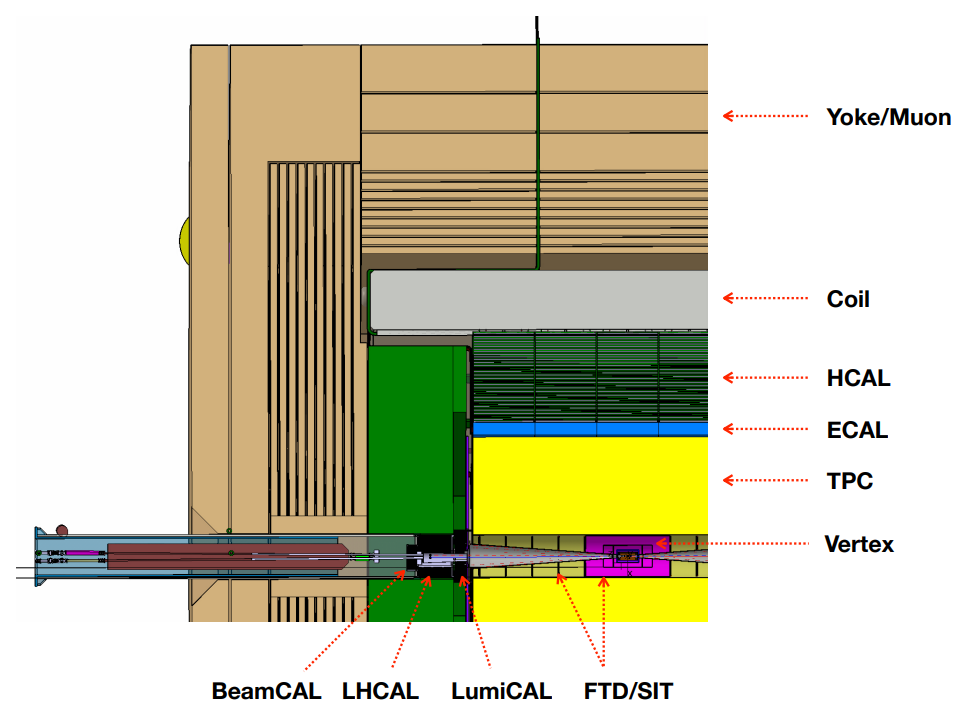}
    
    \caption{Schematic layout of the ILD~\cite{ILDConceptGroup:2020sfq}.}
    \label{fig:ILD_design}
\end{figure}

The detector design is shown in Figure \ref{fig:ILD_design}. Closest to the interaction point is a pixel detector designed for the precise reconstruction of short-lived particle decay vertices. ILD has adopted a layout consisting of three double layers of pixel sensors. Over the past decade, Monolithic Active Pixel Sensor (MAPS) technology has matured significantly, now approaching the required standards in terms of material budget, readout speed, and granularity~\cite{Schambach:2015gga}. Other geometries, for example, involving stitched sensors, are being explored.

For charged particle tracking, ILD uses a hybrid system combining a high-resolution Time Projection Chamber (TPC) with additional strip or pixel detectors. The TPC, roughly 4.6\,m in length and spanning radial distances from 33\,cm to 180\,cm, provides up to 220 three-dimensional points for continuous tracking. The high number of points enables reconstruction of complex topologies, such as long-lived particles. Additional silicon strip or pixel layers placed inside and outside the TPC further refine tracking precision. 

A powerful calorimeter system is essential to meet performance goals, particularly for particle-flow algorithms, which depend on fine-grained sampling. For the electromagnetic part, one option is to use silicon diode sensors with $5 \times 5$ mm$^2$ pads, capable of sampling electromagnetic showers up to 30 times. For the hadronic calorimeter, ILD is exploring two technologies. One uses silicon photomultiplier sensors on scintillator tiles, which offer analogue readout with $3 \times 3$\,cm$^2$ tiles~\cite{Simon:2009tke}. The other relies on resistive plate chambers with finer $1 \times 1$\,cm$^2$ granularity, though it provides only limited 2-bit amplitude information~\cite{Laktineh:2010zsa}. 

Differences in the operation of linear and circular colliders present distinct challenges for the detector. For instance, ILC operates in a bunch-train mode, with trains of 1--2 thousand bunches separated by hundreds of nanoseconds, repeating at 5 or 10\,Hz, while FCC-ee operates continuously, with collisions occurring every 20\,ns. Additionally, FCC-ee anticipates significant operation at the $Z$-pole energy, placing different constraints on the detector, particularly in its innermost regions. Adapting the calorimeter systems to the FCC's operation mode presents challenges in readout bandwidth, power consumption, and cooling, which must be addressed without compromising the compactness and granularity of the system.

In the very forward region, three specialised calorimeter systems are planned. LumiCal, a fine-sampling silicon-tungsten calorimeter, is dedicated to measuring electrons from Bhabha scattering and determining the integrated luminosity. LHCAL (Luminosity Hadronic CALorimeter) extends the reach of the endcap calorimeter system closer to the beamline, closing the gap between the electromagnetic calorimeter endcap and LumiCal. BeamCal, positioned further downstream from the interaction points, provides fast feedback on the beam position. As the inner detector region for FCC-ee differs significantly from the linear-collider environment, the layout of the forward calorimetry requires re-optimisation.

\section{Other detector concepts}
A major challenge for detectors at the Muon Collider is the mitigation of machine-induced backgrounds~\cite{InternationalMuonCollider:2025sys}. To address this, the detector design incorporates typically two cone-shaped tungsten shields, known as nozzles, positioned along the beam line between the interaction point and the outer edges of the detector. The nozzles absorb high-energy particles produced by beam interactions. However, as these particles interact with the shielding material, they might escape the nozzles. The result is a residual background flux composed of low-energy particles that can still reach the sensitive detector components. The presence of the nozzles also reduces the angular coverage close to the beamline. This poses additional challenges for the detector's design and operation.

Two distinct detector concepts have been developed for the Muon Collider: MUSIC (MUon System for Interesting Collisions) and MAIA (Muon Accelerator Instrumented Apparatus)~\cite{Casarsa:MUSIC, MAIA:2025hzm}. Both designs share a common layout, consisting of a cylindrical detector measuring about 11.4 meters in length and 12.8 meters in diameter. Despite differences in several components, the two concepts follow a similar multi-purpose detector architecture, incorporating a full set of subsystems typical for collider experiments. These include an all-silicon tracking system, an electromagnetic calorimeter, a hadronic calorimeter, and an outer muon system. A superconducting solenoid provides the magnetic field necessary for measuring the momentum of charged particles within the tracking volume. The primary differences between MUSIC and MAIA lie in the placement of the solenoid and the choice of electromagnetic calorimeter technology. In the MAIA design, the solenoid is located outside the tracking system, whereas in MUSIC it is placed between the electromagnetic and hadronic calorimeters. As for the calorimetry, MAIA employs a silicon-tungsten electromagnetic calorimeter, while MUSIC uses a semihomogeneous crystal-based technology.

Asymmetric colliders, such as $\mu$Tristan, present other challenges. The asymmetry between the energies of the two colliding beams, 30\,GeV and 1\,TeV or even 50\,GeV and 3\,TeV, results in the final-state particles being strongly boosted in the direction of the higher-energy $\mu^+$ beam. Consequently, these particles appear at small polar angles, typically within a few to ten degrees in the laboratory frame~\cite{Hamada:2022mua}. The detector must be designed to preserve sensitivity at such small angles. While this requirement introduces additional complexity, it should not be too severe, given that, for instance, the LHCb detector currently operating achieves coverage down to polar angles of approximately 15 milliradians~\cite{LHCb:2008vvz}. Another consideration is the background arising from the decay products of the muon beams, the challenge shared with the Muon Collider's detector. In particular, the $\mu^+$ beam generates a significant flux of positrons and photons, and effective shielding is necessary to protect the interaction region from these particles.

Even greater challenges are anticipated at FCC-hh. The experimental environment at FCC-hh would feature extreme conditions, including up to 1000 proton-proton pile-up events, the presence of highly boosted objects, and radiation levels reaching up to $10^{18}$ hadrons per square centimetre~\cite{Mangano:2022ukr}. These factors impose severe demands on detector technology and data acquisition systems. Although a conceptual detector design already exists for performance studies, the long timescale before potential collisions means that the current specifications will significantly evolve.

\chapter{Structure functions in Whizard}
In this appendix, Sindarin files for using structure functions in \whizard are given.

\section{EVA}
\label{app:sinEVA}

The revisited EVA implementation was made available in the \textsc{Whizard} release 3.1.7. Below, an example \textsc{Sindarin} file to run \whizard for $e^+ \mu^- \to HH + X$ in the EVA framework is presented:
\begin{verbatim}
model = SM
sqrts = 10 TeV
beams = "e+", "mu-" => ewa
$ewa_mode = "default" #Other options: "log", "log_pt", "legacy"
ewa_x_min =  2*mW/sqrts
scale = eval 0.25*M [H, H] #Scale corresponding to sqrt(s-hat)/4

process procEWA = "W+", "W-" => H, H
integrate (procEWA)
\end{verbatim}

\section{EWPDF}
\label{app:sinEWPDF}
The new EWPDF implementation will be made available in an official release version in 2026. Below, an example \textsc{Sindarin} file to run \whizard for $e^+ \mu^- \to HH + X$ in the EWPDF framework is presented:

\begin{verbatim}
model = SM
sqrts = 10 TeV
beams = "e+", "mu-" => lhapdf_ew
scale = eval 0.25*M [H, H] #Scale corresponding to sqrt(s-hat)/4

process procEWPDF = "W+", "W-" => H, H
integrate (procEWPDF)
\end{verbatim}

\bibliographystyle{JHEP}
\bibliography{refs}

@article{ParticleDataGroup:2024cfk,
    author = "Navas, S. and others",
    collaboration = "Particle Data Group",
    title = "{Review of particle physics}",
    doi = "10.1103/PhysRevD.110.030001",
    journal = "Phys. Rev. D",
    volume = "110",
    number = "3",
    pages = "030001",
    year = "2024"
}

@article{Irles:2023ojs,
    author = {Irles, A. and P\"oschl, R. and Richard, F.},
    title = "{Experimental methods and prospects on the measurement of electroweak $b$ and $c$-quark observables at the ILC operating at 250 GeV}",
    eprint = "2306.11413",
    archivePrefix = "arXiv",
    primaryClass = "hep-ex",
    reportNumber = "ILD-PHYS-PUB--2023-001, IFIC/23-38",
    month = "6",
    year = "2023"
}

@inproceedings{Fuster:2021ekh,
    author = "Fuster, Juan and Irles, Adrian and Rodrigo, Germ\'an and Tairafune, Seidai and Vos, Marcel and Yamamoto, Hitoshi and Yonamine, Ryo",
    title = "{Prospects for the measurement of the $b$-quark mass at the ILC}",
    booktitle = "{International Workshop on Future Linear Colliders}",
    eprint = "2104.09924",
    archivePrefix = "arXiv",
    primaryClass = "hep-ex",
    reportNumber = "IFIC/21-10, ILD-PHYS-PUB-2021-001",
    month = "4",
    year = "2021"
}

@article{Bilokin:2017lco,
    author = {Bilokin, S. and P\"oschl, R. and Richard, F.},
    title = "{Measurement of b quark EW couplings at ILC}",
    eprint = "1709.04289",
    archivePrefix = "arXiv",
    primaryClass = "hep-ex",
    reportNumber = "LAL-17-052",
    month = "9",
    year = "2017"
}

@article{Zhu:2023xpk,
    author = "Zhu, Yongfeng and Liang, Hao and Wang, Yuexin and Qu, Huilin and Zhou, Chen and Ruan, Manqi",
    title = "{ParticleNet and its application on CEPC jet flavor tagging}",
    eprint = "2309.13231",
    archivePrefix = "arXiv",
    primaryClass = "hep-ex",
    doi = "10.1140/epjc/s10052-024-12475-5",
    journal = "Eur. Phys. J. C",
    volume = "84",
    number = "2",
    pages = "152",
    year = "2024"
}

@article{Blekman:2024wyf,
    author = "Blekman, Freya and Canelli, Florencia and De Moor, Alexandre and Gautam, Kunal and Ilg, Armin and Macchiolo, Anna and Ploerer, Eduardo",
    title = "{Tagging more quark jet flavours at FCC-ee at 91 GeV with a transformer-based neural network}",
    eprint = "2406.08590",
    archivePrefix = "arXiv",
    primaryClass = "hep-ex",
    reportNumber = "PUBDB-2024-01826, DESY/PUBDB-2024-01826",
    doi = "10.1140/epjc/s10052-025-13785-y",
    journal = "Eur. Phys. J. C",
    volume = "85",
    number = "2",
    pages = "165",
    year = "2025"
}

@inproceedings{Albert:2022mpk,
    author = "Albert, Alexander and others",
    title = "{Strange quark as a probe for new physics in the Higgs sector}",
    booktitle = "{Snowmass 2021}",
    eprint = "2203.07535",
    archivePrefix = "arXiv",
    primaryClass = "hep-ex",
    reportNumber = "Report-no: ILD-PHYS-PROC-2022-001",
    month = "3",
    year = "2022"
}

@article{Liang:2023wpt,
    author = "Liang, Hao and Zhu, Yongfeng and Wang, Yuexin and Che, Yuzhi and Zhou, Chen and Qu, Huilin and Ruan, Manqi",
    title = "{Jet-Origin Identification and Its Application at an Electron-Positron Higgs Factory}",
    eprint = "2310.03440",
    archivePrefix = "arXiv",
    primaryClass = "hep-ex",
    doi = "10.1103/PhysRevLett.132.221802",
    journal = "Phys. Rev. Lett.",
    volume = "132",
    number = "22",
    pages = "221802",
    year = "2024"
}

@article{Tagami:2024gtc,
    author = "Tagami, Risako and Suehara, Taikan and Ishino, Masaya",
    title = "{Application of Particle Transformer to quark flavor tagging in the ILC project}",
    eprint = "2410.11322",
    archivePrefix = "arXiv",
    primaryClass = "hep-ex",
    doi = "10.1051/epjconf/202431503011",
    journal = "EPJ Web Conf.",
    volume = "315",
    pages = "03011",
    year = "2024"
}

@article{DELPHI:1995okj,
    author = "Abreu, P. and others",
    collaboration = "DELPHI",
    title = "{Study of prompt photon production in hadronic Z0 decays}",
    reportNumber = "CERN-PPE-95-101",
    doi = "10.1007/s002880050001",
    journal = "Z. Phys. C",
    volume = "69",
    pages = "1--14",
    year = "1995"
}

@article{DELPHI:1991utk,
    author = "Abreu, P. and others",
    collaboration = "DELPHI",
    title = "{Study of final state photons in hadronic Z0 decay and limits on new phenomena}",
    reportNumber = "CERN-PPE-91-174",
    doi = "10.1007/BF01559732",
    journal = "Z. Phys. C",
    volume = "53",
    pages = "555--566",
    year = "1992"
}

@article{L3:1992ukp,
    author = "Adriani, O. and others",
    collaboration = "L3",
    title = "{Determination of quark electroweak couplings from direct photon production in hadronic Z decays}",
    reportNumber = "CERN-PPE-92-209",
    doi = "10.1016/0370-2693(93)90735-Z",
    journal = "Phys. Lett. B",
    volume = "301",
    pages = "136--144",
    year = "1993"
}

@article{L3:1992kcg,
    author = "Adriani, O. and others",
    collaboration = "L3",
    title = "{Isolated hard photon emission in hadronic Z0 decays}",
    reportNumber = "CERN-PPE-92-131",
    doi = "10.1016/0370-2693(92)91205-N",
    journal = "Phys. Lett. B",
    volume = "292",
    pages = "472--484",
    year = "1992"
}

@article{OPAL:2003gdu,
    author = "Abbiendi, G. and others",
    collaboration = "OPAL",
    title = "{Measurement of the partial widths of the Z into up and down type quarks}",
    eprint = "hep-ex/0312043",
    archivePrefix = "arXiv",
    reportNumber = "CERN-EP-2003-082",
    doi = "10.1016/j.physletb.2004.02.046",
    journal = "Phys. Lett. B",
    volume = "586",
    pages = "167--182",
    year = "2004"
}

@article{Bedeschi:2022rnj,
    author = "Bedeschi, Franco and Gouskos, Loukas and Selvaggi, Michele",
    title = "{Jet flavour tagging for future colliders with fast simulation}",
    eprint = "2202.03285",
    archivePrefix = "arXiv",
    primaryClass = "hep-ex",
    doi = "10.1140/epjc/s10052-022-10609-1",
    journal = "Eur. Phys. J. C",
    volume = "82",
    number = "7",
    pages = "646",
    year = "2022"
}

@article{Ntounis:2025itg,
    author = "Ntounis, Dimitrios and Gouskos, Loukas and Vernieri, Caterina",
    title = "{Evaluating the Impact of Detector Design on Jet Flavor Tagging for Future Colliders}",
    eprint = "2501.16584",
    archivePrefix = "arXiv",
    primaryClass = "physics.ins-det",
    month = "1",
    year = "2025"
}

@article{Kalinowski:2020lhp,
    author = "Kalinowski, J. and Kotlarski, W. and Sopicki, P. and Zarnecki, A. F.",
    title = "{Simulating hard photon production with WHIZARD}",
    eprint = "2004.14486",
    archivePrefix = "arXiv",
    primaryClass = "hep-ph",
    reportNumber = "CLICdp-Pub-2020-001",
    doi = "10.1140/epjc/s10052-020-8149-6",
    journal = "Eur. Phys. J. C",
    volume = "80",
    number = "7",
    pages = "634",
    year = "2020"
}

@article{Sjostrand:2014zea,
    author = {Sj\"ostrand, Torbj\"orn and Ask, Stefan and Christiansen, Jesper R. and Corke, Richard and Desai, Nishita and Ilten, Philip and Mrenna, Stephen and Prestel, Stefan and Rasmussen, Christine O. and Skands, Peter Z.},
    title = "{An introduction to PYTHIA 8.2}",
    eprint = "1410.3012",
    archivePrefix = "arXiv",
    primaryClass = "hep-ph",
    reportNumber = "LU-TP-14-36, MCNET-14-22, CERN-PH-TH-2014-190, FERMILAB-PUB-14-316-CD, DESY-14-178, SLAC-PUB-16122",
    doi = "10.1016/j.cpc.2015.01.024",
    journal = "Comput. Phys. Commun.",
    volume = "191",
    pages = "159--177",
    year = "2015"
}

@article{Sjostrand:2006za,
    author = "Sjostrand, Torbjorn and Mrenna, Stephen and Skands, Peter Z.",
    title = "{PYTHIA 6.4 Physics and Manual}",
    eprint = "hep-ph/0603175",
    archivePrefix = "arXiv",
    reportNumber = "FERMILAB-PUB-06-052-CD-T, LU-TP-06-13",
    doi = "10.1088/1126-6708/2006/05/026",
    journal = "JHEP",
    volume = "05",
    pages = "026",
    year = "2006"
}

@article{Sjostrand:2007gs,
    author = "Sjostrand, Torbjorn and Mrenna, Stephen and Skands, Peter Z.",
    title = "{A Brief Introduction to PYTHIA 8.1}",
    eprint = "0710.3820",
    archivePrefix = "arXiv",
    primaryClass = "hep-ph",
    reportNumber = "CERN-LCGAPP-2007-04, LU-TP-07-28, FERMILAB-PUB-07-512-CD-T",
    doi = "10.1016/j.cpc.2008.01.036",
    journal = "Comput. Phys. Commun.",
    volume = "178",
    pages = "852--867",
    year = "2008"
}

@article{Jadach:1999vf,
    author = "Jadach, S. and Ward, B. F. L. and Was, Z.",
    title = "{The Precision Monte Carlo event generator K K for two fermion final states in e+ e- collisions}",
    eprint = "hep-ph/9912214",
    archivePrefix = "arXiv",
    reportNumber = "DESY-99-106, CERN-TH-99-235, UTHEP-99-08-01",
    doi = "10.1016/S0010-4655(00)00048-5",
    journal = "Comput. Phys. Commun.",
    volume = "130",
    pages = "260--325",
    year = "2000"
}

@article{Jadach:2013aha,
    author = "Jadach, S. and Ward, B. F. L. and Was, Z.",
    title = "{KK MC 4.22: Coherent exclusive exponentiation of electroweak corrections for $f\overline{f} \to f'\overline{f}'$ at the LHC and muon colliders}",
    eprint = "1307.4037",
    archivePrefix = "arXiv",
    primaryClass = "hep-ph",
    reportNumber = "BU-HEPP-13-03, IFJPAN-IV-2013-13",
    doi = "10.1103/PhysRevD.88.114022",
    journal = "Phys. Rev. D",
    volume = "88",
    number = "11",
    pages = "114022",
    year = "2013"
}

@article{Jadach:2022mbe,
    author = "Jadach, S. and Ward, B. F. L. and Was, Z. and Yost, S. A. and Siodmok, A.",
    title = "{Multi-photon Monte Carlo event generator KKMCee for lepton and quark pair production in lepton colliders}",
    eprint = "2204.11949",
    archivePrefix = "arXiv",
    primaryClass = "hep-ph",
    reportNumber = "IFJPAN-IV-2022-6, BU-HEPP-22-02, MCnet-22",
    doi = "10.1016/j.cpc.2022.108556",
    journal = "Comput. Phys. Commun.",
    volume = "283",
    pages = "108556",
    year = "2023"
}

@inproceedings{White:2015dxj,
    author = "White, Andrew P.",
    collaboration = "SiD Consortium",
    title = "{The SiD Detector for the International Linear Collider}",
    booktitle = "{Meeting of the APS Division of Particles and Fields}",
    eprint = "1511.00134",
    archivePrefix = "arXiv",
    primaryClass = "physics.ins-det",
    reportNumber = "DPF2015-152",
    month = "10",
    year = "2015"
}

@article{ILDConceptGroup:2020sfq,
    author = "Abramowicz, Halina and others",
    collaboration = "ILD Concept Group",
    title = "{International Large Detector: Interim Design Report}",
    eprint = "2003.01116",
    archivePrefix = "arXiv",
    primaryClass = "physics.ins-det",
    reportNumber = "DESY-20-034, KEK 2019-57",
    month = "3",
    year = "2020"
}

@article{Einhaus:2023npl,
    author = "Einhaus, Ulrich",
    collaboration = "ILD",
    title = "{The International Large Detector (ILD) for a future electron-positron collider: Status and Plans}",
    eprint = "2311.09181",
    archivePrefix = "arXiv",
    primaryClass = "hep-ex",
    reportNumber = "ILD-PHYS-PROC-2023-012",
    doi = "10.22323/1.449.0543",
    journal = "PoS",
    volume = "EPS-HEP2023",
    pages = "543",
    year = "2024"
}

@techreport{Robson:2920434,
      author        = "Robson, Aidan and Leonidopoulos, Christos",
      title         = "{The ECFA Higgs/Electroweak/Top Factory Study}",
        number = {CERN-2025-005},
      institution  = {CERN},
      year          = "2024",
      url           = "https://cds.cern.ch/record/2920434",
}

@article{Gorishnii:1990vf,
    author = "Gorishnii, S. G. and Kataev, A. L. and Larin, S. A.",
    title = "{The $O(\alpha^{3}_{s})$-corrections to $\sigma_{tot}(e^{+}e^{-}\rightarrow hadrons)$ and $\Gamma(\tau^{-} \rightarrow \nu_{\tau} + hadrons)$ in QCD}",
    reportNumber = "UM-TH-91-01",
    doi = "10.1016/0370-2693(91)90149-K",
    journal = "Phys. Lett. B",
    volume = "259",
    pages = "144--150",
    year = "1991"
}

@article{Surguladze:1990tg,
    author = "Surguladze, Levan R. and Samuel, Mark A.",
    title = "{Total hadronic cross-section in e+ e- annihilation at the four loop level of perturbative QCD}",
    reportNumber = "OSU-RN-250",
    doi = "10.1103/PhysRevLett.66.560",
    journal = "Phys. Rev. Lett.",
    volume = "66",
    pages = "560--563",
    year = "1991",
    note = "[Erratum: Phys.Rev.Lett. 66, 2416 (1991)]"
}

@article{Kramer:1991yc,
    author = "Kramer, G. and Lampe, B.",
    title = "{QCD corrections to final state photon Bremsstrahlung in e- e- annihilation}",
    reportNumber = "DESY-91-078",
    doi = "10.1016/0370-2693(91)90191-R",
    journal = "Phys. Lett. B",
    volume = "269",
    pages = "401--406",
    year = "1991"
}

@techreport{Kirkby:1992xm,
    author = "Kirkby, David P.",
    title = "{QCD corrections to isolated hard photon production in e+ e- annihilation}",
    reportNumber = "CALT-68-1822",
    month = "11",
    year = "1992"
}

@article{Kunszt:1992np,
    author = "Kunszt, Zoltan and Trocsanyi, Zoltan",
    title = "{QCD corrections to photon production in association with hadrons in e+ e- annihilation}",
    eprint = "hep-ph/9207232",
    archivePrefix = "arXiv",
    reportNumber = "ETH-TH-92-26",
    doi = "10.1016/0550-3213(93)90104-W",
    journal = "Nucl. Phys. B",
    volume = "394",
    pages = "139--168",
    year = "1993"
}

@article{Mattig:1993jv,
    author = "Mattig, Peter and Spiesberger, H. and Zeuner, W.",
    title = "{On the comparison of matrix element calculations of O (alpha alpha-s) with the measurement of photon emission in hadronic Z0 decays}",
    reportNumber = "CERN-PPE-93-119",
    doi = "10.1007/BF01558390",
    journal = "Z. Phys. C",
    volume = "60",
    pages = "613--632",
    year = "1993"
}

@article{Kilian:2003pc,
    author = "Kilian, W.",
    title = "{Electroweak symmetry breaking: The bottom-up approach}",
    doi = "10.1007/b97367",
    journal = "Springer Tracts Mod. Phys.",
    volume = "198",
    pages = "1--113",
    year = "2003"
}

@book{Peskin:1995ev,
    author = "Peskin, Michael E. and Schroeder, Daniel V.",
    title = "{An Introduction to quantum field theory}",
    doi = "10.1201/9780429503559",
    isbn = "978-0-201-50397-5, 978-0-429-50355-9, 978-0-429-49417-8",
    publisher = "Addison-Wesley",
    address = "Reading, USA",
    year = "1995"
}

@article{Vysotskii:2018eic,
    author = "Vysotskii, Mikhail I. and Zhemchugov, Evgeny",
    title = "{Equivalent photons in proton and ion{\textemdash}ion collisions at the LHC}",
    eprint = "1806.07238",
    archivePrefix = "arXiv",
    primaryClass = "hep-ph",
    doi = "10.3367/UFNe.2018.07.038389",
    journal = "Phys. Usp.",
    volume = "62",
    number = "9",
    pages = "910--919",
    year = "2019"
}

@article{Budnev:1975poe,
    author = "Budnev, V. M. and Ginzburg, I. F. and Meledin, G. V. and Serbo, V. G.",
    title = "{The Two photon particle production mechanism. Physical problems. Applications. Equivalent photon approximation}",
    doi = "10.1016/0370-1573(75)90009-5",
    journal = "Phys. Rept.",
    volume = "15",
    pages = "181--281",
    year = "1975"
}

@article{Brodsky:1971ud,
    author = "Brodsky, Stanley J. and Kinoshita, Toichiro and Terazawa, Hidezumi",
    title = "{Two Photon Mechanism of Particle Production by High-Energy Colliding Beams}",
    reportNumber = "SLAC-REPRINT-1971-007",
    doi = "10.1103/PhysRevD.4.1532",
    journal = "Phys. Rev. D",
    volume = "4",
    pages = "1532--1557",
    year = "1971"
}

@article{Engel:1995yda,
    author = "Engel, R. and Ranft, J.",
    title = "{Hadronic photon-photon interactions at high-energies}",
    eprint = "hep-ph/9509373",
    archivePrefix = "arXiv",
    reportNumber = "ENSLAPP-A-540-95, SI-95-08",
    doi = "10.1103/PhysRevD.54.4244",
    journal = "Phys. Rev. D",
    volume = "54",
    pages = "4244--4262",
    year = "1996"
}

@article{Frixione:1993yw,
    author = "Frixione, Stefano and Mangano, Michelangelo L. and Nason, Paolo and Ridolfi, Giovanni",
    title = "{Improving the Weizsacker-Williams approximation in electron - proton collisions}",
    eprint = "hep-ph/9310350",
    archivePrefix = "arXiv",
    reportNumber = "CERN-TH-7032-93, GEF-TH-18-93",
    doi = "10.1016/0370-2693(93)90823-Z",
    journal = "Phys. Lett. B",
    volume = "319",
    pages = "339--345",
    year = "1993"
}

@article{vonWeizsacker:1934nji,
    author = "von Weizsacker, C. F.",
    title = "{Radiation emitted in collisions of very fast electrons}",
    doi = "10.1007/BF01333110",
    journal = "Z. Phys.",
    volume = "88",
    pages = "612--625",
    year = "1934"
}

@article{Williams:1935dka,
    author = "Williams, E. J.",
    title = "{Correlation of certain collision problems with radiation theory}",
    journal = "Kong. Dan. Vid. Sel. Mat. Fys. Med.",
    volume = "13N4",
    pages = "1--50",
    year = "1935"
}

@article{Fermi:1924tc,
    author = "Fermi, E.",
    title = "{On the Theory of the impact between atoms and electrically charged particles}",
    doi = "10.1007/BF03184853",
    journal = "Z. Phys.",
    volume = "29",
    pages = "315--327",
    year = "1924"
}

@article{Cahn:1983ip,
    author = "Cahn, R. N. and Dawson, Sally",
    title = "{Production of Very Massive Higgs Bosons}",
    reportNumber = "LBL-16976",
    doi = "10.1016/0370-2693(84)91180-8",
    journal = "Phys. Lett. B",
    volume = "136",
    pages = "196",
    year = "1984",
    note = "[Erratum: Phys.Lett.B 138, 464 (1984)]"
}

@article{Dawson:1984gx,
    author = "Dawson, Sally",
    title = "{The Effective W Approximation}",
    reportNumber = "LBL-17497",
    doi = "10.1016/0550-3213(85)90038-0",
    journal = "Nucl. Phys. B",
    volume = "249",
    pages = "42--60",
    year = "1985"
}

@article{Kane:1984bb,
    author = "Kane, Gordon L. and Repko, W. W. and Rolnick, W. B.",
    title = "{The Effective W+-, Z0 Approximation for High-Energy Collisions}",
    reportNumber = "Print-84-0432 (MICHIGAN STATE)",
    doi = "10.1016/0370-2693(84)90105-9",
    journal = "Phys. Lett. B",
    volume = "148",
    pages = "367--372",
    year = "1984"
}

@article{Lindfors:1985yp,
    author = "Lindfors, J.",
    title = "{Distribution Functions for Heavy Vector Bosons Inside Colliding Particle Beams}",
    reportNumber = "UCR-TH-84-3",
    doi = "10.1007/BF01413605",
    journal = "Z. Phys. C",
    volume = "28",
    pages = "427",
    year = "1985"
}

@article{Kleiss:1986xp,
    author = "Kleiss, R. and Stirling, W. James",
    editor = "Loken, S. C.",
    title = "{Anomalous High-energy Behavior in Boson Fusion}",
    reportNumber = "CERN-TH-4451/86",
    doi = "10.1016/0370-2693(86)91081-6",
    journal = "Phys. Lett. B",
    volume = "182",
    pages = "75",
    year = "1986"
}

@article{Johnson:1987tj,
    author = "Johnson, P. W. and Olness, Fredrick I. and Tung, Wu-Ki",
    title = "{The Effective Vector Boson Method for High-energy Collisions}",
    reportNumber = "IIT-TH/87-03",
    doi = "10.1103/PhysRevD.36.291",
    journal = "Phys. Rev. D",
    volume = "36",
    pages = "291",
    year = "1987"
}

@article{Gunion:1986gm,
    author = "Gunion, J. F. and Kalinowski, J. and Tofighi-Niaki, A.",
    title = "{Exact F F $\to$ F F $W W$ Calculation for the Charged Current Sector and Comparison With the Effective $W$ Approximation}",
    reportNumber = "UCD-86-19",
    doi = "10.1103/PhysRevLett.57.2351",
    journal = "Phys. Rev. Lett.",
    volume = "57",
    pages = "2351",
    year = "1986"
}

@article{Kunszt:1987tk,
    author = "Kunszt, Zoltan and Soper, Davison E.",
    title = "{On the Validity of the Effective $W$ Approximation}",
    reportNumber = "OITS-343",
    doi = "10.1016/0550-3213(88)90673-6",
    journal = "Nucl. Phys. B",
    volume = "296",
    pages = "253--289",
    year = "1988"
}

@article{Abbasabadi:1988ja,
    author = "Abbasabadi, A. and Repko, W. W. and Dicus, Duane A. and Vega, Roberto",
    title = "{Comparison of Exact and Effective Gauge Boson Calculations for Gauge Boson Fusion Processes}",
    reportNumber = "MSUTH-88/002, DOE-ER40200-139",
    doi = "10.1103/PhysRevD.38.2770",
    journal = "Phys. Rev. D",
    volume = "38",
    pages = "2770",
    year = "1988"
}

@article{Kuss:1995yv,
    author = "Kuss, I. and Spiesberger, H.",
    title = "{Luminosities for vector boson - vector boson scattering at high-energy colliders}",
    eprint = "hep-ph/9507204",
    archivePrefix = "arXiv",
    reportNumber = "BI-TP-95-25",
    doi = "10.1103/PhysRevD.53.6078",
    journal = "Phys. Rev. D",
    volume = "53",
    pages = "6078--6093",
    year = "1996"
}

@article{Kuss:1996ww,
    author = "Kuss, I.",
    title = "{Improved effective vector boson approximation for hadron hadron collisions}",
    eprint = "hep-ph/9608453",
    archivePrefix = "arXiv",
    reportNumber = "BI-TP-96-41",
    doi = "10.1103/PhysRevD.55.7165",
    journal = "Phys. Rev. D",
    volume = "55",
    pages = "7165--7182",
    year = "1997"
}

@article{Borel:2012by,
    author = "Borel, Pascal and Franceschini, Roberto and Rattazzi, Riccardo and Wulzer, Andrea",
    title = "{Probing the Scattering of Equivalent Electroweak Bosons}",
    eprint = "1202.1904",
    archivePrefix = "arXiv",
    primaryClass = "hep-ph",
    reportNumber = "PP-012-005",
    doi = "10.1007/JHEP06(2012)122",
    journal = "JHEP",
    volume = "06",
    pages = "122",
    year = "2012"
}

@article{Costantini:2020stv,
    author = "Costantini, Antonio and De Lillo, Federico and Maltoni, Fabio and Mantani, Luca and Mattelaer, Olivier and Ruiz, Richard and Zhao, Xiaoran",
    title = "{Vector boson fusion at multi-TeV muon colliders}",
    eprint = "2005.10289",
    archivePrefix = "arXiv",
    primaryClass = "hep-ph",
    reportNumber = "CP3-20-20, MCNET-20-12, VBSCAN-PUB-03-20",
    doi = "10.1007/JHEP09(2020)080",
    journal = "JHEP",
    volume = "09",
    pages = "080",
    year = "2020"
}

@article{Ruiz:2021tdt,
    author = "Ruiz, Richard and Costantini, Antonio and Maltoni, Fabio and Mattelaer, Olivier",
    title = "{The Effective Vector Boson Approximation in high-energy muon collisions}",
    eprint = "2111.02442",
    archivePrefix = "arXiv",
    primaryClass = "hep-ph",
    reportNumber = "CP3-21-59, IFJPAN-IV-2021-17, MCNET-21-13, VBSCAN-PUB-07-21",
    doi = "10.1007/JHEP06(2022)114",
    journal = "JHEP",
    volume = "06",
    pages = "114",
    year = "2022"
}

@article{Bigaran:2025rvb,
    author = "Bigaran, Innes and Ruiz, Richard",
    title = "{Weak bosons as partons below 10 TeV partonic center-of-momentum}",
    eprint = "2502.07878",
    archivePrefix = "arXiv",
    primaryClass = "hep-ph",
    reportNumber = "IFJPAN-IV-2025-2, COMETA-2025-1, FERMILAB-PUB-25-0030-T",
    month = "2",
    year = "2025"
}

@article{Frixione:2025wsv,
    author = "Frixione, Stefano and Maltoni, Fabio and Pagani, Davide and Zaro, Marco",
    title = "{Double neutral-current corrections to NLO electroweak leptonic cross sections}",
    eprint = "2506.10732",
    archivePrefix = "arXiv",
    primaryClass = "hep-ph",
    reportNumber = "TIFF-UNIMI-2025-11, IRMP-CP3-25-22, COMETA-2025-20",
    month = "6",
    year = "2025"
}

@article{Frixione:2025guf,
    author = "Frixione, Stefano and Maltoni, Fabio and Pagani, Davide and Zaro, Marco",
    title = "{Precision phenomenology at multi-TeV muon colliders}",
    eprint = "2506.10733",
    archivePrefix = "arXiv",
    primaryClass = "hep-ph",
    reportNumber = "TIF-UNIMI-2025-12, IRMP-CP3-25-21, COMETA-2025-21",
    month = "6",
    year = "2025"
}

@article{Accomando:2006mc,
    author = "Accomando, Elena and Ballestrero, Alessandro and Belhouari, Aissa and Maina, Ezio",
    title = "{Isolating Vector Boson Scattering at the LHC: Gauge cancellations and the Equivalent Vector Boson Approximation vs complete calculations}",
    eprint = "hep-ph/0608019",
    archivePrefix = "arXiv",
    reportNumber = "DFTT-12-2006",
    doi = "10.1103/PhysRevD.74.073010",
    journal = "Phys. Rev. D",
    volume = "74",
    pages = "073010",
    year = "2006"
}

@article{Han:2020uid,
    author = "Han, Tao and Ma, Yang and Xie, Keping",
    title = "{High energy leptonic collisions and electroweak parton distribution functions}",
    eprint = "2007.14300",
    archivePrefix = "arXiv",
    primaryClass = "hep-ph",
    reportNumber = "PITT-PACC 2006",
    doi = "10.1103/PhysRevD.103.L031301",
    journal = "Phys. Rev. D",
    volume = "103",
    number = "3",
    pages = "L031301",
    year = "2021"
}

@article{Han:2021kes,
    author = "Han, Tao and Ma, Yang and Xie, Keping",
    title = "{Quark and gluon contents of a lepton at high energies}",
    eprint = "2103.09844",
    archivePrefix = "arXiv",
    primaryClass = "hep-ph",
    reportNumber = "PITT-PACC-2104",
    doi = "10.1007/JHEP02(2022)154",
    journal = "JHEP",
    volume = "02",
    pages = "154",
    year = "2022"
}

@article{Gribov:1972ri,
    author = "Gribov, V. N. and Lipatov, L. N.",
    title = "{Deep inelastic e p scattering in perturbation theory}",
    reportNumber = "IPTI-381-71",
    journal = "Sov. J. Nucl. Phys.",
    volume = "15",
    pages = "438--450",
    year = "1972"
}

@article{Altarelli:1977zs,
    author = "Altarelli, Guido and Parisi, G.",
    title = "{Asymptotic Freedom in Parton Language}",
    reportNumber = "LPTENS-77-6",
    doi = "10.1016/0550-3213(77)90384-4",
    journal = "Nucl. Phys. B",
    volume = "126",
    pages = "298--318",
    year = "1977"
}

@article{Dokshitzer:1977sg,
    author = "Dokshitzer, Yuri L.",
    title = "{Calculation of the Structure Functions for Deep Inelastic Scattering and e+ e- Annihilation by Perturbation Theory in Quantum Chromodynamics.}",
    journal = "Sov. Phys. JETP",
    volume = "46",
    pages = "641--653",
    year = "1977"
}

@article{Boos:1997gw,
    author = "Boos, E. and He, H. J. and Kilian, W. and Pukhov, A. and Yuan, C. P. and Zerwas, P. M.",
    title = "{Strongly interacting vector bosons at TeV e+ e- linear colliders}",
    eprint = "hep-ph/9708310",
    archivePrefix = "arXiv",
    reportNumber = "DESY-96-256",
    doi = "10.1103/PhysRevD.57.1553",
    journal = "Phys. Rev. D",
    volume = "57",
    pages = "1553",
    year = "1998"
}

@article{Boos:1999kj,
    author = "Boos, E. and He, H. J. and Kilian, W. and Pukhov, A. and Yuan, C. P. and Zerwas, P. M.",
    title = "{Strongly interacting vector bosons at TeV e+- e- linear colliders: Addendum}",
    eprint = "hep-ph/9908409",
    archivePrefix = "arXiv",
    reportNumber = "DESY-99-111, TTP-99-34, MSUHEP-90815",
    doi = "10.1103/PhysRevD.61.077901",
    journal = "Phys. Rev. D",
    volume = "61",
    pages = "077901",
    year = "2000"
}

@article{Chen:2016wkt,
    author = "Chen, Junmou and Han, Tao and Tweedie, Brock",
    title = "{Electroweak Splitting Functions and High Energy Showering}",
    eprint = "1611.00788",
    archivePrefix = "arXiv",
    primaryClass = "hep-ph",
    reportNumber = "PITT-PACC-1602",
    doi = "10.1007/JHEP11(2017)093",
    journal = "JHEP",
    volume = "11",
    pages = "093",
    year = "2017"
}

@article{Bauer:2017isx,
    author = "Bauer, Christian W. and Ferland, Nicolas and Webber, Bryan R.",
    title = "{Standard Model Parton Distributions at Very High Energies}",
    eprint = "1703.08562",
    archivePrefix = "arXiv",
    primaryClass = "hep-ph",
    reportNumber = "CAVENDISH-HEP-17-03, CERN-TH-2017-067",
    doi = "10.1007/JHEP08(2017)036",
    journal = "JHEP",
    volume = "08",
    pages = "036",
    year = "2017"
}

@article{Bauer:2018xag,
    author = "Bauer, Christian W. and Provasoli, Davide and Webber, Bryan R.",
    title = "{Standard Model Fragmentation Functions at Very High Energies}",
    eprint = "1806.10157",
    archivePrefix = "arXiv",
    primaryClass = "hep-ph",
    reportNumber = "Cavendish-HEP-18/12",
    doi = "10.1007/JHEP11(2018)030",
    journal = "JHEP",
    volume = "11",
    pages = "030",
    year = "2018"
}

@article{Bauer:2018arx,
    author = "Bauer, Christian W. and Webber, Bryan R.",
    title = "{Polarization Effects in Standard Model Parton Distributions at Very High Energies}",
    eprint = "1808.08831",
    archivePrefix = "arXiv",
    primaryClass = "hep-ph",
    reportNumber = "Cavendish-HEP-18/13",
    doi = "10.1007/JHEP03(2019)013",
    journal = "JHEP",
    volume = "03",
    pages = "013",
    year = "2019"
}

@article{Ciafaloni:2005fm,
    author = "Ciafaloni, Paolo and Comelli, Denis",
    title = "{Electroweak evolution equations}",
    eprint = "hep-ph/0505047",
    archivePrefix = "arXiv",
    doi = "10.1088/1126-6708/2005/11/022",
    journal = "JHEP",
    volume = "11",
    pages = "022",
    year = "2005"
}

@article{Garosi:2023bvq,
    author = "Garosi, Francesco and Marzocca, David and Trifinopoulos, Sokratis",
    title = "{LePDF: Standard Model PDFs for high-energy lepton colliders}",
    eprint = "2303.16964",
    archivePrefix = "arXiv",
    primaryClass = "hep-ph",
    doi = "10.1007/JHEP09(2023)107",
    journal = "JHEP",
    volume = "09",
    pages = "107",
    year = "2023"
}

@article{Marzocca:2024ica,
    author = "Marzocca, David and Stanzione, Alfredo",
    title = "{On the impact of the mixed Z/{\ensuremath{\gamma}} PDF at muon colliders}",
    eprint = "2408.13191",
    archivePrefix = "arXiv",
    primaryClass = "hep-ph",
    doi = "10.1007/JHEP03(2025)171",
    journal = "JHEP",
    volume = "03",
    pages = "171",
    year = "2025"
}

@article{ATLAS:2012yve,
    author = "Aad, Georges and others",
    collaboration = "ATLAS",
    title = "{Observation of a new particle in the search for the Standard Model Higgs boson with the ATLAS detector at the LHC}",
    eprint = "1207.7214",
    archivePrefix = "arXiv",
    primaryClass = "hep-ex",
    reportNumber = "CERN-PH-EP-2012-218",
    doi = "10.1016/j.physletb.2012.08.020",
    journal = "Phys. Lett. B",
    volume = "716",
    pages = "1--29",
    year = "2012"
}

@article{CMS:2012qbp,
    author = "Chatrchyan, Serguei and others",
    collaboration = "CMS",
    title = "{Observation of a New Boson at a Mass of 125 GeV with the CMS Experiment at the LHC}",
    eprint = "1207.7235",
    archivePrefix = "arXiv",
    primaryClass = "hep-ex",
    reportNumber = "CMS-HIG-12-028, CERN-PH-EP-2012-220",
    doi = "10.1016/j.physletb.2012.08.021",
    journal = "Phys. Lett. B",
    volume = "716",
    pages = "30--61",
    year = "2012"
}

@article{ATLAS:2015yey,
    author = "Aad, Georges and others",
    collaboration = "ATLAS, CMS",
    title = "{Combined Measurement of the Higgs Boson Mass in $pp$ Collisions at $\sqrt{s}=7$ and 8 TeV with the ATLAS and CMS Experiments}",
    eprint = "1503.07589",
    archivePrefix = "arXiv",
    primaryClass = "hep-ex",
    reportNumber = "ATLAS-HIGG-2014-14, CMS-HIG-14-042, CERN-PH-EP-2015-075",
    doi = "10.1103/PhysRevLett.114.191803",
    journal = "Phys. Rev. Lett.",
    volume = "114",
    pages = "191803",
    year = "2015"
}

@article{ATLAS:2016neq,
    author = "Aad, Georges and others",
    collaboration = "ATLAS, CMS",
    title = "{Measurements of the Higgs boson production and decay rates and constraints on its couplings from a combined ATLAS and CMS analysis of the LHC pp collision data at $ \sqrt{s}=7 $ and 8 TeV}",
    eprint = "1606.02266",
    archivePrefix = "arXiv",
    primaryClass = "hep-ex",
    reportNumber = "CERN-EP-2016-100, ATLAS-HIGG-2015-07, CMS-HIG-15-002",
    doi = "10.1007/JHEP08(2016)045",
    journal = "JHEP",
    volume = "08",
    pages = "045",
    year = "2016"
}

@article{ATLAS:2013dos,
    author = "Aad, Georges and others",
    collaboration = "ATLAS",
    title = "{Measurements of Higgs boson production and couplings in diboson final states with the ATLAS detector at the LHC}",
    eprint = "1307.1427",
    archivePrefix = "arXiv",
    primaryClass = "hep-ex",
    reportNumber = "CERN-PH-EP-2013-103",
    doi = "10.1016/j.physletb.2014.05.011",
    journal = "Phys. Lett. B",
    volume = "726",
    pages = "88--119",
    year = "2013",
    note = "[Erratum: Phys.Lett.B 734, 406--406 (2014)]"
}

@article{ATLAS:2022vkf,
    author = "Aad, Georges and others",
    collaboration = "ATLAS",
    title = "{A detailed map of Higgs boson interactions by the ATLAS experiment ten years after the discovery}",
    eprint = "2207.00092",
    archivePrefix = "arXiv",
    primaryClass = "hep-ex",
    reportNumber = "CERN-EP-2022-057",
    doi = "10.1038/s41586-022-04893-w",
    journal = "Nature",
    volume = "607",
    number = "7917",
    pages = "52--59",
    year = "2022",
    note = "[Erratum: Nature 612, E24 (2022)]"
}

@article{ATLAS:2020fzp,
    author = "Aad, Georges and others",
    collaboration = "ATLAS",
    title = "{A search for the dimuon decay of the Standard Model Higgs boson with the ATLAS detector}",
    eprint = "2007.07830",
    archivePrefix = "arXiv",
    primaryClass = "hep-ex",
    reportNumber = "CERN-EP-2020-117",
    doi = "10.1016/j.physletb.2020.135980",
    journal = "Phys. Lett. B",
    volume = "812",
    pages = "135980",
    year = "2021"
}

@article{ATLAS:2020rej,
    author = "Aad, Georges and others",
    collaboration = "ATLAS",
    title = "{Higgs boson production cross-section measurements and their EFT interpretation in the $4\ell $ decay channel at $\sqrt{s}=$13 TeV with the ATLAS detector}",
    eprint = "2004.03447",
    archivePrefix = "arXiv",
    primaryClass = "hep-ex",
    reportNumber = "CERN-EP-2020-034",
    doi = "10.1140/epjc/s10052-020-8227-9",
    journal = "Eur. Phys. J. C",
    volume = "80",
    number = "10",
    pages = "957",
    year = "2020",
    note = "[Erratum: Eur.Phys.J.C 81, 29 (2021), Erratum: Eur.Phys.J.C 81, 398 (2021)]"
}

@article{ATLAS:2020ior,
    author = "Aad, Georges and others",
    collaboration = "ATLAS",
    title = "{$CP$ Properties of Higgs Boson Interactions with Top Quarks in the $t\bar{t}H$ and $tH$ Processes Using $H \rightarrow \gamma\gamma$ with the ATLAS Detector}",
    eprint = "2004.04545",
    archivePrefix = "arXiv",
    primaryClass = "hep-ex",
    reportNumber = "CERN-EP-2020-046",
    doi = "10.1103/PhysRevLett.125.061802",
    journal = "Phys. Rev. Lett.",
    volume = "125",
    number = "6",
    pages = "061802",
    year = "2020"
}

@article{CMS:2022dwd,
    author = "Tumasyan, Armen and others",
    collaboration = "CMS",
    title = "{A portrait of the Higgs boson by the CMS experiment ten years after the discovery.}",
    eprint = "2207.00043",
    archivePrefix = "arXiv",
    primaryClass = "hep-ex",
    reportNumber = "CMS-HIG-22-001, CERN-EP-2022-039",
    doi = "10.1038/s41586-022-04892-x",
    journal = "Nature",
    volume = "607",
    number = "7917",
    pages = "60--68",
    year = "2022",
    note = "[Erratum: Nature 623, (2023)]"
}

@article{CMS:2020xrn,
    author = "Sirunyan, Albert M and others",
    collaboration = "CMS",
    title = "{A measurement of the Higgs boson mass in the diphoton decay channel}",
    eprint = "2002.06398",
    archivePrefix = "arXiv",
    primaryClass = "hep-ex",
    reportNumber = "CMS-HIG-19-004, CERN-EP-2020-004",
    doi = "10.1016/j.physletb.2020.135425",
    journal = "Phys. Lett. B",
    volume = "805",
    pages = "135425",
    year = "2020"
}

@article{CMS:2022ley,
    author = "Tumasyan, Armen and others",
    collaboration = "CMS",
    title = "{Measurement of the Higgs boson width and evidence of its off-shell contributions to ZZ production}",
    eprint = "2202.06923",
    archivePrefix = "arXiv",
    primaryClass = "hep-ex",
    reportNumber = "CMS-HIG-21-013, CERN-EP-2021-272",
    doi = "10.1038/s41567-022-01682-0",
    journal = "Nature Phys.",
    volume = "18",
    number = "11",
    pages = "1329--1334",
    year = "2022"
}

@article{CMS:2014fzn,
    author = "Khachatryan, Vardan and others",
    collaboration = "CMS",
    title = "{Precise determination of the mass of the Higgs boson and tests of compatibility of its couplings with the standard model predictions using proton collisions at 7 and 8 $\,\text {TeV}$}",
    eprint = "1412.8662",
    archivePrefix = "arXiv",
    primaryClass = "hep-ex",
    reportNumber = "CMS-HIG-14-009, CERN-PH-EP-2014-288",
    doi = "10.1140/epjc/s10052-015-3351-7",
    journal = "Eur. Phys. J. C",
    volume = "75",
    number = "5",
    pages = "212",
    year = "2015"
}

@article{CMS:2013fjq,
    author = "Chatrchyan, Serguei and others",
    collaboration = "CMS",
    title = "{Measurement of the Properties of a Higgs Boson in the Four-Lepton Final State}",
    eprint = "1312.5353",
    archivePrefix = "arXiv",
    primaryClass = "hep-ex",
    reportNumber = "CMS-HIG-13-002, CERN-PH-EP-2013-220",
    doi = "10.1103/PhysRevD.89.092007",
    journal = "Phys. Rev. D",
    volume = "89",
    number = "9",
    pages = "092007",
    year = "2014"
}

@article{CMS:2018nsn,
    author = "Sirunyan, A. M. and others",
    collaboration = "CMS",
    title = "{Observation of Higgs boson decay to bottom quarks}",
    eprint = "1808.08242",
    archivePrefix = "arXiv",
    primaryClass = "hep-ex",
    reportNumber = "CMS-HIG-18-016, CERN-EP-2018-223",
    doi = "10.1103/PhysRevLett.121.121801",
    journal = "Phys. Rev. Lett.",
    volume = "121",
    number = "12",
    pages = "121801",
    year = "2018"
}

@article{ATLAS:2014zve,
    author = "Aad, Georges and others",
    collaboration = "ATLAS",
    title = "{Search for direct production of charginos, neutralinos and sleptons in final states with two leptons and missing transverse momentum in $pp$ collisions at $\sqrt{s} =$ 8 TeV with the ATLAS detector}",
    eprint = "1403.5294",
    archivePrefix = "arXiv",
    primaryClass = "hep-ex",
    reportNumber = "CERN-PH-EP-2014-037",
    doi = "10.1007/JHEP05(2014)071",
    journal = "JHEP",
    volume = "05",
    pages = "071",
    year = "2014"
}

@article{ATLAS:2014jxt,
    author = "Aad, Georges and others",
    collaboration = "ATLAS",
    title = "{Search for squarks and gluinos with the ATLAS detector in final states with jets and missing transverse momentum using $\sqrt{s}=8$ TeV proton--proton collision data}",
    eprint = "1405.7875",
    archivePrefix = "arXiv",
    primaryClass = "hep-ex",
    reportNumber = "CERN-PH-EP-2014-093",
    doi = "10.1007/JHEP09(2014)176",
    journal = "JHEP",
    volume = "09",
    pages = "176",
    year = "2014"
}

@article{ATLAS:2019lff,
    author = "Aad, Georges and others",
    collaboration = "ATLAS",
    title = "{Search for electroweak production of charginos and sleptons decaying into final states with two leptons and missing transverse momentum in $\sqrt{s}=13$ TeV $pp$ collisions using the ATLAS detector}",
    eprint = "1908.08215",
    archivePrefix = "arXiv",
    primaryClass = "hep-ex",
    reportNumber = "CERN-EP-2019-106",
    doi = "10.1140/epjc/s10052-019-7594-6",
    journal = "Eur. Phys. J. C",
    volume = "80",
    number = "2",
    pages = "123",
    year = "2020"
}

@article{ATLAS:2011yrs,
    author = "Aad, Georges and others",
    collaboration = "ATLAS",
    title = "{Search for squarks and gluinos using final states with jets and missing transverse momentum with the ATLAS detector in $\sqrt{s}=7$ TeV proton-proton collisions}",
    eprint = "1109.6572",
    archivePrefix = "arXiv",
    primaryClass = "hep-ex",
    reportNumber = "CERN-PH-EP-2011-145",
    doi = "10.1016/j.physletb.2012.02.051",
    journal = "Phys. Lett. B",
    volume = "710",
    pages = "67--85",
    year = "2012"
}

@article{ATLAS:2019lng,
    author = "Aad, Georges and others",
    collaboration = "ATLAS",
    title = "{Searches for electroweak production of supersymmetric particles with compressed mass spectra in $\sqrt{s}=$ 13 TeV $pp$ collisions with the ATLAS detector}",
    eprint = "1911.12606",
    archivePrefix = "arXiv",
    primaryClass = "hep-ex",
    reportNumber = "CERN-EP-2019-242",
    doi = "10.1103/PhysRevD.101.052005",
    journal = "Phys. Rev. D",
    volume = "101",
    number = "5",
    pages = "052005",
    year = "2020"
}

@article{CMS:2014ewg,
    author = "Khachatryan, Vardan and others",
    collaboration = "CMS",
    title = "{Searches for Electroweak Production of Charginos, Neutralinos, and Sleptons Decaying to Leptons and W, Z, and Higgs Bosons in pp Collisions at 8 TeV}",
    eprint = "1405.7570",
    archivePrefix = "arXiv",
    primaryClass = "hep-ex",
    reportNumber = "CMS-SUS-13-006, CERN-PH-EP-2014-098",
    doi = "10.1140/epjc/s10052-014-3036-7",
    journal = "Eur. Phys. J. C",
    volume = "74",
    number = "9",
    pages = "3036",
    year = "2014"
}

@article{CMS:2011bul,
    author = "Chatrchyan, Serguei and others",
    collaboration = "CMS",
    title = "{Search for Supersymmetry at the LHC in Events with Jets and Missing Transverse Energy}",
    eprint = "1109.2352",
    archivePrefix = "arXiv",
    primaryClass = "hep-ex",
    reportNumber = "CERN-PH-EP-2011-138, CMS-SUS-11-003",
    doi = "10.1103/PhysRevLett.107.221804",
    journal = "Phys. Rev. Lett.",
    volume = "107",
    pages = "221804",
    year = "2011"
}

@article{CMS:2011xek,
    author = "Khachatryan, Vardan and others",
    collaboration = "CMS",
    title = "{Search for Supersymmetry in pp Collisions at 7 TeV in Events with Jets and Missing Transverse Energy}",
    eprint = "1101.1628",
    archivePrefix = "arXiv",
    primaryClass = "hep-ex",
    reportNumber = "CERN-PH-EP-2010-084, CMS-SUS-10-003",
    doi = "10.1016/j.physletb.2011.03.021",
    journal = "Phys. Lett. B",
    volume = "698",
    pages = "196--218",
    year = "2011"
}

@article{CMS:2013cgf,
    author = "Chatrchyan, Serguei and others",
    collaboration = "CMS",
    title = "{Search for Supersymmetry in Hadronic Final States with Missing Transverse Energy Using the Variables $\alpha_T$ and b-Quark Multiplicity in $pp$ collisions at $\sqrt{s}= 8$ TeV}",
    eprint = "1303.2985",
    archivePrefix = "arXiv",
    primaryClass = "hep-ex",
    reportNumber = "CMS-SUS-12-028, CERN-PH-EP-2013-037",
    doi = "10.1140/epjc/s10052-013-2568-6",
    journal = "Eur. Phys. J. C",
    volume = "73",
    number = "9",
    pages = "2568",
    year = "2013"
}

@article{LHCb:2021auc,
    author = "Aaij, Roel and others",
    collaboration = "LHCb",
    title = "{Study of the doubly charmed tetraquark $T_{cc}^{+}$}",
    eprint = "2109.01056",
    archivePrefix = "arXiv",
    primaryClass = "hep-ex",
    reportNumber = "CERN-EP-2021-169, LHCb-PAPER-2021-032",
    doi = "10.1038/s41467-022-30206-w",
    journal = "Nature Commun.",
    volume = "13",
    number = "1",
    pages = "3351",
    year = "2022"
}

@article{LHCb:2021vvq,
    author = "Aaij, Roel and others",
    collaboration = "LHCb",
    title = "{Observation of an exotic narrow doubly charmed tetraquark}",
    eprint = "2109.01038",
    archivePrefix = "arXiv",
    primaryClass = "hep-ex",
    reportNumber = "CERN-EP-2021-165, LHCb-PAPER-2021-031",
    doi = "10.1038/s41567-022-01614-y",
    journal = "Nature Phys.",
    volume = "18",
    number = "7",
    pages = "751--754",
    year = "2022"
}

@article{LHCb:2015yax,
    author = "Aaij, Roel and others",
    collaboration = "LHCb",
    title = "{Observation of $J/\psi p$ Resonances Consistent with Pentaquark States in $\Lambda_b^0 \to J/\psi K^- p$ Decays}",
    eprint = "1507.03414",
    archivePrefix = "arXiv",
    primaryClass = "hep-ex",
    reportNumber = "CERN-PH-EP-2015-153, LHCB-PAPER-2015-029",
    doi = "10.1103/PhysRevLett.115.072001",
    journal = "Phys. Rev. Lett.",
    volume = "115",
    pages = "072001",
    year = "2015"
}

@article{LHCb:2019kea,
    author = "Aaij, Roel and others",
    collaboration = "LHCb",
    title = "{Observation of a narrow pentaquark state, $P_c(4312)^+$, and of two-peak structure of the $P_c(4450)^+$}",
    eprint = "1904.03947",
    archivePrefix = "arXiv",
    primaryClass = "hep-ex",
    reportNumber = "LHCb-PAPER-2019-014 CERN-EP-2019-058",
    doi = "10.1103/PhysRevLett.122.222001",
    journal = "Phys. Rev. Lett.",
    volume = "122",
    number = "22",
    pages = "222001",
    year = "2019"
}

@article{LHCb:2015gmp,
    author = "Aaij, Roel and others",
    collaboration = "LHCb",
    title = "{Measurement of the ratio of branching fractions $\mathcal{B}(\bar{B}^0 \to D^{*+}\tau^{-}\bar{\nu}_{\tau})/\mathcal{B}(\bar{B}^0 \to D^{*+}\mu^{-}\bar{\nu}_{\mu})$}",
    eprint = "1506.08614",
    archivePrefix = "arXiv",
    primaryClass = "hep-ex",
    reportNumber = "CERN-PH-EP-2015-150, LHCB-PAPER-2015-025, CERN-PH-EP-2015-150, LHCb-PAPER-2015-025",
    doi = "10.1103/PhysRevLett.115.111803",
    journal = "Phys. Rev. Lett.",
    volume = "115",
    number = "11",
    pages = "111803",
    year = "2015",
    note = "[Erratum: Phys.Rev.Lett. 115, 159901 (2015)]"
}

@article{LHCb:2021trn,
    author = "Aaij, Roel and others",
    collaboration = "LHCb",
    title = "{Test of lepton universality in beauty-quark decays}",
    eprint = "2103.11769",
    archivePrefix = "arXiv",
    primaryClass = "hep-ex",
    reportNumber = "LHCb-PAPER-2021-004, CERN-EP-2021-042",
    doi = "10.1038/s41567-023-02095-3",
    journal = "Nature Phys.",
    volume = "18",
    number = "3",
    pages = "277--282",
    year = "2022",
    note = "[Addendum: Nature Phys. 19, (2023)]"
}

@article{LHCb:2019hip,
    author = "Aaij, Roel and others",
    collaboration = "LHCb",
    title = "{Search for lepton-universality violation in $B^+\to K^+\ell^+\ell^-$ decays}",
    eprint = "1903.09252",
    archivePrefix = "arXiv",
    primaryClass = "hep-ex",
    reportNumber = "LHCb-PAPER-2019-009, CERN-EP-2019-043, LHCb-PAPER-2019-009 CERN-EP-2019-043",
    doi = "10.1103/PhysRevLett.122.191801",
    journal = "Phys. Rev. Lett.",
    volume = "122",
    number = "19",
    pages = "191801",
    year = "2019"
}

@article{LHCb:2012skj,
    author = "Aaij, R and others",
    collaboration = "LHCb",
    title = "{First Evidence for the Decay $B_s^0 \to \mu^+ \mu^-$}",
    eprint = "1211.2674",
    archivePrefix = "arXiv",
    primaryClass = "hep-ex",
    reportNumber = "CERN-PH-EP-2012-335, LHCB-PAPER-2012-043",
    doi = "10.1103/PhysRevLett.110.021801",
    journal = "Phys. Rev. Lett.",
    volume = "110",
    number = "2",
    pages = "021801",
    year = "2013"
}

@article{LHCb:2015svh,
    author = "Aaij, Roel and others",
    collaboration = "LHCb",
    title = "{Angular analysis of the $B^{0} \to K^{*0} \mu^{+} \mu^{-}$ decay using 3 fb$^{-1}$ of integrated luminosity}",
    eprint = "1512.04442",
    archivePrefix = "arXiv",
    primaryClass = "hep-ex",
    reportNumber = "CERN-PH-EP-2015-314, LHCB-PAPER-2015-051",
    doi = "10.1007/JHEP02(2016)104",
    journal = "JHEP",
    volume = "02",
    pages = "104",
    year = "2016"
}

@article{LHCb:2014vgu,
    author = "Aaij, Roel and others",
    collaboration = "LHCb",
    title = "{Test of lepton universality using $B^{+}\rightarrow K^{+}\ell^{+}\ell^{-}$ decays}",
    eprint = "1406.6482",
    archivePrefix = "arXiv",
    primaryClass = "hep-ex",
    reportNumber = "CERN-PH-EP-2014-140, LHCB-PAPER-2014-024",
    doi = "10.1103/PhysRevLett.113.151601",
    journal = "Phys. Rev. Lett.",
    volume = "113",
    pages = "151601",
    year = "2014"
}

@article{LHCb:2017avl,
    author = "Aaij, R. and others",
    collaboration = "LHCb",
    title = "{Test of lepton universality with $B^{0} \rightarrow K^{*0}\ell^{+}\ell^{-}$ decays}",
    eprint = "1705.05802",
    archivePrefix = "arXiv",
    primaryClass = "hep-ex",
    reportNumber = "LHCB-PAPER-2017-013, CERN-EP-2017-100",
    doi = "10.1007/JHEP08(2017)055",
    journal = "JHEP",
    volume = "08",
    pages = "055",
    year = "2017"
}

@article{ALICE:2016fzo,
    author = "Adam, Jaroslav and others",
    collaboration = "ALICE",
    title = "{Enhanced production of multi-strange hadrons in high-multiplicity proton-proton collisions}",
    eprint = "1606.07424",
    archivePrefix = "arXiv",
    primaryClass = "nucl-ex",
    reportNumber = "CERN-EP-2016-153",
    doi = "10.1038/nphys4111",
    journal = "Nature Phys.",
    volume = "13",
    pages = "535--539",
    year = "2017"
}

@article{ALICE:2013osk,
    author = "Abelev, Betty Bezverkhny and others",
    collaboration = "ALICE",
    title = "{Centrality, rapidity and transverse momentum dependence of $J/\psi$ suppression in Pb-Pb collisions at $\sqrt{s_{\rm NN}}$=2.76 TeV}",
    eprint = "1311.0214",
    archivePrefix = "arXiv",
    primaryClass = "nucl-ex",
    reportNumber = "CERN-PH-EP-2013-203",
    doi = "10.1016/j.physletb.2014.05.064",
    journal = "Phys. Lett. B",
    volume = "734",
    pages = "314--327",
    year = "2014"
}

@article{ALICE:2022wpn,
    author = "Acharya, Shreyasi and others",
    collaboration = "ALICE",
    title = "{The ALICE experiment: a journey through QCD}",
    eprint = "2211.04384",
    archivePrefix = "arXiv",
    primaryClass = "nucl-ex",
    reportNumber = "CERN-EP-2022-227",
    doi = "10.1140/epjc/s10052-024-12935-y",
    journal = "Eur. Phys. J. C",
    volume = "84",
    number = "8",
    pages = "813",
    year = "2024"
}

@article{ALICE:2016ccg,
    author = "Adam, Jaroslav and others",
    collaboration = "ALICE",
    title = "{Anisotropic flow of charged particles in Pb-Pb collisions at $\sqrt{s_{\rm NN}}=5.02$ TeV}",
    eprint = "1602.01119",
    archivePrefix = "arXiv",
    primaryClass = "nucl-ex",
    reportNumber = "CERN-EP-2016-018",
    doi = "10.1103/PhysRevLett.116.132302",
    journal = "Phys. Rev. Lett.",
    volume = "116",
    number = "13",
    pages = "132302",
    year = "2016"
}

@article{ALICE:2015vxz,
    author = "Adam, Jaroslav and others",
    collaboration = "ALICE",
    title = "{Transverse momentum dependence of D-meson production in Pb-Pb collisions at $ \sqrt{{\mathrm{s}}_{\mathrm{NN}}}=$ 2.76  TeV}",
    eprint = "1509.06888",
    archivePrefix = "arXiv",
    primaryClass = "nucl-ex",
    reportNumber = "CERN-PH-EP-2015-252",
    doi = "10.1007/JHEP03(2016)081",
    journal = "JHEP",
    volume = "03",
    pages = "081",
    year = "2016"
}

@article{ALICE:2015mdb,
    author = "Adam, Jaroslav and others",
    collaboration = "ALICE",
    title = "{Measurement of jet quenching with semi-inclusive hadron-jet distributions in central Pb-Pb collisions at $ \sqrt{s_{\mathrm{NN}}}=2.76 $ TeV}",
    eprint = "1506.03984",
    archivePrefix = "arXiv",
    primaryClass = "nucl-ex",
    reportNumber = "CERN-PH-EP-2015-136",
    doi = "10.1007/JHEP09(2015)170",
    journal = "JHEP",
    volume = "09",
    pages = "170",
    year = "2015"
}

@article{ALICE:2008ngc,
    author = "Aamodt, K. and others",
    collaboration = "ALICE",
    title = "{The ALICE experiment at the CERN LHC}",
    doi = "10.1088/1748-0221/3/08/S08002",
    journal = "JINST",
    volume = "3",
    pages = "S08002",
    year = "2008"
}

@article{LHCb:2008vvz,
    author = "Alves, Jr., A. Augusto and others",
    collaboration = "LHCb",
    title = "{The LHCb Detector at the LHC}",
    reportNumber = "LHCb-DP-2008-001",
    doi = "10.1088/1748-0221/3/08/S08005",
    journal = "JINST",
    volume = "3",
    pages = "S08005",
    year = "2008"
}

@article{ATLAS:2008xda,
    author = "Aad, G. and others",
    collaboration = "ATLAS",
    title = "{The ATLAS Experiment at the CERN Large Hadron Collider}",
    doi = "10.1088/1748-0221/3/08/S08003",
    journal = "JINST",
    volume = "3",
    pages = "S08003",
    year = "2008"
}

@article{CMS:2008xjf,
    author = "Chatrchyan, S. and others",
    collaboration = "CMS",
    title = "{The CMS Experiment at the CERN LHC}",
    doi = "10.1088/1748-0221/3/08/S08004",
    journal = "JINST",
    volume = "3",
    pages = "S08004",
    year = "2008"
}

@misc{EPPSU-website,
  title = "{European Strategy for Particle Physics}",
  howpublished = "{\url{https://europeanstrategy.cern/}}",
  note = "{Accessed: 2025-08-24}"
}

@report{Butler:2023glv,
    author = "Butler, Joel N. and others",
    title = "{Report of the 2021 U.S. Community Study on the Future of Particle Physics (Snowmass 2021)}",
    reportNumber = "FERMILAB-CONF-23-008, SLAC-PUB-17717",
    doi = "10.2172/1922503",
    month = "1",
    year = "2023"
}

@article{P5:2023wyd,
    author = "Asai, Shoji and others",
    collaboration = "P5",
    title = "{Exploring the Quantum Universe: Pathways to Innovation and Discovery in Particle Physics}",
    eprint = "2407.19176",
    archivePrefix = "arXiv",
    primaryClass = "hep-ex",
    doi = "10.2172/2368847",
    month = "12",
    year = "2023"
}

@proceedings{Donaldson:1983mq,
    editor = "Donaldson, R. and Gustafson, H. Richard and Paige, F. E.",
    title = "{Proceedings, 1982 DPF Summer Study on Elementary Particle Physics and Future Facilities (Snowmass 82)}: {Snowmass, Colorado, June 28-July 16,1982}",
    year = "1983"
}

@techreport{CERN-ESU-001,
  title        = {Strategy Brochure 2006},
  institution  = {CERN},
  number       = {CERN-ESU-001},
  address      = {Geneva},
  year         = {2006},
  url          = {https://cds.cern.ch/record/2690118},
author = "Strategy Group"
}

@techreport{CERN-ESU-015,
      title         = "{2020 Update of the European Strategy for Particle Physics
                       (Brochure)}",
        institution  = {CERN},
      number  = "CERN-ESU-015",
      address       = "Geneva",
      year          = "2020",
      url           = "https://cds.cern.ch/record/2721370",
      doi           = "10.17181/CERN.JSC6.W89E",
author = "Strategy Group"
}

@techreport{ESC-E-106,
      title         = "{2013 Update of the European Strategy for Particle Physics}",
        institution  = {CERN},
      number  = "esc-e-106",
      address       = "Geneva",
      year          = "2013",
      url           = "https://cds.cern.ch/record/1567258/files/esc-e-106.pdf",
author = "Strategy Group"
}

@article{ILC:2007oiw,
    author = "Aarons, Gerald and others",
    editor = "Brau, James and Okada, Yasuhiro and Walker, Nicholas",
    collaboration = "ILC",
    title = "{ILC Reference Design Report Volume 1 - Executive Summary}",
    eprint = "0712.1950",
    archivePrefix = "arXiv",
    primaryClass = "physics.acc-ph",
    reportNumber = "FERMILAB-DESIGN-2007-03, FERMILAB-PUB-07-794-E",
    month = "8",
    year = "2007"
}

@article{Behnke:2013xla,
    author = "Behnke, Ties and Brau, James E. and Foster, Brian and Fuster, Juan and Harrison, Mike and Paterson, James McEwan and Peskin, Michael and Stanitzki, Marcel and Walker, Nicholas and Yamamoto, Hitoshi",
    title = "{The International Linear Collider Technical Design Report - Volume 1: Executive Summary}",
    eprint = "1306.6327",
    archivePrefix = "arXiv",
    primaryClass = "physics.acc-ph",
    reportNumber = "ILC-REPORT-2013-040, ANL-HEP-TR-13-20, BNL-100603-2013-IR, IRFU-13-59, CERN-ATS-2013-037, COCKCROFT-13-10, CLNS-13-2085, DESY-13-062, FERMILAB-TM-2554, IHEP-AC-ILC-2013-001, INFN-13-04-LNF, JAI-2013-001, JINR-E9-2013-35, JLAB-R-2013-01, KEK-REPORT-2013-1, KNU-CHEP-ILC-2013-1, LLNL-TR-635539, SLAC-R-1004, ILC-HIGRADE-REPORT-2013-003",
    month = "6",
    year = "2013"
}

@report{CEPC-SPPCStudyGroup:2015csa,
    author = "Ahmad, Muhammd and others",
    title = "{CEPC-SPPC Preliminary Conceptual Design Report. 1. Physics and Detector}",
    reportNumber = "IHEP-CEPC-DR-2015-01, IHEP-TH-2015-01, IHEP-EP-2015-01",
    month = "3",
    year = "2015"
}

@article{CEPCStudyGroup:2023quu,
    author = "Abdallah, Waleed and others",
    collaboration = "CEPC Study Group",
    title = "{CEPC Technical Design Report: Accelerator}",
    eprint = "2312.14363",
    archivePrefix = "arXiv",
    primaryClass = "physics.acc-ph",
    reportNumber = "IHEP-CEPC-DR-2023-01, IHEP-AC-2023-01",
    doi = "10.1007/s41605-024-00463-y",
    journal = "Radiat. Detect. Technol. Methods",
    volume = "8",
    number = "1",
    pages = "1--1105",
    year = "2024",
    note = "[Erratum: Radiat.Detect.Technol.Methods 9, 184--192 (2025)]"
}

@article{ILC:2007bjz,
    author = "Aarons, Gerald and others",
    editor = "Djouadi, Abdelhak and Lykken, Joseph and Moenig, Klaus and Okada, Yasuhiro and Oreglia, Mark and Yamashita, Satoru",
    collaboration = "ILC",
    title = "{International Linear Collider Reference Design Report Volume 2: Physics at the ILC}",
    eprint = "0709.1893",
    archivePrefix = "arXiv",
    primaryClass = "hep-ph",
    reportNumber = "SLAC-R-975, FERMILAB-DESIGN-2007-04, FERMILAB-PUB-07-795-E",
    month = "9",
    year = "2007"
}

@article{Adolphsen:2013jya,
    author = "Adolphsen, Chris and others",
    title = "{The International Linear Collider Technical Design Report - Volume 3.I: Accelerator {\textbackslash}{\&} in the Technical Design Phase}",
    eprint = "1306.6353",
    archivePrefix = "arXiv",
    primaryClass = "physics.acc-ph",
    reportNumber = "ILC-REPORT-2013-040, ANL-HEP-TR-13-20, BNL-100603-2013-IR, IRFU-13-59, CERN-ATS-2013-037, COCKCROFT-13-10, CLNS-13-2085, DESY-13-062, FERMILAB-TM-2554, IHEP-AC-ILC-2013-001, INFN-13-04-LNF, JAI-2013-001, JINR-E9-2013-35, JLAB-R-2013-01, KEK-REPORT-2013-1, KNU-CHEP-ILC-2013-1, LLNL-TR-635539, SLAC-R-1004, ILC-HIGRADE-REPORT-2013-003",
    month = "6",
    year = "2013"
}

@article{Adolphsen:2013kya,
    author = "Adolphsen, Chris and others",
    title = "{The International Linear Collider Technical Design Report - Volume 3.II: Accelerator Baseline Design}",
    eprint = "1306.6328",
    archivePrefix = "arXiv",
    primaryClass = "physics.acc-ph",
    reportNumber = "ILC-REPORT-2013-040, ANL-HEP-TR-13-20, BNL-100603-2013-IR, IRFU-13-59, CERN-ATS-2013-037, COCKCROFT-13-10, CLNS-13-2085, DESY-13-062, FERMILAB-TM-2554, IHEP-AC-ILC-2013-001, INFN-13-04-LNF, JAI-2013-001, JINR-E9-2013-35, JLAB-R-2013-01, KEK-REPORT-2013-1, KNU-CHEP-ILC-2013-1, LLNL-TR-635539, SLAC-R-1004, ILC-HIGRADE-REPORT-2013-003",
    month = "6",
    year = "2013"
}

@article{Behnke:2013lya,
    author = "Abramowicz, Halina and others",
    editor = "Behnke, Ties and Brau, James E. and Burrows, Philip N. and Fuster, Juan and Peskin, Michael and Stanitzki, Marcel and Sugimoto, Yasuhiro and Yamada, Sakue and Yamamoto, Hitoshi",
    title = "{The International Linear Collider Technical Design Report - Volume 4: Detectors}",
    eprint = "1306.6329",
    archivePrefix = "arXiv",
    primaryClass = "physics.ins-det",
    reportNumber = "ILC-REPORT-2013-040, ANL-HEP-TR-13-20, BNL-100603-2013-IR, IRFU-13-59, CERN-ATS-2013-037, COCKCROFT-13-10, CLNS-13-2085, DESY-13-062, FERMILAB-TM-2554, IHEP-AC-ILC-2013-001, INFN-13-04-LNF, JAI-2013-001, JINR-E9-2013-35, JLAB-R-2013-01, KEK-REPORT-2013-1, KNU-CHEP-ILC-2013-1, LLNL-TR-635539, SLAC-R-1004, ILC-HIGRADE-REPORT-2013-003",
    month = "6",
    year = "2013"
}

@inproceedings{budker1969accelerators,
  author       = {A. M. Budker},
  title        = {Accelerators and Colliding Beams},
  booktitle    = {Proceedings of the 7th International Conference on High Energy Accelerators},
  year         = {1969},
  address      = {Yerevan, USSR},
  note         = {}
}

@article{MICE:2019jkl,
    author = "Bogomilov, M. and others",
    collaboration = "MICE",
    title = "{Demonstration of cooling by the Muon Ionization Cooling Experiment}",
    eprint = "1907.08562",
    archivePrefix = "arXiv",
    primaryClass = "physics.acc-ph",
    reportNumber = "RAL-P-2019-003",
    doi = "10.1038/s41586-020-1958-9",
    journal = "Nature",
    volume = "578",
    number = "7793",
    pages = "53--59",
    year = "2020"
}

@article{Aicheler:2012bya,
    author = "Aicheler, M and Burrows, P and Draper, M and Garvey, T and Lebrun, P and Peach, K and Phinney, N and Schmickler, H and Schulte, D and Toge, N",
    title = "{A Multi-TeV Linear Collider Based on CLIC Technology}: {CLIC Conceptual Design Report}",
    reportNumber = "CERN-2012-007, SLAC-R-985, KEK-Report-2012-1, PSI-12-01, JAI-2012-001",
    doi = "10.5170/CERN-2012-007",
    month = "10",
    year = "2012"
}

@article{Linssen:2012hp,
    author = "Linssen, Lucie and Miyamoto, Akiya and Stanitzki, Marcel and Weerts, Harry",
    title = "{Physics and Detectors at CLIC: CLIC Conceptual Design Report}",
    eprint = "1202.5940",
    archivePrefix = "arXiv",
    primaryClass = "physics.ins-det",
    reportNumber = "CERN-2012-003, ANL-HEP-TR-12-01, DESY-12-008, KEK-REPORT-2011-7",
    doi = "10.5170/CERN-2012-003",
    month = "2",
    year = "2012"
}

@article{Lebrun:2012hj,
    author = "Lebrun, P. and Linssen, L. and Lucaci-Timoce, A. and Schulte, D. and Simon, F. and Stapnes, S. and Toge, N. and Weerts, H. and Wells, J.",
    title = "{The CLIC Programme: Towards a Staged e+e- Linear Collider Exploring the Terascale : CLIC Conceptual Design Report}",
    eprint = "1209.2543",
    archivePrefix = "arXiv",
    primaryClass = "physics.ins-det",
    reportNumber = "CERN-2012-005, ANL-HEP-TR-12-51, KEK-REPORT-2012-2, MPP-2012-115",
    doi = "10.5170/CERN-2012-005",
    month = "9",
    year = "2012"
}

@article{FCC:2018byv,
    author = "Abada, A. and others",
    collaboration = "FCC",
    title = "{FCC Physics Opportunities}: {Future Circular Collider Conceptual Design Report Volume 1}",
    reportNumber = "CERN-ACC-2018-0056",
    doi = "10.1140/epjc/s10052-019-6904-3",
    journal = "Eur. Phys. J. C",
    volume = "79",
    number = "6",
    pages = "474",
    year = "2019"
}

@article{FCC:2018evy,
    author = "Abada, A. and others",
    collaboration = "FCC",
    title = "{FCC-ee: The Lepton Collider}: {Future Circular Collider Conceptual Design Report Volume 2}",
    reportNumber = "CERN-ACC-2018-0057",
    doi = "10.1140/epjst/e2019-900045-4",
    journal = "Eur. Phys. J. ST",
    volume = "228",
    number = "2",
    pages = "261--623",
    year = "2019"
}

@article{FCC:2018vvp,
    author = "Abada, A. and others",
    collaboration = "FCC",
    title = "{FCC-hh: The Hadron Collider}: {Future Circular Collider Conceptual Design Report Volume 3}",
    reportNumber = "CERN-ACC-2018-0058",
    doi = "10.1140/epjst/e2019-900087-0",
    journal = "Eur. Phys. J. ST",
    volume = "228",
    number = "4",
    pages = "755--1107",
    year = "2019"
}

@article{ILC:2013jhg,
    author = "Baer, Howard and others",
    collaboration = "ILC",
    title = "{The International Linear Collider Technical Design Report - Volume 2: Physics}",
    eprint = "1306.6352",
    archivePrefix = "arXiv",
    primaryClass = "hep-ph",
    reportNumber = "ILC-REPORT-2013-040, ANL-HEP-TR-13-20, BNL-100603-2013-IR, IRFU-13-59, CERN-ATS-2013-037, COCKCROFT-13-10, CLNS-13-2085, DESY-13-062, FERMILAB-TM-2554, IHEP-AC-ILC-2013-001, INFN-13-04-LNF, JAI-2013-001, JINR-E9-2013-35, JLAB-R-2013-01, KEK-REPORT-2013-1, KNU-CHEP-ILC-2013-1, LLNL-TR-635539, SLAC-R-1004, ILC-HIGRADE-REPORT-2013-003",
    month = "6",
    year = "2013"
}

@article{CEPCStudyGroup:2018rmc,
    author = "CEPC Study Group",
    title = "{CEPC Conceptual Design Report: Volume 1 - Accelerator}",
    eprint = "1809.00285",
    archivePrefix = "arXiv",
    primaryClass = "physics.acc-ph",
    reportNumber = "IHEP-CEPC-DR-2018-01, IHEP-AC-2018-01",
    month = "9",
    year = "2018"
}

@article{CEPCStudyGroup:2018ghi,
    author = "Dong, Mingyi and others",
    editor = "Guimar{\~a}es da Costa, Jo{\~a}o Barreiro and others",
    collaboration = "CEPC Study Group",
    title = "{CEPC Conceptual Design Report: Volume 2 - Physics {\&} Detector}",
    eprint = "1811.10545",
    archivePrefix = "arXiv",
    primaryClass = "hep-ex",
    reportNumber = "IHEP-CEPC-DR-2018-02, IHEP-EP-2018-01, IHEP-TH-2018-01",
    month = "11",
    year = "2018"
}

@misc{ESPPinputs:2024,
  title        = {Input to the European Particle Physics Strategy Update 2024‑2026: Overview},
  howpublished = {\url{https://indico.cern.ch/event/1439855/overview}},
  note         = {Accessed: 2025‑09‑30},
  year         = {2025}
}

@inproceedings{Koratzinos:2015fya,
    author = "Koratzinos, Michael and others",
    title = "{The FCC-ee study: Progress and challenges}",
    booktitle = "{6th International Particle Accelerator Conference}",
    eprint = "1506.00918",
    archivePrefix = "arXiv",
    primaryClass = "physics.acc-ph",
    reportNumber = "IPAC-2015-TUPTY060",
    doi = "10.18429/JACoW-IPAC2015-TUPTY060",
    pages = "TUPTY060",
    year = "2015"
}

@article{FCC:2025lpp,
    author = "Benedikt, M. and others",
    collaboration = "FCC",
    title = "{Future Circular Collider Feasibility Study Report: Volume 1, Physics, Experiments, Detectors}",
    eprint = "2505.00272",
    archivePrefix = "arXiv",
    primaryClass = "hep-ex",
    reportNumber = "CERN-FCC-PHYS-2025-0002",
    doi = "10.17181/CERN.9DKX.TDH9",
    month = "4",
    year = "2025"
}

@article{FCC:2025uan,
    author = "Zimmermann, F. and others",
    editor = "Benedikt, M.",
    collaboration = "FCC",
    title = "{Future Circular Collider Feasibility Study Report: Volume 2, Accelerators, Technical Infrastructure and Safety}",
    eprint = "2505.00274",
    archivePrefix = "arXiv",
    primaryClass = "physics.acc-ph",
    reportNumber = "CERN-FCC-ACC-2025-0004",
    doi = "10.17181/CERN.EBAY.7W4X",
    month = "4",
    year = "2025"
}

@article{FCC:2025jtd,
    author = "Benedikt, M. and others",
    collaboration = "FCC",
    title = "{Future Circular Collider Feasibility Study Report: Volume 3, Civil Engineering, Implementation and Sustainability}",
    eprint = "2505.00273",
    archivePrefix = "arXiv",
    primaryClass = "physics.acc-ph",
    reportNumber = "CERN-FCC-ACC-2025-0003",
    doi = "10.17181/CERN.I26X.V4VF",
    month = "3",
    year = "2025"
}

@techreport{Benedikt:2025kwi,
    author = "Benedikt, Michael and others",
    title = "{FCC Integrated Programme Stage 1: The FCC-ee}",
    reportNumber = "CERN-FCC-ACC-2025-0006",
    doi = "10.17181/CERN.SLLK.DA6C",
    month = "3",
    year = "2025",
institution = "CERN"
}

@report{KEKTsukuba:1995urg,
    title = "{KEKB B factory design report}",
author = "Toge, N. and others",
    reportNumber = "KEK-REPORT-95-7, KEK-REPORT-95-07, KEK-REPORT-95-007, KEK-95-7",
    month = "6",
    year = "1995"
}

@article{Zhang:2009ze,
    author = "Zhang, Chuang",
    collaboration = "BEPC II Team",
    title = "{BEPC II: Construction and commissioning}",
    doi = "10.1088/1674-1137/33/S2/016",
    journal = "Chin. Phys. C",
    volume = "33S2",
    pages = "60--64",
    year = "2009"
}

@article{Akai:2018mbz,
    author = "Akai, Kazunori and Furukawa, Kazuro and Koiso, Haruyo",
    collaboration = "SuperKEKB",
    title = "{SuperKEKB Collider}",
    eprint = "1809.01958",
    archivePrefix = "arXiv",
    primaryClass = "physics.acc-ph",
    reportNumber = "KEK Preprint 2018-28",
    doi = "10.1016/j.nima.2018.08.017",
    journal = "Nucl. Instrum. Meth. A",
    volume = "907",
    pages = "188--199",
    year = "2018"
}

@article{Zobov:2010zza,
    author = "Zobov, M. and others",
    title = "{Test of crab-waist collisions at DAFNE Phi factory}",
    doi = "10.1103/PhysRevLett.104.174801",
    journal = "Phys. Rev. Lett.",
    volume = "104",
    pages = "174801",
    year = "2010"
}

@article{Zhou:2023ygh,
    author = "Zhou, Demin and Ohmi, Kazuhito and Funakoshi, Yoshihiro and Ohnishi, Yukiyoshi and Zhang, Yuan",
    title = "{Simulations and experimental results of beam-beam effects in SuperKEKB}",
    eprint = "2306.02681",
    archivePrefix = "arXiv",
    primaryClass = "physics.acc-ph",
    doi = "10.1103/PhysRevAccelBeams.26.071001",
    journal = "Phys. Rev. Accel. Beams",
    volume = "26",
    number = "7",
    pages = "071001",
    year = "2023"
}

@article{Assmann:1998qb,
    author = "Assmann, R. and others",
    title = "{Calibration of center-of-mass energies at LEP-1 for precise measurements of Z properties}",
    reportNumber = "CERN-EP-98-040, CERN-EP-98-40, CERN-SL-98-012, CERN-SL-98-12",
    doi = "10.1007/s100529801030",
    journal = "Eur. Phys. J. C",
    volume = "6",
    pages = "187--223",
    year = "1999"
}

@article{RevModPhys.81.1229,
  title = {Physics of laser-driven plasma-based electron accelerators},
  author = {Esarey, E. and Schroeder, C. B. and Leemans, W. P.},
  journal = {Rev. Mod. Phys.},
  volume = {81},
  issue = {3},
  pages = {1229--1285},
  numpages = {0},
  year = {2009},
  month = {Aug},
  publisher = {American Physical Society},
  doi = {10.1103/RevModPhys.81.1229},
  url = {https://link.aps.org/doi/10.1103/RevModPhys.81.1229}
}

@article{LinearColliderVision:2025hlt,
    author = "Atti{\'e}, D. and others",
    collaboration = "Linear Collider Vision",
    title = "{A Linear Collider Vision for the Future of Particle Physics}",
    eprint = "2503.19983",
    archivePrefix = "arXiv",
    primaryClass = "hep-ex",
    reportNumber = "FERMILAB-PUB-25-0216-CSAID-TD",
    month = "3",
    year = "2025"
}

@article{LinearCollider:2025lya,
    author = "Subba, A. and others",
    collaboration = "Linear Collider",
    title = "{The Linear Collider Facility (LCF) at CERN}",
    eprint = "2503.24049",
    archivePrefix = "arXiv",
    primaryClass = "hep-ex",
    reportNumber = "DESY-25-054, FERMILAB-PUB-25-0239-CSAID",
    month = "3",
    year = "2025"
}

@article{Bambade:2019fyw,
    author = "Bambade, Philip and others",
    title = "{The International Linear Collider: A Global Project}",
    eprint = "1903.01629",
    archivePrefix = "arXiv",
    primaryClass = "hep-ex",
    reportNumber = "DESY 19-037, DESY-19-037, FERMILAB-FN-1067-PPD, IFIC/19-10, IRFU-19-10,
  JLAB-PHY-19-2854, KEK Preprint 2018-92, JLAB-PHY-19-2854, KEK
  Preprint 2018-92, LAL/RT 19-001, PNNL-SA-142168,
  SLAC-PUB-17412, SLAC-PUB-17412",
    month = "3",
    year = "2019"
}

@article{Moortgat-Pick:2005jsx,
    author = "Moortgat-Pick, G. and others",
    title = "{The role of polarized positrons and electrons in revealing fundamental interactions at the linear collider}",
    eprint = "hep-ph/0507011",
    archivePrefix = "arXiv",
    reportNumber = "CERN-PH-TH-2005-036, DCPT-04-100, DESY-05-059, FERMILAB-PUB-05-060-T, IPPP-04-50, KEK-2005-16, PRL-TH-05-01, SHEP-05-03, SLAC-PUB-11087",
    doi = "10.1016/j.physrep.2007.12.003",
    journal = "Phys. Rept.",
    volume = "460",
    pages = "131--243",
    year = "2008"
}

@article{Belomestnykh:2023naf,
    author = "Belomestnykh, S. and others",
    title = "{Superconducting radio frequency linear collider HELEN}",
    reportNumber = "FERMILAB-PUB-23-411-AD-PPD-SQMS-TD",
    doi = "10.1088/1748-0221/18/09/P09039",
    journal = "JINST",
    volume = "18",
    number = "09",
    pages = "P09039",
    year = "2023"
}

@article{Belomestnykh:2023uon,
    author = "Belomestnykh, S. and others",
    title = "{HELEN: Traveling Wave SRF Linear Collider Higgs Factory}",
    eprint = "2307.06248",
    archivePrefix = "arXiv",
    primaryClass = "physics.acc-ph",
    reportNumber = "FERMILAB-CONF-23-183-AD-PPD-SQMS-TD",
    doi = "10.18429/JACoW-IPAC2023-MOPL098",
    journal = "JACoW",
    volume = "IPAC2023",
    pages = "MOPL098",
    year = "2023"
}

@article{Vernieri:2022fae,
    author = "Vernieri, Caterina and others",
    title = "{Strategy for Understanding the Higgs Physics: The Cool Copper Collider}",
    eprint = "2203.07646",
    archivePrefix = "arXiv",
    primaryClass = "hep-ex",
    reportNumber = "SLAC-PUB-17661, FERMILAB-CONF-22-311-AD-PPD-SCD-SQMS-TD",
    doi = "10.1088/1748-0221/18/07/P07053",
    journal = "JINST",
    volume = "18",
    number = "07",
    pages = "P07053",
    year = "2023"
}

@article{Andorf:2025dnn,
    author = "Andorf, Matthew B. and others",
    title = "{ESPPU INPUT: C$^3$ within the ''Linear Collider Vision''}",
    eprint = "2503.20829",
    archivePrefix = "arXiv",
    primaryClass = "physics.acc-ph",
    reportNumber = "FERMILAB-PUB-25-0290-CSAID",
    month = "3",
    year = "2025"
}

@article{Cros:2019tns,
    author = "Cros, B. and Muggli, P.",
    collaboration = "ALEGRO",
    title = "{ALEGRO input for the 2020 update of the European Strategy}",
    eprint = "1901.08436",
    archivePrefix = "arXiv",
    primaryClass = "physics.acc-ph",
    month = "1",
    year = "2019"
}

@article{HALHF:2025hqj,
    author = "Adli, Erik and others",
    collaboration = "HALHF",
    title = "{HALHF: a hybrid, asymmetric, linear Higgs factory using plasma- and RF-based acceleration}",
    eprint = "2503.19880",
    archivePrefix = "arXiv",
    primaryClass = "physics.acc-ph",
    month = "3",
    year = "2025"
}

@article{Foster:2025gsj,
    author = "Foster, Brian and others",
    title = "{Proceedings of the Erice workshop: A new baseline for the hybrid, asymmetric, linear Higgs factory HALHF}",
    eprint = "2501.11072",
    archivePrefix = "arXiv",
    primaryClass = "physics.acc-ph",
    doi = "10.1016/j.physo.2025.100261",
    journal = "Phys. Open",
    volume = "23",
    pages = "100261",
    year = "2025"
}

@article{Caldwell:2025hsa,
    author = "Caldwell, Allen and Farmer, John and Lopes, Nelson and Pukhov, Alexander and Willeke, Ferdinand and Wilson, Thomas",
    title = "{Proton-Driven Plasma Wakefield Acceleration for Future HEP Colliders}",
    eprint = "2503.21669",
    archivePrefix = "arXiv",
    primaryClass = "physics.acc-ph",
    month = "3",
    year = "2025"
}

@article{Adli:2025swq,
    author = "Adli, Erik and others",
    title = "{The Compact Linear e$^+$e$^-$ Collider (CLIC)}",
    eprint = "2503.24168",
    archivePrefix = "arXiv",
    primaryClass = "physics.acc-ph",
    month = "3",
    year = "2025"
}

@article{InternationalMuonCollider:2025sys,
    author = "Accettura, Carlotta and others",
    collaboration = "International Muon Collider",
    title = "{The Muon Collider}",
    eprint = "2504.21417",
    archivePrefix = "arXiv",
    primaryClass = "physics.acc-ph",
    reportNumber = "FERMILAB-PUB-25-0309-AD-PPD-T",
    month = "4",
    year = "2025"
}

@article{Accettura:2023ked,
    author = "Accettura, Carlotta and others",
    title = "{Towards a muon collider}",
    eprint = "2303.08533",
    archivePrefix = "arXiv",
    primaryClass = "physics.acc-ph",
    reportNumber = "FERMILAB-PUB-23-123-AD-PPD-T",
    doi = "10.1140/epjc/s10052-023-11889-x",
    journal = "Eur. Phys. J. C",
    volume = "83",
    number = "9",
    pages = "864",
    year = "2023",
    note = "[Erratum: Eur.Phys.J.C 84, 36 (2024)]"
}

@article{Neuffer:2024dka,
    author = "Neuffer, David and Stratakis, Diktys and Eldred, Jeffery and Nagaitsev, Sergei",
    title = "{Adaptation of the Fermilab proton source to support new muon facilities}",
    reportNumber = "FERMILAB-CONF-24-0230-AD",
    doi = "10.18429/JACoW-IPAC2024-TUPC41",
    journal = "JACoW",
    volume = "IPAC2024",
    pages = "TUPC41",
    year = "2024"
}

@article{Abe:2019thb,
    author = "Abe, M. and others",
    title = "{A New Approach for Measuring the Muon Anomalous Magnetic Moment and Electric Dipole Moment}",
    eprint = "1901.03047",
    archivePrefix = "arXiv",
    primaryClass = "physics.ins-det",
    doi = "10.1093/ptep/ptz030",
    journal = "PTEP",
    volume = "2019",
    number = "5",
    pages = "053C02",
    year = "2019"
}

@article{Hamada:2022mua,
    author = "Hamada, Yu and Kitano, Ryuichiro and Matsudo, Ryutaro and Takaura, Hiromasa and Yoshida, Mitsuhiro",
    title = "{$\mu$TRISTAN}",
    eprint = "2201.06664",
    archivePrefix = "arXiv",
    primaryClass = "hep-ph",
    reportNumber = "KEK-TH-2385",
    doi = "10.1093/ptep/ptac059",
    journal = "PTEP",
    volume = "2022",
    number = "5",
    pages = "053B02",
    year = "2022"
}

@article{Belle-II:2010dht,
    author = "Abe, T. and others",
    collaboration = "Belle-II",
    title = "{Belle II Technical Design Report}",
    eprint = "1011.0352",
    archivePrefix = "arXiv",
    primaryClass = "physics.ins-det",
    reportNumber = "KEK-REPORT-2010-1",
    month = "11",
    year = "2010"
}

@article{Roney:2021pwz,
    author = "Roney, J. Michael",
    collaboration = "Belle II SuperKEKB e- Polarization Upgrade Working Group",
    title = "{Upgrading SuperKEKB with polarized $e^-$ beams}",
    doi = "10.22323/1.390.0699",
    journal = "PoS",
    volume = "ICHEP2020",
    pages = "699",
    year = "2021"
}

@techreport{Benedikt:2928941,
      author        = "Benedikt, Michael and others",
      title         = "{FCC Integrated Programme Stage 2: The FCC-hh}",
      institution   = "CERN",
      reportNumber  = "CERN-FCC-ACC-2025-0007",
      address       = "Geneva",
      year          = "2025",
      month = "3",
      url           = "https://cds.cern.ch/record/2928941",
      doi           = "10.17181/CERN.M6OU.VG88",
}

@techreport{ZurbanoFernandez:2020cco,
    author = "Zurbano Fernandez, I. and others",
    editor = {B{\'e}jar Alonso, I. and Br{\"u}ning, O. and Fessia, P. and Rossi, L. and Tavian, L. and Zerlauth, M.},
    title = "{High-Luminosity Large Hadron Collider (HL-LHC): Technical design report}",
    reportNumber = "CERN-2020-010",
    doi = "10.23731/CYRM-2020-0010",
    volume = "10/2020",
    month = "12",
    year = "2020",
institution = "CERN"
}

@article{ILD:2025yhd,
    author = "Abramowicz, H. and others",
    collaboration = "ILD",
    title = "{The ILD Detector: A Versatile Detector for an Electron-Positron Collider at Energies up to 1 TeV}",
    eprint = "2506.06030",
    archivePrefix = "arXiv",
    primaryClass = "hep-ex",
    month = "6",
    year = "2025"
}

@article{Schambach:2015gga,
    author = "Schambach, Joachim and others",
    editor = "Doyle, Barney L. and Glass, Gary A. and McDaniel, Floyd D. and Wang, Yongqiang and Parriott, Carley R.",
    title = "{A MAPS Based Micro-Vertex Detector for the STAR Experiment}",
    doi = "10.1016/j.phpro.2015.05.067",
    journal = "Phys. Procedia",
    volume = "66",
    pages = "514--519",
    year = "2015"
}

@article{Simon:2009tke,
    author = "Simon, Frank",
    editor = "Behnke, Ties and Mnich, Joachim",
    collaboration = "CALICE",
    title = "{Beam Test Results with Highly Granular Hadron Calorimeters for the ILC}",
    eprint = "1002.1012",
    archivePrefix = "arXiv",
    primaryClass = "physics.ins-det",
    reportNumber = "MPP-2010-19",
    doi = "10.3204/DESY-PROC-2010-04/P44",
    journal = "Conf. Proc. C",
    volume = "0908171",
    pages = "539--541",
    year = "2009"
}

@inproceedings{Laktineh:2010zsa,
    author = "Laktineh, I.",
    collaboration = "CALICE",
    title = "{Construction of a technological semi-digital Hadronic calorimeter prototype for ILC}",
    booktitle = "{2010 IEEE Nuclear Science Symposium, Medical Imaging Conference, and 17th Room Temperature Semiconductor Detectors Workshop}",
    doi = "10.1109/NSSMIC.2010.5874085",
    pages = "1800--1803",
    year = "2010"
}

@inproceedings{Casarsa:MUSIC,
author = {Casarsa, Massimo},
year = {2024},
month = {12},
pages = {1108},
title = {Detector performance for low- and high-momentum particles in 10 TeV muon collisions},
doi = {10.22323/1.476.1108},
booktitle = "42nd International Conference on High Energy Physics"
}

@article{MAIA:2025hzm,
    author = "Bell, Charles and others",
    collaboration = "MAIA",
    title = "{MAIA: A new detector concept for a 10 TeV muon collider}",
    eprint = "2502.00181",
    archivePrefix = "arXiv",
    primaryClass = "physics.ins-det",
    month = "1",
    year = "2025"
}

@report{Mangano:2022ukr,
    author = "Aleksa, M. and others",
    editor = "Mangano, Michelangelo L. and Riegler, Werner",
    title = "{Conceptual design of an experiment at the FCC-hh, a future 100 TeV hadron collider}",
    reportNumber = "CERN-2022-002",
    doi = "10.23731/CYRM-2022-002",
    volume = "2/2022",
    month = "11",
    year = "2022",
institution = "CERN"
}

@article{Bahr:2008pv,
    author = "Bahr, M. and others",
    title = "{Herwig++ Physics and Manual}",
    eprint = "0803.0883",
    archivePrefix = "arXiv",
    primaryClass = "hep-ph",
    reportNumber = "CERN-PH-TH-2008-038, CAVENDISH-HEP-08-03, KA-TP-05-2008, DCPT-08-22, IPPP-08-11, CP3-08-05",
    doi = "10.1140/epjc/s10052-008-0798-9",
    journal = "Eur. Phys. J. C",
    volume = "58",
    pages = "639--707",
    year = "2008"
}

@article{Bellm:2015jjp,
    author = "Bellm, Johannes and others",
    title = "{Herwig 7.0/Herwig++ 3.0 release note}",
    eprint = "1512.01178",
    archivePrefix = "arXiv",
    primaryClass = "hep-ph",
    reportNumber = "CERN-PH-TH-2015-289, MAN-HEP-2015-15, IFJPAN-IV-2015-13, KA-TP-18-2015, DCPT-15-142, MCNET-15-28, IPPP-15-71, HERWIG-2015-01",
    doi = "10.1140/epjc/s10052-016-4018-8",
    journal = "Eur. Phys. J. C",
    volume = "76",
    number = "4",
    pages = "196",
    year = "2016"
}

@article{Bewick:2023tfi,
    author = "Bewick, Gavin and others",
    title = "{Herwig 7.3 release note}",
    eprint = "2312.05175",
    archivePrefix = "arXiv",
    primaryClass = "hep-ph",
    reportNumber = "CERN-TH-2023-223, HERWIG-2023-01, KA-TP-28-2023, MCnet-23-19, IPPP/23/66",
    doi = "10.1140/epjc/s10052-024-13211-9",
    journal = "Eur. Phys. J. C",
    volume = "84",
    number = "10",
    pages = "1053",
    year = "2024"
}

@article{Alwall:2011uj,
    author = "Alwall, Johan and Herquet, Michel and Maltoni, Fabio and Mattelaer, Olivier and Stelzer, Tim",
    title = "{MadGraph 5 : Going Beyond}",
    eprint = "1106.0522",
    archivePrefix = "arXiv",
    primaryClass = "hep-ph",
    reportNumber = "FERMILAB-PUB-11-448-T",
    doi = "10.1007/JHEP06(2011)128",
    journal = "JHEP",
    volume = "06",
    pages = "128",
    year = "2011"
}

@article{Alwall:2014hca,
    author = "Alwall, J. and Frederix, R. and Frixione, S. and Hirschi, V. and Maltoni, F. and Mattelaer, O. and Shao, H. -S. and Stelzer, T. and Torrielli, P. and Zaro, M.",
    title = "{The automated computation of tree-level and next-to-leading order differential cross sections, and their matching to parton shower simulations}",
    eprint = "1405.0301",
    archivePrefix = "arXiv",
    primaryClass = "hep-ph",
    reportNumber = "CERN-PH-TH-2014-064, CP3-14-18, LPN14-066, MCNET-14-09, ZU-TH-14-14",
    doi = "10.1007/JHEP07(2014)079",
    journal = "JHEP",
    volume = "07",
    pages = "079",
    year = "2014"
}

@article{Frederix:2018nkq,
    author = "Frederix, R. and Frixione, S. and Hirschi, V. and Pagani, D. and Shao, H. -S. and Zaro, M.",
    title = "{The automation of next-to-leading order electroweak calculations}",
    eprint = "1804.10017",
    archivePrefix = "arXiv",
    primaryClass = "hep-ph",
    reportNumber = "Nikhef/2018-015, TUM-HEP-1138/18, NIKHEF-2018-015, TUM-HEP-1138-18",
    doi = "10.1007/JHEP11(2021)085",
    journal = "JHEP",
    volume = "07",
    pages = "185",
    year = "2018",
    note = "[Erratum: JHEP 11, 085 (2021)]"
}

@article{Bierlich:2022pfr,
    author = "Bierlich, Christian and others",
    title = "{A comprehensive guide to the physics and usage of PYTHIA 8.3}",
    eprint = "2203.11601",
    archivePrefix = "arXiv",
    primaryClass = "hep-ph",
    reportNumber = "LU-TP 22-16, MCNET-22-04, FERMILAB-PUB-22-227-SCD",
    doi = "10.21468/SciPostPhysCodeb.8",
    journal = "SciPost Phys. Codeb.",
    volume = "2022",
    pages = "8",
    year = "2022"
}

@article{Gleisberg:2003xi,
    author = "Gleisberg, Tanju and Hoeche, Stefan and Krauss, Frank and Schalicke, Andreas and Schumann, Steffen and Winter, Jan-Christopher",
    title = "{SHERPA 1. alpha: A Proof of concept version}",
    eprint = "hep-ph/0311263",
    archivePrefix = "arXiv",
    reportNumber = "CERN-TH-2003-284",
    doi = "10.1088/1126-6708/2004/02/056",
    journal = "JHEP",
    volume = "02",
    pages = "056",
    year = "2004"
}

@article{Sherpa:2019gpd,
    author = "Bothmann, Enrico and others",
    collaboration = "Sherpa",
    title = "{Event Generation with Sherpa 2.2}",
    eprint = "1905.09127",
    archivePrefix = "arXiv",
    primaryClass = "hep-ph",
    reportNumber = "FERMILAB-PUB-19-218-T, SLAC-PUB-17433, IPPP/19/42, MCNET-19-11",
    doi = "10.21468/SciPostPhys.7.3.034",
    journal = "SciPost Phys.",
    volume = "7",
    number = "3",
    pages = "034",
    year = "2019"
}

@article{Sherpa:2024mfk,
    author = "Bothmann, Enrico and others",
    collaboration = "Sherpa",
    title = "{Event generation with Sherpa 3}",
    eprint = "2410.22148",
    archivePrefix = "arXiv",
    primaryClass = "hep-ph",
    reportNumber = "IPPP/24/67, LTH-1385, FERMILAB-PUB-24-0748-T, ZU-TH 51/24, MCNET-24-17, CERN-TH-2024-171",
    doi = "10.1007/JHEP12(2024)156",
    journal = "JHEP",
    volume = "12",
    pages = "156",
    year = "2024"
}

@article{Moretti:2001zz,
    author = "Moretti, Mauro and Ohl, Thorsten and Reuter, Jurgen",
    editor = "Behnke, Ties and Bertolucci, S. and Heuer, Rolf D. and Miller, David and Richard, Francois and Settles, Ron and Telnov, V. and Zerwas, Peter",
    title = "{O'Mega: An Optimizing matrix element generator}",
    eprint = "hep-ph/0102195",
    archivePrefix = "arXiv",
    reportNumber = "IKDA-2001-06, LC-TOOL-2001-040",
    pages = "1981--2009",
    month = "2",
    year = "2001"
}

@article{Kilian:2007gr,
    author = "Kilian, Wolfgang and Ohl, Thorsten and Reuter, Jurgen",
    title = "{WHIZARD: Simulating Multi-Particle Processes at LHC and ILC}",
    eprint = "0708.4233",
    archivePrefix = "arXiv",
    primaryClass = "hep-ph",
    reportNumber = "DESY-11-126, EDINBURGH-2010-36, FR-PHENO-2010-037, SI-HEP-2010-18",
    doi = "10.1140/epjc/s10052-011-1742-y",
    journal = "Eur. Phys. J. C",
    volume = "71",
    pages = "1742",
    year = "2011"
}

@article{Reuter:2024dvz,
    author = {Reuter, J{\"u}rgen and Bredt, Pia and H{\"o}fer, Marius and Kilian, Wolfgang and Kreher, Nils and L{\"o}schner, Maximilian and M{\k{e}}ka{\l}a, Krzysztof and Ohl, Thorsten and Striegl, Tobias and {\.Z}arnecki, Aleksander Filip},
    title = "{New developments in the Whizard event generator}",
    eprint = "2412.06605",
    archivePrefix = "arXiv",
    primaryClass = "hep-ph",
    reportNumber = "DESY-24-188, KA-TP-22-2024, P3H-24-091, SI-HEP-2024-25",
    doi = "10.1051/epjconf/202431501020",
    journal = "EPJ Web Conf.",
    volume = "315",
    pages = "01020",
    year = "2024"
}

@article{Blondel:2019qlh,
    author = "Blondel, Alain and Freitas, Ayres and Gluza, Janusz and Riemann, Tord and Heinemeyer, Sven and Jadach, Stanislaw and Janot, Patrick",
    title = "{Theory Requirements and Possibilities for the FCC-ee and other Future High Energy and Precision Frontier Lepton Colliders}",
    eprint = "1901.02648",
    archivePrefix = "arXiv",
    primaryClass = "hep-ph",
    month = "1",
    year = "2019"
}

@techreport{deBlas:2944678,
      author        = "de Blas, Jorge and others",
      title         = "{Physics Briefing Book: Input for the 2026 update of the
                       European Strategy for Particle Physics}",
        institution  = {CERN},
      number  = "CERN-ESU-2025-001",
      address       = "Geneva",
      year          = "2025",
      url           = "https://cds.cern.ch/record/2944678",
      doi           = "10.17181/CERN.35CH.2O2P",
}

@article{GoSam:2014iqq,
    author = "Cullen, Gavin and others",
    collaboration = "GoSam",
    title = "{G$\scriptsize{O}$S$\scriptsize{AM}$-2.0: a tool for automated one-loop calculations within the Standard Model and beyond}",
    eprint = "1404.7096",
    archivePrefix = "arXiv",
    primaryClass = "hep-ph",
    reportNumber = "DESY-14-061, MPP-2014-145",
    doi = "10.1140/epjc/s10052-014-3001-5",
    journal = "Eur. Phys. J. C",
    volume = "74",
    number = "8",
    pages = "3001",
    year = "2014"
}

@article{Braun:2025afl,
    author = {Braun, Jens and Campillo Aveleira, Benjamin and Heinrich, Gudrun and H{\"o}fer, Marius and Jones, Stephen P. and Kerner, Matthias and Lang, Jannis and Magerya, Vitaly},
    title = "{One-Loop Calculations in Effective Field Theories with GoSam-3.0}",
    eprint = "2507.23549",
    archivePrefix = "arXiv",
    primaryClass = "hep-ph",
    reportNumber = "KA-TP-21-2025, P3H-25-051, IPPP/25/50, CERN-TH-2025-134",
    month = "7",
    year = "2025"
}

@article{Cascioli:2011va,
    author = "Cascioli, Fabio and Maierhofer, Philipp and Pozzorini, Stefano",
    title = "{Scattering Amplitudes with Open Loops}",
    eprint = "1111.5206",
    archivePrefix = "arXiv",
    primaryClass = "hep-ph",
    reportNumber = "ZU-TH-23-11, LPN11-66",
    doi = "10.1103/PhysRevLett.108.111601",
    journal = "Phys. Rev. Lett.",
    volume = "108",
    pages = "111601",
    year = "2012"
}

@article{Buccioni:2019sur,
    author = {Buccioni, Federico and Lang, Jean-Nicolas and Lindert, Jonas M. and Maierh{\"o}fer, Philipp and Pozzorini, Stefano and Zhang, Hantian and Zoller, Max F.},
    title = "{OpenLoops 2}",
    eprint = "1907.13071",
    archivePrefix = "arXiv",
    primaryClass = "hep-ph",
    reportNumber = "IPPP/19/62, FR-PHENO-2019-12, PSI-PR-19-15, ZU-TH 37/19",
    doi = "10.1140/epjc/s10052-019-7306-2",
    journal = "Eur. Phys. J. C",
    volume = "79",
    number = "10",
    pages = "866",
    year = "2019"
}

@article{Actis:2016mpe,
    author = "Actis, Stefano and Denner, Ansgar and Hofer, Lars and Lang, Jean-Nicolas and Scharf, Andreas and Uccirati, Sandro",
    title = "{RECOLA: REcursive Computation of One-Loop Amplitudes}",
    eprint = "1605.01090",
    archivePrefix = "arXiv",
    primaryClass = "hep-ph",
    reportNumber = "ICCUB-16-020",
    doi = "10.1016/j.cpc.2017.01.004",
    journal = "Comput. Phys. Commun.",
    volume = "214",
    pages = "140--173",
    year = "2017"
}

@article{Feynman:1969ej,
    author = "Feynman, Richard P.",
    editor = "Brown, L. M.",
    title = "{Very high-energy collisions of hadrons}",
    reportNumber = "PRINT-69-2817",
    doi = "10.1103/PhysRevLett.23.1415",
    journal = "Phys. Rev. Lett.",
    volume = "23",
    pages = "1415--1417",
    year = "1969"
}

@article{Buckley:2014ana,
    author = {Buckley, Andy and Ferrando, James and Lloyd, Stephen and Nordstr{\"o}m, Karl and Page, Ben and R{\"u}fenacht, Martin and Sch{\"o}nherr, Marek and Watt, Graeme},
    title = "{LHAPDF6: parton density access in the LHC precision era}",
    eprint = "1412.7420",
    archivePrefix = "arXiv",
    primaryClass = "hep-ph",
    reportNumber = "GLAS-PPE-2014-05, MCNET-14-29, IPPP-14-111, DCPT-14-222",
    doi = "10.1140/epjc/s10052-015-3318-8",
    journal = "Eur. Phys. J. C",
    volume = "75",
    pages = "132",
    year = "2015"
}

@article{Ohl:1996fi,
    author = "Ohl, Thorsten",
    title = "{CIRCE version 1.0: Beam spectra for simulating linear collider physics}",
    eprint = "hep-ph/9607454",
    archivePrefix = "arXiv",
    reportNumber = "IKDA-96-13",
    doi = "10.1016/S0010-4655(96)00167-1",
    journal = "Comput. Phys. Commun.",
    volume = "101",
    pages = "269--288",
    year = "1997"
}

@article{Andersson:1983ia,
    author = "Andersson, Bo and Gustafson, G. and Ingelman, G. and Sjostrand, T.",
    title = "{Parton Fragmentation and String Dynamics}",
    reportNumber = "LU-TP-83-10",
    doi = "10.1016/0370-1573(83)90080-7",
    journal = "Phys. Rept.",
    volume = "97",
    pages = "31--145",
    year = "1983"
}

@article{Webber:1983if,
    author = "Webber, B. R.",
    title = "{A QCD Model for Jet Fragmentation Including Soft Gluon Interference}",
    reportNumber = "CERN-TH-3713",
    doi = "10.1016/0550-3213(84)90333-X",
    journal = "Nucl. Phys. B",
    volume = "238",
    pages = "492--528",
    year = "1984"
}

@article{Sjostrand:1984ic,
    author = "Sjostrand, Torbjorn",
    title = "{Jet Fragmentation of Nearby Partons}",
    reportNumber = "DESY-T-84-01",
    doi = "10.1016/0550-3213(84)90607-2",
    journal = "Nucl. Phys. B",
    volume = "248",
    pages = "469--502",
    year = "1984"
}

@article{Sjostrand:1986hx,
    author = "Sjostrand, Torbjorn and Bengtsson, Mats",
    title = "{The Lund Monte Carlo for Jet Fragmentation and e+ e- Physics. Jetset Version 6.3: An Update}",
    reportNumber = "LU-TP-86-22",
    doi = "10.1016/0010-4655(87)90054-3",
    journal = "Comput. Phys. Commun.",
    volume = "43",
    pages = "367",
    year = "1987"
}

@inproceedings{Reuter:2025imz,
    author = {Reuter, J{\"u}rgen},
    title = "{Monte Carlo Event Generators}",
    eprint = "2509.21611",
    archivePrefix = "arXiv",
    primaryClass = "hep-ph",
    reportNumber = "DESY-25-130",
    month = "9",
    year = "2025",
booktitle = "Encyclopedia of Particle Physics"
}

@inproceedings{Sjostrand:2009ad,
    author = "Sjostrand, Torbjorn",
    title = "{Monte Carlo Tools}",
    booktitle = "{65th Scottish Universities Summer School in Physics: LHC Physics}",
    eprint = "0911.5286",
    archivePrefix = "arXiv",
    primaryClass = "hep-ph",
    reportNumber = "LU-TP-09-31, MCNET-09-17",
    doi = "10.1201/b11865-14",
    pages = "309--339",
    month = "11",
    year = "2009"
}

@inproceedings{Hoche:2014rga,
    author = {H{\"o}che, Stefan},
    title = "{Introduction to parton-shower event generators}",
    booktitle = "{Theoretical Advanced Study Institute in Elementary Particle Physics}: {Journeys Through the Precision Frontier: Amplitudes for Colliders}",
    eprint = "1411.4085",
    archivePrefix = "arXiv",
    primaryClass = "hep-ph",
    reportNumber = "SLAC-PUB-16160",
    doi = "10.1142/9789814678766_0005",
    pages = "235--295",
    year = "2015"
}

@article{Skands:2014pea,
    author = "Skands, Peter and Carrazza, Stefano and Rojo, Juan",
    title = "{Tuning PYTHIA 8.1: the Monash 2013 Tune}",
    eprint = "1404.5630",
    archivePrefix = "arXiv",
    primaryClass = "hep-ph",
    reportNumber = "CERN-PH-TH-2014-069, MCNET-14-08, OUTP-14-05P",
    doi = "10.1140/epjc/s10052-014-3024-y",
    journal = "Eur. Phys. J. C",
    volume = "74",
    number = "8",
    pages = "3024",
    year = "2014"
}

@article{Fieg:2023kld,
    author = {Fieg, Max and Kling, Felix and Schulz, Holger and Sj{\"o}strand, Torbj{\"o}rn},
    title = "{Tuning pythia for forward physics experiments}",
    eprint = "2309.08604",
    archivePrefix = "arXiv",
    primaryClass = "hep-ph",
    reportNumber = "DESY-23-133",
    doi = "10.1103/PhysRevD.109.016010",
    journal = "Phys. Rev. D",
    volume = "109",
    number = "1",
    pages = "016010",
    year = "2024"
}

@article{Nason:2004rx,
    author = "Nason, Paolo",
    title = "{A New method for combining NLO QCD with shower Monte Carlo algorithms}",
    eprint = "hep-ph/0409146",
    archivePrefix = "arXiv",
    reportNumber = "BICOCCA-FT-04-11",
    doi = "10.1088/1126-6708/2004/11/040",
    journal = "JHEP",
    volume = "11",
    pages = "040",
    year = "2004"
}

@article{Frixione:2007vw,
    author = "Frixione, Stefano and Nason, Paolo and Oleari, Carlo",
    title = "{Matching NLO QCD computations with Parton Shower simulations: the POWHEG method}",
    eprint = "0709.2092",
    archivePrefix = "arXiv",
    primaryClass = "hep-ph",
    reportNumber = "BICOCCA-FT-07-9, GEF-TH-21-2007",
    doi = "10.1088/1126-6708/2007/11/070",
    journal = "JHEP",
    volume = "11",
    pages = "070",
    year = "2007"
}

@article{Heinemeyer:2021rgq,
    author = {Heinemeyer, Sven and Jadach, Stanislaw and Reuter, J{\"u}rgen},
    title = "{Theory requirements for SM Higgs and EW precision physics at the FCC-ee}",
    eprint = "2106.11802",
    archivePrefix = "arXiv",
    primaryClass = "hep-ph",
    reportNumber = "DESY-21-100",
    doi = "10.1140/epjp/s13360-021-01875-1",
    journal = "Eur. Phys. J. Plus",
    volume = "136",
    number = "9",
    pages = "911",
    year = "2021"
}

@article{Kleiss:2020rcg,
    author = "Kleiss, Ronald and Verheyen, Rob",
    title = "{Collinear electroweak radiation in antenna parton showers}",
    eprint = "2002.09248",
    archivePrefix = "arXiv",
    primaryClass = "hep-ph",
    doi = "10.1140/epjc/s10052-020-08510-w",
    journal = "Eur. Phys. J. C",
    volume = "80",
    number = "10",
    pages = "980",
    year = "2020"
}

@article{Brooks:2021kji,
    author = "Brooks, Helen and Skands, Peter and Verheyen, Rob",
    title = "{Interleaved resonance decays and electroweak radiation in the Vincia parton shower}",
    eprint = "2108.10786",
    archivePrefix = "arXiv",
    primaryClass = "hep-ph",
    doi = "10.21468/SciPostPhys.12.3.101",
    journal = "SciPost Phys.",
    volume = "12",
    number = "3",
    pages = "101",
    year = "2022"
}

@article{Wilson:1974sk,
    author = "Wilson, Kenneth G.",
    editor = "Taylor, J. C.",
    title = "{Confinement of Quarks}",
    reportNumber = "CLNS-262",
    doi = "10.1103/PhysRevD.10.2445",
    journal = "Phys. Rev. D",
    volume = "10",
    pages = "2445--2459",
    year = "1974"
}

@article{FlavourLatticeAveragingGroupFLAG:2024oxs,
    author = "Aoki, Y. and others",
    collaboration = "Flavour Lattice Averaging Group (FLAG)",
    title = "{FLAG Review 2024}",
    eprint = "2411.04268",
    archivePrefix = "arXiv",
    primaryClass = "hep-lat",
    reportNumber = "CERN-TH-2024-192, FERMILAB-PUB-24-0785-T",
    month = "11",
    year = "2024"
}

@article{Sjostrand:1982fn,
    author = "Sjostrand, Torbjorn",
    title = "{The Lund Monte Carlo for Jet Fragmentation}",
    reportNumber = "LU TP 82-3",
    doi = "10.1016/0010-4655(82)90175-8",
    journal = "Comput. Phys. Commun.",
    volume = "27",
    pages = "243",
    year = "1982"
}

@article{Sjostrand:1982am,
    author = "Sjostrand, Torbjorn",
    title = "{The Lund Monte Carlo for e+ e- Jet Physics}",
    reportNumber = "LU TP-82-7",
    doi = "10.1016/0010-4655(83)90041-3",
    journal = "Comput. Phys. Commun.",
    volume = "28",
    pages = "229",
    year = "1983"
}

@article{Sjostrand:1985ys,
    author = "Sjostrand, Torbjorn",
    title = "{The Lund Monte Carlo for Jet Fragmentation and e+ e- Physics: Jetset Version 6.2}",
    reportNumber = "LU-TP-85-10",
    doi = "10.1016/0010-4655(86)90096-2",
    journal = "Comput. Phys. Commun.",
    volume = "39",
    pages = "347--407",
    year = "1986"
}

@article{Bengtsson:1982jr,
    author = "Bengtsson, H. U.",
    title = "{The Lund Monte Carlo for High $p_T$ Physics}",
    reportNumber = "LU-TP-82-15",
    doi = "10.1016/0010-4655(84)90018-3",
    journal = "Comput. Phys. Commun.",
    volume = "31",
    pages = "323",
    year = "1984"
}

@article{Bengtsson:1984yx,
    author = "Bengtsson, H. U. and Ingelman, G.",
    title = "{The Lund Monte Carlo for High $p_T$ Physics}",
    reportNumber = "LU TP 84-3, CERN-TH-3820",
    doi = "10.1016/0010-4655(85)90003-7",
    journal = "Comput. Phys. Commun.",
    volume = "34",
    pages = "251",
    year = "1985"
}

@article{Bengtsson:1987kr,
    author = "Bengtsson, Hans-Uno and Sjostrand, Torbjorn",
    title = "{The Lund Monte Carlo for Hadronic Processes: Pythia Version 4.8}",
    reportNumber = "LU-TP-87-3, UCLA-87-001",
    doi = "10.1016/0010-4655(87)90036-1",
    journal = "Comput. Phys. Commun.",
    volume = "46",
    pages = "43",
    year = "1987"
}

@article{Sjostrand:1993yb,
    author = "Sjostrand, Torbjorn",
    title = "{High-energy physics event generation with PYTHIA 5.7 and JETSET 7.4}",
    reportNumber = "CERN-TH-7111-93",
    doi = "10.1016/0010-4655(94)90132-5",
    journal = "Comput. Phys. Commun.",
    volume = "82",
    pages = "74--90",
    year = "1994"
}

@article{Sjostrand:2000wi,
    author = "Sjostrand, Torbjorn and Eden, Patrik and Friberg, Christer and Lonnblad, Leif and Miu, Gabriela and Mrenna, Stephen and Norrbin, Emanuel",
    title = "{High-energy physics event generation with PYTHIA 6.1}",
    eprint = "hep-ph/0010017",
    archivePrefix = "arXiv",
    reportNumber = "LU-TP-00-30",
    doi = "10.1016/S0010-4655(00)00236-8",
    journal = "Comput. Phys. Commun.",
    volume = "135",
    pages = "238--259",
    year = "2001"
}

@inproceedings{Boos:2001cv,
    author = "Boos, E. and others",
    title = "{Generic User Process Interface for Event Generators}",
    booktitle = "{2nd Les Houches Workshop on Physics at TeV Colliders}",
    eprint = "hep-ph/0109068",
    archivePrefix = "arXiv",
    reportNumber = "FERMILAB-CONF-01-496-T",
    month = "9",
    year = "2001"
}

@article{Alwall:2006yp,
    author = "Alwall, J. and others",
    title = "{A Standard format for Les Houches event files}",
    eprint = "hep-ph/0609017",
    archivePrefix = "arXiv",
    reportNumber = "FERMILAB-PUB-06-337-T, CERN-LCGAPP-2006-03",
    doi = "10.1016/j.cpc.2006.11.010",
    journal = "Comput. Phys. Commun.",
    volume = "176",
    pages = "300--304",
    year = "2007"
}

@article{Andersen:2014efa,
    author = "Andersen, J. R. and others",
    title = "{Les Houches 2013: Physics at TeV Colliders: Standard Model Working Group Report}",
    eprint = "1405.1067",
    archivePrefix = "arXiv",
    primaryClass = "hep-ph",
    month = "5",
    year = "2014"
}

@inproceedings{Whalley:2005nh,
    author = "Whalley, M. R. and Bourilkov, D. and Group, R. C.",
    title = "{The Les Houches accord PDFs (LHAPDF) and LHAGLUE}",
    booktitle = "{HERA and the LHC: A Workshop on the Implications of HERA and LHC Physics (Startup Meeting, CERN, 26-27 March 2004; Midterm Meeting, CERN, 11-13 October 2004)}",
    eprint = "hep-ph/0508110",
    archivePrefix = "arXiv",
    pages = "575--581",
    month = "8",
    year = "2005"
}

@article{Dobbs:2001ck,
    author = "Dobbs, Matt and Hansen, Jorgen Beck",
    title = "{The HepMC C++ Monte Carlo event record for High Energy Physics}",
    reportNumber = "ATL-SOFT-2000-001",
    doi = "10.1016/S0010-4655(00)00189-2",
    journal = "Comput. Phys. Commun.",
    volume = "134",
    pages = "41--46",
    year = "2001"
}

@article{Branco:2011iw,
    author = "Branco, G. C. and Ferreira, P. M. and Lavoura, L. and Rebelo, M. N. and Sher, Marc and Silva, Joao P.",
    title = "{Theory and phenomenology of two-Higgs-doublet models}",
    eprint = "1106.0034",
    archivePrefix = "arXiv",
    primaryClass = "hep-ph",
    doi = "10.1016/j.physrep.2012.02.002",
    journal = "Phys. Rept.",
    volume = "516",
    pages = "1--102",
    year = "2012"
}

@article{Sjostrand:2004ef,
    author = "Sjostrand, T. and Skands, Peter Z.",
    title = "{Transverse-momentum-ordered showers and interleaved multiple interactions}",
    eprint = "hep-ph/0408302",
    archivePrefix = "arXiv",
    reportNumber = "LU-TP-04-29",
    doi = "10.1140/epjc/s2004-02084-y",
    journal = "Eur. Phys. J. C",
    volume = "39",
    pages = "129--154",
    year = "2005"
}

@article{Christiansen:2014kba,
    author = {Christiansen, Jesper Roy and Sj{\"o}strand, Torbj{\"o}rn},
    title = "{Weak Gauge Boson Radiation in Parton Showers}",
    eprint = "1401.5238",
    archivePrefix = "arXiv",
    primaryClass = "hep-ph",
    reportNumber = "LU-TP-14-02, MCNET-14-01",
    doi = "10.1007/JHEP04(2014)115",
    journal = "JHEP",
    volume = "04",
    pages = "115",
    year = "2014"
}

@article{Ilten:2012zb,
    author = "Ilten, Philip",
    editor = "Hayasaka, Kiyoshi and Iijima, Toru",
    title = "{Tau Decays in Pythia 8}",
    eprint = "1211.6730",
    archivePrefix = "arXiv",
    primaryClass = "hep-ph",
    doi = "10.1016/j.nuclphysbps.2014.09.019",
    journal = "Nucl. Phys. B Proc. Suppl.",
    volume = "253-255",
    pages = "77--80",
    year = "2014"
}

@misc{whizard-manual,
  author       = {{WHIZARD Collaboration}},
  title        = {{WHIZARD Manual, Version 3.1}},
  year         = {2025},
  howpublished = {\url{https://whizard.hepforge.org/manual/}},
  note         = {Accessed: 2025-10-07},
}

@article{ECFADESYLCPhysicsWorkingGroup:2001igx,
    author = "Aguilar-Saavedra, J. A. and others",
    collaboration = "ECFA/DESY LC Physics Working Group",
    title = "{TESLA: The Superconducting electron positron linear collider with an integrated x-ray laser laboratory. Technical design report. Part 3. Physics at an e+ e- linear collider}",
    eprint = "hep-ph/0106315",
    archivePrefix = "arXiv",
    reportNumber = "SLAC-REPRINT-2001-002, DESY-01-011, DESY-01-011C, DESY-TESLA-2001-23, DESY-TESLA-FEL-2001-05, ECFA-2001-209",
    month = "3",
    year = "2001"
}

@article{Ohl:1998jn,
    author = "Ohl, Thorsten",
    title = "{Vegas revisited: Adaptive Monte Carlo integration beyond factorization}",
    eprint = "hep-ph/9806432",
    archivePrefix = "arXiv",
    reportNumber = "IKDA-98-15",
    doi = "10.1016/S0010-4655(99)00209-X",
    journal = "Comput. Phys. Commun.",
    volume = "120",
    pages = "13--19",
    year = "1999"
}

@article{Schulte:1999tx,
    author = "Schulte, D.",
    title = "{Beam-beam simulations with GUINEA-PIG}",
    reportNumber = "CERN-PS-99-014-LP, CERN-PS-99-14-LP, CLIC-NOTE-387, CERN-CLIC-NOTE-387",
    year = "1999"
}

@article{Rimbault:2007wfy,
    author = "Rimbault, C. and Bambade, P. and Dadoun, O. and Le Meur, G. and Touze, F. and del Alabau, M. C. and Schulte, D.",
    editor = "Petit-Jean-Genaz, C.",
    title = "{GUINEA PIG++ : An Upgraded Version of the Linear Collider Beam Beam Interaction Simulation Code GUINEA PIG}",
    reportNumber = "PAC07-THPMN010",
    doi = "10.1109/PAC.2007.4440556",
    journal = "Conf. Proc. C",
    volume = "070625",
    pages = "2728",
    year = "2007"
}

@article{Bonvicini:1989ci,
    author = "Bonvicini, Giovanni and Gero, E. and Frey, R. and Koska, Wayne A. and Field, R. C. and Phinney, N. and Minten, A.",
    title = "{First Observation of Beamstrahlung}",
    reportNumber = "SLAC-PUB-4856, UM-HE-89-5",
    doi = "10.1103/PhysRevLett.62.2381",
    journal = "Phys. Rev. Lett.",
    volume = "62",
    pages = "2381",
    year = "1989"
}

@article{Ohl:2000pr,
    author = "Ohl, Thorsten",
    editor = "Para, A. and Fisk, H. E.",
    title = "{O'Mega {\textbackslash}{\&} WHIZARD: Monte Carlo event generator generation for future colliders}",
    eprint = "hep-ph/0011287",
    archivePrefix = "arXiv",
    reportNumber = "IKDA-2000-31",
    doi = "10.1063/1.1394396",
    journal = "AIP Conf. Proc.",
    volume = "578",
    number = "1",
    pages = "638--641",
    year = "2001"
}

@inproceedings{Stienemeier:2021cse,
    author = {Stienemeier, Pascal and Bra{\ss}, Simon and Bredt, Pia and Kilian, Wolfgang and Kreher, Nils and Ohl, Thorsten and Reuter, J{\"u}rgen and Rothe, Vincent and Striegl, Tobias},
    title = "{WHIZARD 3.0: Status and News}",
    booktitle = "{International Workshop on Future Linear Colliders}",
    eprint = "2104.11141",
    archivePrefix = "arXiv",
    primaryClass = "hep-ph",
    reportNumber = "DESY-21-050",
    month = "4",
    year = "2021"
}

@article{kibble2009englert,
  author       = {Kibble, Tom W.~B.},
  title        = {Englert–Brout–Higgs–Guralnik–Hagen–Kibble mechanism},
  journal      = {Scholarpedia},
  volume       = {4},
  number       = {1},
  pages        = {6441},
  year         = {2009},
  doi          = {10.4249/scholarpedia.6441},
  url          = {https://www.scholarpedia.org/article/Englert-Brout-Higgs-Guralnik-Hagen-Kibble_mechanism}
}

@article{Englert:1964et,
    author = "Englert, F. and Brout, R.",
    editor = "Taylor, J. C.",
    title = "{Broken Symmetry and the Mass of Gauge Vector Mesons}",
    doi = "10.1103/PhysRevLett.13.321",
    journal = "Phys. Rev. Lett.",
    volume = "13",
    pages = "321--323",
    year = "1964"
}

@article{Higgs:1964pj,
    author = "Higgs, Peter W.",
    editor = "Taylor, J. C.",
    title = "{Broken Symmetries and the Masses of Gauge Bosons}",
    doi = "10.1103/PhysRevLett.13.508",
    journal = "Phys. Rev. Lett.",
    volume = "13",
    pages = "508--509",
    year = "1964"
}

@article{Guralnik:1964eu,
    author = "Guralnik, G. S. and Hagen, C. R. and Kibble, T. W. B.",
    editor = "Taylor, J. C.",
    title = "{Global Conservation Laws and Massless Particles}",
    doi = "10.1103/PhysRevLett.13.585",
    journal = "Phys. Rev. Lett.",
    volume = "13",
    pages = "585--587",
    year = "1964"
}

@article{Nambu:1960tm,
    author = "Nambu, Yoichiro",
    editor = "Taylor, J. C.",
    title = "{Quasiparticles and Gauge Invariance in the Theory of Superconductivity}",
    doi = "10.1103/PhysRev.117.648",
    journal = "Phys. Rev.",
    volume = "117",
    pages = "648--663",
    year = "1960"
}

@article{Anderson:1963pc,
    author = "Anderson, Philip W.",
    editor = "Taylor, J. C.",
    title = "{Plasmons, Gauge Invariance, and Mass}",
    doi = "10.1103/PhysRev.130.439",
    journal = "Phys. Rev.",
    volume = "130",
    pages = "439--442",
    year = "1963"
}

@article{Horiguchi:2013wra,
    author = "Horiguchi, Tomohiro and Ishikawa, Akimasa and Suehara, Taikan and Fujii, Keisuke and Sumino, Yukinari and Kiyo, Yuichiro and Yamamoto, Hitoshi",
    title = "{Study of top quark pair production near threshold at the ILC}",
    eprint = "1310.0563",
    archivePrefix = "arXiv",
    primaryClass = "hep-ex",
    month = "10",
    year = "2013"
}

@article{Canetti:2012vf,
    author = "Canetti, Laurent and Drewes, Marco and Shaposhnikov, Mikhail",
    title = "{Sterile Neutrinos as the Origin of Dark and Baryonic Matter}",
    eprint = "1204.3902",
    archivePrefix = "arXiv",
    primaryClass = "hep-ph",
    reportNumber = "TTK-12-08, CAS-KITPC-ITP-315",
    doi = "10.1103/PhysRevLett.110.061801",
    journal = "Phys. Rev. Lett.",
    volume = "110",
    number = "6",
    pages = "061801",
    year = "2013"
}

@article{Caputo:2018zky,
    author = "Caputo, Andrea and Hernandez, Pilar and Rius, Nuria",
    title = "{Leptogenesis from oscillations and dark matter}",
    eprint = "1807.03309",
    archivePrefix = "arXiv",
    primaryClass = "hep-ph",
    doi = "10.1140/epjc/s10052-019-7083-y",
    journal = "Eur. Phys. J. C",
    volume = "79",
    number = "7",
    pages = "574",
    year = "2019"
}

@article{Gninenko:2012anz,
    author = "Gninenko, S. N. and Gorbunov, D. S. and Shaposhnikov, M. E.",
    title = "{Search for GeV-scale sterile neutrinos responsible for active neutrino oscillations and baryon asymmetry of the Universe}",
    eprint = "1301.5516",
    archivePrefix = "arXiv",
    primaryClass = "hep-ph",
    doi = "10.1155/2012/718259",
    journal = "Adv. High Energy Phys.",
    volume = "2012",
    pages = "718259",
    year = "2012"
}

@article{Mekala:2022cmm,
    author = {M{\k{e}}ka{\l}a, Krzysztof and Reuter, J{\"u}rgen and {\.Z}arnecki, Aleksander Filip},
    title = "{Heavy neutrinos at future linear e$^{+}$e$^{-}$ colliders}",
    eprint = "2202.06703",
    archivePrefix = "arXiv",
    primaryClass = "hep-ph",
    reportNumber = "DESY-22-029",
    doi = "10.1007/JHEP06(2022)010",
    journal = "JHEP",
    volume = "06",
    pages = "010",
    year = "2022"
}

@article{Mekala:2023diu,
    author = {M{\k{e}}ka{\l}a, Krzysztof and Reuter, J{\"u}rgen and {\.Z}arnecki, Aleksander Filip},
    title = "{Optimal search reach for heavy neutral leptons at a muon collider}",
    eprint = "2301.02602",
    archivePrefix = "arXiv",
    primaryClass = "hep-ph",
    doi = "10.1016/j.physletb.2023.137945",
    journal = "Phys. Lett. B",
    volume = "841",
    pages = "137945",
    year = "2023"
}

@article{Mekala:2023kzo,
    author = {M{\k{e}}ka{\l}a, Krzysztof and Reuter, J{\"u}rgen and {\.Z}arnecki, Aleksander Filip},
    title = "{Discriminating Majorana and Dirac heavy neutrinos at lepton colliders}",
    eprint = "2312.05223",
    archivePrefix = "arXiv",
    primaryClass = "hep-ph",
    reportNumber = "DESY-23-210",
    doi = "10.1007/JHEP03(2024)075",
    journal = "JHEP",
    volume = "03",
    pages = "075",
    year = "2024"
}

@article{Kalinowski:2021tyr,
    author = "Kalinowski, Jan and Kotlarski, Wojciech and Mekala, Krzysztof and Sopicki, Pawel and Zarnecki, Aleksander Filip",
    title = "{Sensitivity of future linear $\hbox {e}^+\hbox {e}^-$ colliders to processes of dark matter production with light mediator exchange}",
    eprint = "2107.11194",
    archivePrefix = "arXiv",
    primaryClass = "hep-ph",
    doi = "10.1140/epjc/s10052-021-09758-6",
    journal = "Eur. Phys. J. C",
    volume = "81",
    number = "10",
    pages = "955",
    year = "2021"
}

@article{Korshynska:2024suh,
    author = {Korshynska, Kateryna and L{\"o}schner, Maximilian and Marinichenko, Mariia and M{\k{e}}ka{\l}a, Krzysztof and Reuter, J{\"u}rgen},
    title = "{Z{\textquoteright} boson mass reach and model discrimination at muon colliders}",
    eprint = "2402.18460",
    archivePrefix = "arXiv",
    primaryClass = "hep-ph",
    reportNumber = "DESY-24-028",
    doi = "10.1140/epjc/s10052-024-12892-6",
    journal = "Eur. Phys. J. C",
    volume = "84",
    number = "6",
    pages = "568",
    year = "2024"
}

@article{ATLAS:2016gzy,
    author = "Aaboud, Morad and others",
    collaboration = "ATLAS",
    title = "{Search for resonances in diphoton events at $\sqrt{s}$=13 TeV with the ATLAS detector}",
    eprint = "1606.03833",
    archivePrefix = "arXiv",
    primaryClass = "hep-ex",
    reportNumber = "CERN-EP-2016-120",
    doi = "10.1007/JHEP09(2016)001",
    journal = "JHEP",
    volume = "09",
    pages = "001",
    year = "2016"
}

@article{ATLAS:2019erb,
    author = "Aad, Georges and others",
    collaboration = "ATLAS",
    title = "{Search for high-mass dilepton resonances using 139 fb$^{-1}$ of $pp$ collision data collected at $\sqrt{s}=$13 TeV with the ATLAS detector}",
    eprint = "1903.06248",
    archivePrefix = "arXiv",
    primaryClass = "hep-ex",
    reportNumber = "CERN-EP-2019-030",
    doi = "10.1016/j.physletb.2019.07.016",
    journal = "Phys. Lett. B",
    volume = "796",
    pages = "68--87",
    year = "2019"
}

@article{ATLAS:2019fgd,
    author = "Aad, Georges and others",
    collaboration = "ATLAS",
    title = "{Search for new resonances in mass distributions of jet pairs using 139 fb$^{-1}$ of $pp$ collisions at $\sqrt{s}=13$ TeV with the ATLAS detector}",
    eprint = "1910.08447",
    archivePrefix = "arXiv",
    primaryClass = "hep-ex",
    reportNumber = "CERN-EP-2019-162",
    doi = "10.1007/JHEP03(2020)145",
    journal = "JHEP",
    volume = "03",
    pages = "145",
    year = "2020"
}

@article{CMS:2019gwf,
    author = "Sirunyan, Albert M and others",
    collaboration = "CMS",
    title = "{Search for high mass dijet resonances with a new background prediction method in proton-proton collisions at $\sqrt{s} =$ 13 TeV}",
    eprint = "1911.03947",
    archivePrefix = "arXiv",
    primaryClass = "hep-ex",
    reportNumber = "CMS-EXO-19-012, CERN-EP-2019-222",
    doi = "10.1007/JHEP05(2020)033",
    journal = "JHEP",
    volume = "05",
    pages = "033",
    year = "2020"
}

@article{CMS:2021ctt,
    author = "Sirunyan, Albert M and others",
    collaboration = "CMS",
    title = "{Search for resonant and nonresonant new phenomena in high-mass dilepton final states at $ \sqrt{s} $ = 13 TeV}",
    eprint = "2103.02708",
    archivePrefix = "arXiv",
    primaryClass = "hep-ex",
    reportNumber = "CMS-EXO-19-019, CERN-EP-2021-026",
    doi = "10.1007/JHEP07(2021)208",
    journal = "JHEP",
    volume = "07",
    pages = "208",
    year = "2021"
}

@article{ATLAS:2020fry,
    author = "Aad, Georges and others",
    collaboration = "ATLAS",
    title = "{Search for heavy diboson resonances in semileptonic final states in pp collisions at $\sqrt{s}=13$ TeV with the ATLAS detector}",
    eprint = "2004.14636",
    archivePrefix = "arXiv",
    primaryClass = "hep-ex",
    reportNumber = "CERN-EP-2020-049",
    doi = "10.1140/epjc/s10052-020-08554-y",
    journal = "Eur. Phys. J. C",
    volume = "80",
    number = "12",
    pages = "1165",
    year = "2020"
}

@article{CMS:2021klu,
    author = "Tumasyan, Armen and others",
    collaboration = "CMS",
    title = "{Search for heavy resonances decaying to WW, WZ, or WH boson pairs in the lepton plus merged jet final state in proton-proton collisions at $\sqrt{s}$ = 13 TeV}",
    eprint = "2109.06055",
    archivePrefix = "arXiv",
    primaryClass = "hep-ex",
    reportNumber = "CMS-B2G-19-002, CERN-EP-2021-159",
    doi = "10.1103/PhysRevD.105.032008",
    journal = "Phys. Rev. D",
    volume = "105",
    number = "3",
    pages = "032008",
    year = "2022"
}

@article{Alvarez:2020yim,
    author = "Alvarez, Ezequiel and Est\'evez, Mariel and Sand\'a Seoane, Rosa Mar\'\i{}a",
    title = "{Z'-explorer: A simple tool to probe Z' models against LHC data}",
    eprint = "2005.05194",
    archivePrefix = "arXiv",
    primaryClass = "hep-ph",
    reportNumber = "ICAS-050-20",
    doi = "10.1016/j.cpc.2021.108144",
    journal = "Comput. Phys. Commun.",
    volume = "269",
    pages = "108144",
    year = "2021"
}

@article{Lozano:2021zbu,
    author = "Lozano, Victor Martin and Seoane, Rosa Maria Sanda and Zurita, Jose",
    title = "{Z'-explorer 2.0: Reconnoitering the dark matter landscape}",
    eprint = "2109.13194",
    archivePrefix = "arXiv",
    primaryClass = "hep-ph",
    reportNumber = "DESY 21-146, ICAS 066/21, IFIC/21-33",
    doi = "10.1016/j.cpc.2023.108729",
    journal = "Comput. Phys. Commun.",
    volume = "288",
    pages = "108729",
    year = "2023"
}

@inbook{Cvetic:1995zs,
    author = "Cvetic, Mirjam and Godfrey, Stephen",
    title = "{Discovery and identification of extra gauge bosons}",
    booktitle = "{Electroweak symmetry breaking and new physics at the TeV scale}",
    eprint = "hep-ph/9504216",
    archivePrefix = "arXiv",
    reportNumber = "OCIP-C-95-2, UPR-648-T",
    doi = "10.1142/9789812830265_0007",
    pages = "383--415",
    month = "3",
    year = "1995",
publisher = "World Scientific"
}

@article{Leike:1996pj,
    author = "Leike, A. and Riemann, S.",
    title = "{$Z^\prime$ search in $e^{+} e^{-}$ annihilation}",
    eprint = "hep-ph/9607306",
    archivePrefix = "arXiv",
    reportNumber = "DESY-96-111, LMU-02-96",
    doi = "10.1007/s002880050477",
    journal = "Z. Phys. C",
    volume = "75",
    pages = "341--348",
    year = "1997"
}

@article{Leike:1998wr,
    author = "Leike, A.",
    title = "{The Phenomenology of extra neutral gauge bosons}",
    eprint = "hep-ph/9805494",
    archivePrefix = "arXiv",
    reportNumber = "LMU-03-98",
    doi = "10.1016/S0370-1573(98)00133-1",
    journal = "Phys. Rept.",
    volume = "317",
    pages = "143--250",
    year = "1999"
}

@inproceedings{Godfrey:2005pm,
    author = "Godfrey, Stephen and Kalyniak, Pat and Tomkins, Alexander",
    title = "{Distinguishing between models with extra gauge bosons at the ILC}",
    booktitle = "{2005 International Linear Collider Physics and Detector Workshop and 2nd ILC Accelerator Workshop}",
    eprint = "hep-ph/0511335",
    archivePrefix = "arXiv",
    reportNumber = "SNOWMASS-2005-ALCPG0108",
    month = "11",
    year = "2005"
}

@article{Osland:2009dp,
    author = "Osland, P. and Pankov, A. A. and Tsytrinov, A. V.",
    title = "{Identification of extra neutral gauge bosons at the International Linear Collider}",
    eprint = "0912.2806",
    archivePrefix = "arXiv",
    primaryClass = "hep-ph",
    doi = "10.1140/epjc/s10052-010-1272-z",
    journal = "Eur. Phys. J. C",
    volume = "67",
    pages = "191--204",
    year = "2010"
}

@article{Andreev:2012cj,
    author = "Andreev, V. V. and Moortgat-Pick, G. and Osland, P. and Pankov, A. A. and Paver, N.",
    title = "{Discriminating Z' from Anomalous Trilinear Gauge Coupling Signatures in e+e- \textbackslash{}to W+W- at ILC with Polarized Beams}",
    eprint = "1205.0866",
    archivePrefix = "arXiv",
    primaryClass = "hep-ph",
    reportNumber = "DESY-12-016, CERN-PH-TH-2012-077",
    doi = "10.1140/epjc/s10052-012-2147-2",
    journal = "Eur. Phys. J. C",
    volume = "72",
    pages = "2147",
    year = "2012"
}

@article{Han:2013mra,
    author = "Han, Tao and Langacker, Paul and Liu, Zhen and Wang, Lian-Tao",
    title = "{Diagnosis of a New Neutral Gauge Boson at the LHC and ILC for Snowmass 2013}",
    eprint = "1308.2738",
    archivePrefix = "arXiv",
    primaryClass = "hep-ph",
    month = "8",
    year = "2013"
}

@inproceedings{Godfrey:2013eta,
    author = "Godfrey, Stephen and Martin, Travis",
    title = "{Z' Discovery Reach at Future Hadron Colliders: A Snowmass White Paper}",
    booktitle = "{Snowmass 2013}: {Snowmass on the Mississippi}",
    eprint = "1309.1688",
    archivePrefix = "arXiv",
    primaryClass = "hep-ph",
    month = "9",
    year = "2013"
}

@article{Kapukchyan:2013hfa,
    author = "Kapukchyan, David and Tait, Tim M. P.",
    title = "{Sensitivity of a Future High Energy $e^+ e^-$ Collider to $Z^\prime$ Bosons}",
    eprint = "1312.3377",
    archivePrefix = "arXiv",
    primaryClass = "hep-ph",
    doi = "10.1088/0954-3899/41/7/075011",
    journal = "J. Phys. G",
    volume = "41",
    pages = "075011",
    year = "2014"
}

@article{Pankov:2017dkv,
    author = "Pankov, A. . A. . and Tsytrinov, A. . V.",
    title = "{Model identification of new heavy $Z^\prime$ bosons at ILC with polarized beams}",
    doi = "10.1088/1742-6596/938/1/012059",
    journal = "J. Phys. Conf. Ser.",
    volume = "938",
    number = "1",
    pages = "012059",
    year = "2017"
}

@article{Gulov:2017uqa,
    author = "Gulov, Alexey and Moroz, Yaroslav",
    title = "{Optimal observables for $Z'$ models in annihilation leptonic processes}",
    eprint = "1711.02853",
    archivePrefix = "arXiv",
    primaryClass = "hep-ph",
    doi = "10.1103/PhysRevD.98.115014",
    journal = "Phys. Rev. D",
    volume = "98",
    number = "11",
    pages = "115014",
    year = "2018"
}

@article{Jenkins:2017jig,
    author = "Jenkins, Elizabeth E. and Manohar, Aneesh V. and Stoffer, Peter",
    title = "{Low-Energy Effective Field Theory below the Electroweak Scale: Operators and Matching}",
    eprint = "1709.04486",
    archivePrefix = "arXiv",
    primaryClass = "hep-ph",
    doi = "10.1007/JHEP03(2018)016",
    journal = "JHEP",
    volume = "03",
    pages = "016",
    year = "2018",
    note = "[Erratum: JHEP 12, 043 (2023)]"
}

@article{Kling:2025zsb,
    author = {Kling, Felix and Ma, Yang and M{\k{e}}ka{\l}a, Krzysztof and Reuter, J{\"u}rgen and Tabrizi, Zahra},
    title = "{Non-Standard Neutrino Interactions at a Muon Collider Neutrino Detector}",
    eprint = "2508.00761",
    archivePrefix = "arXiv",
    primaryClass = "hep-ph",
    reportNumber = "DESY 25-112, IRMP-CP3-25-27, PITT-PACC-2506, UCI-TR-2025-15",
    month = "8",
    year = "2025"
}

@article{Blondel:2025kbl,
    author = "Blondel, Alain and Grojean, Christophe and Janot, Patrick and Rajagopalan, Srini and Wilkinson, Guy",
    title = "{The FCC integrated programme: a physics manifesto}",
    eprint = "2504.02634",
    archivePrefix = "arXiv",
    primaryClass = "hep-ex",
    month = "4",
    year = "2025"
}

@article{CLICdp:2018esa,
    author = "Abramowicz, H. and others",
    collaboration = "CLICdp",
    title = "{Top-Quark Physics at the CLIC Electron-Positron Linear Collider}",
    eprint = "1807.02441",
    archivePrefix = "arXiv",
    primaryClass = "hep-ex",
    reportNumber = "CLICdp-Pub-2018-003, CLICDP-PUB-2018-003",
    doi = "10.1007/JHEP11(2019)003",
    journal = "JHEP",
    volume = "11",
    pages = "003",
    year = "2019"
}

@article{Defranchis:2025auz,
    author = "Defranchis, Matteo M. and de Blas, Jorge and Mehta, Ankita and Selvaggi, Michele and Vos, Marcel",
    title = "{A detailed study on the prospects for a $ \textrm{t}\overline{\textrm{t}} $ threshold scan in e$^{+}$e$^{-}$ collisions}",
    eprint = "2503.18713",
    archivePrefix = "arXiv",
    primaryClass = "hep-ph",
    doi = "10.1007/JHEP11(2025)020",
    journal = "JHEP",
    volume = "11",
    pages = "020",
    year = "2025"
}

@article{Ake:2023xcz,
    author = "Ake, Daniel and Bouzas, Antonio O. and Larios, F.",
    title = "{Top Quark Flavor Changing Couplings at a Muon Collider}",
    eprint = "2311.09488",
    archivePrefix = "arXiv",
    primaryClass = "hep-ph",
    doi = "10.1155/2024/2038180",
    journal = "Adv. High Energy Phys.",
    volume = "2024",
    pages = "2038180",
    year = "2024"
}

@article{Han:2024gan,
    author = "Han, Tao and Liu, Da and Wang, Si",
    title = "{Top quark electroweak dipole moment at a high energy muon collider}",
    eprint = "2410.11015",
    archivePrefix = "arXiv",
    primaryClass = "hep-ph",
    reportNumber = "PITT-PACC-2406",
    doi = "10.1103/PhysRevD.111.035015",
    journal = "Phys. Rev. D",
    volume = "111",
    number = "3",
    pages = "035015",
    year = "2025"
}

@article{Fujii:2017vwa,
    author = "Fujii, Keisuke and others",
    title = "{Physics Case for the 250 GeV Stage of the International Linear Collider}",
    eprint = "1710.07621",
    archivePrefix = "arXiv",
    primaryClass = "hep-ex",
    reportNumber = "DESY-17-155, KEK-PREPRINT-2017-31, LAL-17-059, SLAC-PUB-17161",
    month = "10",
    year = "2017"
}

@article{deBlas:2025xhe,
    author = "de Blas, J. and Goncalves, A. and Miralles, V. and Reina, L. and Silvestrini, L. and Valli, M.",
    title = "{Constraining new physics effective interactions via a global fit of electroweak, Drell-Yan, Higgs, top, and flavour observables}",
    eprint = "2507.06191",
    archivePrefix = "arXiv",
    primaryClass = "hep-ph",
    month = "7",
    year = "2025"
}

@article{Blondel:2014bra,
    author = "Blondel, Alain and Graverini, E. and Serra, N. and Shaposhnikov, M.",
    editor = "Aguilar-Ben{\'\i}tez, M and Fuster, J and Mart{\'\i}-Garc{\'\i}a, S and Santamar{\'\i}a, A",
    collaboration = "FCC-ee study Team",
    title = "{Search for Heavy Right Handed Neutrinos at the FCC-ee}",
    eprint = "1411.5230",
    archivePrefix = "arXiv",
    primaryClass = "hep-ex",
    doi = "10.1016/j.nuclphysbps.2015.09.304",
    journal = "Nucl. Part. Phys. Proc.",
    volume = "273-275",
    pages = "1883--1890",
    year = "2016"
}

@article{Romao:2012pq,
    author = "Romao, Jorge C. and Silva, Joao P.",
    title = "{A resource for signs and Feynman diagrams of the Standard Model}",
    eprint = "1209.6213",
    archivePrefix = "arXiv",
    primaryClass = "hep-ph",
    doi = "10.1142/S0217751X12300256",
    journal = "Int. J. Mod. Phys. A",
    volume = "27",
    pages = "1230025",
    year = "2012"
}

@article{tHooft:1971qjg,
    author = "'t Hooft, Gerard",
    editor = "Taylor, J. C.",
    title = "{Renormalizable Lagrangians for Massive Yang-Mills Fields}",
    doi = "10.1016/0550-3213(71)90139-8",
    journal = "Nucl. Phys. B",
    volume = "35",
    pages = "167--188",
    year = "1971"
}

@article{Cabibbo:1963yz,
    author = "Cabibbo, Nicola",
    title = "{Unitary Symmetry and Leptonic Decays}",
    doi = "10.1103/PhysRevLett.10.531",
    journal = "Phys. Rev. Lett.",
    volume = "10",
    pages = "531--533",
    year = "1963"
}

\end{document}